\documentclass[11pt,a4paper]{article}
\usepackage[latin,english]{babel}
\usepackage[utf8]{inputenc}
\usepackage[T1]{fontenc}
\usepackage{eufrak}
\usepackage{geometry}
\usepackage{amsmath}
\usepackage{bbm}
\usepackage{amssymb}
\usepackage{amsbsy}
\usepackage{amsthm}
\usepackage{makeidx}
\usepackage{mathtools}
\usepackage{mathabx, mathrsfs, dsfont}
\usepackage{comment}
\usepackage{cases}
\usepackage{braket}
\usepackage[toc,page]{appendix}
\usepackage{dirtytalk}
\usepackage{pdfsync}
\usepackage{graphicx}
\usepackage{epstopdf}
\usepackage{todonotes}
\usepackage{fancyhdr}
\usepackage{math}
\usepackage{cite}
\usepackage{color}
\usepackage{enumitem}
\usepackage{subcaption}
\usepackage{cancel}
\usepackage[colorlinks=true, pdfstartview=FitV, urlcolor=blue, citecolor=red, linkcolor=blue]{hyperref}
\usepackage{tikz}
\usetikzlibrary{decorations.markings, arrows.meta}
\usepackage{pgfplots} \pgfplotsset{compat=1.18}

\newcommand{\pmtwo}[4]{\begin{pmatrix}#1 & #2 \\ #3 & #4 \end{pmatrix}}
\renewcommand{\vector}[2]{\begin{pmatrix}#1 \\ #2 \end{pmatrix}}
\DeclareMathOperator{\res}{Res}
\DeclareMathOperator{\Ai}{Ai}
\newcommand{\fW}{\mathfrak{W}}
\newcommand{\fw}{\mathfrak{w}}

\renewcommand{\Re}{\operatorname{Re}}
\renewcommand{\Im}{\operatorname{Im}}
\newcommand{\dd}{\xi}

\numberwithin{equation}{section}
\setlist[enumerate]{align=left}

\tikzset{
    every picture/.style={line width=0.75pt}, 
    arrow rev/.style={
        postaction={
            decorate,
            decoration={
                markings,
                mark=at position #1 with {\arrow{Latex[reversed, length=7pt, width=5pt]}} 
            }
        }
    },
    arrow fwd/.style={
        postaction={
            decorate,
            decoration={
                markings,
                mark=at position #1 with {\arrow{Latex[length=7pt, width=5pt]}} 
            }
        }
    }
}

\title{Global Asymptotics of a Toda Soliton Condensate with Random Scattering Data }
\usepackage{authblk}    \author[1]{Guido Mazzuca} \author[2]{Mateusz Piorkowski} \affil[1]{Department of Mathematics, Tulane University, New Orleans, LA, USA} \affil[2]{Department of Mathematical Sciences, Durham University, Durham, UK}
\date{\today}

\begin{document}
\maketitle
\begin{abstract}
    We study a dense $N$-soliton condensate for the infinite Toda lattice in the  limit $N \to \infty$. In contrast to previous mathematical studies of soliton gases where the norming constants are assumed to be deterministic or infinitesimal, we consider initial scattering data where both the discrete eigenvalues and the norming constants are independent random variables. The norming constants $\gamma_j$ are allowed to be arbitrarily large under a mild moment condition $\mathbb{E}[\gamma_j^\nu] < \infty$ ($\nu > 1$). Our main result is divided into two parts. First, we develop a probabilistic Riemann--Hilbert framework to show that the system self-averages: despite large, independent fluctuations in the initial data, the random Toda lattice variables converge deterministically to an effective continuum model at a polynomial rate $\mathcal{O}(N^{-\delta})$ for $0 < \delta < 1/2$. Second, by performing the Deift--Zhou nonlinear steepest descent analysis of the continuum problem, we obtain the complete leading-order macroscopic description of the lattice across the entire space-time plane. The solution decomposes into five distinct finite-gap regimes, characterized explicitly by modulated elliptic and genus-2 waves with an error of order $\mathcal{O}(1/\ln N)$. In contrast to the straight characteristic rays typical of standard soliton gas limits, the interfaces separating these five macroscopic regimes in the space-time plane are genuinely non-linear curves; in particular, the boundary of the modulated genus-2 region is completely non-linear. Finally, under our assumptions,  we establish the Nonlinear Dispersion Relations for the Toda soliton condensate and prove that the macroscopic ensemble averages are determined by the soliton gas density of states.
\end{abstract}
\tableofcontents

\section{Introduction and main results}
\subsection{Background and motivation}
Solitons are fundamental, localized nonlinear structures that arise in completely integrable Hamiltonian systems and nonlinear dispersive wave equations, such as the Korteweg--de Vries (KdV) equation \cite{KdV1895}, the nonlinear Schr\"odinger (NLS) equation \cite{Ablowitz1974}, and the Toda lattice \cite{Toda1967}. Since the foundational discoveries of Gardner, Greene, Kruskal, and Miura \cite{GGKM1967} and Zabusky and Kruskal \cite{ZK1965}, an $N$-soliton solution has enjoyed a natural and compelling particle-like interpretation. In the asymptotic regimes $t \to \pm\infty$, an $N$-soliton solution decomposes into a collection of $N$ well-separated, isolated traveling waves, each propagating with a distinct, invariant velocity determined by its associated spectral parameter. When two solitons collide, they interact elastically, emerging from the interaction zone with their asymptotic profiles and velocities completely intact, experiencing only an explicit spatial displacement, known as a two-body phase shift.

When the number of solitons $N$ is macroscopic and their spatial density is sufficiently high that their individual localized profiles continuously overlap, the classical particle-like picture of isolated traveling waves breaks down. In this dense regime, individual solitons can no longer be visually distinguished within the wave field; instead, their nonlinear interactions give rise to a fluctuating and turbulent macroscopic state. To describe the statistical mechanics of such dense multi-soliton ensembles, V.~E.~Zakharov \cite{Zakharov1971} proposed the innovative framework of a ``soliton gas''. In Zakharov's kinetic theory, the ensemble is modeled as a continuous thermodynamic gas of quasi-particles whose macroscopic motion is governed by the cumulative effect of pairwise collision phase shifts. This heuristic vision was formalized by El \cite{El2003} and El and Kamchatnov \cite{ElKamchatnov2005}, who derived integro-differential kinetic equations governing the spectral density of solitons and their effective propagation velocities in dense gases.

Motivated by the kinetic theory, an extensive mathematical program dedicated to the analysis of soliton gases in continuous integrable PDEs has been pioneered in \cite{GGJM, Girotti2023SolitonGas, Girotti2026} using the Inverse Scattering Transform (IST) and the Deift--Zhou steepest descent method for Riemann--Hilbert problems \cite{Deift1993,Deift1997,DZ03}. In their foundational work on the KdV equation \cite{GGJM}, the authors rigorously established the emergence of a modulated elliptic wave from a dense soliton gas as $N \to \infty$. In \cite{Girotti2023SolitonGas}, the interaction of a single tracer soliton with a dense modified KdV (mKdV) soliton gas was analyzed, proving that the tracer's average velocity satisfies an analogue of Zakharov's kinetic velocity equation. More recently, in the context of the focusing NLS equation, a law of large numbers and a central limit theorem were established for random sets of solitons \cite{Girotti2026}, and universality for extreme rogue-wave formations was proven in \cite{Girotti2025,Gkogkou2025}.

Despite these remarkable advances, a fundamental limitation in the existing mathematical literature on soliton gases is the treatment of norming constants. In the physical Inverse Scattering Transform, an $N$-soliton solution is completely parametrized by $2N$ spectral quantities: the discrete eigenvalues $\{\zeta_j\}_{j=1}^N$, which determine the soliton amplitudes and velocities, and the norming constants $\{\gamma_j\}_{j=1}^N$, which fix the initial spatial positions (phases) of the individual solitons. In previous rigorous Riemann--Hilbert studies of soliton gases, the norming constants are assumed to be \textbf{deterministic} (for instance, set identically to $1$, or deterministically interpolated by a smooth profile of the eigenvalues as in \cite{Girotti2026,GGJM,Gkogkou2025soliton}) and/or of order $\cO(N^{-1})$, where $N$ is the (large) number of solitons. In any realistic random soliton gas, however, one cannot expect the norming constants to be deterministic or infinitesimal: the initial spatial centers of the solitons must be allowed to fluctuate randomly throughout the lattice. One of the objectives of this paper is the \textit{resolution of this randomness}. In particular,  we study a dense $N$-soliton condensate where both the discrete eigenvalues $\{\zeta_j\}_{j=1}^N$ and the norming constants $\{\gamma_j\}_{j=1}^N$ are independent random variables.

The inclusion of random norming constants in our work is further motivated by profound developments in the past decade in the study of integrable systems in \emph{thermal equilibrium} driven by the emergence of \textbf{Generalized Hydrodynamics (GHD)} \cite{CastroAlvaredo2016, Bertini2016}. Originating in the study of non-equilibrium quantum and classical integrable many-body systems, GHD provides a systematic framework for Euler-scale transport by incorporating the full infinite hierarchy of conservation laws \cite{Bonnemain2022, DoyonLecutureNotes}. In GHD, solitons are identified as the fundamental quasi-particles carrying ballistic transport, and the kinetic equations of soliton gases emerge as the hydrodynamic Euler equations of the underlying field theory.

In the context of the classical Toda lattice, a first breakthrough was achieved by Herbert Spohn \cite{Spohn2019}. Starting from Generalized Gibbs Ensembles and using a transfer operator method, Spohn was able to characterize the density of states for the Toda lattice in thermal equilibrium, see also \cite{Mazzuca2022, Guionnet2022,grava2026largedeviationsperiodictoda,Mazzuca2024}. Then, employing the collision rate ansatz, he derived the hydrodynamic Euler equations for the Toda chain. Furthermore, in \cite{Spohn2020,Spohn2020Toda} he investigated dynamical current fluctuations in the Toda lattice, formulating the seminal conjecture that at thermal equilibrium, the two-point space-time correlation functions decay ballistically ($t^{-1}$) along characteristic rays, exhibiting universal scaling profiles dictated by the effective velocity of the quasi-particles. This conjecture was numerically verified in \cite{Mazzuca2023}.

Spohn's conjectures catalyzed an intense interest in establishing these hydrodynamic limits from microscopic first principles. A crowning mathematical breakthrough in this direction was recently achieved in a remarkable series of works by Amol Aggarwal \cite{Aggarwal2025Scattering, Aggarwal2025Velocities}, who analyzed the classical Toda chain at thermal equilibrium, establishing the asymptotic scattering relation and proving the law of large numbers for effective quasi-particle velocities. Most recently, Aggarwal and Nicoletti \cite{AggarwalNicoletti2026} developed the corresponding equilibrium fluctuation theory under diffusive scaling, rigorously proving the ballistic decay of correlation functions conjectured by Spohn.

We remark that while the results of Aggarwal establish the hydrodynamic description of the Toda lattice at \textbf{thermal equilibrium}, the present paper investigates an entirely different physical and mathematical regime: an \textbf{out-of-equilibrium} dense $N$-soliton gas condensate as $N \to \infty$. Rather than sampling the microscopic particles from an invariant spatial Gibbs measure, we consider a coherent ensemble of $N$ solitons where randomness is introduced directly at the level of the \textit{scattering data} (the discrete eigenvalues and norming constants). Over macroscopic space-time scales, this out-of-equilibrium condensate does not remain stationary or homogeneous; instead, it undergoes several spatio-temporal phase transitions, generating macroscopic multi-phase oscillatory waves and modulated elliptic structures.

\begin{quote}
     Motivated by these developments, the primary objective of the present work is  to analyze the \textbf{long time behavior} of a dense $N$-soliton condensate for the infinite Toda lattice with random scattering data; specifically, both the discrete eigenvalues $\{\zeta_j\}_{j=1}^N$ and the norming constants $\{\gamma_j\}_{j=1}^N$ are \textbf{independent random variables}.
    We show that this random soliton ensemble converges in the large $N$-limit to a deterministic profile which displays spatio-temporal (non-linear) phase transitions and is given by modulated (hyper)elliptic wave solutions. 
\end{quote}

It is worth mentioning that the norming constants $\gamma_j$ are subject only to a mild moment condition $\E[\gamma_j^\nu] < \infty$ ($\nu > 1$), and in particular are allowed to be arbitrarily large. We achieve these objectives in two steps. First, by developing a probabilistic small-norm Riemann--Hilbert framework in the same spirit as \cite{Gkogkou2025,Girotti2026}, we prove that despite independent stochastic fluctuations in the initial scattering data, the macroscopic state self-averages: the random physical lattice variables converge deterministically to the solution of an effective continuum problem with a rate $\mathcal{O}(N^{-\delta})$ for some $\delta \in (0, 1/2)$, holding uniformly in $L^1$, and almost surely if $\nu > 2$. In our analysis, the self-averaging is achieved through conjugation with the $g$-function of the model problem. This mechanism allows us to consider larger fluctuations in the initial data and the inclusion of random norming constants.

Second, through a non-linear steepest descent analysis of the continuum problem, we show that the leading-order macroscopic profile decomposes across spacetime into five distinct finite-gap regimes described explicitly by modulated elliptic (genus $1$) and genus-$2$ waves, with an asymptotic error of order $\mathcal{O}((\ln N)^{-1})$ arising from the local parametrices. We observe slow modulation on the order of $\cO(\ln N)$ in both space and time, and the emergence of nonlinear phase boundaries; see Figures \ref{fig:phase_diagram} and \ref{fig:evolution1}. This is in stark contrast with the usual \emph{linear} phase diagram that emerges in which the waveform depends \emph{only} on the ratio of the space and time variables (i.e., one slowly varying parameter), and the boundaries are given by straight rays going through the origin. See for instance \cite{bertola2026arbitrarygenusdarksolitongases}, where the authors observe the same type of genus transition, but the phase boundaries are linear functions. 

We remark that the $g$-function of the model problem plays a major role in both results: on the one hand, it allows us to obtain the asymptotics of the continuum model problem; on the other hand, it allows us to include random scattering data, see Remark \ref{rem:jump_difference}. Finally, the $g$-function also provides a direct bridge to the theory of Soliton Gases and Generalized Hydrodynamics. Specifically, by representing the $g$-function as a logarithmic potential, we derive the \textit{Nonlinear Dispersion Relations} (NDR) \cite{El2003,ElTovbis2020,Congy2023} for the Toda soliton condensate and prove that the macroscopic ensemble averages of the Flaschka variables are determined directly by the soliton gas density of states. In the physics literature such relations are typically deduced via heuristic phase-shift arguments or formal thermodynamic limits. Our probabilistic framework establishes them rigorously from random scattering data.

\subsection{The model}
 The infinite Toda lattice \cite{Toda1967} is the Hamiltonian system described by the following Hamiltonian

 \begin{equation}
	H_{T}(\boldsymbol{p},\boldsymbol{q}) = \sum_{j\in \mathbb{Z}}\left[ \frac{p_j^2}{2} + V_T(q_{j+1}-q_{j} )\right]\ , \quad  V_T(x) =  e^{- x} +  x - 1\, ,
	\label{toda}
	\end{equation}
	$p_j,q_j\in \mathbb{R}$ (we will follow the notational conventions of \cite{KRGER2009}). Its equations of motion take the form
	\begin{equation}
	\label{Todaeq}
	\frac{\mathrm{d} q_j}{\mathrm{d} t} \equiv \dot{q}_j=\dfrac{\partial H_T}{\partial p_j}=p_j,\quad \frac{\mathrm{d} p_j}{\mathrm{d} t}\equiv \dot{p}_j=-\dfrac{\partial H_T}{\partial q_j}=V_T'(q_{j+1}-q_{j}) -V_T'(q_{j}-q_{j-1}), \;\;j\in \mathbb{Z}\, .
	\end{equation}

The remarkable property of this model is that it is integrable, specifically its equations of motion are equivalent to the following Lax pair due to Flaschka \cite{Flaschka1974} and Manakov \cite{manakov1975}

\begin{equation}\label{eq:Lax_Eq}
    \dot{L} = [P,L] =PL - LP\,,
\end{equation}
where $L$ is the symmetric Jacobi Operator 
\begin{equation}
\label{eq:Lax_operator}
    L = \begin{pmatrix}
        \ddots & \ddots&\ddots \\
        &a_{-1} &b_{0} & a_0 \\
        &&a_0 & b_1 & a_1\\
        &&& \ddots & \ddots & \ddots
    \end{pmatrix}
\end{equation}
$P$ reads

\begin{equation}
    P = \begin{pmatrix}
        \ddots & \ddots&\ddots \\
        &-a_{-1} &0 & a_0 \\
        &&-a_0 & 0 & a_1\\
        &&& \ddots & \ddots & \ddots
    \end{pmatrix}\,.
\end{equation}
and 

\begin{equation}
    b_j = -\frac{1}{2}p_j\quad a_j = \frac{1}{2}e^{-\frac{1}{2}(q_{j+1} - q_j)}\,.
\end{equation}
In fact the Lax pair equation \eqref{eq:Lax_Eq} is equivalent to the following time evolutions
\begin{equation}
   \begin{split}
    \dot{b}_j &= 2(a_j^2 - a_{j-1}^2)
    \\
    \dot{a}_j &= a_j(b_{j+1} - b_j)
\end{split}, \qquad j \in \mathbb Z,  
\end{equation}
which are equivalent to our original Hamiltonian evolution equations \eqref{Todaeq}.

As a consequence of the Lax pair representation, the solution of this system can be computed explicitly applying the Deift--Zhou nonlinear steepest descent method, see \cite{Kamvissis1993,Krger2008,Teschl1999}. Specifically, consider the space 

\begin{equation}
\label{eq:func_space}
    \mathfrak{S} = \left\{ \{a_j,b_j\}_{j\in \mathbb{Z}} \, :\, \sum_{j\in \mathbb{Z}} |j| \left(\bigg\lvert a_j-\frac{1}{2}\bigg\rvert + |b_j|\right) < \infty \right\}\,.
\end{equation}
The solution $(a_j(t),b_j(t))$ of the initial value problem starting in $\mathfrak{S}$ remains in $\mathfrak{S}$ for all time, see \cite[Lem.~13.7]{Teschl1999}, and can be recovered for all $j,t$ through a $(j,t)$-dependent Riemann-Hilbert Problem (RHP) \cite{Kamvissis1993,KRGER2009}. In the two papers that we mentioned, two different RHP are given. In \cite{Kamvissis1993}, the RHP that the author study is a matrix valued RHP, but it is valid only in the situation when the operator $L$ \eqref{eq:Lax_operator} does not have discrete spectrum, i.e. the spectrum of the operator is the interval $(-1,1)$. In \cite{KRGER2009}, the considered RHP is a vector-valued RHP, but in this way the authors can also handle the discrete spectrum. Since we are going to consider the situation with discrete spectrum, we consider the RHP in \cite[Thm.~3.3]{KRGER2009}. 

\begin{rhp}
\label{rhp:toda}
    Let $\{\zeta_j\}_{j=1}^{N}\in\mathbb{R}$ such that $0<|\zeta_j|<1$, and $\{\gamma_j\}_{j=1}^{N}\in\mathbb{C}$. 
    We look for a vector $m^{(1)}(z)=(m^{(1)}_1(z),m^{(1)}_2(z))$ such that the following conditions are satisfied

    \begin{itemize}
        \item $m^{(1)}(z)$ is analytic in $\mathbb{C}\setminus\{\zeta_j,\zeta_j^{-1}\}$
        \item the pole condition,
        \begin{align}
        \label{eq:res_zeta_j}
                &\mathrm{Res}_{\zeta_j}m^{(1)}(z) = \lim_{z\to\zeta_j}m^{(1)}(z)\begin{pmatrix}
                    0 & 0 \\
                    -\zeta_j\gamma_je^{t\Phi(\zeta_j)} & 0
                \end{pmatrix},\\
                \label{eq:res_zeta_j_inv}
                &\mathrm{Res}_{\zeta^{-1}_j}m^{(1)}(z) = \lim_{z\to\zeta^{-1}_j}m^{(1)}(z)\begin{pmatrix}
                    0 & 
                    \zeta^{-1}_j\gamma_je^{t\Phi(\zeta_j)}\\ 0 & 0
                \end{pmatrix},
        \end{align}
        with $\Phi(z;n,t) = z-z^{-1} + 2 \frac{n}{t}\ln(z)$,
        \item the symmetry condition 
        \begin{equation}
            m^{(1)}(z^{-1})=m^{(1)}(z)\sigma_1,\quad \sigma_1=\begin{pmatrix}
                0 & 1\\1&0
            \end{pmatrix},
        \end{equation}
        \item and the normalization:  $ m^{(1)}(0)= (m^{(1)}_1,m^{(1)}_2)$ such that $m^{(1)}_1\cdot m^{(1)}_2 =1 $, $m_1^{(1)}>0$.
        
    \end{itemize}
\end{rhp}

We call $\{\zeta_j,\zeta_j^{-1}\}_{j=1}^N$ the \textit{poles} and the $\{\gamma_j\}_{j=1}^N$ the \textit{norming constants}.
Given the solution to this RHP, then it is possible to explicitly characterize the solution $(a_j(t),b_j(t))$ by means of the following recovery formula found in \cite[Lem.~3.4]{KRGER2009}.

\begin{lemma}
\label{lem:recovery}
    The function $m^{(1)}(z)$ defined in the RHP \ref{rhp:toda} satisfies the following asymptotic as $z\to 0$

    \begin{equation}
        \label{eq:recovery}
        m^{(1)}(z,n,t) = \Big(A(n,t)(1-2B(n-1,t)z),\; A(n,t)^{-1}(1+2B(n,t)z) \Big) + \mathcal{O}(z^{2})\,,
    \end{equation}
    where

    \begin{equation}
    \label{eq:infity_prod_sum}
        A(n,t) = \prod_{j=n}^\infty 2a_j(t) \qquad B(n,t) = -\sum_{j=n+1}^\infty b_j(t)\,.
    \end{equation}
\end{lemma}

\begin{remark}
    We notice that $A(n,t),B(n,t)$ in \eqref{eq:infity_prod_sum} are well defined, since we assume that the initial data is in the space $\mathfrak{S}$ \eqref{eq:func_space}.
\end{remark} 
In this paper, we study a fully \textbf{random soliton gas} (i.e.~$N \to \infty$) for the Infinite Toda lattice in the \textbf{scaling regime} $n, t = \cO(\ln N)$. This is the natural scaling under which we observe nonlinear wave modulation in both space \emph{and} time. Specifically, we work under the following assumptions:

\begin{assumptions}[Random $N$-Soliton Condensate]\label{ass:condensate_assumptions}
Let $N \in \mathbb{N}$ denote the number of solitons. We prescribe the scattering data as follows:
\begin{enumerate}[leftmargin=*]
    \item \textbf{Random Eigenvalues:} The discrete eigenvalues $\{\zeta_j\}_{j=1}^N$ are independent and identically distributed (i.i.d.) random variables supported on a fixed compact interval $[\alpha, \beta] \subset (0, 1)$, with $0 < \alpha < \beta < 1$. The common probability density $\rho(s)$ is analytic in an open neighborhood of $[\alpha, \beta]$ and satisfies strict positivity $0 < \rho(s) < \infty$ for $s \in [\alpha, \beta]$.
    \item \textbf{Random Norming Constants:} The norming constants $\{\gamma_j\}_{j=1}^N$ are i.i.d.\ positive random variables, independent of $\{\zeta_j\}_{j=1}^N$, satisfying a moment condition of order $\nu > 1$:
    \begin{equation}
        \mu_1 \coloneqq \E[\gamma_1] \in (0, \infty), \qquad \mu_\nu \coloneqq \E[\gamma_1^\nu] < \infty.
    \end{equation}
    \item \textbf{Spatio-Temporal Logarithmic Scaling:} The macroscopic spacetime coordinates $(n, t) \in \Z \times \R$ scale with the number of solitons according to
    \begin{equation}
        n = \eta \ln N + \cO(1), \qquad t = \tau \ln N,
        \label{eq:intro_scaling}
    \end{equation}
    for fixed \textit{slow} parameters $(\eta, \tau) \in \R^2$.
\end{enumerate}
\end{assumptions}

\begin{remark}\label{rem:soliton_direction}
    Under Assumptions \ref{ass:condensate_assumptions}, the discrete eigenvalues $\{\zeta_j\}_{j=1}^N$ are supported in $(0, 1)$. In the Toda lattice, a single soliton with eigenvalue $\zeta \in (0, 1)$ propagates with velocity
    \begin{equation}\label{eq:bare_velocity}
        v_0(\zeta) = \frac{\zeta^{-1} - \zeta}{2\ln \zeta} < 0,
    \end{equation}
    meaning that all solitons travel to the left. We emphasize that we do not restrict the temporal parameter to be positive; the evolution is well-defined for all $\tau \in \mathbb{R}$. Under time-reversal $\tau \mapsto -\tau$, the dynamics is isomorphic to reflecting the eigenvalues $\zeta_j \mapsto -\zeta_j$ into $(-1, 0)$, which reverses the direction of propagation so that all solitons travel to the right.
\end{remark}

\begin{remark}
    The strict inequality $0 < \rho(z) < \infty$ on the compact support $[\alpha, \beta]$ ensures that the background density does not vanish or blow up at the spectral edges. Analytically, this non-vanishing property guarantees that the local parametrices at the branch points are build out of Bessel functions of order $0$.  If instead $\rho(s)$ would have power type behavior $(\, \cdot\,)^p$, $p > -1$, at one of the spectral edges, then one would have to use different local parametrix. Despite that, we strongly believe that our analysis and final results about the asymptotic of the Toda lattice would be essentially unchanged.
\end{remark}

\begin{remark}\label{rem:log_scaling_mechanism}
    The logarithmic scaling $n = \eta \ln N$ and $t = \tau \ln N$ in \eqref{eq:intro_scaling} is the unique scaling regime where the nonlinear effects from the accumulation of solitons and their time evolution are of the same order, producing non-trivial macroscopic dynamics. Analytically, this becomes clear by looking at the off-diagonal entries in the Riemann--Hilbert problem jump matrices \eqref{eq:jump_mhat}, e.g.:
    \begin{equation}\label{eq:phase_balance_intro}
        e^{t\Phi(z;\eta,\tau)}\underbrace{\sum_{j=1}^N \frac{\zeta_j\gamma_j}{z-\zeta_j}}_{=\, \cO(N)} = \cO\left(e^{\tau \ln N \Phi(z; \eta, \tau)}N\right) = \cO(e^{\ln N (\tau \Phi(z; \eta, \tau) + 1)}), \qquad \tau = \cO(1).
    \end{equation}
    We see that the phase varies linearly in $\ln(N)$, which is exactly the same rate as the density of discrete poles $N \rho(s)$ growths, giving us a crucial $+1$ contribution in the exponential above; this is the mechanism that creates a very rich and interesting dynamics.
\end{remark}

\subsection{Summary of the main results}
\begin{figure}[!htbp]
    \centering
    \includegraphics[width=\linewidth]{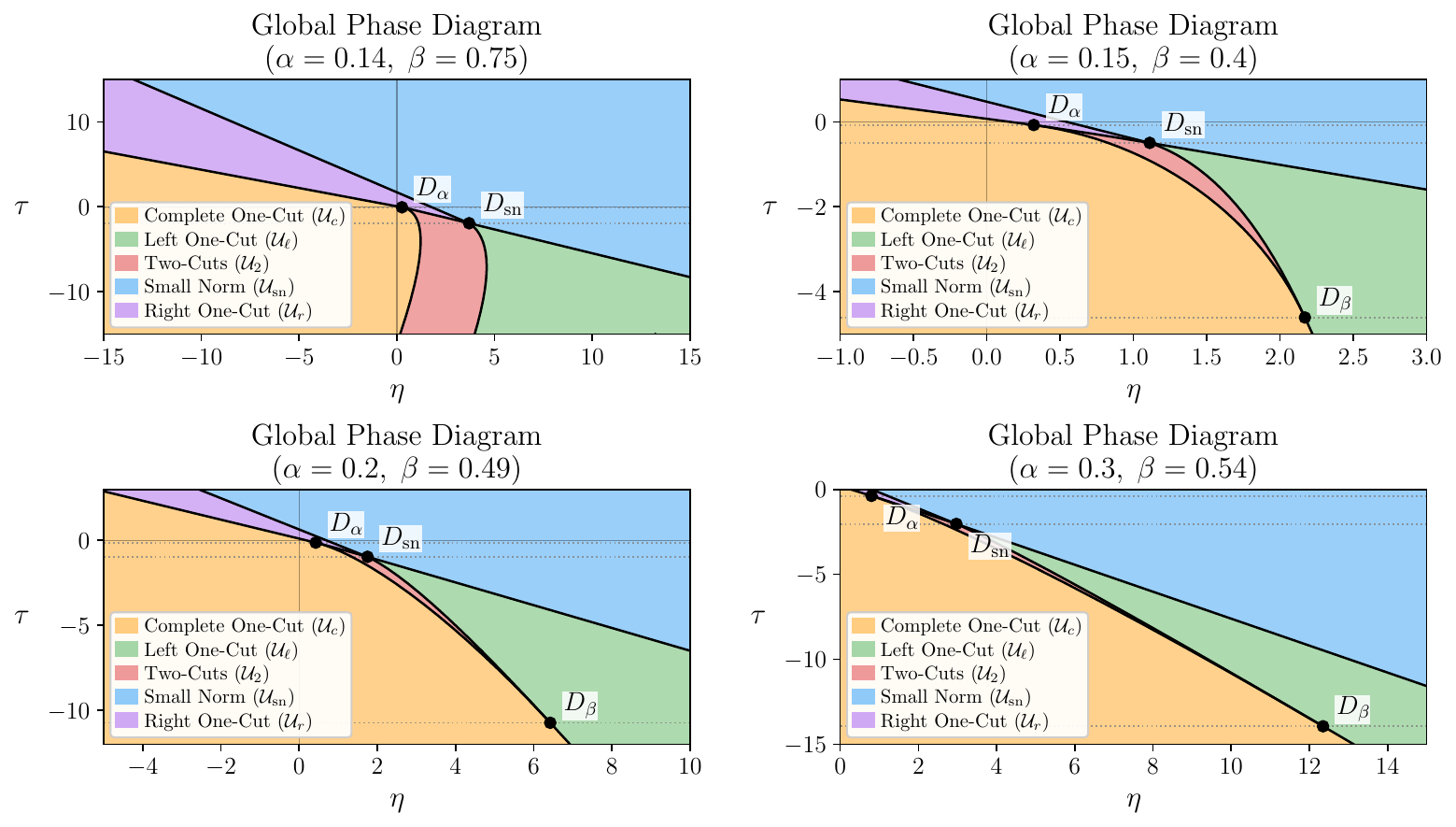}
    \caption{Global phase diagram in the macroscopic $(\eta, \tau)$-plane for four distinct spectral configurations $(\alpha, \beta)$. The colors demarcate the five macroscopic regimes: complete one-cut $\mathcal{U}_c$, modulated left one-cut $\mathcal{U}_\ell$, modulated right one-cut $\mathcal{U}_r$, modulated two-cuts $\mathcal{U}_2$, and small-norm vacuum $\mathcal{U}_{\mathrm{sn}}$.}
    \label{fig:phase_diagram}
\end{figure}

Under Assumptions \ref{ass:condensate_assumptions}, we establish that the leading-order asymptotic behavior of the Flaschka variables $(a_n(t), b_n(t))$ in the logarithmic scaling limit is \textbf{completely deterministic and universal}, i.e. it depends mildly on the distribution of $\{\boldsymbol{\zeta},\boldsymbol{\gamma}\}=\{\zeta_j,\gamma_j\}_{j=1}^N$. In fact, the leading order asymptotics only depends on the support $[\alpha, \beta]$ of the density $\rho(s)$. The  $(\eta,\tau)$-plane is partitioned into five   regions according to the genus of the limiting finite-gap solutions (see Figure \ref{fig:phase_diagram}):
\begin{enumerate}[leftmargin=*]
    \item \textbf{Small-Norm/Equilibrium Regime($\mathcal{U}_{\mathrm{sn}}$):} In this domain, the lattice relaxes to its equilibrium:
    \begin{equation}
        a_n(t) = \frac{1}{2} + \mathcal{O}(N^{-\delta}), \qquad b_n(t) = \mathcal{O}(N^{-\delta}), \qquad \text{for some }\, \delta > 0.
    \end{equation}
    \item \textbf{Complete One-Cut Regime ($\mathcal{U}_c$):} In this region, the solution is described by an elliptic wave whose spectral band coincides with the entire interval $[\alpha, \beta]$. At leading order, the variables $a_n(t)$ and $b_n(t)$ are given in terms of Jacobi theta functions $\theta_3$ and modulated background fields $a_{\mathrm{bg}}(\eta, \tau), b_{\mathrm{bg}}(\eta, \tau)$.
    \item \textbf{Modulated Left One-Cut Regime ($\mathcal{U}_\ell$):} In this regime, the solution is represented by a modulated genus-$1$ elliptic wave; in particular, only the right branch point is modulated according a non-linear Whitham equation.
    \item \textbf{Modulated Right One-Cut Regime ($\mathcal{U}_r$):} In this set, the solution is represented by a modulated genus-$1$ elliptic wave; in particular, only the left branch point is modulated according a non-linear Whitham equation.
    \item \textbf{Modulated Two-Cuts Regime ($\mathcal{U}_2$):} Here the solution is a modulated genus-$2$ hyperelliptic wave, where we have $2$ modulated branch-points, evolving according non-linear Whitham equations
\end{enumerate}

\begin{figure}[htbp]
    \centering
    \includegraphics[width=\linewidth]{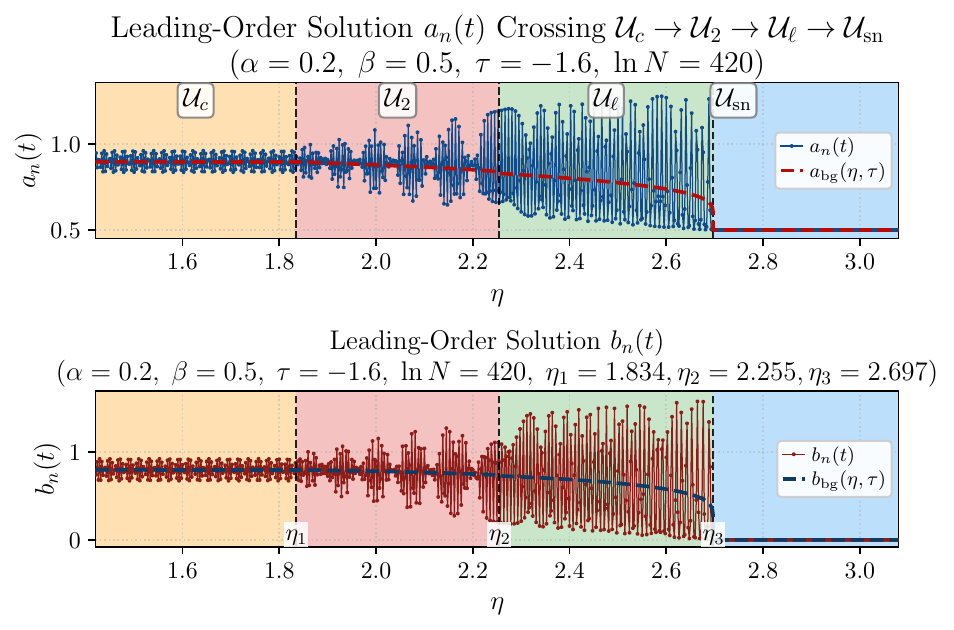}
    \includegraphics[width=\linewidth]{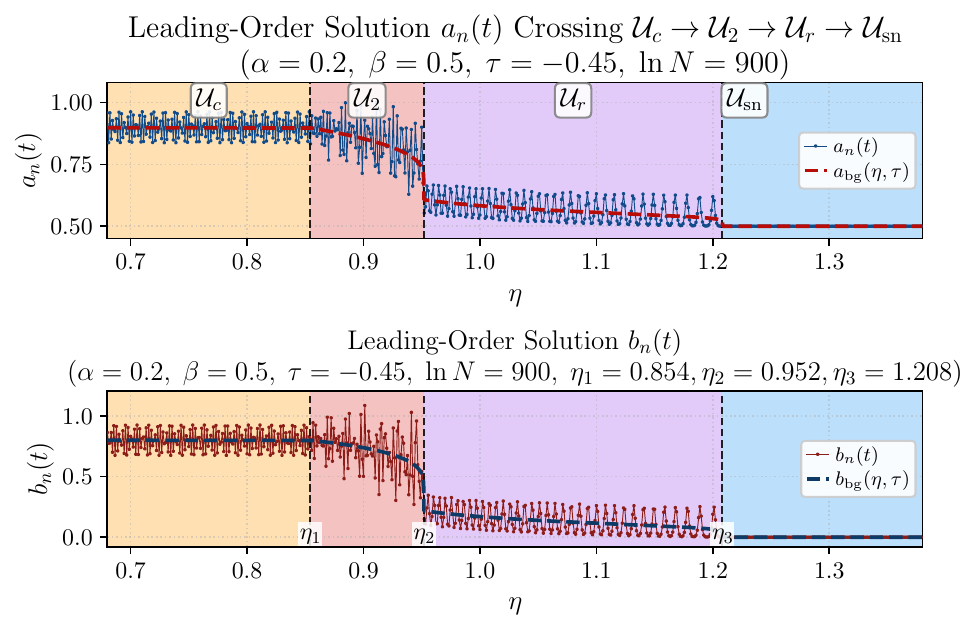}
    \caption{Spatio-temporal evolution of the Flaschka variables $a_n(t)$ and $b_n(t)$ for $\tau = -1.6$ and $\tau = -0.45$. The background shading corresponds to the macroscopic regimes identified in Figure \ref{fig:phase_diagram}. }
    \label{fig:evolution1}
\end{figure}

\begin{figure}
    \centering
    \includegraphics[width=\linewidth]{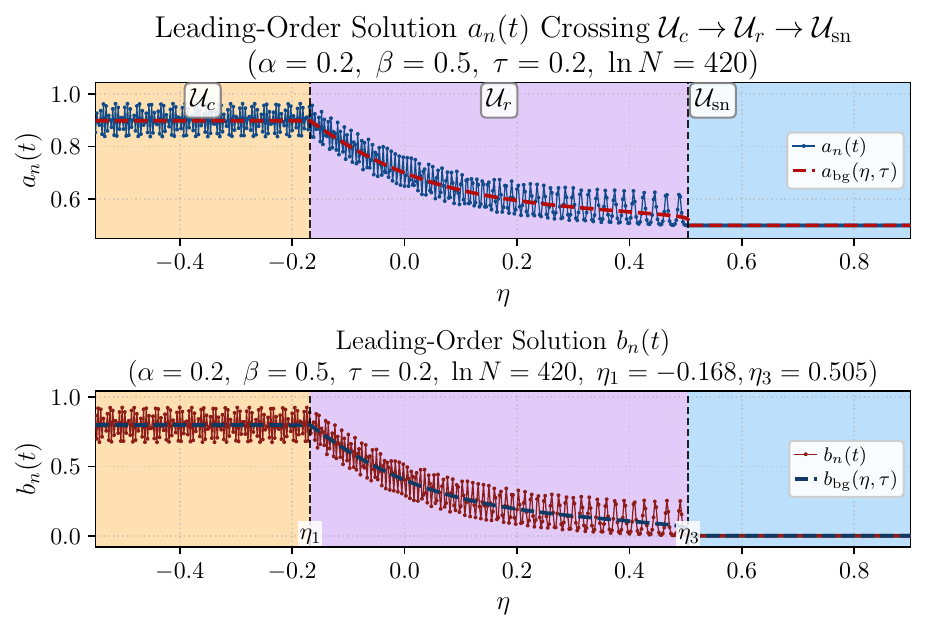}
    \caption{Spatio-temporal evolution of the Flaschka variables $a_n(t)$ and $b_n(t)$ for $\tau = 0.2$. The background shading corresponds to the macroscopic regimes identified in Figure \ref{fig:phase_diagram}.}
    \label{fig:evolution2}
\end{figure}

A pictorial representation of these macroscopic phases across the $(\eta, \tau)$-plane is displayed in Figure \ref{fig:phase_diagram}, and representative profile snapshots are reported in Figure \ref{fig:evolution1}-\ref{fig:evolution2}. As $N \to \infty$ give us a series of \emph{different} solitons solutions, our results should be interpreted as follows: choose $N$ very large but fixed. Then for each $t = \cO(\ln N)$ fixed the $N$-soliton solution we study can be approximated by slowly modulated waves of genus $0$, $1$ or  $2$ in different deterministic regions which have length of order $\cO(\ln N)$, see the aforementioned snapshots in Figure \ref{fig:evolution1}-\ref{fig:evolution2}. Additionally, as $t$ varies on the order of $\cO(\ln N)$, the wave profile undergoes a slow modulation in time, best visualized in Figure \ref{fig:phase_diagram}.

Before detailing the strategy of proof and the structure of the paper, we emphasize several key aspects that distinguish our analysis from the existing literature:

\begin{remark}\label{rem:intro_random_norming}
    In the majority of the mathematical literature on soliton gases (see  \cite{Girotti2023SolitonGas, GGJM, ElTovbis2020,Gkogkou2025soliton,Girotti2026}), the norming constants are predominantly treated as deterministic, and/or constrained to be asymptotically small. In our work, the norming constants $\{\gamma_j\}_{j=1}^N$ are \textbf{independent, identically distributed positive random variables}. In particular, they can be \textbf{arbitrarily large or close to zero}; subject only to a mild moment condition $\mathbb{E}[\gamma_1^\nu] < \infty$ with $\nu > 1$. Together with the random discrete eigenvalues $\{\zeta_j\}_{j=1}^N$, our results demonstrate that despite microscopic stochasticity in both amplitudes and norming constants, the Flaschka variables converge to a universal, deterministic profile.
\end{remark}

\begin{remark}\label{rem:intro_actual_small_norm}
     One of the main differences with previous work on soliton gases is that we perform a \textbf{global small-norm analysis} that rigorously relate the finite-$N$ system to its continuum limit in the \textbf{long time} asymptotic regime. In earlier studies of soliton gases-- such as \cite{GGJM,Girotti2023SolitonGas}---the authors formulate a Riemann--Hilbert model problem with continuous jumps and investigate its long-time asymptotic behavior, without having small-norm control connecting this model problem back to the original $N$-soliton system. In \cite{Gkogkou2025soliton}, the authors did perform a small-norm analysis connecting to the continuous problem, but with deterministic eigenvalues and norming constant, and for $x,t$ bounded. In contrast, in the present paper we construct an explicit matrix-valued small-norm problem that rigorously establishes uniform control between the discrete random problem $m^{(2)}(z)$ and the continuum deterministic problem $q(z)$ on a high-probability set $\mathfrak{W}_N$ which is valid in the long-time asymptotic regime.
\end{remark}

\begin{remark}\label{rem:intro_discrete_shocks}
    To the best of our knowledge, this work provides the \textbf{first complete asymptotic description of a dense soliton gas on a discrete integrable lattice}. Due to the logarithmic scaling, the boundaries $\Sigma_{\mathrm{phase}}$ between the various asymptotic regions are genuinely non-linear curves rather than self-similar straight rays. Moreover, our phase diagram exhibits a novel critical point $D_{\mathrm{sn}}$ where the genus-$2$ solution transitions to the equilibrium one, without passing through a genus-$1$ phase.
\end{remark}

\begin{remark}
    In \cite{Congy2023}, the authors conjectured that any dense soliton gas for the KdV equation should be described, at leading order, by some finite-genus elliptic/hyper-elliptic wave. Our result can be interpreted as a first step toward proving this conjecture in the context of the infinite Toda lattice. 
\end{remark}

\subsection{Idea of the proof}
We now describe the main ideas of our proof. Specifically, we first simplify our problem by introducing an \textit{approximate and fully deterministic} Riemann-Hilbert problem, which is the one we actually analyze and solve explicitly upto a uniform $\cO\left((\ln N)^{-1}\right)$-error.
Following a standard procedure \cite{Girotti2023SolitonGas}, for each realization of the random variables $\{\zeta_j,\gamma_j\}_{j=1}^N$,  we trade the poles in the RHP \ref{rhp:toda} for jumps along some contours. Consider a curve $\sigma$ that encloses all the poles $\zeta_1,\ldots,\zeta_N$ and does not intersect the unit circumference. Then one can define the following transformation
\begin{equation}
    m^{(2)}(z) = m^{(1)}(z)J_{m^{(2)}}(z)
    \end{equation}
    where
    \begin{equation}
    \label{eq:jump_mhat}
    J_{m^{(2)}}(z)=\begin{cases}
      \begin{pmatrix}
            1 & 0 \\
            e^{t\Phi(z,\eta,\tau)}\sum_{j=1}^N \frac{\zeta_j\gamma_j}{z-\zeta_j} & 1
        \end{pmatrix} \quad& z \in \mathring{\sigma} \\\\[5pt]
        \begin{pmatrix}
            1 &-ze^{-t\Phi(z,\eta,\tau)}\sum_{j=1}^N \frac{\gamma_j}{z-\zeta^{-1}_j}\\
            0 & 1
        \end{pmatrix} \quad &z^{-1} \in \mathring{\sigma} \\\\[5pt]
        \begin{pmatrix}
            1 & 0 \\
            0 & 1
        \end{pmatrix} \quad& \text{ otherwise } \\
    \end{cases}\,.
\end{equation}
One can check that 
\begin{equation}
    J_{m^{(2)}}(z^{-1}) = \sigma_1 J_{m^{(2)}}(z)\sigma_1\,, \qquad \sigma_1 = \begin{pmatrix}
        0 & 1 \\
        1&0
    \end{pmatrix}\,,
\end{equation}
and therefore $m^{(2)}(z^{-1})=m^{(2)}(z)\sigma_1$. Moreover, $m^{(2)}(z)$ does not have any poles, so it solves the following RHP

\begin{rhp}
\label{rhp:m_hat}
    The function $m^{(2)}(z) = (m^{(2)}_1(z)\quad  m^{(2)}_2(z))$ satisfies the following conditions

    \begin{itemize}
        \item $m^{(2)}(z)$ is analytic in $\mathbb{C}\setminus\{ \sigma,\sigma^{-1}\}$
        \item $m^{(2)}_+(z)=m^{(2)}_-(z) J_{m^{(2)}}(z)$ for $z\in \sigma,\sigma^{-1}$, which are oriented counterclockwise and $J_{m^{(2)}}(z)$ is defined in \eqref{eq:jump_mhat}
        \item $ m^{(2)}(z^{-1})=m^{(2)}(z)\sigma_1$
        \item $m^{(2)}_1(0)\cdot m^{(2)}_2(0) = 1$, $m_1^{(2)}(0)>0$
    \end{itemize}
\end{rhp}

We notice that the two contours $\sigma,\sigma^{-1}$ are such that there exists a constant $d>0$ independent of $N$, and depending only on the contour $\sigma$ such that

\begin{equation}
    |\zeta_j - z|>d \qquad \forall\, z\in \sigma, \,\forall\,j=1,\ldots,N\,.
\end{equation}
Therefore, the following limits hold for the off-diagonal elements of the jump matrix $J_{m^{(2)}}(z)$ in view of the Law of Large Numbers for all fixed $z\in\sigma\cup \sigma^{-1}$:
\begin{equation}
  \frac{1}{N}\sum_{j=1}^N \frac{\zeta_j\gamma_j}{z-\zeta_j} \xrightarrow[N\to\infty]{\text{a.s}}  \mu_1\int_\alpha^\beta \frac{s\rho(s)}{z-s}\di s \,, \qquad 
    \frac{1}{N}\sum_{j=1}^N \frac{\gamma_j}{z-\zeta_j^{-1}} \xrightarrow[N\to\infty]{\text{a.s}}  \mu_1\int_{\alpha}^{\beta} \frac{\rho(s)}{z-s^{-1}} \di s\,.
\end{equation}

\begin{remark}
    The rate of convergence to these limits will play a major role in our analysis, but, at this stage, we do not need it to specify it. See Section \ref{sec:L1AS} for details.
\end{remark}

This leads us to consider the following RHP for a vector $ q(z) = ( q_1(z)\quad q_2(z))$ 

\begin{rhp}\label{RHP:q}
    We seek for a vector $ q(z) = (q_1(z)\quad q_2(z))$  such that 
    \begin{itemize}
        \item $q(z)$ is analytic in $\mathbb{C}\setminus\{ \sigma,\sigma^{-1}\}$
        \item $q_+(z)=q_-(z) J_{q}(z)$ for $z\in \sigma,\sigma^{-1}$, which are oriented counterclockwise and $J_{q}(z)$ is defined as
        
      \begin{equation}
  	\label{eq:jump_q}
  	J_{q}(z)=\begin{cases}
  		\begin{pmatrix}
  			1 & 0 \\
  			N\mu_1 e^{t\Phi(z,\eta,\tau)}\int_\alpha^\beta \frac{s\rho(s)}{z-s}\mathrm{d} s& 1
  		\end{pmatrix} \quad &z \in \sigma \\\\[5pt]
  		\begin{pmatrix}
  			1 &-N\mu_1 ze^{-t\Phi(z,\eta,\tau)} \int_{\alpha}^{\beta} \frac{\rho(s)}{z-s^{-1}} \mathrm{d} s\\
  			0 & 1
  		\end{pmatrix} \quad &z^{-1}\in\sigma 
  	\end{cases}\,.
  \end{equation}
        \item $q(z^{-1})=q(z)\sigma_1$
        \item $\lim_{z\to 0} q_1(z)\cdot q_2(z) = 1$, $q_1(0)>0$
    \end{itemize}
\end{rhp}
In particular, the following holds:

\begin{lemma}
\label{lem:recovery_q}
    The function $q(z) \equiv q(z;\eta,\tau)$ defined in RHP \ref{RHP:q} satisfies the following asymptotic as $z\to 0$:
    \begin{equation}
        \label{eq:recovery_q}
        q(z;\eta,\tau) = \Big(A^{(q)}(\eta,\tau)\big(1-2B^{(q)}(\eta-\ln(N)^{-1},\tau)z\big),\; A^{(q)}(\eta,\tau)^{-1}\big(1+2B^{(q)}(\eta,\tau)z\big) \Big) + \mathcal{O}(z^{2})\,.
    \end{equation}
    Defining the continuum Flaschka variables
    \begin{equation}
    \label{eq:recovery_q_fields}
        a^{(q)}(\eta,\tau) \coloneqq \frac{1}{2}\frac{A^{(q)}(\eta,\tau)}{A^{(q)}(\eta+\ln(N)^{-1},\tau)}\,, \qquad b^{(q)}(\eta,\tau) \coloneqq B^{(q)}(\eta,\tau)-B^{(q)}(\eta-\ln(N)^{-1},\tau)\,,
    \end{equation}
    and evaluating them along the lattice coordinates $n = \eta\ln N$ and $t = \tau\ln N$ via
    \begin{equation}
    \label{eq:recovery_q_lattice}
        a_n^{(q)}(t) \coloneqq a^{(q)}\left(\frac{n}{\ln N}, \frac{t}{\ln N}\right) = a^{(q)}(\eta,\tau)\,, \qquad b_n^{(q)}(t) \coloneqq b^{(q)}\left(\frac{n}{\ln N}, \frac{t}{\ln N}\right) = b^{(q)}(\eta,\tau)\,,
    \end{equation}
    the resulting infinite sequence $\{a_n^{(q)}(t), b_n^{(q)}(t)\}_{n\in\mathbb{Z}}$ satisfies the infinite Toda lattice equations \eqref{Todaeq}.
\end{lemma}
In Appendix \ref{App:Fred}, we show that RHP \ref{RHP:q} has a unique solution via a vanishing lemma and Fredholm theory. Lemma \ref{lem:recovery_q} follows from an explicit calculation using the Lax pair formulation \eqref{eq:Lax_Eq} for the Toda lattice and the unique solvability of RHP \ref{RHP:q}.

The RHP \ref{RHP:q} is the one that we consider and solve to obtain the leading order asymptotic of the soliton gas that we are studying. Our main result will then follow from small norm analysis \cite{DZ03,PT22} combined with some probabilistic estimate in the same spirit of \cite{Gkogkou2025,Girotti2026}. More specifically, let $\nu > 1$ be as in Assumption \ref{ass:condensate_assumptions} point 2, and define
\begin{equation}
\label{eq:kappa_nu}
    \kappa_\nu := \min\left(\nu - 1,\, \frac{\nu}{2}\right) = \begin{cases}
        \nu - 1\,, & \text{if } 1 < \nu \leq 2\,, \\[4pt]
        \dfrac{\nu}{2}\,, & \text{if } \nu > 2\,.
    \end{cases}
\end{equation}
For some fixed $\varepsilon_\nu \in \left(0, \frac{\kappa_\nu}{\nu}\right)$, we define the \textit{good set} $\mathfrak{W}_N\subset (\alpha,\beta)^N\times\R_+^{N}$ as 
\begin{align}\label{eq:good_set}
&\fW_N = \left\{(\zeta_1,\ldots,\zeta_N,\gamma_1,\ldots,\gamma_N )\in (\alpha,\beta)^N\times\R_+^{N} \;\left\vert\;\sup_{z\in\sigma}\left\vert \frac{1}{N}\sum_{j=1}^N \frac{\zeta_j\gamma_j}{z-\zeta_j} -\mu_1\int_\alpha^\beta \frac{s\rho(s)}{z-s}\di s \right\vert\right. \leq N^{-\varepsilon_\nu} \right\}\,,
\end{align}
which alternatively can be written as
\begin{align*}
   \fW_N = \left\{(\zeta_1,\ldots,\zeta_N,\gamma_1,\ldots,\gamma_N )\in (\alpha,\beta)^N\times\R_+^{N} \;\left\vert\;\sup_{z\in\sigma^{-1}}\left\vert \frac{1}{N}\sum_{j=1}^N \frac{z\gamma_j}{z-\zeta^{-1}_j} -z\mu_1\int_\alpha^\beta \frac{\rho(s)}{z-s^{-1}}\di s \right\vert\right. \leq N^{-\varepsilon_\nu} \right\}\,.
\end{align*}
\begin{remark}\label{rem:jump_difference}
Notice that the on the good set $\fW_N$ we only have the estimate 
\begin{align*}
    \Vert J_{m^{(2)}} - J_q \Vert_{L^\infty(\sigma \cup \sigma^{-1})} \leq N^{1-\varepsilon_\nu} \to +\infty, \quad \text{for} \ N \to \infty 
\end{align*}
as $\varepsilon_\nu < 1/2$. In particular the jump matrices of the two problems are, in general, \emph{not} close to each other in the $L^\infty$-norm, which is usually required for a small norm analysis, despite not being necessary. However, we will see that the natural small norm problem for $m^{(2)}(z)Q^{-1}(z)$, where $Q(z)$ is a matrix-valued solution of RHP \ref{RHP:q}, nonetheless has the small norm property due to the \emph{smoothing effect} of the $g$-function, see Section \ref{Sect:SNGoodSet}, in particular \eqref{eq:Jdelta}. This implies that the $\nu^{th}$-moment bound (for any $\nu > 1$) on the norming constants is sufficient for our analysis, see Assumption \ref{ass:condensate_assumptions}.  
\end{remark}

\begin{remark}
    We reported both expressions for $\fW_N$ as we need to bound the difference of the off-diagonal entries of the jump matrices of $m^{(2)}(z),q(z)$ on $\sigma$ \emph{and} $\sigma^{-1}$; both entries have the same numerical value under the symmetry $z \to z^{-1}$, which implies that the two expressions for the good set $\fW_N$ are equivalent.
\end{remark}
For $(\zeta_1,\ldots,\zeta_N,\gamma_1,\ldots,\gamma_N )\in\fW_N$, as a byproduct of our Riemann-Hilbert analysis, we deduce that
\begin{theorem}
\label{thm:m_2_q_approx}
    Under Assumptions \ref{ass:condensate_assumptions}, the RHP \ref{RHP:q} has a unique solution. Furthermore, for $(\zeta_1,\ldots,\zeta_N,\gamma_1,\ldots,\gamma_N )\in\fW_N$ and $(\eta, \tau)$ in any compact subset of the interior of each macroscopic regime ($\mathbb{R}^2 \setminus \Sigma_{\mathrm{phase}}$, where $\Sigma_{\mathrm{phase}}$ denotes the finite union of the phase transition boundary curves in the $(\eta, \tau)$-plane) and any $\delta\in (0,\varepsilon_\nu)\subset\left(0,\frac{\kappa_\nu}{\nu}\right)$, the following holds
    \begin{equation}
        \label{eq:small_approx}
	    m^{(1)}_1(z) = m^{(2)}_1(z) =   q_1(z)(1 + \mathcal{O}(N^{-\delta})),
    \end{equation}
    locally, uniformly for $z$ in a neighborhood of $0$.
\end{theorem}
 The proof of this result is given in Section \ref{sec:prob_estimate} following the approach of Deift--Zhou \cite{Deift1993, Deift1997,DZ03}.
We analyze the asymptotic behavior of $q(z)$ and in Sections \ref{sec:small_norm}-\ref{sec:complete_one_cut}-\ref{sec:two_cuts}-\ref{sec:one_cut_modulated_regions} we obtain the leading order behavior of $a_n(t),\, b_n(t)$ (equiv.~of $a_n^{(q)}(t),\, b_n^{(q)}(t)$) which holds with high probability and denote it by $\fa(\eta,\tau),\fb(\eta,\tau)$. To \textit{promote} this convergence to $L^1$-convergence (and, for $\nu>2$, almost sure convergence) on the full probability space, we use the following Lemma, whose proof can be found in Section \ref{sec:prob_estimate}:
\begin{lemma}
\label{lem:measure_bad_set}
    Under Assumptions \ref{ass:condensate_assumptions}, for any $\nu > 1$ and $\varepsilon_\nu \in \left(0, \frac{\kappa_\nu}{\nu}\right)$, there exists a positive constant $c_{\textrm{bad}}$ independent of $N$, depending only on $\sigma,\nu,\mu_\nu$, such that
    \begin{equation}
        \mathbb{P}(\fW_N^c) = |\fW_N^c| \leq c_{\textrm{bad}} N^{-\kappa_\nu + \nu\varepsilon_\nu} = \begin{cases}
            c_{\textrm{bad}} N^{1-\nu+\nu\varepsilon_\nu}\,, & \text{if } 1 < \nu \leq 2\,, \\[4pt]
            c_{\textrm{bad}} N^{-\frac{\nu}{2}+\nu\varepsilon_\nu}\,, & \text{if } \nu > 2\,.
        \end{cases}
    \end{equation}
\end{lemma}
This, together with some a priori estimates on $\norm{a_n(t)}_\infty,\norm{b_n(t)}_\infty$ \cite{Teschl1999}, allows us to deduce the following (here $\di\mathbb{P}(\boldsymbol{\zeta}, \boldsymbol{\gamma})$ denotes the joint probability measure of $\lbrace \zeta_j, \gamma_j\rbrace_{j=1}^N$):
\begin{equation}
\label{eq:split_mean_val}
\begin{split}
    &\mathbb{E}\left[\left\vert a_n(t,\zeta_1,\ldots,\zeta_N,\gamma_1,\ldots,\gamma_N) - \fa(\eta,\tau)\right\vert\right] \\& = \int_{\fW_N} \left\vert a_n(t,\zeta_1,\ldots,\zeta_N,\gamma_1,\ldots,\gamma_N) - \fa(\eta,\tau)\right\vert\di \mathbb{P}(\boldsymbol{\zeta}, \boldsymbol{\gamma}) 
    \\
    &\hspace{130pt}+ \int_{\fW_N^c} \left\vert a_n(t,\zeta_1,\ldots,\zeta_N,\gamma_1,\ldots,\gamma_N) - \fa(\eta,\tau)\right\vert\di\mathbb{P}(\boldsymbol{\zeta}, \boldsymbol{\gamma})\\
    &\leq \int_{\fW_N} \left\vert a_n(t,\zeta_1,\ldots,\zeta_N,\gamma_1,\ldots,\gamma_N) - \fa(\eta,\tau)\right\vert \di\mathbb{P}(\boldsymbol{\zeta}, \boldsymbol{\gamma}) + \mathcal{O}\left(N^{-\kappa_\nu+\nu\varepsilon_\nu}\right)\,,
    \end{split}
\end{equation}
(notice that $-\kappa_\nu+\nu\varepsilon_\nu < 0$ by the choice of $\varepsilon_\nu$) where we explicitly write the dependence of $a_n(t)$ from the random variables $\zeta_1,\ldots,\zeta_N,\gamma_1,\ldots,\gamma_N$; an analogous estimate holds for $\fb(\eta,\tau)$.
Furthermore, when $\nu > 2$, we have $\kappa_\nu = \frac{\nu}{2} > 1$, so choosing $\varepsilon_\nu \in \left(0, \frac{1}{2} - \nu^{-1}\right)$ ensures $-\frac{\nu}{2} + \nu\varepsilon_\nu < -1$, therefore there exists a constant $c_\fW>0$ such that
\begin{equation}
    \sum_{N=1}^\infty\mathbb{P}(\fW_N^c) \leq c_\fW \sum_{N=1}^\infty N^{-\frac{\nu}{2} +\nu\varepsilon_\nu} < +\infty\,.
\end{equation}
Combining these probabilistic estimates with the Riemann--Hilbert analysis and the Borel--Cantelli lemma, we deduce our main results. We present them in two different parts: first, a strong approximation of the random lattice variables $a_n(t),b_n(t)$ by the deterministic continuum Flaschka variables $a_n^{(q)}(t)$ and $b_n^{(q)}(t)$; second, the explicit asymptotic representation of these variables across the phase diagram in terms of modulated finite-gap solutions.

\begin{theorem}
\label{thm:poly_continuum_approx}
Consider the infinite Toda lattice \eqref{toda} under Assumption \ref{ass:condensate_assumptions}. Let $a^{(q)}(\eta,\tau)$ and $b^{(q)}(\eta,\tau)$ denote the deterministic continuum Flaschka variables extracted from the unique solution $q(z)$ of RHP \ref{RHP:q} via Lemma \ref{lem:recovery_q}, and let $a_n^{(q)}(t)$ and $b_n^{(q)}(t)$ denote their lattice evaluations \eqref{eq:recovery_q_lattice}. 
Then for any $(\eta, \tau)$ in any compact subset of the interior of each macroscopic regime ($\mathbb{R}^2 \setminus \Sigma_{\mathrm{phase}}$) and for any $\delta \in (0, \varepsilon_\nu) \subset \left(0, \frac{\kappa_\nu}{\nu}\right)$, there exists a deterministic constant $C_0 > 0$ such that:
\begin{enumerate}
    \item (Uniform Convergence on the Good Set $\fW_N$)
    \begin{equation}
    \label{eq:good_set_poly_bound}
        \sup_{(\boldsymbol{\zeta},\boldsymbol{\gamma})\in \fW_N} \left( |a_n(t) - a_n^{(q)}(t)| + |b_n(t) - b_n^{(q)}(t)| \right) \leq C_0 N^{-\delta}\,.
    \end{equation}
    \item ($L^1(\mathbb{P})$ Convergence) for any $0<\delta_1 < \min(\delta, \kappa_\nu - \nu\varepsilon_\nu)$,
    \begin{equation}
        \mathbb{E}\left[ |a_n(t) - a_n^{(q)}(t)| \right] \leq C_0 N^{-\delta_1}\,, \qquad \mathbb{E}\left[ |b_n(t) - b_n^{(q)}(t)| \right] \leq C_0 N^{-\delta_1}\,.
    \end{equation}
    \item (Almost Sure Convergence) if $\nu>2$, for $\mathbb{P}$-almost every realization of the scattering data $\{(\zeta_j, \gamma_j)\}_{j=1}^\infty$, as $N \to \infty$,
    \begin{equation}
    \label{eq:almost_sure_poly}
        |a_n(t) - a_n^{(q)}(t)| = \mathcal{O}(N^{-\delta})\,, \qquad |b_n(t) - b_n^{(q)}(t)| = \mathcal{O}(N^{-\delta})
    \end{equation}
\end{enumerate}
\end{theorem}

\begin{theorem}
\label{thm:main_prob}
Consider the infinite Toda lattice \eqref{toda} under Assumption \ref{ass:condensate_assumptions}. For $(\eta, \tau)$ in any compact subset of the interior of each macroscopic regime ($\mathbb{R}^2 \setminus \Sigma_{\mathrm{phase}}$), there exist explicit deterministic finite-gap wave $\fa(\eta,\tau)$ and $\fb(\eta,\tau)$ such that:
\begin{enumerate}
    \item ($L^1(\mathbb{P})$ convergence) If $\nu > 1$, then
    \begin{equation}
        \lim_{N\to\infty} \mathbb{E}\left[\left\vert a_n(t) - \fa(\eta,\tau)\right\vert\right] =0 \,,\qquad
        \lim_{N\to\infty} \mathbb{E}\left[\left\vert b_n(t) - \fb(\eta,\tau)\right\vert\right] =0\,.
    \end{equation}
    \item (Almost sure convergence) If in addition $\nu > 2$, then for $\mathbb{P}$-almost every realization of the sequence of i.i.d.\ scattering data $\{(\zeta_j, \gamma_j)\}_{j=1}^\infty$, as $N \to \infty$,
    \begin{equation}
    \label{eq:almost_sure_conv}
        a_n(t) = \fa(\eta,\tau) + \mathcal{O}\left(\frac{1}{\ln N}\right) \,,\qquad
        b_n(t) = \fb(\eta,\tau) + \mathcal{O}\left(\frac{1}{\ln N}\right) \,.
    \end{equation}
\end{enumerate}
\end{theorem}

\begin{remark}\label{rem:explicit_solutions_roadmap}
The explicit finite-gap waveforms $\fa(\eta,\tau)$ and $\fb(\eta,\tau)$ appearing in Theorem \ref{thm:main_prob} are the leading-order asymptotic limits of the continuum variables $a_n^{(q)}(t)$ and $b_n^{(q)}(t)$, derived via the nonlinear steepest descent analysis of RHP \ref{RHP:q}. The reader will find their explicit closed-form expressions across the subsequent sections. Specifically in Corollary \ref{cor:small_norm_sol}, and Theorem \ref{thm:complete_one_cut_asymptotics}-\ref{thm:modulated_two_cuts_asymptotics}-\ref{thm:left_one_cut_asymp}-\ref{thm:right_one_cut_asymp}.
Regarding nonlinear dispersion relations and ensemble averages, we state an informal version of our main results below; formal statements can be found in Section \ref{sec:NDR}.
\end{remark}
\begin{theorem}\label{Thm:NDR}
    Under Assumptions \ref{ass:condensate_assumptions}, in the large $N$-limit the continuum solution $a_n^{(q)}(\eta, \tau)$, $b_n^{(q)}(\eta, \tau)$ can be described by a continuous density of states $u(\lambda, \eta, \tau)$ and a flux density $v(\lambda, \eta, \tau)$, with $\lambda$ in an explicit $(\eta, \tau)$-dependent subset $\Sigma(\eta, \tau) \subseteq(\alpha, \beta)$. These densities satisfy the Toda nonlinear dispersion relations:
    \begin{equation}
    \int_{\Sigma}\ln\left(\left\vert\frac{z-\lambda}{1-\lambda z}\right\vert\right)u(\lambda,\eta,\tau)\di\lambda = \ln(z)\,,\qquad \int_{\Sigma}\ln\left(\left\vert\frac{z-\lambda}{1-\lambda z}\right\vert\right)v (\lambda,\eta,\tau)\di\lambda= \frac{z^{-1}-z}{2}\,.
    \end{equation}
    As a consequence we obtain the Toda soliton gas kinetic equation \eqref{eq:KineticEq} and explicit formulas for ensemble averages \eqref{eq:a_avg}, \eqref{eq:b_avg}.
\end{theorem}

\begin{remark}
    It is worth mentioning that the ensemble averages are slowly varying in space and time.
\end{remark}
The paper is organized as follows: in Section \ref{sec:model_problems}, we solve two model Riemann--Hilbert problems that provide the algebraic building blocks for the global parametrices. In Sections \ref{sec:small_norm}--\ref{sec:one_cut_modulated_regions}, we perform the Deift--Zhou steepest descent analysis of RHP \ref{RHP:q} across each macroscopic regime, obtaining the explicit asymptotic formulas for $a_n^{(q)}(t)$ and $b_n^{(q)}(t)$ summarized in Remark \ref{rem:explicit_solutions_roadmap}. In Section \ref{sec:prob_estimate}, we establish the small-norm approximation on the good set $\fW_N$, prove Lemma \ref{lem:measure_bad_set}, and complete the proofs of Theorem \ref{thm:poly_continuum_approx} and Theorem \ref{thm:main_prob}. In Section \ref{sec:NDR} we prove Theorem \ref{Thm:NDR} and compare our formulas with analogous formulas for the KdV equation. Finally, Appendix \ref{Appendix} contains technical results on existence, uniqueness, and small-norm theory for vector Riemann--Hilbert problems, and a proof of the unique solvability of RHPs \ref{RHP:q}, \ref{RHP:q_hat}.

Before starting with our analysis, we perform a final transformation

\begin{equation}
    \wh q(z) = \begin{cases}
        q(z)\begin{pmatrix}
  			1 & 0 \\
  			-N\mu_1 e^{t\Phi(z,\eta,\tau)}\int_\alpha^\beta \frac{s\rho(s)}{z-s}\mathrm{d} s& 1
  		\end{pmatrix} \qquad &z\in \mathring{\sigma}\\\\
        q(z) \begin{pmatrix}
  			1 &zN\mu_1 e^{-t\Phi(z,\eta,\tau)} \int_{\beta^{-1}}^{\alpha^{-1}} \frac{\rho(s^{-1})}{(z-s)s^2} \mathrm{d} s\\
  			0 & 1
  		\end{pmatrix} \qquad &z^{-1}\in \mathring{\sigma}\\\\
        q(z) \qquad &\text{otherwise}
    \end{cases}
\end{equation}
where by $\mathring{\sigma}$ we indicate the finite portion of the complex plane surrounded by $\sigma$.
Then, it follows that $\wh q(z)$ satisfies the following RHP

\begin{rhp}
\label{RHP:q_hat}
      We seek for a vector $ \wh q(z) = (\wh q_1(z)\quad \wh q_2(z))$  such that 
    \begin{itemize}
        \item $\wh q(z)$ is analytic in $\mathbb{C}\setminus\{(\alpha,\beta)\cup(\beta^{-1},\alpha^{-1}) \}$
        \item $\wh q_+(z)=\wh q_-(z) J_{\wh q}(z)$ for $z\in (\alpha,\beta)\cup(\beta^{-1},\alpha^{-1})$, which are oriented from left to right and $J_{\wh q}(z)$ is defined as
        
      \begin{equation}
  	\label{eq:jump_q_hat}
  	J_{\wh q}(z)=\begin{cases}
  		\begin{pmatrix}
  			1 & 0 \\
  			2\pi iN\mu_1 e^{t\Phi(z,\eta,\tau)}z\rho(z)& 1
  		\end{pmatrix} \quad &z \in (\alpha,\beta) \\\\[5pt]
  		\begin{pmatrix}
  			1 &-2\pi i \mu_1 Ne^{-t\Phi(z,\eta,\tau)}z^{-1}\rho(z^{-1})\\
  			0 & 1
  		\end{pmatrix} \quad &z\in (\beta^{-1},\alpha^{-1}) 
  	\end{cases}\,.
  \end{equation}
        \item $\wh q(z)$ can have at most logarithmic singularities near $\lbrace \alpha^{\pm1}, \beta^{\pm 1} \rbrace$
        \item $\wh q(z^{-1})=\wh q(z)\sigma_1$
        \item $\lim_{z\to 0} \wh q_1(z)\cdot \wh q_2(z) = 1$, $\wh q_1(0)>0$
    \end{itemize}
\end{rhp}
and we notice that solving RHP \ref{RHP:q_hat} is equivalent to solving RHP \ref{RHP:q}.

\begin{remark}
    We notice that a similar RHP was considered in \cite{Gkogkou2025soliton} for the focusing NLS equation. The main difference is that in \cite{Gkogkou2025soliton} the authors considered $(x,t)$ in a compact set of the plane; here we set $n = \eta \ln  N$ (the analogue of the space variable $x$) , $t = \tau \ln  N$, where $N$ is the number of solitons.
\end{remark}

\section{Two model problems}
\label{sec:model_problems}

In order to construct the solution to the RHP \ref{RHP:q}, we need the explicit solution to the following vector model problems.
Here and for the remaining part of the section $0<\ta<\tb<\tc<\td<1$. We start with the $1$-cut, i.e.~genus $1$, case.
\begin{rhp}\label{OuterRHP_1cut}
    We seek for a vector $k(z) = (k_1(z)\quad k_2(z))$  such that 
      \begin{itemize}
        \item $k(z)$ is analytic in $\C\setminus \left(\ta,\ta^{-1}\right)$, with at most a $(z - \tp)^{-1/4}$ blow-up near the branch points $\tp \in \lbrace \ta^{\pm 1}, \tb^{\pm 1} \rbrace$
        \item $k_+(z)=k_-(z) J_{k}(z)$ for $z\in \left(\ta,\ta^{-1}\right) $, which is oriented from left to right  and $J_{k}(z)$ is defined as 
    \begin{equation}
    J_{k}(z) = \begin{cases}
    \begin{pmatrix}
        0 & i \\ i & 0
    \end{pmatrix}
    \quad &z \in \left(\ta,\tb\right)\\\\[5pt]
    \begin{pmatrix}
        e^{2\pi i \Delta} & 0 \\ 
        0 & e^{-2\pi i \Delta}
    \end{pmatrix}
    \quad &z \in \left(\tb,\frac{1}{\tb}\right)
     \\\\[5pt]
    \begin{pmatrix}
        0 & -i \\ -i & 0
    \end{pmatrix}  \qquad &z \in \left(\frac{1}{\tb},\frac{1}{\ta}\right)
\end{cases}
\end{equation}
where $\Delta\in\R$
        \item $k(z^{-1})=\pm k(z)\sigma_1$
		\item $k_1(0) = 1$
\end{itemize}
If $k(z^{-1}) = k(z) \sigma_1$ we will refer to $k$ as the \emph{symmetric} vector solution, while if $k(z^{-1}) = -k(z) \sigma_1$ we call it the \emph{antisymmetric} vector solution. Only in the latter case we will allow for an additional pole at $z = 1$:
\begin{itemize}
    \item the antisymmetric vector solution $k(z)$ (i.e., $k(z^{-1}) = -k(z)\sigma_1$) has at most a simple pole at $z = 1$. 
\end{itemize}
\end{rhp}
The reason for allowing an extra pole in the antisymmetric solution is that otherwise an antisymmetric solution satisfying all the other conditions might not exist, see the paragraph following \eqref{Eq:AntiSymE}. It will be clear from the small-norm analysis that the presece of this extra pole is not a real problem, see Appendix \ref{App:SmallNorm}. This approach (or similar ones) have been used in earlier work which use the theta-functions outer parametrix (see \cite[Sect.~3.4]{GGJM}).  In the present paper, we follow \cite{PT22} to construct both the symmetric and antisymmetric solutions.

The following model problem will be used in the two-cuts region and it is a genus $3$ generalization of Riemann--Hilbert problem \ref{OuterRHP_1cut}.
\begin{rhp}
\label{OuterRHP_2cuts}
    We seek a piecewise analytic row vector $k(z) = \big(k_1(z), \; k_2(z)\big)$  such that 
    \begin{itemize}
        \item $k(z)$ is analytic for $z \in \mathbb{C} \setminus \left\{ \left(\ta,\ta^{-1}\right)\right\}$  with at most a $(z - \tp)^{-1/4}$ blow-up near the branch points $\tp \in \lbrace \ta^{\pm 1}, \tb^{\pm 1}, \tc^{\pm 1}, \td^{\pm 1} \rbrace$
        \item it satisfies the jump condition $k_+(z) = k_-(z) J_k(z)$ for $z \in \left(\ta,\ta^{-1}\right)$. The jump matrix $J_k(z)$ is defined piecewise:
\begin{equation}
    J_k(z) = 
    \begin{cases}
        \begin{pmatrix} 0 & i \\ i & 0 \end{pmatrix}, & z \in (\ta, \tb) \cup (\tc, \td) \\[15pt]
        \begin{pmatrix} e^{2\pi i \Delta_1} & 0 \\ 0 & e^{-2\pi i \Delta_1} \end{pmatrix}, & z \in (\tb, \tc) \cup (\tc^{-1}, \tb^{-1}) \\[15pt]
        \begin{pmatrix} e^{2\pi i \Delta_2} & 0 \\ 0 & e^{-2\pi i \Delta_2} \end{pmatrix}, & z \in (\td, \td^{-1}) \\[15pt]
        \begin{pmatrix} 0 & -i \\ -i & 0 \end{pmatrix}, & z \in (\td^{-1}, \tc^{-1}) \cup (\tb^{-1}, \ta^{-1})
    \end{cases}
\end{equation}
where $\Delta_1, \Delta_2 \in \mathbb{R}$,

\item $  k(z^{-1}) = \pm \, k(z) \sigma_1$,

    \item the antisymmetric vector solution $k(z)$ (i.e., $k(z^{-1}) = -k(z)\sigma_1$) has at most a simple pole at $z = 1$,
\item $k_1(0) = 1$.
\end{itemize}
\end{rhp}

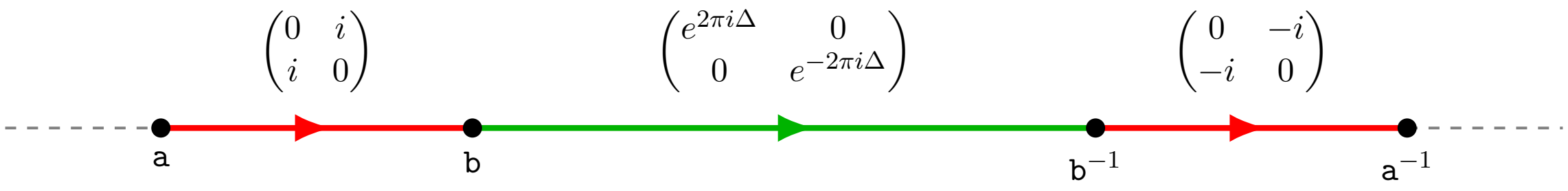
\begin{figure}
    \centering
    \begin{tikzpicture}[scale=1.5, every node/.style={scale=0.85}]
    % Define coordinates based on 0 < a < b < 1
    % 1/a > 1/b > b > a
    \coordinate (a) at (4,0);
    \coordinate (b) at (2,0);
    \coordinate (binv) at (-2,0);
    \coordinate (ainv) at (-4,0);
    
    % Draw reference real axis (dashed)
    \draw[dashed, gray] (-5,0) -- (5,0);
    
    % Styles for colored segments with arrows pointing right
    \tikzset{
        band/.style={red, ultra thick, decoration={markings, mark=at position 0.55 with {\arrow{Latex}}}, postaction={decorate}},
        gap/.style={green!70!black, ultra thick, decoration={markings, mark=at position 0.55 with {\arrow{Latex}}}, postaction={decorate}}
    }
    
    % Draw Segments
    \draw[band] (ainv) -- (binv);
    \draw[gap] (binv) -- (b);
    \draw[band] (b) -- (a);
    
    % Nodes/Labels for the points
    \filldraw (ainv) circle (1.5pt) node[below=3pt] {$\ta$};
    \filldraw (binv) circle (1.5pt) node[below=3pt] {$\tb$};
    \filldraw (b) circle (1.5pt) node[below=3pt] {$\tb^{-1}$};
    \filldraw (a) circle (1.5pt) node[below=3pt] {$\ta^{-1}$};
    
    % Matrices above the segments
    \node[above=6pt] at (-3,0) {$\begin{pmatrix} 0 & i \\ i & 0 \end{pmatrix}$};
    \node[above=6pt] at (0,0)  {$\begin{pmatrix} e^{2\pi i \Delta} & 0 \\ 0 & e^{-2\pi i \Delta} \end{pmatrix}$};
    \node[above=6pt] at (3,0)  {$\begin{pmatrix} 0 & -i \\ -i & 0 \end{pmatrix}$};
    
\end{tikzpicture}
    \caption{Jumps of the model problem \ref{OuterRHP_1cut}}
    \label{fig:RHP_model1}
\end{figure}

\begin{figure}
    \centering
    \begin{tikzpicture}[scale=1.4, every node/.style={scale=0.75}]
    % Coordinates mapping out: 1/a > 1/d > 1/c > 1/b > b > c > d > a
    \coordinate (a) at (5.2,0);
    \coordinate (d) at (4.0,0);
    \coordinate (c) at (2.8,0);
    \coordinate (b) at (1.4,0);
    \coordinate (binv) at (-1.4,0);
    \coordinate (cinv) at (-2.8,0);
    \coordinate (dinv) at (-4.0,0);
    \coordinate (ainv) at (-5.2,0);
    
    % Draw reference real axis (dashed)
    \draw[dashed, gray] (-6,0) -- (6,0);
    
    % Styles for colored segments with arrows pointing right
    \tikzset{
        band/.style={red, ultra thick, decoration={markings, mark=at position 0.6 with {\arrow{Latex}}}, postaction={decorate}},
        gap/.style={green!70!black, ultra thick, decoration={markings, mark=at position 0.6 with {\arrow{Latex}}}, postaction={decorate}}
    }
    
    % Draw Segments
    \draw[band] (ainv) -- (dinv);
    \draw[gap] (dinv) -- (cinv);
    \draw[band] (cinv) -- (binv);
    \draw[gap] (binv) -- (b);
    \draw[band] (b) -- (c);
    \draw[gap] (c) -- (d);
    \draw[band] (d) -- (a);
    
    % Point labels
    \filldraw (ainv) circle (1.2pt) node[below=3pt] {$\ta$};
    \filldraw (dinv) circle (1.2pt) node[below=3pt] {$\tb$};
    \filldraw (cinv) circle (1.2pt) node[below=3pt] {$\tc$};
    \filldraw (binv) circle (1.2pt) node[below=3pt] {$\td$};
    \filldraw (b) circle (1.2pt) node[below=3pt] {$\td^{-1}$};
    \filldraw (c) circle (1.2pt) node[below=3pt] {$\tc^{-1}$};
    \filldraw (d) circle (1.2pt) node[below=3pt] {$\tb^{-1}$};
    \filldraw (a) circle (1.2pt) node[below=3pt] {$\ta^{-1}$};
    
    % Matrices above the segments
    \node[above=5pt] at (-4.6,0) {$\begin{pmatrix} 0 & i \\ i & 0 \end{pmatrix}$};
    \node[above=5pt] at (-3.4,0) {$\begin{pmatrix} e^{2\pi i \Delta_1} & 0 \\ 0 & e^{-2\pi i \Delta_1} \end{pmatrix}$};
    \node[above=5pt] at (-2.1,0) {$\begin{pmatrix} 0 & i \\ i & 0 \end{pmatrix}$};
    \node[above=5pt] at (0,0)     {$\begin{pmatrix} e^{2\pi i \Delta_2} & 0 \\ 0 & e^{-2\pi i \Delta_2} \end{pmatrix}$};
    \node[above=5pt] at (1.9,0)  {$\begin{pmatrix} 0 & -i \\ -i & 0 \end{pmatrix}$};
    \node[above=5pt] at (3.3,0)  {$\begin{pmatrix} e^{2\pi i \Delta_1} & 0 \\ 0 & e^{-2\pi i \Delta_1} \end{pmatrix}$};
    \node[above=5pt] at (4.7,0)  {$\begin{pmatrix} 0 & -i \\ -i & 0 \end{pmatrix}$};
    
\end{tikzpicture}
    \caption{Jumps of the model problem \ref{OuterRHP_2cuts}}
    \label{fig:RHP_model2}
\end{figure}
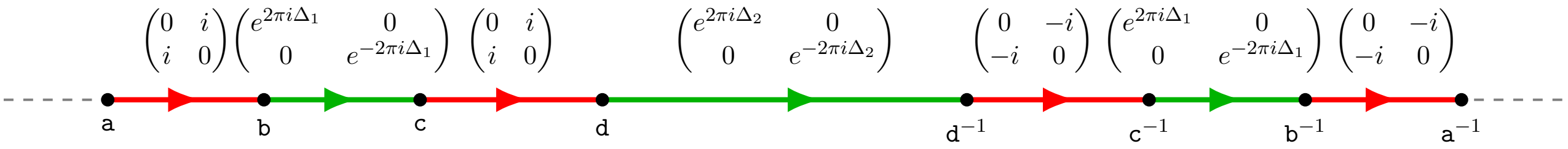

Again, the $\pm$ sign gives rise to the \textit{symmetric} and \textit{antisymmetric} solutions to the previous RHP. Both are required in order to construct a suitable global parametrix, see Sections \ref{sec:complete_one_cut}, \ref{sec:two_cuts}, and \ref{sec:one_cut_modulated_regions}. The previous RHPs are displayed in Figure \ref{fig:RHP_model1}-\ref{fig:RHP_model2}.

Now, we construct an explicit solution to the model RHPs in terms of Riemann and Jacobi Theta functions.
\subsection{Explicit solution to the model problem \ref{OuterRHP_1cut}}

To write an explicit solution, we introduced the two-sheeted Riemann surface $\mathcal R = \mathcal R(\ta,\tb)$ with cuts on $(\ta,\tb) \cup (\tb^{-1}, \ta^{-1})$ and a canonical basis of cycles $\bf a$, $\bf b$. The $\bf a$-cycle starts from the point $\tb^{-1}$, goes to $\tb$ on the upper sheet, and returns on the lower sheet. The $\bf b$-cycle surrounds the cut $[\tb^{-1}, \ta^{-1}]$ on the upper sheet in the counterclockwise direction. The Abel map for that surface is given by
\begin{align}
\label{eq:abel_map_1cut}
    \cA: \mathcal R \to \mathbb C / \Lambda, \qquad p \mapsto \Gamma\int_{\ta^{-1}}^p \frac{1}{w(z)} \frac{dz}{z}\,,
\end{align}
with $\Gamma = \Big(\int_{\bf a} \frac{1}{w(z)} \frac{dz}{z} \Big)^{-1}$ and $w(z) =\sqrt{(z+1/z-\ta-1/\ta)(z+1/z-\tb - 1/\tb)}$ with $w$ being analytic in $\mathbb C \setminus ((\ta, \tb) \cup (\tb^{-1},\ta^{-1}))$ and $w(z) = z + \cO(1)$ as $z\to\infty$. 

It follows that $w_\pm(z) \in  i \mathbb R_\pm$ for $z \in (\tb^{-1},\ta^{-1})$, $w_\pm(z) \in i \mathbb R_\mp$ for $z \in (\ta, \tb)$ and $w(z) < 0$ for $z \in (\tb, \tb^{-1})$. Note that $w(z^{-1}) = w(z)$. In particular, taking $p^{\pm 1}\not\in(\ta,\tb)\cup(\tb^{-1},\ta^{-1})$ to denote a point on the same sheet, we have
\begin{align*}
    \cA(p) = \Gamma\int_{\ta^{-1}}^p \frac{1}{w(z)} \frac{dz}{z} = \Gamma\int_{\ta}^{p^{-1}} \frac{1}{w(y^{-1})}\Big(-\frac{dy}{y}\Big) = -\cA(p^{-1}) + \cA(\ta) =  -A(p^{-1}) + \frac{1}{2}\,,
\end{align*}(cf.~\cite[p.~11]{PT22}). Next we define
\begin{equation}
\label{eq:B_period_g1}
    B = \Gamma\int_{\bf b} \frac{1}{w(z)}\frac{dz}{z} \in i \mathbb R_+, \qquad \Lambda = \mathbb Z \oplus B \mathbb Z.
\end{equation}
Then the following holds:

\begin{proposition}\label{Prop:Abel}
    Define the Abel map $\cA(p)$ as in \eqref{eq:abel_map_1cut}, then the following holds
    \begin{enumerate}
    \item $\cA(\ta^{-1}) = 0$, $\cA(\ta) = \frac{1}{2}$;
    \item $\cA(\tb^{-1}) = B/2$, $\cA(\tb) = 1/2 + B/2$;
    \item $\cA(1^{(1)}) = 1/4 + B/2$, $\cA(1^{(2)}) = 3/4 + B/2$;
    \item $\cA(\infty^{(1)}) \in (0,1/4)$, $\cA(0^{(1)}) \in (1/4, 1/2)$, $\cA(\infty^{(2)}) \in (3/4, 1)$, $\cA(0^{(2)}) \in (1/2, 3/4)$;
    \item $\cA(p^{(1)}) = - \cA(p^{(2)})$;
    \item $\cA(p) = -\cA(p^{-1}) + 1/2$, where $p, p^{-1}$ are on the same sheet
\end{enumerate}
where every equality is understood modulo $\Lambda$ and the superscript denote the first vs.~the second sheet.
\end{proposition}
\begin{proof}
    All the properties are obtained from direct computations; regarding the zeros and infinities, note that $\cA(\infty^{(1)}) + \cA(0^{(1)}) = 1/2$ by the last item. Moreover, as the integrand $w(z) < 0$ for $z < 0$ on the upper sheet, we have that $\cA(0^{(1)}) - \cA(\infty^{(1)}) = \Gamma\int_{-\infty}^0 \frac{1}{w(z)} \frac{dz}{z} > 0$. It follows that $\cA(\infty^{(1)}) \in (0,1/4)$ and $\cA(0^{(1)}) \in (1/4,1/2)$. The claims for $\cA(\infty^{(2)})$, $\cA(0^{(2)})$ follow by symmetry.
\end{proof}

To solve the outer parametrix problem, we first define 
\begin{equation}
\label{eq:n_gamma_def_1_cut}
    n(z) = \gamma(z)k(z), \qquad \text{with} \quad \gamma(z) =  \frac{\left(z + \frac{1}{z} - \ta - \frac{1}{\ta}\right)^{1/4}}{\left(z + \frac{1}{z} - \tb - \frac{1}{\tb}\right)^{1/4}}, 
\end{equation}
where $\gamma(z)$ has branch cuts on $[\ta, \tb] \cup [\tb^{-1}, \ta^{-1}]$ and $\gamma(\infty) = 1$. It follows that $\gamma_{+}(z) = -i\gamma_{-}(z)$ for $z \in (\ta,\tb)$, $\gamma_{+}(z) = i\gamma_{-}(z)$ for $z \in (\tb^{-1},\ta^{-1})$, $\gamma(0) = \gamma(\infty) = 1$ and $1/\gamma(\tb) = 1/\gamma(\tb^{-1}) = 0$ (the last point will allow us to impose pole conditions at $\tb^{\pm 1}$). It follows that $n(z)$ will satisfy exactly the same R-H problem as $k(z)$, except that the jump matrices are replaced with permutation matrices $\sigma_1 = \begin{pmatrix}
    0 & 1
    \\
    1 & 0
\end{pmatrix}$. In particular, we have a \emph{scalar} Riemann--Hilbert problem for the function $F(p)$ with $n(z) = (F(z^{(1)}), F(z^{(2)}))$. As we are in the genus $1$ case the Abel map $\mathcal A$ defines a bi-holomorphism between $\mathcal R$ and the torus $\C / \Lambda$ (see e.g.~\cite[Cor.~A.7]{Teschl1999}), we can formulate the Riemann-Hilbert problem in $\C / \Lambda$. Define $G(\mathcal A(p)) = F(p)$. Then in the new coordinates the jump condition becomes $G_+(x-B/2) = G_-(x+B/2)\exp(-2\pi i \Delta)$ on the cycle $x + B/2 \in \mathbb C / \Lambda$, $x \in \mathbb R$, oriented from left to right (i.e.~the $\bf a$-cycle). Note that $x + B/2 \equiv x - B/2 \ \, \text{mod} \ \Lambda$ making this a genuine jump condition; our choice for using both $x \pm B/2$ will become clear below. The minus in the exponent comes from the fact that the orientation of the $\bf a$-cycle is opposite to the orientation of the jump contour on the first sheet, but agrees on the second.

We can switch perspective again and define the \emph{entire} function $E(A)$ such that $E(A) \equiv G(A)$ for $\text{Im}(A) \in (-B/2, B/2)$. The jump condition for $G$ now becomes a monodromy condition $E(x-\frac{B}{2}) = E(x+\frac{B}{2})\exp(-2\pi i \Delta)$, which by analyticity of $E$ actually holds for \emph{all} $x \in \C$. We now summarize the properties of $E$:
\begin{rhp}\label{TorusRHP}
    We seek for a meromorphic function $E = E(A)$  such that 
      \begin{itemize}
        \item \emph{Analyticity}: $E(A)$ is meromorphic in $\mathbb C$ with its only poles at $\frac{B}{2} + \Lambda \cup \frac{1}{2} + \frac{B}{2} + \Lambda$
        \item \emph{Monodromy}: $E(A + B) = E(A)\exp(2\pi i \Delta)$

        \item \emph{(Anti-)Symmetry}: $E(A) = \pm E(A + \frac{1}{2})$
		\item \emph{Normalization}: $E(\mathcal A(0^{(1)})) = 1$
\end{itemize}
\end{rhp}

This is identical to the 1-gap KdV model Riemann--Hilbert problem studied in \cite{PT22}. The unique symmetric solution is given by (cf.~\cite[Lem.~3.8]{PT22})
\begin{align}\label{eq:FormEs}
    E_s(A) = c(\Delta) \frac{\theta_3(A - \frac{\Delta}{2} + \frac{1}{2}| B)\theta_3(A - \frac{\Delta}{2}| B)}{\theta_3(A + \frac{1}{2}| B)\theta_3(A| B)} = c'(\Delta) \frac{\theta_3(2A - \Delta  + \frac{1}{2}| 2B)}{\theta_3(2A + \frac{1}{2}| 2B)}
\end{align}
where the constants $c$, $c'$ are chosen such that the normalization condition at $\mathcal A(0^{(1)}) \in \left(\frac{1}{4}, \frac{1}{2}\right)$ is satisfied. Note that this is always possible, as the zeros of $\theta(A|B)$ are located at $\frac{1}{2} + \frac{B}{2} +\Lambda$ with $B \in i \mathbb R_+$. However, $\Delta \in \mathbb R$ by assumption (in our application $\Delta = (\ln(N)\zeta(\eta, \tau) + \omega(\eta, \tau))/(2\pi i)$ with $\zeta(\eta, \tau), \omega(\eta, \tau) \in i \mathbb R$). 
\begin{remark}
    The second equality in \eqref{eq:FormEs} can be viewed as a simplification using $\theta_3$-functions on the genus $1$ Riemann surface $\widehat \cR$ with period matrix $2B$. This simplification is only possible for the symmetric solution. In fact there is a natural double cover map $\mathcal R \to \widehat \cR$ given by $A \mapsto 2A$ on the Jacobian, and the symmetry condition $E_s(A) = E_s(A+\frac{1}{2})$ becomes the ordinary real periodicity condition on $\widehat \cR$. A similar simplification will also be possible for the genus $3$ RHP studied next. In this case, the corresponding Riemann surface $\wh \cR$ will be of genus $2$, giving us a modulated genus $2$ wave in the two-cut region, despite the two-cut model RHP problem being related to a genus $3$ Riemann surface. See Section \ref{Sect:genus2Red} for details.  
\end{remark}

It follows that the solution of the outer parametrix problem is given by
\begin{align*}
    k_s(z) &= \frac{1}{\gamma(z)} \Big(E_s(\mathcal A(z)), \ E_s(-\mathcal A(z)) \Big) 
    \\
    &= \frac{c'(\Delta)}{\gamma(z)}\Big(\frac{\theta_3(2\mathcal A(z) - \Delta  + \frac{1}{2}| 2B)}{\theta_3(2\mathcal A(z) + \frac{1}{2}| 2B)}, \frac{\theta_3(-2\mathcal A(z) - \Delta  + \frac{1}{2}| 2B)}{\theta_3(-2\mathcal A(z) + \frac{1}{2}| 2B)} \Big),
\end{align*}
where the path of integration in $\mathcal A(z)$ is in $\mathbb C \setminus (\ta, \ta^{-1})$ and therefore is restricted to be on the upper-sheet (therefore the minus in the second entry to account for the second sheet). 

In order to apply the small norm theory and prove that the solution to the RHP \ref{rhp:m_hat} converges to the one of the model problem, we need to find an \textit{anti-symmetric} solution to the previous RHP, i.e.~$E_a(A) = - E_a(A + \frac{1}{2})$. Again according to \cite[Lem.~3.8]{PT22} this is given by
\begin{align}\label{Eq:AntiSymE}
    E_a(A) = c(\Delta)\frac{\theta_3(A - \frac{\Delta}{2} + \frac{1}{2} - \frac{B}{2}| B)\theta_3(A - \frac{\Delta}{2} + \frac{B}{2}| B)}{\theta_3(A + \frac{1}{2}| B)\theta_3(A| B)}, 
\end{align}
(the normalization constant $c(\Delta)$ might differ from the one above). The corresponding antisymmetric vector solution has the form 
\begin{align*}
    k_a(z) = \frac{1}{\gamma(z)} \Big(E_a(\mathcal A(z)), \ E_a(-\mathcal A(z)) \Big).
\end{align*}
Note that $E_a$ differs from $E_s$ only by the $\pm \frac{B}{2}$ in the numerator (and the normalization constant), which makes it anti-symmetric, see \cite[Lem.~3.8]{PT22}. However, this solution is not normalizable whenever $\mathcal A(0^{(1)}) - \frac{\Delta}{2} = 0 \ \text{or} \ \frac{1}{2} \mod \mathbb Z$ (the zeros of $\theta_3$ are of the form $\frac{1+B}{2} + \Lambda$). As both quantities are real, this \emph{can} occur. In this case, one can instead construct a meromorphic antisymmetric vector solution where we allow the poles to be at $1^{(1)}$, $1^{(2)}$ instead. These are fixed points of the symmetry $z \to 1/z$ and the poles will cancel in the final small-norm analysis. A meromorphic antisymmetric solution is given according to \cite[Sect.~4]{PT22} by
\begin{align}\label{Eq:AntiETilde}
    \widetilde E_a(A) =  \wt c(\Delta) \frac{\theta_3(A +\frac{1}{4}- \frac{\Delta}{2}  -\frac{B}{2}| B)\theta_3(A - \frac{1}{4}-\frac{\Delta}{2} + \frac{B}{2}| B)}{\theta_3(A + \frac{1}{4}| B)\theta_3(A -  \frac{1}{4}| B)} \,,
\end{align}
and as before
\begin{align*}
    \widetilde k_a(z) = \frac{1}{\gamma(z)} \Big(\widetilde E_a(\mathcal A(z)), \ \widetilde E_a(-\mathcal A(z)) \Big). 
\end{align*}
Again, it is not normalizable whenever $\mathcal A(0^{(1)}) - \frac{\Delta}{2} = \pm \frac{1}{4} \ \mod \ \mathbb Z$. Importantly, this case corresponds to different values of $\Delta$, hence one can always construct an anti-symmetric solution with permissible pole structure (in principle one could also take linear combinations of $k_a$, $\widetilde k_a$); the exact form of the meromorphic antisymmetric solution will not matter in the small-norm analysis, see Appendix \ref{App:SmallNorm} for details.

\begin{remark}\label{Rmk:Zeros}
    We see that the zeros of $k_s$, meaning the zeros of either $k_{s,1}$ or $k_{s,2}$, are confined to lie in the gap $(\tb, \tb^{-1})$. On the upper sheet (i.e.~in the first entry $k_{s,1}$) the zero moves to the left with increasing $\Delta$, while on the lower sheet (i.e.~in the second entry $k_{s,2}$) the zero moves to the right. They perform a continuous flow on the $\bf a$-cycle which becomes linear on the Jacobian, (i.e.~the torus $\C / \Lambda)$. Similarly, either one of the antisymmetric solutions has two zeros that evolve linearly on the Jacobian, however the antisymmetry requirement moves the zero by $B/2$ in the vertical direction. Back on the two-sheeted Riemann surface, this means that the zeros evolve on $[\ta^{-1}, \infty] \cup [-\infty, \ta]$ on either sheet, which forms a cycle homotopic to the $\bf a$-cycle. With our choice of normalization, whenever a zero hits the point $0^{(1)}$, the solution will not be normalizable. 
\end{remark}
The topology of the zeros also plays an important role in the proof of linear independence that we will consider next.

For the purposes of the small-norm analysis, it remains to check that there is \emph{at least one} antisymmetric solution $\ell_a(z)$ (can be $k_a$, or $\widetilde k_a$), which is linearly independent of $k_s(z)$ for all $z$. Define the determinant $d(z) = \det\begin{pmatrix}
    k_s(z)
    \\
    \ell_a(z)
\end{pmatrix}$. As all jump matrices have determinant $1$, $d(z)$ defines a meromorphic in $\C$ with only a possible \emph{simple} pole at $z = 1$. We use the symmetries of the two vector solutions to write
\begin{align*}
    d(z) = \det \begin{pmatrix}
        k_{s,1}(z) & k_{s,1}(z^{-1})
        \\
        \ell_{a,1}(z) & -\ell_{a,1}(z^{-1})
    \end{pmatrix} = -\big(k_{s,1}(z)\ell_{a,1}(z^{-1}) + \ell_{a,1}(z)k_{s,1}(z^{-1})\big).
\end{align*}
This shows that $d(z) = d(z^{-1})$, which, by performing a Laurent expansion around $z=1$, excludes the possibility of the simple pole at $z = 1$. It follows that $d(z) \equiv const.$, hence to show linear independence it is enough to show $d(z) \not = 0$ for one $z \in \mathbb C$. 

\textbf{Case 1}: Assume $k_{s, 1}(1) = k_{s,2}(1) \not = 0$. In that case there will be a $z_0 \in [\tb, \tb^{-1}] \setminus \lbrace 1 \rbrace$ with $k_{s,2}(z_0) = 0 = k_{s,1}(z_0^{-1})$ (it will be unique unless $z_0 = \tb^{\pm 1}$). Let us assume that $z_0 \in (\tb, \tb^{-1}) \setminus \lbrace 1 \rbrace$. It follows that 
\begin{align*}
    d(z_0) = k_{s,1}(z_0)\ell_{a,2}(z_0).
\end{align*}
We have that $k_{s,1}(z_0) \not = 0$ because $z_0^{-1} (\not = z_0)$ will be the unique zero of $k_{s,1}$ on $(\tb, \tb^{-1})$, and $\ell_{a,2}(z_0) \not = 0$ because both antisymmetric solutions have zeros confined to $[\ta^{-1}, \infty] \cup [-\infty, \ta]$ (on either sheet), see Remark \ref{Rmk:Zeros} above. 

In the special case $z_0 = \tb^{\pm 1}$ we have that $k_{s,1}(z) =\frac{1}{\gamma(z)} = k_{s,2}(z)$ (as $\frac{\Delta}{2} \equiv 0, \frac{1}{2}  \mod \Lambda$). It follows that $d(z) = -\frac{1}{\gamma(1)}(\ell_{a,1}(z^{-1})+\ell_{a,1}(z))$. If we set $\ell_a = k_a$ we immediately see that $d(1) \not = 0$ again due to the zero structure of $k_a$ (this argument also works for $z_0 \not = \tb^{\pm1}, 1$). Meanwhile, if we choose $\ell_a = \widetilde k_a$ we see that $d(\tb^{\pm 1}) = 0$ as all the entries of the matrix will vanish at that point.

\textbf{Case 2}: Now consider the case $k_{s,1}(1) = k_{s,2}(1) = 0$ which happens exactly for $\frac{\Delta}{2} = \pm \frac{1}{4} \ \mod \ \Z$. In this case $d(1) = 0$ if $\ell_a = k_a$, as $k_a$ is bounded at $z =1$. Meanwhile, setting $\ell_a = \widetilde k_a$, we obtain
\begin{align*}
    d(1) = 2 k_{s,1}'(1)\res \widetilde k_a(z)|_{z=1} \not = 0,
\end{align*}
which is nonzero as $k_{s,j}(1)$ can have only a simple zero at any given point (as the zeros of $\theta_3$ are simple), and $\res \widetilde k_a(z)|_{z=1} \not = 0$ follows from the nonvanishing of the numerator in \eqref{Eq:AntiETilde}, which can be shown as in the previous case.

We summarize our findings in the following proposition:

\begin{proposition}\label{prop:model_1cut_solutions}

Let $\cA(z)$ be the Abel map defined in \eqref{eq:abel_map_1cut}, let $B \in i\R_+$ be the $\mathbf{b}$-cycle period, and $\Delta\in \R$. Define the function $\gamma(z)$ as in \eqref{eq:n_gamma_def_1_cut}.

Then, the symmetric solution ($k_s(z^{-1})= + k_s(z)\sigma_1$) to the model RHP \ref{OuterRHP_1cut} is given by
\begin{equation}
    k_s(z) = \gamma(z)^{-1} \Big(E_s(\cA(z)), \ E_s(-\cA(z)) \Big)\ ,
\end{equation} 
where the path of integration for $\cA(z)$ is restricted to the upper sheet in $\C \setminus (\ta, \ta^{-1})$, and $E_s(A)$ is the elliptic function
\begin{equation}
    E_s(A) = c'(\Delta) \frac{\theta_3(2A - \Delta  + \frac{1}{2}| 2B)}{\theta_3(2A + \frac{1}{2}| 2B)}\ ,
\end{equation} 
here $\theta_3(z\vert B)$ is the Jacobi theta-$3$ function \cite[Ch.~20]{dlmf},
with the constant $c'(\Delta)$ chosen to satisfy the normalization $E_s(\cA(0^{(1)})) = 1$.

Furthermore, there exists an antisymmetric solution $\ell_a(z)$ to the RHP \ref{OuterRHP_1cut} satisfying $\ell_a(z^{-1}) = -\ell_a(z)\sigma_1$, which is appropriately normalized and linearly independent from $k_s(z)$.
Depending on the value of $\Delta$, $\ell_a(z)$ takes one of the following forms:
\begin{enumerate}
    \item If $\Delta \neq 2\cA(0^{(1)}) \mod \Lambda$ and $\Delta \neq \frac{1}{2} \mod \Z$, we can choose $\ell_a(z) = k_a(z)$, where
    \begin{equation}
        k_a(z) = \gamma(z)^{-1} \Big(E_a(\cA(z)), \ E_a(-\cA(z)) \Big)\ ,
    \end{equation} 
    and 
    \begin{equation}
        E_a(A) = c(\Delta) \frac{\theta_3(A - \frac{\Delta}{2} + \frac{1}{2} - \frac{B}{2}| B)\theta_3(A - \frac{\Delta}{2} + \frac{B}{2}| B)}{\theta_3(A + \frac{1}{2}| B)\theta_3(A| B)}\ .
    \end{equation} 

    \item If $\Delta \neq 2\cA(0^{(1)}) + \frac{1}{2} \mod \Lambda$ and $\Delta \not = 0 \mod \Z$, then we can choose $\ell_a(z) = \widetilde k_a(z)$, where
    \begin{equation}
        \widetilde k_a(z) = \gamma(z)^{-1} \Big(\widetilde E_a(\cA(z)), \ \widetilde E_a(-\cA(z)) \Big)\ ,
    \end{equation} 
    and
    \begin{equation}
        \widetilde E_a(A) = \widetilde c(\Delta) \frac{\theta_3(A +\frac{1}{4}- \frac{\Delta}{2}  -\frac{B}{2}| B)\theta_3(A - \frac{1}{4}-\frac{\Delta}{2} + \frac{B}{2}| B)}{\theta_3(A + \frac{1}{4}| B)\theta_3(A -  \frac{1}{4}| B)}\ .
    \end{equation} 
\end{enumerate}
\end{proposition}

\begin{remark}\label{rmk:UnifBoundPara}
    Notice that as $\mathcal A(0^{(1)}) \in (\frac{1}{4}, \frac{1}{2})$, the conditions excluding case 1.~and case 2.~are disjoint and in fact a finite distance in $\Delta$ of each other. Hence, we can choose $\ell_a(z)$ such that 
    \begin{align*}
      c < \left|\det\begin{pmatrix}
        k_s(z)
        \\
        \ell_a(z)
    \end{pmatrix}\right| = \left|\frac{1}{2}\det \begin{pmatrix}
        k_s(z) + \ell_a(z)
        \\
        k_s(z) - \ell_a(z)
    \end{pmatrix} \right| < C  \,,
    \end{align*}
    (cf.~\eqref{eq:MAnsatz}) for some fixed constants $c, C>0$, i.e.~the matrix solution and its inverse are uniformly bounded away from the branch points and the poles at $z = 1$ (the same is true for the two-cut outer parametrix constructed in the following subsection). This fact is used implicitly in the small norm analysis in each region.
\end{remark}
\subsection{Explicit solution to the model problem \ref{OuterRHP_2cuts}}
\label{Sect:ExplSol2}

In this section, we construct an explicit solution to the model problem \ref{OuterRHP_2cuts}. We follow the same ideas as in the previous section, but in this case we find the solution in terms of \emph{Riemann} Theta functions \cite[21.2]{dlmf}.

As in the previous section, we define a new function $n(z)$ 

\begin{equation}
    n(z) = \gamma(z)k(z)\,,\qquad  \gamma(z) = \frac{\left(z+z^{-1} - \ta - \ta^{-1}\right)^{1/4}\left(z+z^{-1} - \tc - \tc^{-1}\right)^{1/4}}{\left(z+z^{-1} - \tb - \tb^{-1}\right)^{1/4}\left(z+z^{-1} - \td - \td^{-1}\right)^{1/4}}\,.
\end{equation}

We consider the canonical branch-cuts, therefore $\gamma(z)$ is analytic in $\mathbb{C} \setminus \Sigma$, where $\Sigma$ is the union of the four branch-cuts $\Sigma = (\ta,\tb)\cup(\tc,\td)\cup (\tb^{-1},\ta^{-1})\cup(\td^{-1},\tc^{-1})$. By grouping inverse pairs, it immediately follows that $\gamma(z^{-1}) = \gamma(z)$. 

Because the argument of the fourth root changes sign across the cuts, the boundary values satisfy $\gamma_+(z) = -i\gamma_-(z)$ for $z \in (\ta,\tb) \cup (\tc, \td)$ and $\gamma_+(z) = i\gamma_-(z)$ for $z \in (\td^{-1},\tc^{-1})\cup(\tb^{-1},\ta^{-1})$. In view of these properties, we conclude that $n(z)$ solves the same RHP as $k(z)$ except that the off-diagonal jump matrices become $\sigma_1$.

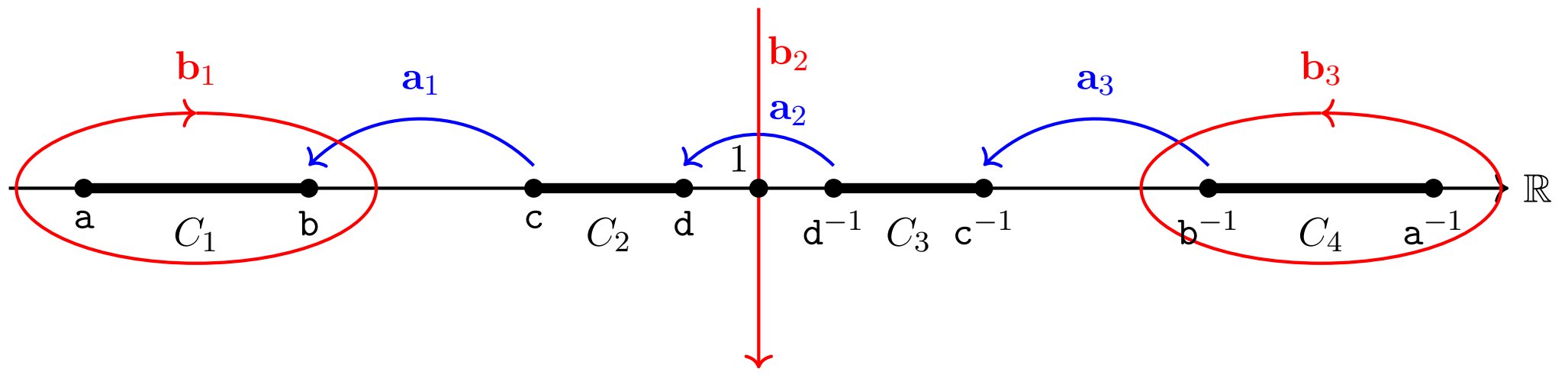
\begin{figure}[htbp]
    \centering
    \begin{tikzpicture}[scale=1.3, every node/.style={scale=0.9}]
        % Real axis
        \draw[thick, ->] (-5, 0) -- (5, 0) node[right] {$\mathbb{R}$};
        
        % Cuts
        \draw[line width=2.5pt, black] (-4.5, 0) -- (-3, 0) node[midway, below=3pt, black] {$C_1$};
        \draw[line width=2.5pt, black] (-1.5, 0) -- (-0.5, 0) node[midway, below=3pt, black] {$C_2$};
        \draw[line width=2.5pt, black] (0.5, 0) -- (1.5, 0) node[midway, below=3pt, black] {$C_3$};
        \draw[line width=2.5pt, black] (3, 0) -- (4.5, 0) node[midway, below=3pt, black] {$C_4$};

        % A-cycles
        \draw[->, thick, blue] (-1.5, 0.15) to[out=135, in=45] (-3, 0.15); \node[blue] at (-2.25, 0.7) {$\mathbf{a}_1$};
        \draw[->, thick, blue] (0.5, 0.15) to[out=135, in=45] (-0.5, 0.15); \node[blue] at (0.2, 0.5) {$\mathbf{a}_2$};
        \draw[->, thick, blue] (3, 0.15) to[out=135, in=45] (1.5, 0.15); \node[blue] at (2.25, 0.7) {$\mathbf{a}_3$};

        % B-cycles (Fully visible, counterclockwise)
        \draw[<-, thick, red] (-3.75, 0.5) arc (90:450:1.2 and 0.5); \node[red] at (-3.75, 0.8) {$\mathbf{b}_1$};
        \draw[->, thick, red] (3.75, 0.5) arc (90:450:1.2 and 0.5); \node[red] at (3.75, 0.8) {$\mathbf{b}_3$};
        \draw[->, thick, red] (0, 1.2) -- (0, -1.2); \node[red] at (0.2, 0.9) {$\mathbf{b}_2$};
             % Points
        \foreach \x/\l in {-4.5/\ta, -3/\tb, -1.5/\tc, -0.5/\td, 0.5/\td^{-1}, 1.5/\tc^{-1}, 3/\tb^{-1}, 4.5/\ta^{-1}} {
            \filldraw (\x,0) circle (1.5pt) node[below=2pt] {$\l$};
        }
        \filldraw (0,0) circle (1.5pt) node[above=0.5pt] {$1 \quad$};
    \end{tikzpicture}
    \caption{The branch cuts (black) and the homology basis.}
    \label{fig:cycles_2cuts}
\end{figure}

Following the same ideas as before, we consider the hyperelliptic Riemann surface $\mathcal{R}(\ta,\tb,\tc,\td)\equiv\cR$ of genus $g=3$ with cuts on $\Sigma$ and a canonical homology base of cycles $\lbrace \mathbf{a}_j,\mathbf{b}_j\rbrace_{j=1}^3$, which are depicted in Fig.~\ref{fig:cycles_2cuts} (the $\bf a$-cycles return to the right branch point on the lower sheet).

Introduce now $\vec{\Delta}$ and the permutation matrix $\Pi$: 
\begin{align*}
    \vec{\Delta} = \begin{pmatrix}
        \Delta_1
        \\
        \Delta_2
        \\
        \Delta_1
    \end{pmatrix}, \qquad \Pi = \begin{pmatrix}
        0 & 0 & 1
        \\
        0 & 1 & 0
        \\
        1 & 0  & 0 
    \end{pmatrix}.
\end{align*}
We clearly have $\Pi \vec{\Delta} = \vec{\Delta}$. Consider next the involution $\iota \colon p^{(j)} \mapsto (p^{-1})^{(j+1)}$, where the superscript is understood $\mod$ $2\Z$ (in particular $\iota$ maps the first sheet to the second sheet and vice versa). It is easy to see that (here $\bf a_j, \bf b_j$ are understood as homotopy classes of cycles rather than fixed contours)
\begin{align*}
    \iota_* \bf{a}_1 &= \bf{a}_3, \qquad  \iota_* \bf{a}_2 = \bf{a}_2,
    \\
    \iota_* \bf{b}_1 &= \bf{b}_3, \qquad  \iota_* \bf{b}_2 = \bf{b}_2.
\end{align*}
Consequently for any holomorphic differential $\omega$, we have
\begin{align*}
    \int_{\bf{a}_1} \omega = \int_{\bf{a}_3} \iota^* \omega, \qquad  \int_{\bf{a}_2} \omega = \int_{\bf{a}_2} \iota^* \omega
\end{align*}
where $\iota^* \omega$ denotes the pullback of $\omega$. In particular for the canonical basis of holomorphic differential $\lbrace \omega_j \rbrace_{j =1}^3$ satisfying $\int_{\bf{a}_j} \omega_k = \delta_{jk}$ we conclude that $\omega_3 = \iota^* \omega_1$ and $\omega_2 =  \iota^* \omega_2$, or in vector notation $\iota^*\vec{\omega} = \Pi \vec{\omega}$. In particular we obtain 
\begin{align*}
    B_{11} = \int_{\bf{b}_1} \omega_1 = \int_{\bf{b}_1} \iota^* \omega_3 = \int_{\bf{b}_3} \omega_3 = B_{33}, \quad B_{12} = \int_{\bf{b}_1} \omega_2 = \int_{\bf{b}_1} \iota^* \omega_2 = \int_{\bf{b}_3} \omega_2 = B_{32},
\end{align*}
where $B = (B_{jk})_{j,k=1}^3$ is the period matrix defined as $B_{jk}=\int_{\bf b_j}\omega_k$. As we additionally have the symmetry $B_{jk} = B_{kj}$ (which always holds, see e.g.~\cite[Eq.~(A.13)]{Teschl1999}) it follows that $B$ has the highly symmetric form  
\begin{align}\label{Eq:BSym}
    B = \begin{pmatrix}
        e & g & h
        \\
        g & f & g
        \\
        h & g & e
    \end{pmatrix} \qquad \Longleftrightarrow \qquad B = B^\mathsf{T} \ \text{ and } \ B\Pi = \Pi B.
\end{align}
\begin{remark}
It is known that for any holomorphic differential $\omega$ on a hyperelliptic Riemann surface we have $\pi^* \omega = - \omega$, where $\pi \colon \mathcal R \to \mathcal R$ is the sheet exchange map. In particular, it follows that $\underbrace{\int_{\tb^{-1}}^{\tc^{-1}} \omega}_{\text{int.~on upper sheet}} = \frac{1}{2} \int_{\bf{a}_3} \omega$, and similarly for the other $\bf{a}$ and $\bf{b}$ cycles.
\end{remark}
Having the basis of holomorphic differentials we define Abel map
\begin{align*}
    \mathcal A \colon \mathcal R \to \C^3 / \underbrace{\Z^3 \oplus B \Z^3}_{= \Lambda}, \quad p \mapsto \int_{\ta^{-1}}^p \vec{\omega}, \qquad \vec{\omega} = \begin{pmatrix}
        \omega_1
        \\
        \omega_2
        \\
        \omega_3
    \end{pmatrix}.
\end{align*}
The Abel map $\mathcal A$ no longer defines a biholomorphism (though it is still injective); this fact will however lead to only minor modification compared to the genus $1$ case. 

For later usage, we notice that
\begin{equation}
\label{eq:inversion_prop}
    \mathcal A(\iota(p)) = \int_{\ta^{-1}}^{\iota(p)} \vec{\omega} = \int_{\ta}^{p} \iota^* \vec{\omega} = \Pi(\mathcal A(p) - \mathcal A(\ta)) = \Pi \mathcal A(p) - \frac{1}{2} \begin{pmatrix}
        1
        \\
        1
        \\
        1
    \end{pmatrix}.
\end{equation}
We will use the abbreviation $\vec{1} = (1, \, 1, \, 1)^\mathsf{T}$. One can verify that $\mathcal A(\ta) = \frac{1}{2}\vec{1}$.

Next we introduce the vector of Riemann constants $K \in \C^g$ given by (see e.g.~\cite[Ch.~VI.2]{Farkas1992})
\begin{align*}
    K_j = \frac{1 + B_{jj}}{2} + \sum_{\ell \not = j} \int_{\bf a_\ell}\mathcal A(p)_j\omega_\ell(p).
\end{align*}
Let us now introduce the genus $g$ Riemann Theta function defined by
\begin{align}\label{Eq:ThetaDef}
\Theta(A \, | \, B) = \sum_{\vec{n} \in \Z^g} \exp\Big\lbrace 2 \pi i\Big( \frac{1}{2}\langle \vec{n}, B\vec{n}\rangle + \langle \vec{n}, A \rangle  \Big) \Big\rbrace, \qquad A \in \C^g,
\end{align}
and satisfying
\begin{align}\label{Eq:ThetaPer}
    \Theta(A + \vec{m} + B \vec{n}\, | \, B) = \exp\left( - 2 \pi i\left( \frac{1}{2}\langle \vec{n}, B\vec{n}\rangle + \langle \vec{n}, A \rangle  \right) \right) \Theta(A | B).
\end{align} 
From here we will set $g = 3$ and build \emph{one} symmetric solution and \emph{two} antisymmetric solutions.

\subsubsection{The symmetric genus 3 vector solution}

For the symmetric solution, we make the following ansatz
\begin{align}\label{Eq:ks}
    k_s(z) &= \frac{1}{\gamma(z)} \Big(E_s(\mathcal A(z)), \ E_s(-\mathcal A(z)) \Big), \qquad \text{integration on the upper sheet},
\end{align}
where 
\begin{align}\label{Eq:AnsatzEs}
     E_s(\mathcal{A}(z)) = c(\vec{\Delta}) \frac{\Theta(\mathcal{A}(z)  - \mathcal{A}(D_0) + K\mid B)\Theta(\mathcal{A}(z) - \Pi \mathcal{A}(D_0) + \frac{1}{2}\vec{1} + K\mid B)}{\Theta(\mathcal{A}(z)  - \mathcal{A}(D_\infty) + K\mid B)\Theta(\mathcal{A}(z) - \Pi \mathcal{A}(D_\infty) + \frac{1}{2}\vec{1} + K\mid B)},
\end{align}
and 
\begin{align*}
    D_0 = z_1 + z_2 + q, \qquad D_\infty = \tb  + \td + q.
\end{align*}
Here $z_1 = z_1(\vec{\Delta}) \in \mathcal R$ is a point in the gap $[\tb, \tc]$ (on either sheet), $z_2 = z_2(\vec{\Delta}) \in \mathcal R$ is a point in the gap $[\td, \td^{-1}]$ (on the \emph{upper} sheet) and $q$ is a fixed point in the gap $[\tc^{-1}, \tb^{-1}]$ (again on either sheet). The placement of $q$ guarantees that the divisors $D_0$, $D_\infty$ are nonspecial, see Remark \ref{Rmk:NonSp} below. Observe that for any divisor $D$ from \eqref{eq:inversion_prop} we have
\begin{align}\label{eq:SymDiv}
    \Pi \mathcal A(D) - \frac{1}{2}\vec{1} \equiv \mathcal A(\iota(D)).
\end{align}
We will now use the following basic facts about positive nonspecial divisors $D = q_1 + \dots + q_g$ of degree $g$ on Riemann surfaces of genus $g$ (see e.g.~\cite[Ch.~VI.2]{Farkas1992}):
\begin{align}\label{Eq:AbelInv}
    \Theta(\mathcal A(p) - \mathcal A(D) + K \, | \, B) = 0 \qquad \Longleftrightarrow \qquad p = q_j \text{ for some } j \in \lbrace 1, \dots, g \rbrace.
\end{align}
Specializing to our setting it follows from \eqref{eq:SymDiv}, \eqref{Eq:AbelInv} that $E_s(\mathcal A(z))$ will be a multivalued meromorphic function on $\mathcal R$ with a well-defined zero/pole divisor (due to $\exp\lbrace ... \rbrace \not = 0$ in \eqref{Eq:ThetaPer}) given by
\begin{align*}
\text{div}\big(E_s(\mathcal{A}(z))\big) &= D_0 + \iota(D_0) - D_\infty - \iota(D_\infty)
\\
&= z_1 + z_2 + \iota(z_1) + \iota(z_2) - \tb - \td - \tb^{-1} - \td^{-1},
\end{align*}
 as $q$, $\iota(q)$ cancel. It remains to choose $z_1, z_2$ in the gaps to satisfy the required (diagonal) jump conditions, which are equivalent to
 \begin{align}\label{Eq:EQuasiPer}
     E_s(A + \vec{B}_j) = E_s(A)\exp(2\pi i \Delta_j), \qquad j = 1,2,3, \ \vec{B}_j = \begin{pmatrix}
         B_{1j}
         \\
         B_{2j}
         \\
         B_{3j}
     \end{pmatrix}.
 \end{align}
 From \eqref{Eq:ThetaPer} it follows that the above condition is equivalent to
 \begin{align}\label{Eq:AbelCond}
     \mathcal A(D_0) + \Pi\mathcal A(D_0) - \mathcal A(D_\infty) - \Pi \mathcal A(D_\infty) = (I + \Pi)\big(\mathcal A(\widehat D_0) - \mathcal A(\widehat D_\infty)\big) \equiv \vec{\Delta} \ \mod \Z^3.
 \end{align}
 where $\widehat D_0 = z_1 + z_2$ and $\widehat D_\infty = \tb + \td$. Note the $\mod \, \Z^3$ instead of the $\mod \Lambda$ that one might have expected. The reason is that Riemann Theta function are only $\Z^g$-periodic, hence the choice of representative matters.

 The equation \eqref{Eq:AbelCond} is essentially an Abel condition and can be rephrased as follows. Find $z_1$ in the first gap and $z_2$ in the second gap on the upper sheet such that
 \begin{equation}
 \label{eq:z_1_z_2}
 \begin{split}
   \int_\tb^{z_1} (\omega_1 + \omega_3) + \int_\td^{z_2} (\omega_1 + \omega_3) &\equiv  \Delta_1 \ \mod \, \Z,
     \\
     2\left(\int_\tb^{z_1} \omega_2  + \int_\td^{z_2} \omega_2 \right) &\equiv\Delta_2 \ \mod \, \Z.
\end{split}
 \end{equation}
 We restrict $z_2$ to lie w.l.o.g. on the upper sheet, as we could always replace it with $\iota(z_2)$ which lies still in the second gap, but on the opposite sheet. The exception are the two branch points $\td$, $\td^{-1} = \iota(\td)$, which we identify with each other. With this choice both $z_1$, $z_2$ will lie on topological circles, and the map $\mu \colon (z_1, z_2) \mapsto (\Delta_1, \Delta_2)$ defined by \eqref{eq:z_1_z_2} can be viewed as a continuous map from $S^1 \times S^1$ to itself. It follows from $\int_{\bf a_j} \omega_k = \delta_{jk}$ that if we fix $z_2$ and let $z_1$ go around the first gap once, the image of $\mu$ transforms as $(\Delta_1, \Delta_2) \mapsto (\Delta_1 + 1, \Delta_2)$. Similarly, if we fix $z_1$ and let $z_2$ go from $\td \to \td^{-1}$, the image of $\mu$ transform as $(\Delta_1, \Delta_2) \mapsto (\Delta_1, \Delta_2+1)$. It follows from standard results of algebraic topology that $\mu$ must be surjective, see e.g.~\cite{Hatcher2002}. As we will show below, the map $\mu$ is in fact bijective. 
\begin{lemma}\label{lem:mu}
    The map $\mu \colon (z_1, z_2) \mapsto (\Delta_1, \Delta_2)$ defined above is bijective.
\end{lemma}

\begin{proof}
 We have already shown surjectivity in the previous paragraph. To show injectivity of the map $\mu$, we first have to show that any $k_s(z)$ constructed that way will already be symmetric (rather than antisymmetric). Note that $\widehat k_s(z) = k_s(z^{-1})\sigma_1$ will satisfy the same Riemann--Hilbert problem. Let us define $\widehat M(z) = \begin{pmatrix}
     k_s(z)
     \\
     \widehat k_s(z)
 \end{pmatrix}$. As all the jump matrices have determinant $1$, and the singularities are only of fourth-root type, it follows that the singularities of $\det \widehat M(z)$ are removable, and thus $\det \widehat M(z) \equiv c$. Evaluating at say $z_1$ which we assume w.l.o.g to lie on the first sheet (or be one of the branch points) we obtain
 \begin{align*}
     c=\det \widehat M(z_1) = \det\begin{pmatrix}
         0 & k_{s,2}(z_1)
         \\
         0 & k_{s,1}(z_1^{-1})
     \end{pmatrix} = 0\,,
 \end{align*}
as $k_{s,2}(z_1^{-1}) = 0$ by symmetry. As the zero structure of $k_s$ and $\widehat k_s$ is identical, it already follows that $\widehat k_s(z) = \widehat c k_s(z) = k_s(z^{-1})\sigma_1$ for some constant $\widehat c \not = 0$ (rather than some meromorphic $\widehat c(z)$). Moreover, $k_s(z) = \widehat c^{-1}k_s(z^{-1})\sigma_1 = \widehat c^{-2}k_s(z)$, and so $\widehat c = \pm 1$. To see that $\widehat c = 1$, we observe that $k_{s,1}(-1) = \widehat c k_{s,2}(-1)$. It is enough to show that $\gamma(-1)k_{s,1}(-1)$ and $\gamma(-1)k_{s,2}(-1)$ have the same sign. This is easy to see as $-1$ on either sheet lies on the unbounded cycle formed by the two copies of $[\ta^{-1}, +\infty]\cup [-\infty,\ta]$ on either sheet. The claim follows from the fact that $E_s(\mathcal A(p))$ has no zeros on this cycle by construction (i.e., it has a fixed sign), and $\gamma(-1)k_{s,1}(-1) = E_s(\mathcal A((-1)^{(1)}))$, while   $\gamma(-1)k_{s,2}(-1) = E_s(-\mathcal A((-1)^{(1)})) = E_s(\mathcal A((-1)^{(2)}))$ where $(-1)^{(j)}$ stands for the point $-1$ on the $j^{th}$ sheet. 

 To continue with the proof of injectivity, assume we have two choices $z_1, z_2$ and $\widetilde z_1$, $\widetilde z_2$ with $\mu(z_1, z_2) = \mu(\widetilde z_1, \widetilde z_2)$. Following \eqref{Eq:ks} we obtain two symmetric vector solutions $k_s, \widetilde k_s$ satisfying the same jump conditions given by $(\Delta_1, \Delta_2)$. We define the matrix $M(z) = \begin{pmatrix}
     k_s(z)
     \\
     \widetilde k_s(z)
 \end{pmatrix}$. Just as before it follows that $\det M(z) \equiv c$. Moreover, as both solutions are symmetric by the previous argument we have $M(z^{-1}) = M(z) \sigma_1$, and thus $\det M(z^{-1}) = \det M(z) \det \sigma_1 = - \det M(z)$. We conclude that $\det M(z) \equiv 0$. 

 Next, observe that if $z_1 = \widetilde z_1$, then necessarily $z_2 = \widetilde z_2$. The reason is that a non-trivial holomorphic differential $\omega$ has exactly $2g - 2 = 4$ zeros. As $\int_{\bf a_1} \omega_2 = \int_{\bf a_3} \omega_2 = 0$ and $\omega_2$ is real in all the gaps, it follows that it must have at least one zero in the first and third gap (viewed as cycles on the Riemann surface). However, due to its periodicity the number of zeros must be even, hence we have located all $4$ zeros to the first and third gap. In particular on the second gap no zeros are present meaning that $z_2 \mapsto \int_{\td}^{z_2} \omega_2$ for $z_2 \in[\td, \td^{-1}]$ (on the first sheet) is monotonic, implying that necessarily $z_2 = \widetilde z_2$.

 With this in mind, we can assume $z_1 \not = \widetilde z_1$ and derive a contradiction as follows. Assume for simplicity that $z_1$ is on the first sheet. In that case we have the contradiction
 \begin{align*}
     0 = \det M(z_1) = \begin{pmatrix}
         0 & k_{s,2}(z_1)
         \\
         \widetilde k_{s,1}(z_1) & \widetilde k_{s,2}(z_1)
     \end{pmatrix} = - \widetilde k_{s,1}(z_1) k_{s,2}(z_1) \not = 0
 \end{align*}
 To understand why the above product is nonzero, we should remind the reader that $z_1$ (resp.~$\widetilde z_1$) is by construction the \emph{only} zero of $k_s$ (resp.~$\widetilde k_s$) on the first gap (manifesting as the vanishing of $k_{s,1}$ if $z_1$ is on the first sheet etc.). With this in mind $\widetilde k_{s,1}(z_1) \not = 0$ as $\widetilde z_1 \not = z_1$, and $k_{s,2}(z_1)\not = 0$ because $z_1$ lies by assumption on the first sheet. 

 The final case we have to consider is $z_1 \in \lbrace \tb, \tc \rbrace$ (we assume $z_1 = \tb$, the other case is completely analogous). Then we have for $z \to \tb$ and the same choice of fourth-root in all four entries:
 \begin{align*}
     0 = \det M(z) = \begin{pmatrix}
         c(z-\tb)^{1/4} + l.o.t & c(z -\tb)^{1/4} + l.o.t
         \\
         \widetilde c(z-\tb)^{-1/4} + l.o.t & -\widetilde c(z -\tb)^{-1/4} + l.o.t
     \end{pmatrix} = -2 c \widetilde c.
 \end{align*}
 Here $c \not = 0$ as otherwise $E_s$ would have a \emph{second} order zero there. Hence, necessarily $\widetilde c = 0$, but this is just equivalent to $\widetilde z_1 = \tb$. This proves that $k_s = \widetilde k_s$ and the map $\mu$ is therefore a bijection.
 \end{proof}
 
\begin{remark}\label{Rmk:NonSp}
    By definition, a positive divisor $D = q_1 + \dots + q_n$ on a compact Riemann surface is special if and only if there exists a holomorphic differential $\omega$ such that $\omega(q_j) = 0$ for $j = 1, \dots, n$ \cite[App.~A.3]{Teschl1999}. In the case of a hyperelliptic Riemann surface of genus $g$, it is additionally known that holomorphic differentials have the form $\omega = \frac{p_{g-1}(z)}{R_{2g+2}(z)} dz$ in some appropriate local coordinates, with $R^2_{2g+2}(z)$ being a polynomial of degree $2g+2$, and $p_{g-1}(z)$ a polynomial of degree $\leq g-1$. Here, the zero divisor of $\omega$ is precisely the zero divisor of $p_{g-1}$ on either sheet, i.e.~if $p_{g-1}(z_0) = 0$ then $z_0$ on both the first and second sheet will be zeros of $\omega$.
    
    It follows that a positive degree $g$ divisor is special if and only if it contains a pair of points $z^{(1)}_0$, $z^{(2)}_0$ on the first and second sheet which project to the same point $z_0 \in \C$ (otherwise $p_{g-1}(z)$ above would have to have $g$ \emph{distinct} zeros). In particular, assuming that $q_1, \dots, q_n$ lie on different gaps, the divisor $D$ will always be nonspecial (however plenty of other choices are also possible). This is exactly the setting we have encountered in the construction above.
\end{remark}
 We summarize our findings in the following proposition:
 \begin{proposition}
 \label{prop:sym_sol_2cutss}
     The unique symmetric solution $k_s$ to the Riemann--Hilbert problem \ref{OuterRHP_2cuts} is given by
     \begin{align*}
    k_s(z) &= \frac{1}{\gamma(z)} \Big(E_s(\mathcal A(z)), \ E_s(-\mathcal A(z)) \Big), \qquad \text{integration on the upper sheet},
\end{align*}
where 
\begin{align*}
     E_s(\mathcal{A}(z)) = c(\vec{\Delta}) \frac{\Theta(\mathcal{A}(z)  - \mathcal{A}(D_0) + K\mid B)\Theta(\mathcal{A}(z) - \Pi \mathcal{A}(D_0) + \frac{1}{2}\vec{1} + K\mid B)}{\Theta(\mathcal{A}(z)  - \mathcal{A}(D_\infty) + K\mid B)\Theta(\mathcal{A}(z) - \Pi \mathcal{A}(D_\infty) + \frac{1}{2}\vec{1} + K\mid B)},
\end{align*}
and 
\begin{align*}
    D_0 = z_1 + z_2 + q, \qquad D_\infty = \tb  + \td + q.
\end{align*}
where $q$ is any point on the Riemann surface such that $D_0$, $D_\infty$ are nonspecial, and $z_1$, $z_2$ are located on the first and second gap and are the unique solution of $\mu(z_1, z_2) =(\Delta_1, \Delta_2)$. The constant $c(\vec{\Delta})$ is chosen such that $k_{s,1}(0) = 1$. 
 \end{proposition}
 \begin{proof} 
     We have constructed our solution by finding an entire function $E_s$ satisfying pole, symmetry and quasi-periodicity conditions. It is easy to check that in fact finding such an $E_s$ is \emph{equivalent} to finding a symmetric vector solution $k_s$. Due to the pole condition and the symmetry condition, the only (possible) freedom that remains is finding $z_1$, $z_2$ such that the quasi-periodicity conditions \eqref{Eq:EQuasiPer} are satisfied. We have shown that this problem is always uniquely solvable if we restrict $z_1$, $z_2$ to lie on the first and second gaps (with $z_2$ being on the upper sheet). Clearly if we let $z_1$, $z_2$ be any other points on the Riemann surface, we could use the same argument as in the preceding paragraph to argue for uniqueness. This proves the statement.
 \end{proof}

\subsubsection{The holomorphic antisymmetric genus $3$ vector solution}
For the \emph{holomorphic} antisymmetric solution we proceed similarly as in the case of the symmetric solution and make the ansatz
\begin{align}\label{Eq:as}
    k_a(z) &= \frac{1}{\gamma(z)} \Big(E_a(\mathcal A(z)), \ E_a(-\mathcal A(z)) \Big), \qquad \text{integration on the upper sheet},
\end{align}
where 
\begin{align}\label{Eq:AnsatzEa}
     E_a(\mathcal{A}(z)) = c'(\vec{\Delta}) \frac{\Theta(\mathcal{A}(z)  - \mathcal{A}(D_0) + K \mid B)\Theta(\mathcal{A}(z) - \Pi \mathcal{A}(D_0) + \frac{1}{2}\vec{1} + K+ \vec{B}_2 \mid B)}{\Theta(\mathcal{A}(z)  - \mathcal{A}(D_\infty) + K \mid B)\Theta(\mathcal{A}(z) - \Pi \mathcal{A}(D_\infty) + \frac{1}{2}\vec{1} + K \mid B)}.
\end{align}
The reason for the inclusion of the period vector $\vec{B}_2$ in one of the $\Theta$ functions will become apparent in a moment. We choose the pole divisor $D_\infty = \tb + \td + q$ as before to guarantee holomorphicity, and for the zero divisor we choose $D_0 = z_1 + z_\infty + q$, where $z_1$ is restricted to the first gap, while $z_\infty$ is restricted to the \emph{unbounded} gap (or cycle) consisting of $[\ta^{-1}, +\infty]\cup [-\infty,\ta]$ on either sheet glued together at $\ta^{\pm 1}$ (as $\iota(z_\infty)$ will be on the same unbounded cycle but opposite sheet, we can w.l.o.g.~assume that it is located on the first sheet). We will denote this cycle by $\bf a_\infty$ and orient it such that it goes from $\ta^{-1}$ to $+\infty$ on the upper sheet. Note that the union of the three cycles $\bf a_1 \cup \bf a_2 \cup \bf a_3$ is homologous to the cycle  $\bf a_\infty$. The Abel condition reads now (notice that additional $\vec{B_2}$)
\begin{align*}
    (I + \Pi)\big(\mathcal A(\widehat D_0) - \mathcal A(\widehat D_\infty)\big) + \vec{B_2} \equiv \vec{\Delta} \quad \mod \Z^3,
\end{align*}
 with $\widehat D_0 = z_\infty + z_1$ and $\widehat D_\infty = \tb + \td$. We remind the reader that the above equation is $\mod \Z^3$, i.e., apart from finding the correct divisors, we also have to choose the correct path of integration defining $\mathcal A(\widehat D_0)$, $\mathcal A(\widehat D_\infty)$.
 
 Written in terms of integrals, the Abel condition is equivalent to
 \begin{align}\label{Eq:AbelCond2}
 \begin{split}
     \int_\tb^{z_1} (\omega_1 + \omega_3) + \int_\td^{z_\infty} (\omega_1 + \omega_3) + B_{12} &\equiv  \Delta_1  \quad \mod \Z,
     \\
      2\Big(\int_\tb^{z_1} \omega_2 + \int_\td^{z_\infty} \omega_2\Big) + B_{22} &\equiv \Delta_2  \quad \mod \Z .
 \end{split}
\end{align}
 Observe that $\int_\td^{z_\infty} (\omega_1 + \omega_3) = \int_{\td}^{\ta} (\omega_1 + \omega_3) + \int_{\ta}^{z_\infty} (\omega_1 + \omega_3)  \mod \Lambda$. By deforming the cycle $\bf b_2$ to surround the interval $[\ta, \td]$ on the upper sheet in the clockwise direction, and using the fact that $\omega_{j,+} = \omega_{j,-}$ on the cuts, it follows that
\begin{align*}
    \underbrace{\int_{\td}^{\ta} (\omega_1 + \omega_3)}_{\text{integration in the upper-half plane}} \hspace{-30pt}= -\frac{1}{2} \int_{\bf b_2} (\omega_1 + \omega_3) + \int_\tc^\tb (\omega_1 + \omega_3) &=-\frac{1}{2} \int_{\bf b_2} (\omega_1 + \omega_3)  + \frac{1}{2}\int_{\bf a_1} (\omega_1 + \omega_3) 
    \\
    &= -\frac{1}{2} (B_{21} + B_{23}) + \frac{1}{2} 
    \\
    &= - B_{12} + \frac{1}{2} \quad \text{by } \, \eqref{Eq:BSym}.
\end{align*} 
Similarly
\begin{align*}
    \underbrace{\int_{\td}^{\ta} \omega_2}_{\text{integration in the upper-half plane}} \hspace{-45pt}=-\frac{1}{2} \int_{\bf b_2} \omega_2  + \frac{1}{2}\int_{\bf a_1} \omega_2 = -\frac{1}{2} B_{22}.
\end{align*} 
It follows that \eqref{Eq:AbelCond2} can be rewritten as
 \begin{align}\label{Eq:AbelCond11}
 \begin{split}
     \int_\tb^{z_1} (\omega_1 + \omega_3) + \int_\ta^{z_\infty} (\omega_1 + \omega_3) + \frac{1}{2} &\equiv  \Delta_1  \quad \mod \Z,
     \\
      2\Big(\int_\tb^{z_1} \omega_2 + \int_\ta^{z_\infty} \omega_2\Big) &\equiv  \Delta_2  \quad \mod \Z.
 \end{split}
\end{align}
If we now introduce $\widetilde \Delta_1 = \Delta_1 - \Delta_2$, $\widetilde \Delta_2 = \Delta_2$, it follows that the mapping $\widetilde \mu \colon (z_1, z_\infty) \mapsto (\widetilde \Delta_1, \widetilde \Delta_2)$ can be viewed as a continuous mapping from $S^1 \times S^1$ to itself, with identical topological properties as in the case of the symmetric solution. We conclude that $\widetilde \mu$ is surjective. 

However, it is impossible to normalize $E_a$ such that $E_a(\mathcal A(0^{(1)})) = 1$ if and only if $z_\infty = 0$ (we have observed the same phenomenon for the antisymmetric genus $1$ solution). In the following, we assume $z_\infty \not = 0$. To show that the resulting $k_a$ is in fact antisymmetric, i.e.~$k_a(z^{-1}) = -k_a(z) \sigma_1$, it is again sufficient to compare the signs of $\gamma(-1) k_{a,1}(-1) = E_a(\mathcal A((-1)^{(1)}))$ and $\gamma(-1) k_{a,2}(-1) = E_a(-\mathcal A((-1)^{(1)})) = E_a(\mathcal A((-1)^{(2)}))$; the signs need to be different for $k_a$ to be antisymmetric. Note that as $  \left.E_a(\mathcal A(p))\right\vert_{p \in \bf a_\infty}$ is real ($\overline{k_a(\overline{z})}$, $k_a(z)$ satisfy the same Riemann--Hilbert problem and have the same zeros $\Rightarrow$ $\overline{k_a(x)} = k_a(x)$ for $x \in \bf a_\infty$) and has exactly two zeros $z_\infty$ and $\iota(z_\infty)$, and furthermore $(-1)^{(2)} = \iota((-1)^{(1)})$, there will be generically exactly one simple zero between $(-1)^{(1)}$ and $(-1)^{(2)}$ on $\bf a_\infty$ (going in either direction). The exception is of course $z_\infty = (-1)^{(1)}$ for which one can use a limiting argument. It follows that $k_a(z)$ is indeed antisymmetric. 

Knowing the symmetry class, one can show that $k_a(z)$ is the unique \emph{holomorphic} antisymmetric solution following the argument for the symmetric solution (see the proof of Lem.~\ref{lem:mu}). In particular the mapping $\widetilde \mu$ is a bijection. We leave the details to the reader. 

\begin{remark}
    The main difference between the holomorphic symmetric vs.~antisymmetric vector solutions is the location of their zeros. While both solutions always have one zero each on $\bf a_1$ and $\bf a_3$, the symmetric solution will have a pair of zeros on $\bf a_2$, while the antisymmetric solution will have a pair of zeros on $\bf a_\infty$. This is a direct consequence of the inclusion of $\vec{B}_2$ in our ansatz which forces the remaining zero to $\bf a_\infty$ through \eqref{Eq:AbelCond3}. It follows that the antisymmetric solution will not be normalizable whenever one of the zeros on $\bf a_\infty$ hits $0$ on the first sheet (as we require $k_{a,1}(0) = 1$).  This will necessitate the construction of a \emph{meromorphic} antisymmetric solution, which is the content of the next subsection.
\end{remark}

\subsubsection{The meromorphic  antisymmetric genus $3$ vector solution}
The construction of the \emph{meromorphic} antisymmetric solution follows closely the construction of the holomorphic antisymmetric solution with many key arguments being completetly analogous; that is why we will be rather brief. We start with the ansatz

\begin{align}\label{Eq:as2}
    \wt k_a(z) &= \frac{1}{\gamma(z)} \Big(\wt E_a(\mathcal A(z)), \ \wt E_a(-\mathcal A(z)) \Big), \qquad \text{integration on the upper sheet},
\end{align}
where 
\begin{align}\label{Eq:AnsatzEa2}
     \wt E_a(\mathcal{A}(z)) = \wt c(\vec{\Delta}) \frac{\Theta(\mathcal{A}(z)  - \mathcal{A}(D_0) + K| B)\Theta(\mathcal{A}(z) - \Pi \mathcal{A}(D_0) + \frac{1}{2}\vec{1} + K+ \vec{B}_2)}{\Theta(\mathcal{A}(z)  - \mathcal{A}(\wt D_\infty) + K| B)\Theta(\mathcal{A}(z) - \Pi \mathcal{A}(\wt D_\infty) + \frac{1}{2}\vec{1} + K)}.
\end{align}
and
\begin{align*}
    D_0 = z_1 + z_\infty + q, \qquad \wt D_\infty = 1^{(1)} + \tb + q.
\end{align*}
Here $z_1 \in \bf a_1$, $z_\infty \in \bf a_\infty$ (this will imply the antisymmetry as for the holomorphic solution). Note that the choice of pole divisor is different as we replaced $\td$ with $1^{(1)}$ (i.e.~the point $1$ on the first sheet). This induces a shift in the corresponding Abel condition which now reads (cf.~\eqref{Eq:AbelCond11})
 \begin{align}\label{Eq:AbelCond3}
 \begin{split}
     \int_\tb^{z_1} (\omega_1 + \omega_3) + \int_\ta^{z_\infty} (\omega_1 + \omega_3) + \frac{1}{2} &\equiv  \Delta_1 + \underbrace{\int_{\td}^{1^{(1)}}
     (\omega_1 + \omega_3)}_{= \, 0}  \quad \mod \Z,
     \\
      2\Big(\int_\tb^{z_1} \omega_2 + \int_\ta^{z_\infty} \omega_2\Big) &\equiv  \Delta_2 + \underbrace{2\int_{\td}^{1^{(1)}}
     \omega_2}_{=\, -\frac{1}{2}} \quad \mod \Z.
 \end{split}
\end{align}
We see that except for the shift in $\Delta_2$, this Jacobi inversion problem is completely equivalent to the one in \eqref{Eq:AbelCond11}. It follows that we can solve it uniquely, and the critical case $z_\infty = 0^{(1)}$ occurs for disjoint values $\Delta_1$, $\Delta_2$ compared to the previous case. In fact, if for some $(\Delta_1, \Delta_2)$ both \eqref{Eq:AbelCond11}, \eqref{Eq:AbelCond3} lead to $z_\infty = 0^{(1)}$, then by the uniqueness of the stated Jacobi inversion problem we must have $z_1 \not = \wt z_{1}$, where $z_1$ solves \eqref{Eq:AbelCond2} and  $\wt z_{1}$ solves \eqref{Eq:AbelCond3} (in both cases $z_\infty = 0^{(1)}$). This would imply that $\int_{\tb}^{z_1} (\omega_1 + \omega_3) = \int_{\tb}^{\wt z_{1}} (\omega_1 + \omega_3)$. We claim that this is not possible as $z_1 \mapsto \int_{\tb}^{z_1} (\omega_1 + \omega_3)$ is monotonic on the first gap. If it was not the case, then $\omega_1 + \omega_3$ would need to have at least \emph{two} zero on the $\bf{a}_1$-cycle, and by symmetry at least two zeros on the $\bf{a}_3$-cycle. Additionally, it would have to have at least two zeros on the $\bf{a}_2$-cycle, as $\int_{\bf{a}_2} (\omega_1 + \omega_3) = 0$, giving a total of at least six zeros. This contradicts the fact that holomorphic differentials have $2g - 2 = 4$ zeros, where $g$ is the genus of the underlying Riemann surface. 

While the Jacobi inversion problem has a unique solution, and therefore the vector Riemann--Hilbert problem has a unique solution of the form \eqref{Eq:as2}, \eqref{Eq:AnsatzEa2}, there will be many meromorphic solutions with at most fourth-root singularities at the branch point and a simple pole at $1$; in fact any linear combination of $\widetilde k_a(z)$ and the previous holomorphic solution $k_a(z)$ will be a meromorphic antisymmetric vector solution of Riemann--Hilbert problem \ref{OuterRHP_2cuts}.   
\section{The small norm regime}
\label{sec:small_norm}

We first analyze the situation where the RHP \ref{RHP:q_hat} becomes ``trivial'', meaning that the solution is asymptotically the vector $(1,1)$. This is the case where the off-diagonal entries of the jump matrix $J_{\wh q}(z)$ \eqref{eq:jump_q_hat} converge to zero as $N\to\infty$. Therefore, to be in this regime, we require that

\begin{equation}
\label{eq:small_norm_req}
    1+\tau \Phi(z,\eta,\tau) <0\; \forall z\in[\alpha,\beta]\,, \quad 1-\tau \Phi(z,\eta,\tau) <0\; \forall z\in[\beta^{-1},\alpha^{-1}]\,, \quad \ (\eta, \tau) \in \R^2. 
\end{equation}
 By symmetry, it is enough to find $\eta, \tau$ such that $\varphi(z, \eta, \tau) = 1 + \tau \Phi(z, \eta, \tau) < 0$ for all $z \in [\alpha, \beta]$. To see that $\varphi(\alpha, \eta,\tau), \varphi(\beta, \eta, \tau) < 0$ is already sufficient, we need to exclude the possibility of $\varphi(z_*) > 0$ with $z_* \in (\alpha, \beta)$, while $\varphi(\alpha), \varphi(\beta) < 0$. This would imply that $\varphi(z)$ would have a local maximum in $(\alpha, \beta) \subset (0,1)$. Noticing that $\varphi'(z)$ has two zeros $z_1$, $z_2$ with $z_1 z_2 = 1$, we we deduce that there is at most one local maximum $z_1\in(\alpha,\beta)$. However, since $\varphi(\beta) < 0$ by assumption and $\varphi(1) = 1$ for all $\eta, \tau$, it follows that there would have to be other local minima between $z_1$ and $1$, giving us a contradiction. We can therefore deduce that the small norm regime is given by the following set of parameters:

\begin{equation}
\label{eq:Usn}
    \cU_{sn} = \left\{ (\eta,\tau)\in \R\times\R \big\vert \, 1 + \tau(\alpha-\alpha^{-1}) + 2\eta \ln(\alpha) <0 \text{ and } 1 + \tau(\beta-\beta^{-1}) + 2\eta \ln(\beta) <0\right\}\,.
\end{equation}
By a trivial application of the small norm theory one deduces the following

\begin{proposition}\label{Prop:AsymSN}
    Fix $(\eta,\tau)\in \cU_{sn}$, and let $\wh q(z)$ be the solution to the RHP \ref{RHP:q_hat} with this choice of parameters. Then, there exists a constant $\delta>0$ such that

    \begin{equation}
        \norm{\wh q(z) - (1,\;1)}_\infty  = \cO\left(\frac{1}{N^{\delta}}\right)\,,
    \end{equation}
    for $z$ in any compact subset of  $\C\setminus\{(\alpha,\beta)\cup(\beta^{-1},\alpha^{-1})\}$.
\end{proposition}
\begin{proof}
    Let us define $r(z) = \frac{1}{\widehat q_1(0)} \widehat q(z)$. Then $r$ will satisfy exactly the same Riemann--Hilbert problem \ref{RHP:q_hat} except for the last normalization condition which reads $r(0) = (1, \, *)$ (and by symmetry $r(\infty) = (*, \, 1)$). From the small-norm theory in Appendix \ref{App:SmallNorm} it follows that $r$ satisfies a singular integral equation of the form $(1 - C_{\widehat w})r_- = (1, \, 1)$ with $\Vert \widehat w \Vert_{\infty} = \Vert J_{\widehat q} - I \Vert_\infty = O(N^{-\delta})$ for some $\delta > 0$. By the small-norm theory $1- C_{\widehat w}$ is uniformly invertible for $N$ large enough and $\Vert r_- \Vert_2 = O(1)$. From the representation 
    \begin{equation}
           r(z) = (1, \, 1) + \frac{1}{2\pi i} \int r_-(s)w_r(s)  \begin{pmatrix}
        \frac{1}{s-z} - \frac{1}{s} & 0
        \\
        0 & \frac{1}{s-z}
    \end{pmatrix} \di s\,,
    \end{equation}
  and the decay of $\widehat w$ it now follows that $r(z) = (1, \, 1) + O(N^{-\delta})$ for $z\not \in (\alpha,\beta) \cup (\beta^{-1},\alpha^{-1})$. In particular 
  
  \[ r_2(0) = 1 + O(N^{-\delta}) = \frac{\widehat q_2(0)}{\widehat q_1(0)}\] 
  which, together with the normalization $\widehat q_1(0) \widehat q_2(0) = 1$, implies that $\widehat q_1(0), \widehat q_2(0) = 1 + O(N^{-\delta})$ and therefore $\Vert \widehat q(z) - r(z) \Vert = O(N^{-\delta})$, finishing the proof. 
\end{proof}

In particular, as $q(z) = \wh q(z)$ for $z$ close to the origin, this implies that the corresponding Toda lattice solution is the trivial one, i.e. the following holds

\begin{corollary}
\label{cor:small_norm_sol}
 Fix $(\eta,\tau)\in \cU_{sn}$ and $(\eta\pm\ln(N)^{-1},\tau)\in \cU_{sn}$ \eqref{eq:Usn}. Let $q(z)$ be the solution to RHP \ref{RHP:q}. Then there exists a constant $\delta>0$ such that the Toda lattice sequence $\{a_n^{(q)}(t), b_n^{(q)}(t)\}$ recovered via Lemma \ref{lem:recovery_q} has the following asymptotic as $N\to\infty$:

    \begin{equation}
        a_n^{(q)}(t) = \frac{1}{2} + \cO(N^{-\delta})\,,\qquad b_n^{(q)}(t)= \cO(N^{-\delta})\,,
    \end{equation}
    where $n=\eta \ln(N),\, t =\tau \ln(N)$.
\end{corollary}

\begin{remark}\label{Rmk:eta3}
    For the future analysis, the intersection point $D_{\text{sn}}$ of the two lines in the $(\eta,\tau)$-plane $1+ \tau \Phi(\alpha,\eta,\tau)=0$ and $1+ \tau \Phi(\beta,\eta,\tau)=0$ will be important. Its coordinates can be computed explicitly as

    \begin{equation}\label{eq:Dsn}
        D_{\text{sn}}=(\eta_{\text{sn}},\tau_{\text{sn}})=\left(\frac{(\beta -\alpha ) (\alpha  \beta +1)}{2 \left(\alpha ^2-1\right) \beta  \ln (\beta )-2 \alpha  \left(\beta ^2-1\right) \ln (\alpha )},\; \frac{\alpha  \beta  \ln \left(\frac{\alpha }{\beta }\right)}{\left(\alpha ^2-1\right) \beta  \ln (\beta )+\alpha\left(1 -  \beta ^2\right) \ln (\alpha )}\right)\,.
    \end{equation}
    Moreover, we introduce the notation
    \begin{align}
        \eta_3(\tau) = \begin{cases}
              - \frac{1 + \tau(\alpha - \alpha^{-1})}{2\ln(\alpha)}, \quad \text{for} \ \tau \leq \tau_{\text{sn}},
            \\
              - \frac{1 + \tau(\beta - \beta^{-1})}{2\ln(\beta)}, \quad \text{for} \ \tau \geq \tau_{\text{sn}}.
        \end{cases}
    \end{align}
Note that $(\eta_3(\tau), \tau)$ with $\tau \in \R$ is exactly the boundary of the small-norm region. The notation $\eta_3$ will become more clear later on.
\end{remark}

\section{The complete one-cut regime}
\label{sec:complete_one_cut}
We now enter into the main part of the manuscript. In this section, we analyze the RHP \ref{RHP:q_hat} in the regime that we call \textit{complete one-cut}. In this regime, the corresponding solution to the Toda lattice is asymptotically given by an elliptic wave with a fixed period. In particular, we construct the solution to this RHP away from the branch points $\alpha,\beta,\alpha^{-1},\beta^{-1}$ in terms of the solution to the model problem RHP \ref{OuterRHP_1cut}. 

\subsection{The first transformation}

To solve the general RHP \ref{RHP:q_hat}, we define the following transformation
\begin{equation}
      \label{eq:def_transformation_1}
      p(z) = \wh q(z)e^{\sigma_3(\ln(N)g(z) + f(z)) }\,.
\end{equation}
In particular, we require $g(z),f(z)$ to be analytic away from the segment $(\alpha,\alpha^{-1})$. By applying this transformation and defining $\psi(z) = f(z) +\ln(N)g(z)$, one gets the following RHP for $p(z)$

\begin{rhp}
\label{RHP:p}
      We seek for a vector $ p(z) = (p_1(z)\quad p_2(z))$  such that 
    \begin{itemize}
        \item $p(z)$ is analytic in $\C\setminus(\alpha,\alpha^{-1}) $
        \item $p_+(z)=p_-(z) J_{p}(z)$ for $z\in (\alpha,\alpha^{-1})$, which is oriented from left to right and $J_{p}(z)$ is defined as
        
      \begin{equation}
  	\label{eq:jump_p}
  	J_{p}(z)=\begin{cases}
  		\begin{pmatrix}
  			e^{\left(\psi_+(z) - \psi_-(z)\right)} & 0 \\
  			2\pi i N\mu_1e^{t\Phi(z,\eta,\tau) + \psi_+(z)+\psi_-(z)}z \rho(z)& e^{-\left(\psi_+(z) - \psi_-(z)\right)}
  		\end{pmatrix} \quad &z \in (\alpha,\beta) \\\\[5pt]
        \exp\left(\sigma_3\left(\psi_+(z) - \psi_-(z)\right)\right)\qquad & z\in (\beta,\beta^{-1})\\\\[5pt]
  		\begin{pmatrix}
  			e^{\left(\psi_+(z) - \psi_-(z)\right)} &-2\pi iN\mu_1e^{-t\Phi(z,\eta,\tau) -\left(\psi_+(z) + \psi_-(z)\right)}z^{-1}\rho(z^{-1})\\
  			0 & e^{-\left(\psi_+(z) - \psi_-(z)\right)}
  		\end{pmatrix} \quad &z\in (\beta^{-1},\alpha^{-1}) 
  	\end{cases}\,.
  \end{equation}

  \item $p(z^{-1}) = p(z) \sigma_1$,
       
        \item $\lim_{z\to 0} p_1(z)\cdot p_2(z) = 1$, $p_1(0)>0$.
    \end{itemize}
\end{rhp}

In view of Assumptions \ref{ass:condensate_assumptions}, we set $t=\tau\ln(N),\, n = \eta \ln(N)$; therefore, following the same ideas as in \cite{Gkogkou2025soliton}, we require that $g(z),f(z)$ solve the following RHP

\begin{rhp}
      \label{rhp:g_function_one_cut_whole}
      We seek for a scalar function $g(z)$ such that
      \begin{itemize}
            \item $g(z)$ is analytic in $\C\setminus \left(\alpha,\alpha^{-1}\right)$
            \item $g_+(z)+g_-(z)+1 + \tau\Phi(z,\eta,\tau) = 0$ for $z\in (\alpha,\beta)$
            \item $g_+(z)+g_-(z) - 1 + \tau\Phi(z,\eta,\tau) = 0$ for $z\in \left(\frac{1}{\beta},\frac{1}{\alpha}\right)$
            \item $g_+(z) - g_-(z) = \zeta_c(\eta,\tau)$ for $z\in (\beta,\beta^{-1})$
            \item $g(z^{-1}) = -g(z)$
            \item $g(z) = g_\infty + \cO(\frac{1}{z})$ as $z\to\infty$
            \item $g(z) + \frac{1 + \tau\Phi(z,\eta,\tau)}{2} = \cO((z-\alpha)^{1/2})$ as $z\to \alpha$ and $g(z) + \frac{1 + \tau\Phi(z,\eta,\tau)}{2} = \pm \frac{1}{2}\zeta_c +\cO((z-\beta)^{1/2})$ as $z\to \beta$ where $\pm$ indicates that $z \in \mathbb{C}_\pm$.
\end{itemize}
\end{rhp}

 \begin{rhp}
    \label{rhp:f_function_one_cut_whole}
    We seek for a scalar function $f(z)$ such that
    \begin{itemize}
        \item $f(z)$ is analytic in $\C\setminus \left(\alpha,\alpha^{-1}\right)$
        \item $f_+(z)+f_-(z) + \ln(2\pi ) + \ln(z) + \ln(\mu_1\rho(z)) = 0$ for $z\in(\alpha,\beta)$
        \item $f_+(z)+f_-(z)  -\ln(2\pi ) + \ln(z) - \ln(\mu_1\rho(z^{-1})) = 0$ for $z\in(\beta^{-1},\alpha^{-1})$
        \item $f_+(z)-f_-(z) = \omega_c(\eta,\tau)$ for $z\in(\beta,\beta^{-1})$
        \item $f(z^{-1}) = -f(z)$
        \item $f(z) = f_\infty + \cO(\frac{1}{z})$ as $z\to\infty$
    \end{itemize}
\end{rhp}

By standard techniques, one can solve the previous RHP as follows

\begin{proposition}
    \label{prop:g_whole}
    Define the function $g(z)$ as

    \begin{equation}
    \label{eq:g_one_cut_whole}
        g(z) = \frac{1 -\tau \Phi(z,\eta,\tau)}{2} + \int_{\alpha^{-1}}^z \frac{P(s,\eta,\tau)}{s R_1(s)}\di s\,, \quad P(s,\eta,\tau) = \frac{\tau}{4}(s+s^{-1})^{2} + \frac{y_1}{2}(s+s^{-1}) + y_0\,,
    \end{equation}
    where
    \begin{equation}
        R_1(z) = \frac{1}{2}\left(z+\frac{1}{z} - \alpha -\frac{1}{\alpha}\right)^{\frac{1}{2}}\left(z+\frac{1}{z} - \beta -\frac{1}{\beta}\right)^{\frac{1}{2}}, \quad y_1 = \eta - \frac{\tau}{4}\left(\alpha + \alpha^{-1} + \beta + \beta^{-1}\right)\,,
    \end{equation}
    and 
    \begin{equation}
    \label{eq:p_0_whole}
        y_0 = \left(\frac{1}{2} -\int_1^{\beta^{-1}} \frac{\frac{\tau}{4}(s+s^{-1})^{2} + \frac{y_1}{2}(s+s^{-1})}{sR_1(s)}\di s \right)\left( \int_1^{\beta^{-1}}\frac{1}{sR_1(s)}\di s\right)^{-1}\,,
    \end{equation}
    then $g(z)$ solves the RHP \ref{rhp:g_function_one_cut_whole} with

    \begin{equation}\label{eq:zeta_c}
        \zeta_c(\eta,\tau) = 2i\Im(g_+(\beta^{-1})) =  2\int_{\alpha^{-1}}^{\beta^{-1}} \frac{P(s,\eta,\tau)}{sR_{1,+}(s)}\di s\,.
    \end{equation}
\end{proposition}
Note that $y_0$ is chosen precisely such that
\begin{align}\label{eq:IntP}
   \int_\beta^{\beta^{-1}} \frac{P(s, \eta, \tau)}{s R_1(s)} \di s = 2 \int_1^{\beta^{-1}} \frac{P(s, \eta, \tau)}{s R_1(s)} \di s = 1.
\end{align}
This condition guarantees that $g_+(z)+g_-(z)+1 + \tau\Phi(z,\eta,\tau) = 0$ for $z\in (\alpha,\beta)$.
\begin{remark}\label{rem:positivity_R_whole}
    The domain of $R_1(z)$ is $\mathbb C \setminus ([\alpha, \beta] \cup [\beta^{-1}, \alpha^{-1}])$, and we choose the branch of the square root such that $R_1(z) = \frac{z}{2} + O(1)$ at infinity. This fixes $R_1$ uniquely, which then satisfies $R_{1,\pm}(x) \in \pm i \mathbb R$ for $x \in (\beta^{-1}, \alpha^{-1})$ and $R_{1, \pm}(x) \in \mp i \mathbb R$ for $x \in (\alpha, \beta)$ (the orientation of the two intervals is as usual from left to right). Also $R_1(z)$ will be negative on $(\beta, \beta^{-1})$.
\end{remark}

\begin{proposition}\label{prop:f_whole_anti_symmetric}

Define the function $f(z)$ as

\begin{equation}\label{eq:f_one_cut_whole_symmetric}f(z) = \frac{R_1(z)}{2\pi i} \left( \int_{\alpha}^{\beta} \frac{s+z}{2s(s-z)}\frac{H_1(s)}{R_{1+}(s)}\di s + \int_{\beta^{-1}}^{\alpha^{-1}} \frac{s+z}{2s(s-z)}\frac{H_2(s)}{R_{1+}(s)}\di s + \omega_c(\eta,\tau)\int_{\beta}^{\beta^{-1}} \frac{s+z}{ 2s(s-z)}\frac{1}{R_{1}(s)}\di s \right)\,,
\end{equation}

where $R_1(z)$ is defined in Proposition \ref{prop:g_whole}, the jump functions $H_1(s)$ and $H_2(s)$ are given by

\begin{equation}
\label{eq:H_functions}
H_1(s) = -\ln(2\pi) - \ln(s) - \ln(\mu_1\rho(s))\,,\qquad  H_2(s) = \ln(2\pi) - \ln(s) + \ln(\mu_1\rho(s^{-1}))\,.
\end{equation}

 finally the constant $\omega_c(\eta,\tau)$ is defined as:
 
 \begin{equation}\label{eq:omega_whole}
 \omega_c(\eta,\tau) = \frac{ \int_{\alpha}^{\beta} \frac{-H_1(s)}{sR_{1+}(s)}\di s}{\int_{1}^{\beta^{-1}} \frac{1}{sR_{1}(s)}\di s}= \frac{ \int_{\alpha}^{\beta} \frac{\ln(2\pi s\rho(s)\mu_1)}{sR_{1+}(s)}\di s}{ \int_{1}^{\beta^{-1}} \frac{1}{sR_{1}(s)}\di s }\,.\end{equation}
 
 Then $f(z)$ is the unique solution to RHP \ref{rhp:f_function_one_cut_whole}.
 \end{proposition}

The proof is a direct application of the Sokhotski--Plemelj formula. Since the underlying argument is identical to -- and simpler than --the two-cut case treated in detail in Section~\ref{sec:two_cuts} (see the proof of Proposition~\ref{prop:sol_f_two_cuts}), we omit the computation here.
\begin{remark}
\label{rem:elliptic_1}
    For later usage, we notice that one can compute both $  \zeta_c(\eta,\tau), I_\ell= \int_1^{\beta^{-1}}\frac{(s+s^{-1})^\ell}{2^\ell sR_1(s)}\di s$ explicitly in terms of elliptic integrals as

    \begin{align}
        &I_0  = -\tg K(k)\,, \qquad I_1 = -\tg \Big[ \wt\alpha K(k) + (\wt\beta - \wt\alpha) \Pi(n, k) \Big]\,,\\
        & I_2 = -\tg\left[\left(\wt \alpha^2 - \frac{1}{2}(\wt\alpha -\wt\beta)(\wt \alpha-1)\right)K(k) - \frac{1}{2}(\wt \alpha - 1)(\wt \beta+1)E(k)- \frac{1}{2}(\wt \alpha - \wt\beta)(\wt\alpha+\wt\beta)\Pi(n,k)\right],\\
        & \zeta_c(\eta,\tau) =
2i\,\tg\Bigg[\left(y_0(\eta,\tau)+\eta+\frac{\tau}{2}\left(1-\widetilde\alpha\widetilde\beta\right)
\right)K(k_c)+\frac{\tau}{2}(\widetilde\alpha-1)(\widetilde\beta+1)
E(k_c)+\eta(\widetilde\beta-1)\Pi(p,k_c)
\Bigg],
    \end{align}
    where $n = \frac{\wt\beta-1}{\wt \alpha-1}, k^2 = \frac{(\wt \alpha +1)(\wt \beta-1)}{(\wt \alpha-1)(\wt \beta+1)}, \tg = \frac{2}{\sqrt{(\wt\alpha-1)(\wt \beta+1)}}, p=\frac{\widetilde\alpha-\widetilde\beta}{\widetilde\alpha-1},k_c^2=\frac{2(\widetilde\alpha-\widetilde\beta)}{(\widetilde\alpha-1)(\widetilde\beta+1)}$ and $K(k),E(k),\Pi(n,k)$ are the complete elliptic integrals of the first, second and third kind \cite[Ch.~19]{dlmf}. Notice that the previous expression coincides with the one in \cite{Biondini2024}.
    Finally, one can simplify the expression for $y_0$ as

\begin{equation}
\label{eq:y0_elliptic_corrected}
y_0 = -\frac{1}{2 \tg K(k)}-\eta\left[\widetilde\alpha-(\widetilde\alpha-\widetilde\beta)\frac{\Pi(n,k)}{K(k)}\right]-\frac{\tau}{2}
\left[(\widetilde\alpha-\widetilde\beta)-(\widetilde\alpha-1)(\widetilde\beta+1)\frac{E(k)}{K(k)}\right],
\end{equation}
\end{remark}

\begin{remark}
    We notice that the symmetric Cauchy kernel $\frac{z+s}{2s(s-z)}$ was already used in \cite{Egorova2023} to solve an analogous problem. 
\end{remark}

Applying the previous proposition to the RHP \ref{RHP:p}, the jump matrix $J_p(z)$ becomes

\begin{equation}
\label{eq:J_p_dressed}
    	J_{p}(z)=\begin{cases}
  		\begin{pmatrix}
  			e^{\left(2\ln(N)\int_{\alpha}^z \frac{P(s,\eta,\tau)}{sR_{1,+}(s)}\di s + f_+(z) -f_-(z)\right)}& 0 \\
  			i& e^{-\left(2\ln(N)\int_{\alpha}^z \frac{P(s,\eta,\tau)}{sR_{1,+}(s)}\di s + f_+(z) -f_-(z)\right)}
  		\end{pmatrix} \; &z \in (\alpha,\beta) \\\\[5pt]
        \exp\left(\sigma_3\left(\ln(N)\zeta_c(\eta,\tau) + \omega_c(\eta,\tau)\right)\right)\; & z\in (\beta,\beta^{-1})\\\\[5pt]
  		\begin{pmatrix}
  			e^{\left(2\ln(N)\int_{\alpha^{-1}}^z \frac{P(s,\eta,\tau)}{sR_{1,+}(s)}\di s + f_+(z) -f_-(z)\right)} &-i\\
  			0 & e^{-\left(2\ln(N)\int_{\alpha^{-1}}^z \frac{P(s,\eta,\tau)}{sR_{1,+}(s)}\di s + f_+(z) -f_-(z)\right)}
  		\end{pmatrix} \; &z\in (\beta^{-1},\alpha^{-1}) 
  	\end{cases}\,.
  \end{equation}
We are now in position to proceed with the \textit{opening of lenses}.

\subsection{Lenses opening and regime characterization}
In this subsection, we proceed by opening the lenses in the previous RHP, this will allow us to determine where the $g(z),f(z)$ functions are useful to obtain meaningful information about the asymptotic of our problem. We start by noticing that 
\begin{align}
\label{eq:f_factorization 1}
    f_+(z) - f_-(z) = \begin{cases}
        2 f_+(z) + \ln(2 \pi z \mu_1 \rho(z))
        \\
        -2f_-(z) - \ln(2 \pi z \mu_1 \rho(z)) 
    \end{cases}, \quad z \in (\alpha, \beta),
\end{align}
and likewise
\begin{align}
\label{eq:f_factorization 2}
    f_+(z) - f_-(z) = \begin{cases}
        2 f_+(z) - \ln(2 \pi \mu_1 \rho(z^{-1})/z)
        \\
        -2f_-(z) + \ln(2 \pi \mu_1 \rho(z^{-1})/z) 
    \end{cases}, \quad z \in (\beta^{-1}, \alpha^{-1})\,.
\end{align}
Recalling the following factorizations for $2\times 2$ matrices with determinant $1$, see \cite{BilmanMem2024}

\begin{align}
\label{eq:LTL}
\begin{pmatrix} a & b \\ c & d \end{pmatrix} &= \begin{pmatrix} 1 & 0 \\ db^{-1} & 1 \end{pmatrix} \begin{pmatrix} 0 & b \\ -b^{-1} & 0 \end{pmatrix} \begin{pmatrix} 1 & 0 \\ ab^{-1} & 1 \end{pmatrix}, & b &\neq 0, \quad & (\text{``LTL''}), \\
\label{eq:UTU}
\begin{pmatrix} a & b \\ c & d \end{pmatrix} &= \begin{pmatrix} 1 & ac^{-1} \\ 0 & 1 \end{pmatrix} \begin{pmatrix} 0 & -c^{-1} \\ c & 0 \end{pmatrix} \begin{pmatrix} 1 & dc^{-1} \\ 0 & 1 \end{pmatrix}, & c &\neq 0, \quad & (\text{``UTU''}).
\end{align}
Using the factorization ``UTU'', one deduces that on $(\alpha, \beta)$ the following holds:
\begin{equation}
\begin{split}
        &\begin{pmatrix}
  			e^{\left(2\ln(N)\int_{\alpha}^z \frac{P(s,\eta,\tau)}{sR_{1,+}(s)}\di s + f_+(z) -f_-(z)\right)}& 0 \\
  			i& e^{-\left(2\ln(N)\int_{\alpha}^z \frac{P(s,\eta,\tau)}{sR_{1,+}(s)}\di s + f_+(z) -f_-(z)\right)}
  		\end{pmatrix}=\\ &\begin{pmatrix}
        1 &  -ie^{-2 f_-(z) - \ln(2 \pi z \mu_1 \rho(z))  -2\ln(N)\int_{\alpha}^z \frac{P(s,\eta,\tau)}{sR_{1,-}(s)}\di s}
        \\
        0 & 1
    \end{pmatrix}\\
    &\times
    \begin{pmatrix}
        0 & i
        \\
        i & 0
    \end{pmatrix} 
    \begin{pmatrix}
        1 &  -ie^{-2 f_+(z) - \ln(2 \pi z \mu_1 \rho(z))  -2\ln(N)\int_{\alpha}^z \frac{P(s,\eta,\tau)}{sR_{1,+}(s)}\di s}
        \\
        0 & 1
    \end{pmatrix}.
    \end{split}
\end{equation}
By enforcing the factorization ``LTL'', one deduces that for $z\in (\beta^{-1},\alpha^{-1})$

\begin{equation}
\begin{split}&
\begin{pmatrix}
  			e^{\left(2\ln(N)\int_{\alpha^{-1}}^z \frac{P(s,\eta,\tau)}{sR_{1,+}(s)}\di s + f_+(z) -f_-(z)\right)} &-i\\
  			0 & e^{-\left(2\ln(N)\int_{\alpha^{-1}}^z \frac{P(s,\eta,\tau)}{sR_{1,+}(s)}\di s + f_+(z) -f_-(z)\right)}
  		\end{pmatrix} = \\
        &\pmtwo{1}{0}{ie^{\left(2\ln(N)\int_{\alpha^{-1}}^z \frac{P(s,\eta,\tau)}{sR_{1,-}(s)}\di s + 2f_-(z) - \ln(2 \pi \mu_1 \rho(z^{-1})/z)\right)}}{1}\\ & \times  \pmtwo{0}{-i}{-i}{0} \pmtwo{1}{0}{ie^{\left(2\ln(N)\int_{\alpha^{-1}}^z \frac{P(s,\eta,\tau)}{sR_{1,+}(s)}\di s + 2f_+(z) - \ln(2 \pi \mu_1 \rho(z^{-1})/z)\right)}}{1}
\end{split}\,.
\end{equation}

\begin{figure}
    \centering
    \begin{tikzpicture}[
    decoration={markings, mark=at position 0.55 with {\arrow{Latex}}},
    scale=1.5
]

    % Draw the real axis
    \draw[dashed,thin, black, ->] (-4.5, 0) -- (4.5, 0) node[right] {$\mathbb{R}$};

    % Define the coordinates for the branch points
    % Assuming alpha > 1, beta > alpha, so beta^{-1} < alpha^{-1} < 1 < alpha < beta
    \coordinate (B_inv) at (-3, 0);
    \coordinate (A_inv) at (-1, 0);
    \coordinate (A) at (1, 0);
    \coordinate (B) at (3, 0);

    % ==========================================
    % LEFT CUT: 
    % ==========================================
    
    % Main cut
    \draw[thick, red, postaction={decorate}] (B_inv) -- (A_inv) 
        node[midway, above=2pt, black] {$\pmtwo{0}{i}{i}{0}$};
        
    % Upper lens (Increased angles and looseness)
    \draw[thick, blue, postaction={decorate}] (B_inv) to[out=75, in=105, looseness=2] (A_inv);
    % Shifted node Y from 0.75 to 1.6
    \node at (-3, 1.6) {{\small $\pmtwo{1}{{\color{gray}-ie^{-2 f(z) - \ln(2 \pi z \mu_1 \rho(z))  -2\ln(N)\int_{\alpha}^z \frac{P(s,\eta,\tau)}{sR_{1}(s)}\di s}}}{0}{1} $}};
    
    % Lower lens (Increased angles and looseness)
    \draw[thick, blue, postaction={decorate}] (B_inv) to[out=-75, in=-105, looseness=2] (A_inv);
    % Shifted node Y from -0.75 to -1.6
    \node at (-3, -1.6) {{\small $\pmtwo{1}{{\color{gray}-ie^{-2 f(z) - \ln(2 \pi z \mu_1 \rho(z))  -2\ln(N)\int_{\alpha}^z \frac{P(s,\eta,\tau)}{sR_{1}(s)}\di s}}}{0}{1}$}};

    % Branch points (Left Cut)
    \filldraw[black] (B_inv) circle (1.5pt) node[below left] {$\alpha$};
    \filldraw[black] (A_inv) circle (1.5pt) node[below right] {$\beta$};

    % ==========================================
    % RIGHT CUT: (\alpha, \beta)
    % ==========================================

    % Main cut
    \draw[thick, red, postaction={decorate}] (A) -- (B) 
        node[midway, above=2pt, black] {$\pmtwo{0}{-i}{-i}{0}$};
        
    % Upper lens (Increased angles and looseness)
    \draw[thick, blue, postaction={decorate}] (A) to[out=75, in=105, looseness=2] (B);
    % Shifted node Y from 0.75 to 1.6
    \node at (2.5, 1.6) {\small {$\pmtwo{1}{0}{{\color{gray}ie^{\left(2\ln(N)\int_{\alpha^{-1}}^z \frac{P(s,\eta,\tau)}{sR_{1}(s)}\di s + 2f(z) - \ln(2 \pi \mu_1 \rho(z^{-1})/z)\right)}}}{1}$}};
    
    % Lower lens (Increased angles and looseness)
    \draw[thick, blue, postaction={decorate}] (A) to[out=-75, in=-105, looseness=2] (B);
    % Shifted node Y from -0.75 to -1.6
    \node at (2.5, -1.6) {{\small $\pmtwo{1}{0}{{\color{gray}ie^{\left(2\ln(N)\int_{\alpha^{-1}}^z \frac{P(s,\eta,\tau)}{sR_{1}(s)}\di s + 2f(z) - \ln(2 \pi \mu_1 \rho(z^{-1})/z)\right)}}}{1}$}};

    % Branch points (Right Cut)
    \filldraw[black] (A) circle (1.5pt) node[below left] {$\beta^{-1}$};
    \filldraw[black] (B) circle (1.5pt) node[below right] {$\alpha^{-1}$};

    % Optional fixed point at z = 1
    \filldraw[black] (0,0) circle (1pt) node[below] {$1$};

     \draw[thick, green, postaction={decorate}] (A_inv) -- (A) 
        node[midway, above=2pt, black] {$e^{\sigma_3\left(\ln(N)\zeta(\eta,\tau) + \omega\right)}$};

\end{tikzpicture}
    \caption{RHP for $v(z)$ defined in \eqref{eq:v_whole}. Lenses Opening: notice that, in the regime of Lemma \ref{lem:whole_regime}, the terms in gray go to zero as $N$ grows.}
    \label{fig:jump_v_whole}
\end{figure}
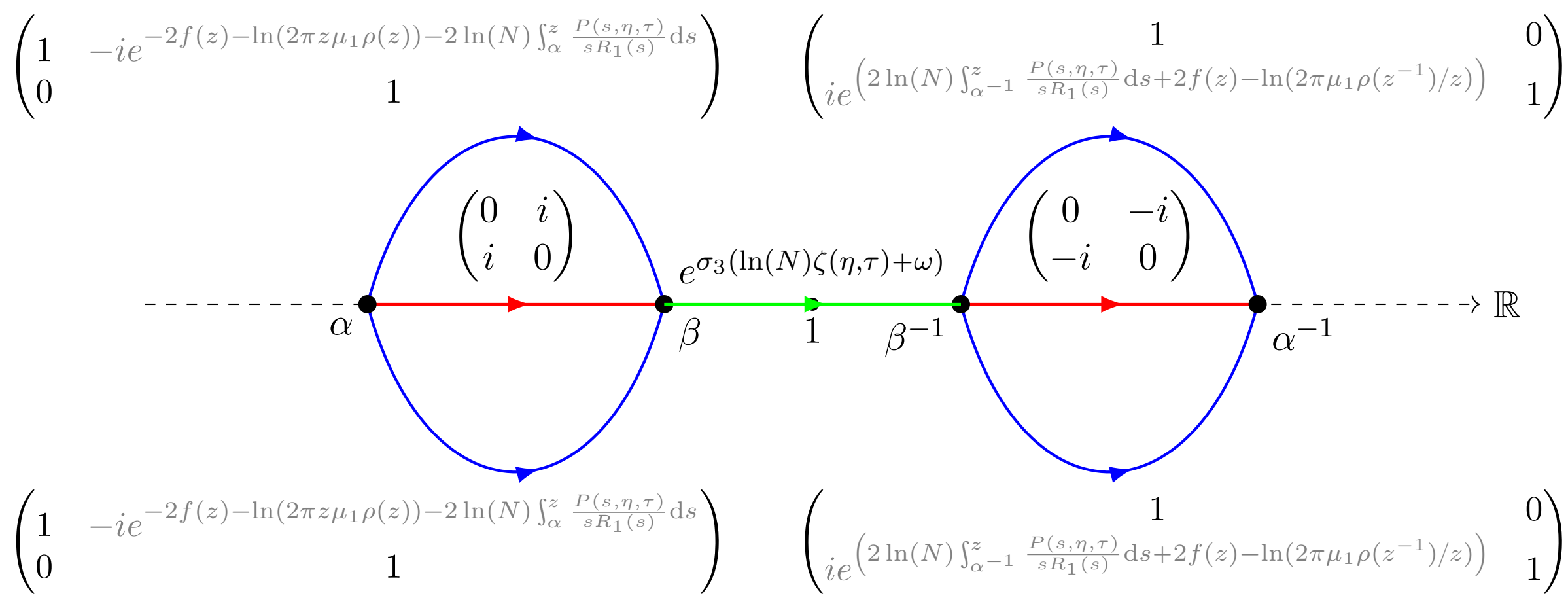
We can now define a new vector function $v(z)$ as follows
\begin{equation}
\label{eq:v_whole}
    v(z) = \begin{cases}
        p(z)\pmtwo{1}{-ie^{-2 f(z) - \ln(2 \pi z \mu_1 \rho(z))  -2\ln(N)\int_{\alpha}^z \frac{P(s,\eta,\tau)}{sR_{1}(s)}\di s}}{0}{1} \quad &z \text{ in the lower lenses below }(\alpha,\beta) \\\\
        p(z)\pmtwo{1}{ie^{-2 f(z) - \ln(2 \pi z \mu_1 \rho(z))  -2\ln(N)\int_{\alpha}^z \frac{P(s,\eta,\tau)}{sR_{1}(s)}\di s}}{0}{1}  \quad &z \text{ in the upper lenses above }(\alpha,\beta) \\\\
        p(z) \pmtwo{1}{0}{ie^{\left(2\ln(N)\int_{\alpha^{-1}}^z \frac{P(s,\eta,\tau)}{sR_{1}(s)}\di s + 2f(z) - \ln(2 \pi \mu_1 \rho(z^{-1})/z)\right)}}{1} \quad &z \text{ in the lower lenses below }(\beta^{-1},\alpha^{-1}) \\\\
        p(z)\pmtwo{1}{0}{-ie^{\left(2\ln(N)\int_{\alpha^{-1}}^z \frac{P(s,\eta,\tau)}{sR_{1}(s)}\di s + 2f(z) - \ln(2 \pi \mu_1 \rho(z^{-1})/z)\right)}}{1}  \quad &z \text{ in the upper lenses above }(\beta^{-1},\alpha^{-1}) \\\\
        p(z) \quad & \text{otherwise}
    \end{cases}\,.
\end{equation}
Then, the vector $v(z)$ satisfies the following conditions

\begin{equation}
    v_+(z) = v_-(z)J_v(z)\,, \quad
    \lim_{z\to 0} v_1(z)v_2(z)=1\,, \quad
     v(z^{-1}) = v(z)\sigma_1\,, \quad v_1(0)>0\,,
\end{equation}
where the jump matrix $J_v(z)$ is as in Figure \ref{fig:jump_v_whole}. 

We now need to identify the set $\cU_c$, where this factorization is useful to obtain the asymptotic behavior of $\wh q(z)$. To fully characterize this set, we need some preparation. First, we introduce the Joukowsky map 

\begin{equation}
\label{eq:Joukowsky}
    z \mapsto x(z) = \frac{1}{2}(z+z^{-1})\,,
\end{equation}
which defines a bijection from $(0,1)$ to $(1, \infty)$, and introduce the notation $\wt \alpha = x(\alpha), \wt\beta = x(\beta)$ etc.
Furthermore, define the polynomial

\begin{equation}
\label{eq:Q_poly}
    Q(x,\eta,\tau)\equiv Q(x) = P(z(x)) =\tau x^2 + y_1 x + y_0\,,
\end{equation}
and its vertex $x_v = - \frac{y_1}{2\tau} = - \frac{\eta}{2\tau} + \frac{\wt \alpha + \wt\beta}{4}$. Note that in these new variables condition \eqref{eq:IntP} is equivalent to
\begin{align}\label{eq:IntQ}
    \int_1^{\wt \beta} \frac{Q(x, \eta, \tau)}{\sqrt{\wt \alpha - x} \sqrt{\wt \beta - x}\sqrt{x^2-1} } \di x = -\frac{1}{2}. 
\end{align}
We can now state the following Lemma that characterizes the set $\cU_c$:

\begin{lemma}
    \label{lem:whole_regime}
    Define the set $\cU_c = \cU_{c}^{(+)}\cup \cU_c^{(-)}$, where

    \begin{align}
    \label{eq:U_cplus}
        &\cU_{c}^{(+)} = \left\{(\eta,\tau)\in \R\times \R_{\geq 0} \big\vert \; Q(\wt\beta)<0 \text{ and } Q(\wt \alpha)<0\right\}\\
        \label{eq:U_cminus}
        &\cU_{c}^{(-)} = \left\{(\eta,\tau)\in \R\times \R_- \big\vert\; Q(x_v)<0 \text{ or } \left( x_v>\wt \alpha \text{ and } Q(\wt\alpha)<0  \right) \text{ or } \left(x_v<\wt \beta \text{ and } Q(\wt\beta)<0\right) \right\}
    \end{align}

    then for all $(\eta,\tau)\in \cU_c$ there are $2$ neighborhoods $V_1,V_2$ of $(\alpha,\beta)$ and $(\beta^{-1},\alpha^{-1})$ respectively such that 

    \begin{equation}
        \Re\left(\int_{\alpha}^z \frac{P(s,\eta,\tau)}{sR_{1}(s)}\di s\right) \geq 0 \quad \forall \, z\in V_1\,, \quad \Re\left(\int_{\alpha^{-1}}^z \frac{P(s,\eta,\tau)}{sR_{1}(s)}\di s \right)\leq 0 \quad \forall \, z\in V_2\,.
    \end{equation}
\end{lemma}

\begin{remark}
    The previous lemma implies that the jump matrices on the outer lenses converge to zero as $N\to\infty$, see Fig. \ref{fig:jump_v_whole}
\end{remark}
For a pictorial representation of the regions $\cU_c$ we refer to Figure \ref{fig:phase_diagram}.
\begin{proof}
    We are going to prove the Lemma only for $z$ in a neighborhood of $(\alpha,\beta)$, the proof in the other case is completely analogous.

    First, we evaluate the integral $\int_{\alpha}^z \frac{P(s,\eta,\tau)}{sR_{1}(s)}\di s$ for $z\to x\in(\alpha,\beta)$ from above, which evaluates to

    \begin{equation}
        \lim_{z\to x^+} \int_{\alpha}^z \frac{P(s,\eta,\tau)}{sR_{1}(s)}\di s = \int_{\alpha}^x \frac{P(s,\eta,\tau)}{sR_{1,+}(s)}\di s\,.
    \end{equation}
    From Remark \ref{rem:positivity_R_whole}, one deduces that $\frac{1}{R_{1,+}(s)}\in i\R_+$ for $s\in(\alpha,\beta)$, and, since we are integrating on a real segment, $\int_{\alpha}^x \frac{P(s,\eta,\tau)}{sR_{1,+}(s)}\di s\in i\R$. Therefore, if $P(s,\eta,\tau)<0$ for all $s\in (\alpha,\beta)$ it would follow that $\Im\big(\int_\alpha^z \frac{P(s,\eta,\tau)}{sR_1(s)} \di s \big)$ is an decreasing function for $z \in (\alpha, \beta)$, which by the Cauchy-Riemann equations would imply the claim in $V_1 \cap \C_+$ for a sufficiently small $V_1$. Similarly we obtain the claim for $V_1 \cap \C_-$ by observing that $R_{1,+}(s) = - R_{1,-}(s)$. Consider the parabola $Q(x)=P(s(x),\eta,\tau)$ \eqref{eq:Q_poly}, where the Joukowsky map is defined in \eqref{eq:Joukowsky}. It is easy to see that

    \begin{equation}\label{Eq:PQequiv}
        P(s,\eta,\tau)<0, \quad  \forall\,s\in(\alpha,\beta)  \quad \Longleftrightarrow \quad Q(x)<0, \quad \forall \,x\in(\wt\beta,\wt\alpha)\,.
    \end{equation}
    For this to happen, one of the following has to be true

    \begin{enumerate}
            \item For $\tau \geq 0$ (upwards parabola or linear function)
        \begin{enumerate}
            \item $Q(\wt \beta)<0$ and $Q(\wt \alpha)<0$
        \end{enumerate}
        \item For $\tau<0$ (downward parabola)
        \begin{enumerate}
            \item $Q(x_v)<0$ where $x_v$ is the vertex of $Q(x)$
            \item $Q(\wt \alpha)<0$ and $x_v>\wt\alpha$
            \item $Q(\wt \beta)<0$ and $x_v<\wt\beta$
        \end{enumerate}
    \end{enumerate}
    These correspond exactly to the definition of $\cU_{c}^{(+)},\cU_{c}^{(-)}$ respectively.
\end{proof}
Applying the previous lemma, we are led to consider as the outer parametrix problem the model problem RHP \ref{OuterRHP_1cut} in its symmetric version, where $\ta \equiv \alpha$, $\tb\equiv \beta$ and $\Delta = (\ln(N)\zeta_c(\eta,\tau) + \omega_c(\eta,\tau))/(2\pi i)$. To approximate $v(z)$, we need to construct a local approximation near the endpoints of the $2$ intervals $(\alpha,\beta),\,(\beta^{-1},\alpha^{-1})$.

\begin{remark}
\label{rem:intersections_one_cut}
As before, the intersection points $D_{\beta},D_\alpha$ in the $(\eta,\tau)$-plane of $Q(\wt\beta)=0,Q(x_v)=0$ and of $Q(\wt\alpha)=0,Q(x_v)=0$ are important for the future analysis, and one can, with a little help of Mathematica, compute their coordinates explicitly as

\begin{equation}
    \begin{split}
        &D_\beta = (\eta_\beta,\tau_\beta)=\left(\frac{\beta  (\alpha  (\beta  (\alpha -3 \beta )-3)+\beta )}{2 \alpha  \left(\left(\beta ^2+1\right)^2 I_0-4 \beta  \left(\beta ^2 I_1+I_1-\beta  I_2\right)\right)},\; \frac{2 \beta ^2}{\left(\beta ^2+1\right)^2 I_0-4 \beta  \left(\beta ^2 I_1+I_1-\beta  I_2\right)}\right)\,,\\ 
        &D_\alpha=(\eta_\alpha,\tau_\alpha)=\left( \frac{\alpha  \left(-3 \left(\alpha ^2+1\right) \beta +\alpha  \beta ^2+\alpha \right)}{2 \beta  \left(\left(\alpha ^2+1\right)^2 I_0-4 \alpha  \left(\alpha ^2 I_1+I_1-\alpha  I_2\right)\right)}, \; \frac{2 \alpha ^2}{\left(\alpha ^2+1\right)^2 I_0-4 \alpha  \left(\alpha ^2 I_1+I_1-\alpha  I_2\right)} \right)\,,
\end{split}
\end{equation}

Note that as $Q(x, \eta_{\gamma}, \tau_\gamma)$ has a double root at $x=\wt \gamma$, with $\gamma \in \lbrace \alpha, \beta \rbrace$ and $\wt \gamma = \frac{1}{2}(\gamma+\gamma^{-1})$ as usual, it follows that $Q(x, \eta_\gamma, \tau_\gamma) = \tau_\gamma(x-\wt \gamma)^2$. As the integral condition \eqref{eq:IntQ} holds, it follows that $\tau_\gamma < 0$ (this was already implied by the fact that $Q(x,\eta,\tau)$ is a downward parabola by \eqref{Eq:PQequiv}). Additionally, as $(x-\wt \alpha)^2 > (x-\wt \beta)^2$ for all $x \in (1, \wt \beta)$, we can conclude (again using \eqref{eq:IntQ}) that $\tau_\beta < \tau_\alpha < 0$.

Next, using $\wt \gamma = - \frac{\eta_\gamma}{2\tau_\gamma} + \frac{\wt \alpha + \wt\beta}{4}$ for $\gamma=\alpha,\beta$, we can also determine the sign of $\eta_\gamma$. In summary we obtain 
\begin{equation}
\label{eq:point_ineq}
   \tau_\beta < \tau_\alpha <0\,, \ \text{and} \ \eta_\alpha>0,\, \ \text{sgn}(\eta_\beta) = \text{sgn}(3\wt\beta-\wt \alpha)\,.
\end{equation} 
\end{remark}
Before we proceed with the construction of the local parametrices, we need a technical Lemma. With a slight abuse of notation, we promote the vertex $x_v$ to $x_v(\eta,\tau)$, hopefully without confusion. Define $\eta_{\tau,\alpha}(\tau)\equiv \eta_{\tau,\alpha}$, $\eta_{\tau,\beta}(\tau)\equiv \eta_{\tau,\beta}$ the two functions $\tau\to\eta_{\tau,\alpha},\tau\to\eta_{\tau,\beta}$ such that $Q(\wt\alpha,\eta_{\tau,\alpha},\tau)=0, Q(\wt\beta,\eta_{\tau,\beta},\tau)=0$ respectively (that $\eta_{\tau, \alpha}, \eta_{\tau, \beta}$ exist and are unique is shown in the proof of Lem.~\ref{lem:building_Uc} below). Define also the function $\eta_{\tau}$  defined for $\tau\in(\tau_\beta,\tau_\alpha)$ mapping $\tau\to\eta_{\tau}$ such that $Q(x_v,\eta_{\tau}, \tau)=0$, we notice that this is not a function  since, in general, the equation $Q(x_v,\eta,\tau)=0$ has two solutions $\eta_1<\eta_2$, to define the function uniquely, we always consider the biggest root. The reason for this choice becomes clear due to the geometric interpretation of the curve $\eta_\tau$, which implies that $\tau \mapsto \eta_\tau$ is a concave function (see the proof of Lem.~\ref{lem:building_Uc} below). Given these definitions, the following holds
\begin{lemma}
\label{lem:building_Uc}
    Fix $0<\alpha<\beta<1$. Let $\eta_{\tau,\alpha}$, $\eta_{\tau,\beta}$ and $\eta_{\tau}$ as before, then the following holds
    
    \begin{equation}
        \cU_{c} = \cU_{c,\alpha}\cup\cU_{c,\beta}\cup\cU_{c,x_v}\,,
    \end{equation}
    where 

    \begin{equation}
        \cU_{c,\alpha} =\left\{(\eta,\tau)\;\vert\; \tau>\tau_\alpha,\, \eta<\eta_{\tau,\alpha}\right\}\,,\qquad  \cU_{c,\beta} =\left\{(\eta,\tau)\;\vert\; \tau<\tau_\beta,\, \eta<\eta_{\tau,\beta}\right\}\,,
    \end{equation}
    \begin{equation}
        \cU_{c,x_v} =\left\{(\eta,\tau)\;\vert\; \tau_\beta<\tau<\tau_\alpha,\, \eta<\eta_{\tau}\right\}\,.
    \end{equation}
\end{lemma}

\begin{proof}
    For a point $x \in [\wt \beta, \wt \alpha]$, we consider the sets 
    \begin{equation}
        H_x = \lbrace (\eta, \tau) \in \R^2 \colon Q(x, \eta, \tau) < 0 \rbrace, \quad T_x = \lbrace (\eta, \tau) \in \R^2 \colon Q(x, \eta, \tau) = 0 \rbrace
    \end{equation}
    which represent translated and rotated open half-planes and their boundaries, respectively, since $Q$ is linear in $\eta$ and $\tau$. It follows directly from its definition that the complete one-cut regime can be written as the intersection of these half-planes:
    \begin{equation}
        \cU_c = \bigcap_{x \in [\wt \beta, \wt \alpha]} H_x\,.
    \end{equation}
    As an intersection of convex sets, $\cU_c$ is itself convex. 
    
    Next, we show that $Q(x, \eta, \tau)$ is a monotonically increasing function in $\eta$ for $x\in [\wt \beta, \wt \alpha]$. Taking the partial derivative yields
    \begin{equation}
        \frac{\partial Q}{\partial \eta} = x - \frac{\int_1^{\beta^{-1}} x(s) \frac{\di s}{sR_1(s)}}{\int_1^{\beta^{-1}}  \frac{\di s}{sR_1(s)}} = x - \langle x(s) \rangle_{(1, \ \beta^{-1})}\,,
    \end{equation}
    where $\langle \, \cdot \,  \rangle_{(1, \ \beta^{-1})}$ denotes a weighted average. As $x(s) \in (1, \wt \beta)$ for $s \in (1, \beta^{-1})$ while $x\in [\wt \beta, \wt \alpha]$, it follows that $\frac{\partial Q}{\partial \eta} > 0$ for $x \geq \wt \beta$.
    
    Therefore, for each fixed $\tau$, there exists a unique threshold $\eta_\tau$ such that 
    \begin{equation}
        (\eta, \tau) \in \cU_c \quad \Longleftrightarrow \quad \eta < \eta_\tau\,.
    \end{equation}
    Because $\cU_c$ is convex, the mapping $\tau \mapsto \eta_\tau$ is concave, and $\cU_c$ simply represents the region under this graph. Furthermore, at the boundary $(\eta_\tau, \tau)$, there must be at least one $x_* \in [\wt \beta, \wt \alpha]$ such that the boundary condition becomes active, meaning $Q(x_*, \eta_\tau, \tau) = 0$, and thus the boundary point lies on $T_{x_*}$.
    
    Assuming this $x_* \in (\wt \beta, \wt \alpha)$ is unique, then $Q(x_*, \eta_\tau, \tau) = 0$ while $Q(x, \eta_\tau, \tau) < 0$ for all other $x \in [\wt \beta, \wt \alpha] \setminus \lbrace x_* \rbrace$. Since $Q(x)$ is a parabola, this isolated maximum at $x_*$ is only possible if the parabola opens downwards ($\tau < 0$) and its vertex $x_v(\eta_\tau, \tau) = -\frac{\eta_\tau}{2\tau} + \frac{\wt \alpha + \wt \beta}{4}$ coincides exactly with $x_*$. 
    
    Solving the equation $Q(x_v(\eta, \tau), \eta, \tau) = 0$ for $x_v \in [\wt \beta, \wt \alpha]$ traces a segment of an ellipse connecting the points $D_\beta$ and $D_\alpha$ for $\tau \in (\tau_\beta, \tau_\alpha)$. By the previous observations, this ellipse segment completely characterizes the boundary $\eta_\tau$ over this interval, defining the subset $\cU_{c,x_v}$.
    
    For $\tau$ outside the interval $(\tau_\beta, \tau_\alpha)$, the vertex $x_v$ falls outside the domain $[\wt \beta, \wt \alpha]$. The boundary is instead fixed by the endpoints of the spatial interval:
    \begin{itemize}
        \item For $\tau \geq \tau_\alpha$, the active constraint is exactly $Q(\wt \alpha, \eta, \tau) < 0$, which restricts $\eta < \eta_{\tau,\alpha}$ and corresponds to the subset $\cU_{c,\alpha}$.
        \item For $\tau \leq \tau_\beta$, the active constraint is $Q(\wt \beta, \eta, \tau) < 0$, which restricts $\eta < \eta_{\tau,\beta}$ and corresponds to the subset $\cU_{c,\beta}$.
    \end{itemize}
    As the graph of $\tau \mapsto \eta_\tau$ is concave and continuously covered by these three segments for all $\tau \in \R$, we conclude that $\cU_c = \cU_{c,\alpha} \cup \cU_{c,\beta} \cup \cU_{c,x_v}$.
\end{proof}
\begin{remark}\label{Rmk:eta1}
    Following Remark \ref{Rmk:eta3} we introduce the simpler notation
    \begin{align*}
        \eta_1(\tau) = \begin{cases}
            \eta_{\tau, \alpha}, \qquad &\text{for} \ \tau \geq \tau_\alpha
            \\
            \eta_{\tau}, \qquad &\text{for} \ \tau_\alpha \geq \tau \geq \tau_\beta
            \\
            \eta_{\tau, \beta}, \qquad &\text{for} \ \tau \leq \tau_\beta.
        \end{cases}
    \end{align*}
\end{remark}

\begin{figure}
    \centering
    \includegraphics[width=0.49\linewidth]{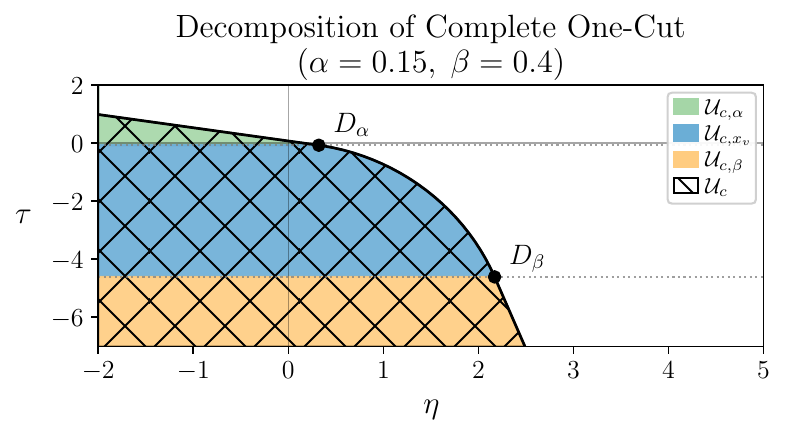}\hfill
    \includegraphics[width=0.49\linewidth]{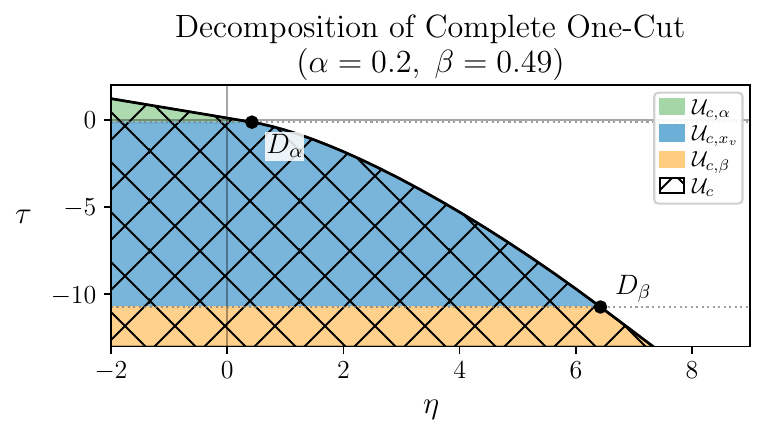}
     \caption{Decomposition of the complete one-cut region $\mathcal{U}_c$ for two distinct spectral parameters $(\alpha, \beta)$.}
    \label{fig:covering_one_cut}
\end{figure}

\begin{remark}
\label{rem:Uc_envelope}
    The set $\cU_c$ can also be interpreted through the geometry of affine lines. Specifically, the boundary of $\cU_c$ forms the envelope of the family of affine lines $T_x = \lbrace (\eta, \tau) \in \R^2 \colon Q(x, \eta, \tau) = 0 \rbrace$ parametrized by $x \in [\wt \beta, \wt \alpha]$. The ellipse segment connecting $D_\alpha$ and $D_\beta$, which acts as the upper boundary for $\cU_{c,x_v}$, is precisely the envelope curve where the lines $T_{x_*}$ are tangent to $\cU_c$. Meanwhile, the linear boundaries corresponding to $\cU_{c,\alpha}$ and $\cU_{c,\beta}$ are simply the specific affine lines $T_{\wt \alpha}$ and $T_{\wt \beta}$. Because the mapping $\tau \mapsto \eta_1$ is concave, these lines naturally extend as the one-sided tangents to the ellipse segment at the intersection points $D_\alpha$ and $D_\beta$.
\end{remark}
To further convince the reader, we report the three different sets in Figure \ref{fig:covering_one_cut} showing that they cover all the set $\cU_c$

Finally, we notice that the previous Lemma implies the following Corollary.

\begin{corollary}
\label{cor:techincal_cor}
    In the same hypothesis and notation as in Lemma \ref{lem:building_Uc}, the following holds

    \begin{equation}
    \begin{split}
                x_v(\eta_1,\tau)>\wt\alpha  \quad &\text{if  }\, 0>\tau>\tau_\alpha\\
                x_v(\eta_{1},\tau)<\wt\alpha  \quad &\text{if  }\, \tau>0\\
                x_v(\eta_{1},\tau)<\wt\beta  \quad &\text{if  }\, \tau<\tau_\beta
    \end{split}
    \end{equation}
\end{corollary}
\subsection{Local parametrices}
\label{subsec:one_cut_local_par}
In this subsection, we construct the local parametrices for the approximate solution. Since the construction is standard, we only construct such parametrix around the point $\alpha^{-1}$, all the others can be constructed in a similar way; furthermore, they do not contribute to the leading order asymptotic of the solution. 

In particular, we focus on a disk $\mathbb{D}(\alpha^{-1},\varepsilon)$ centered at $\alpha^{-1}$ of radius $\varepsilon>0$ small. Then we are left to find a matrix value function $M_{\alpha^{-1}}(z)$ that solves the following RHP ($\gamma_{1,2}$ are defined below)

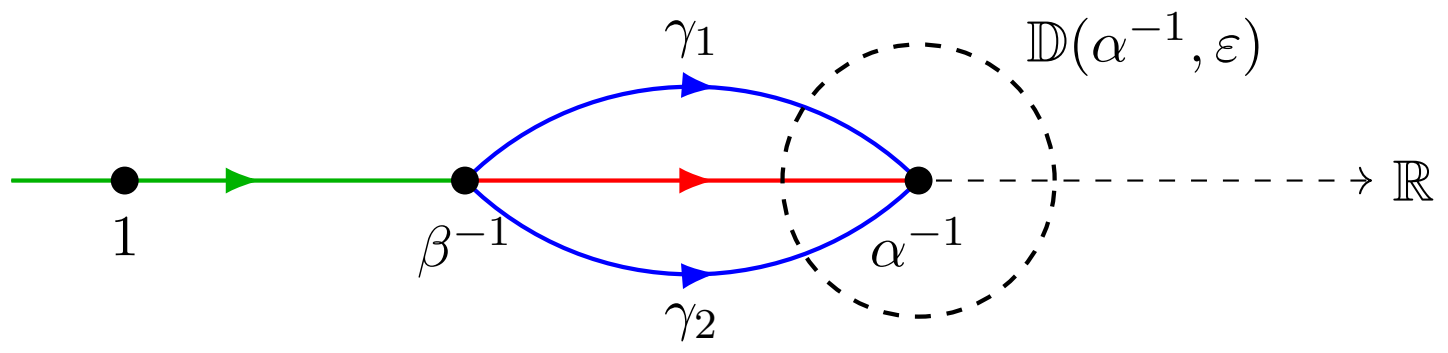
\begin{figure}[htbp]
    \centering
    \begin{tikzpicture}[
        decoration={markings, mark=at position 0.55 with {\arrow{Latex}}},
        scale=1.5
    ]
        % Draw the real axis
        \draw[dashed,thin, black, ->] (-2, 0) -- (4, 0) node[right] {$\mathbb{R}$};

        % Coordinates
        \coordinate (one) at (-1.5, 0);
        \coordinate (beta_inv) at (0, 0);
        \coordinate (alpha_inv) at (2, 0);

        % Gap
        \draw[thick, green!70!black, postaction={decorate}] (-2, 0) -- (beta_inv);
        
        % Cut
        \draw[thick, red, postaction={decorate}] (beta_inv) -- (alpha_inv);
        
        % Lenses
        \draw[thick, blue, postaction={decorate}] (beta_inv) to[out=45, in=135, looseness=1] (alpha_inv);
        \draw[thick, blue, postaction={decorate}] (beta_inv) to[out=-45, in=-135, looseness=1] (alpha_inv);

        % Labels
        \filldraw[black] (one) circle (1.5pt) node[below=3pt] {$1$};
        \filldraw[black] (beta_inv) circle (1.5pt) node[below=3pt] {$\beta^{-1}$};
        \filldraw[black] (alpha_inv) circle (1.5pt) node[below=3pt] {$\alpha^{-1}$};

        % Disk
        \draw[dashed, thick, black] (alpha_inv) circle (0.6);
        \node at (3, 0.6) {$\mathbb{D}(\alpha^{-1},\varepsilon)$};
        
        \node[above] at (1, 0.45) {$\gamma_1$};
        \node[below] at (1, -0.45) {$\gamma_2$};

    \end{tikzpicture}
    \caption{The contour for the local parametrix construction in the complete one-cut case.}
    \label{fig:local_zoom_complete_cut}
\end{figure}

\begin{rhp}
\label{rhp:local_alpha_inv}
    We seek for a matrix valued function $M_{\alpha^{-1}}(z)$ such that

    \begin{enumerate}
        \item $M_{\alpha^{-1}}(z)$ is analytic in $\mathbb{D}(\alpha^{-1},\varepsilon)\setminus\left((\beta^{-1},{\alpha^{-1}}) \cup \gamma_1 \cup \gamma_2\right)$
        \item $M_{\alpha^{-1}}(z)_+=M_{\alpha^{-1}}(z)_- J_{\alpha^{-1}}(z)$ where 

        \begin{equation}\label{eq:Malpha_Jump}
            J_{\alpha^{-1}}(z) = \begin{cases}
                 \pmtwo{1}{0}{ie^{2\ln(N)\int_{\alpha^{-1}}^z \frac{P(s,\eta,\tau)}{sR_{1}(s)}\di s + 2f(z) - \ln(2 \pi \mu_1 \rho(z^{-1})/z)}}{1}\qquad &z\in \gamma_1\cap\mathbb{D}(\alpha^{-1},\varepsilon)\\\\
                 \pmtwo{1}{0}{ie^{2\ln(N)\int_{\alpha^{-1}}^z \frac{P(s,\eta,\tau)}{sR_{1}(s)}\di s + 2f(z) - \ln(2 \pi \mu_1 \rho(z^{-1})/z)}}{1}\qquad &z\in \gamma_2\cap\mathbb{D}(\alpha^{-1},\varepsilon)\\\\
    
    \begin{pmatrix}
        0 & -i
        \\
        -i & 0
    \end{pmatrix} \qquad &z\in (\beta^{-1},\alpha^{-1})\cap\mathbb{D}(\alpha^{-1},\varepsilon)
            \end{cases}
        \end{equation}
    \end{enumerate}
\end{rhp}
Where $\gamma_1,\gamma_2$ are upper and lower lenses connecting $\beta^{-1}$ with $\alpha^{-1}$, the only restriction is that we have $\gamma_1\subset \mathbb{C}_+$ and $\gamma_2\subset \mathbb{C}_-$; for a sketch of the contours see Figure \ref{fig:local_zoom_complete_cut}. We notice that we do not impose any asymptotic condition at the boundary of the domain, this will be fixed later.

First, we notice that 

\begin{equation}
\begin{split}
    &\begin{pmatrix}
        1 & 0 \\
         i e^{2\ln(N)\int_{\alpha^{-1}}^z \frac{P(s,\eta,\tau)}{sR_{1}(s)}\di s + 2f(z) - \ln(2 \pi \mu_1 \rho(z^{-1})/z)} & 1
    \end{pmatrix} \\
    & \quad =\exp\left(-\sigma_3 \left(f(z) - \frac{\ln(2 \pi \mu_1 \rho(z^{-1})/z)}{2}\right)\right) \begin{pmatrix}
        1 &  0 \\
        i e^{2\ln(N)\int_{\alpha^{-1}}^z\frac{P(s,\eta,\tau)}{sR_{1}(s)}\di s} & 1
    \end{pmatrix}\\
    &\qquad \times\exp\left(\sigma_3 \left(f(z) - \frac{\ln(2 \pi \mu_1 \rho(z^{-1})/z)}{2}\right)\right)
\end{split}
\end{equation}
Therefore, the function $N_{\alpha^{-1}}(z) = M_{\alpha^{-1}}(z)\exp\left(-\sigma_3 \left(f(z) - \frac{\ln(2 \pi \mu_1 \rho(z^{-1})/z)}{2}\right)\right)$ solves the following RHP

\begin{rhp}
        We seek for a matrix valued function $N_{\alpha^{-1}}(z)$ such that

    \begin{enumerate}
        \item $N_{\alpha^{-1}}(z)$ is analytic in $\mathbb{D}(\alpha^{-1},\varepsilon)\setminus\left\{(\beta^{-1},{\alpha^{-1}}),\gamma_1, \gamma_2\right\}$
        \item $N_{\alpha^{-1}}(z)_+=N_{\alpha^{-1}}(z)_- \wt J_{\alpha^{-1}}$ where 

        \begin{equation}
            \wt J_{\alpha^{-1}}(z) = \begin{cases}
                \begin{pmatrix}
        1 &  0 \\
        ie^{ 2\ln(N)\int_{\alpha^{-1}}^z\frac{P(s,\eta,\tau)}{sR_{1}(s)}\di s} & 1
    \end{pmatrix}\qquad &z\in \gamma_1\cap\mathbb{D}(\alpha^{-1},\varepsilon)\\\\
    \begin{pmatrix}
        1 &  0 \\
        ie^{ 2\ln(N)\int_{\alpha^{-1}}^z\frac{P(s,\eta,\tau)}{sR_{1}(s)}\di s} & 1
    \end{pmatrix}\qquad &z\in \gamma_2\cap\mathbb{D}(\alpha^{-1},\varepsilon)\\\\
    
    \begin{pmatrix}
        0 & -i
        \\
        -i & 0
    \end{pmatrix} \qquad &z\in (\beta^{-1},\alpha^{-1})\cap\mathbb{D}(\alpha^{-1},\varepsilon)
            \end{cases}
        \end{equation}
    \end{enumerate}
\end{rhp}
Following a standard procedure \cite[Sect.~6]{Kuijlaars-2004}, we define the following conformal map 

\begin{equation}
    \lambda(z) = \frac{1}{4} \left(-\ln N\left(\int_{\alpha^{-1}}^z \frac{P(s,\eta,\tau)}{sR_1(s)}\di s\right)\right)^2\, \quad z \in \mathbb{D}(\alpha^{-1}, \varepsilon).
\end{equation}
Here we use that $P(\alpha^{-1}, \eta, \tau) < 0$ in the complete-cut region, see \eqref{Eq:PQequiv} and Lemma \ref{lem:whole_regime}. In particular we have $\lambda(\alpha^{-1}) = 0$ and $\lambda'(\alpha^{-1}) > 0$. We now choose the two contour $\gamma_1$ and $\gamma_2$ so that $\arg(\lambda(\gamma_1\cap \mathbb{D}(\alpha^{-1},\varepsilon))) \equiv \frac{2\pi}{3}$ and $\arg(\lambda(\gamma_2\cap \mathbb{D}(\alpha^{-1},\varepsilon))) \equiv -\frac{2\pi}{3}$, and  we aim to solve the following RHP in the $\lambda$-plane for the function $\fQ(\lambda)$
\begin{rhp}
        We seek for a matrix valued function $\fQ(\lambda)$ such that

    \begin{enumerate}
        \item $\fQ(\lambda)$ is analytic in $\mathbb{C}\setminus\{(-\infty,0),\wt\gamma_1,\wt\gamma_2\}$
        \item $\fQ_+(\lambda)=\fQ_-(\lambda) J_{\fQ}(\lambda)$ where 

        \begin{equation}
             J_{\fQ}(\lambda) = \begin{cases}
                \begin{pmatrix}
        1 & 0 \\
        i e^{ -4\lambda^{1/2}} & 1
    \end{pmatrix}\qquad &\lambda\in \wt\gamma_1\\\\
                \begin{pmatrix}
        1 & 0 \\
        i e^{ -4\lambda^{1/2}} & 1
    \end{pmatrix}\qquad &\lambda\in \wt\gamma_2\\\\
    
    \begin{pmatrix}
        0 & -i
        \\
        -i & 0
    \end{pmatrix} \qquad &\lambda\in (-\infty,0)
            \end{cases}
        \end{equation}
    \end{enumerate}
\end{rhp}
where the two contour $\wt\gamma_1,\wt\gamma_2$ are defined as

\begin{equation}
    \wt \gamma_1 = \left\{\lambda \in \mathbb{C}_+ \; \vert\; \arg(\lambda)=\frac{2\pi}{3}\right\}\,,\qquad \wt \gamma_2 = \left\{\lambda \in \mathbb{C}_- \; \vert\; \arg(\lambda)=-\frac{2\pi}{3}\right\}\,.
\end{equation}
One can reduce the previous RHP to a constant jump one by the following transformation
\begin{equation}
    \fQ(\lambda) = E(z(\lambda))\Psi_B(\lambda)\exp\left(\sigma_3\left(-2\lambda^{\frac{1}{2}} + i \frac{\pi}{4}\right)\right)\,,
\end{equation}
for some matrix-valued function $E(z)$ analytic in $\mathbb{D}(\alpha^{-1}, \varepsilon)$ which will be specified later. Then $\Psi_B(\lambda)$ solves the following RHP

\begin{rhp}
\label{rhp:psi}
        We seek for a matrix valued function $\Psi_B(\lambda)$ such that

    \begin{enumerate}
        \item $\Psi_B(\lambda)$ is analytic in $\mathbb{C}\setminus\left\{(-\infty,0),\wt\gamma_1,\wt\gamma_2\right\}$
        \item $\Psi_{B,+}(\lambda)=\Psi_{B,-}(\lambda) J_{\Psi_B}(\lambda)$ where 

        \begin{equation}
             J_{\Psi_B}(\lambda) = \begin{cases}
                \pmtwo{1}{0}{1}{1}\qquad &\lambda\in \wt\gamma_1\cup\wt\gamma_2\\\\
        \pmtwo{0}{1}{-1}{0}\qquad &\lambda\in (-\infty,0)
            \end{cases}
        \end{equation}
    \end{enumerate}
\end{rhp}

\begin{figure}[htbp]
    \centering
    \begin{tikzpicture}[
        decoration={markings, mark=at position 0.55 with {\arrow{Latex}}},
        scale=1
    ]
        \coordinate (O) at (0,0);
        
        % Rays
        \draw[thick, red, postaction={decorate}] (-3, 0) -- (O) node[pos=0.4, above=2pt] {$\pmtwo{0}{1}{-1}{0}$};
        
        \draw[thick, blue, postaction={decorate}] (-1.5, 2.6) -- (O) node[pos=0.5, right=4pt] {$\pmtwo{1}{0}{1}{1}$};
        \draw[thick, blue, postaction={decorate}] (-1.5, -2.6) -- (O) node[pos=0.5, right=4pt] {$\pmtwo{1}{0}{1}{1}$};
        
        % Labels
        \filldraw[black] (O) circle (1.5pt) node[right=4pt] {$0$};
        \node[above left] at (-1.5, 2.6) {$\wt\gamma_1$};
        \node[below left] at (-1.5, -2.6) {$\wt\gamma_2$};

    \end{tikzpicture}
    \caption{The contour for the Bessel parametric $\Psi_{B}(\lambda)$}
    \label{fig:bessel_par}
\end{figure}
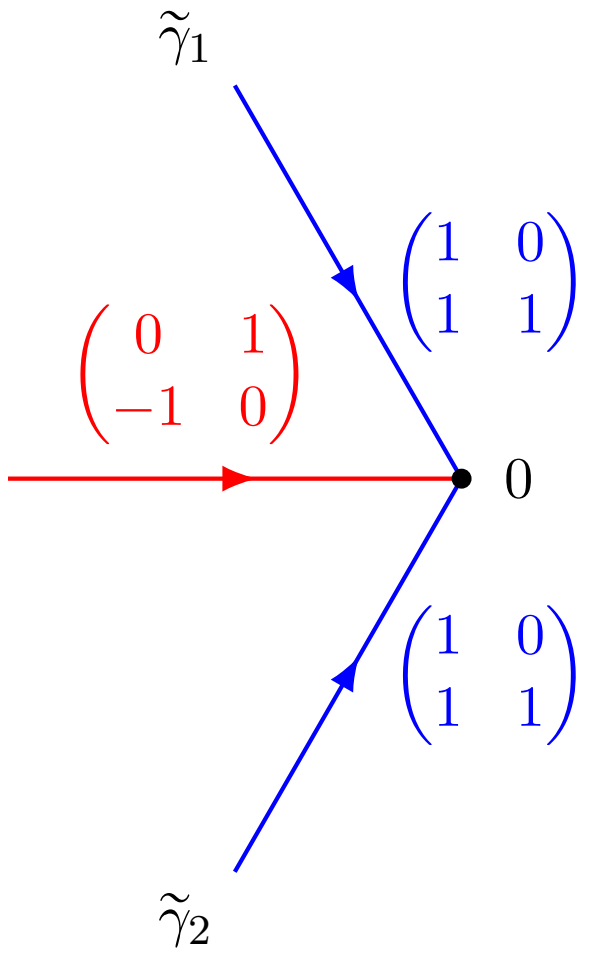

As we have not specified the normalization yet, the above problem has many solutions. The one which will allow us to match the local parametrix with the outer parametrix is explicitly given in \cite[Sect.~6]{Kuijlaars-2004} as 

\begin{equation}
\label{eq:local_explicit}
\Psi_B(\lambda) = \begin{cases}
\begin{pmatrix}
I_{0}(2 \lambda^{1/2}) & \frac{i}{\pi} K_{0}(2 \lambda^{1/2}) \\\\[1pt]
2 \pi i \lambda^{1/2} I_{0}'(2 \lambda^{1/2}) & -2 \lambda^{1/2} K_{0}'(2 \lambda^{1/2})
\end{pmatrix}, \quad & \arg(\lambda) \in (-2\pi/3,2\pi/3) \\\\[1pt]
\begin{pmatrix}
\frac{1}{2} H_{0}^{(1)}(2(-\lambda)^{1/2}) & \frac{1}{2} H_{0}^{(2)}(2(-\lambda)^{1/2}) \\\\[1pt]
\pi \lambda^{1/2} (H_{0}^{(1)})'(2(-\lambda)^{1/2}) & \pi \lambda^{1/2} (H_{0}^{(2)})'(2(-\lambda)^{1/2})
\end{pmatrix}, \quad & \arg(\lambda) \in (2\pi/3,\pi) \\\\[1pt]
\begin{pmatrix}
\frac{1}{2} H_{0}^{(2)}(2(-\lambda)^{1/2}) & -\frac{1}{2} H_{0}^{(1)}(2(-\lambda)^{1/2}) \\\\[1pt]
-\pi \lambda^{1/2} (H_{0}^{(2)})'(2(-\lambda)^{1/2}) & \pi \lambda^{1/2} (H_{0}^{(1)})'(2(-\lambda)^{1/2})
\end{pmatrix}, \quad & \arg(\lambda) \in (-\pi,-2\pi/3)
\end{cases}
\end{equation}
where $H_{0}^{(1)}$ and $H_{0}^{(2)}$ are the Hankel functions and $K_{0}$ and $I_{0}$ are the modified Bessel functions, see \cite[Ch.~10]{dlmf}. In \cite{Kuijlaars-2004}, it is shown that $\Psi_B$ satisfies RHP \ref{rhp:psi}, see Fig. \ref{fig:bessel_par}. We mention that $\Psi_B(\lambda) = \cO(\ln|\lambda|)$ for $\lambda \to 0$, giving us enough regularity for the small norm argument later on.

Given this explicit solution, one can unravel all the transformations and obtain an explicit expression for $M_{\alpha^{-1}}(z)$ as

\begin{equation}
    M_{\alpha^{-1}}(z) = E(z)\Psi_B(\lambda(z))\exp\left(\sigma_3\left(-2\lambda^{\frac{1}{2}}(z) + i\frac{\pi}{4} + f(z) - \frac{\ln(2\pi  \mu_1\rho(z^{-1})/z)}{2}\right)\right)\,,
\end{equation}
where $E(z)$ is an analytic function inside $\mathbb{D}(\alpha^{-1},\varepsilon)$ that we determine by imposing that the jump relation between the local parametrix and the outer parametrix $M(z)$ is small on the boundary of the disk $\mathbb{D}(\alpha^{-1},\varepsilon)$.

The uniform large-$\lambda$ asymptotics of the Bessel functions give us \cite[Eq.~(6.28)]{Kuijlaars-2004}
\begin{equation}\label{eq:PsiBAsym}
    \Psi_B(\lambda) = \left( 2 \pi \lambda^{1/2} \right)^{-\sigma_3/2} \, \frac{1}{\sqrt{2}} \begin{pmatrix}
1 + \mathcal{O}(\lambda^{-1/2}) & i + \mathcal{O}(\lambda^{-1/2})\\
i + \mathcal{O}(\lambda^{-1/2}) & 1 + \mathcal{O}(\lambda^{-1/2})
\end{pmatrix} e^{2 \lambda^{1/2} \sigma_3}.
\end{equation}
With this in mind we set $E(z)$ to be
\begin{align*}
    E(z) =  M(z) \exp\left(-\sigma_3\left( i\frac{\pi}{4} + f(z) - \frac{\ln(2\pi  \mu_1\rho(z^{-1})/z)}{2}\right)\right)\frac{1}{\sqrt{2}}\begin{pmatrix}
        1 & -i
        \\
        -i & 1
    \end{pmatrix}\left(2\pi \lambda^{1/2}(z)\right)^{\sigma_3/2}.
\end{align*}
Using $M_+(z) = M_-(z) \begin{pmatrix}
    0 & -i
    \\
    -i & 0
\end{pmatrix}$ and $\lambda_+^{1/4}(z) = i\lambda_-^{1/4}(z)$ for $z \in (\alpha^{-1} - \varepsilon, \alpha^{-1})$, together with \eqref{eq:f_factorization 2}, it follows that $E(z)$ is holomorphic in $\mathbb{D}(\alpha^{-1}, \varepsilon)$, implying that $M_{\alpha^{-1}}$ satisfies the jump condition \eqref{eq:Malpha_Jump}. Moreover, a direct computation using \eqref{eq:PsiBAsym} gives us
\begin{align*}
    M_{\alpha^{-1}}(z) = M(z)\left(I +  \cO(\lambda^{-1/2}(z))\right) = M(z)\left(I + \cO\left(\frac{1}{\ln(N)}\right)\right)
\end{align*}
as required.

\subsection{The complete one-cut solution}
\label{subsec:complete_one_cut_solution}

With the global outer parametrix $k_s(z)$ explicitly constructed in Section \ref{sec:model_problems} and the local Bessel parametrices handling the fourth-root branch point singularities as in Section \ref{subsec:one_cut_local_par}, we are now in the position to extract the leading-order asymptotics for the Toda lattice variables $a_n(t)$ and $b_n(t)$.

\subsubsection{Global parametrix and small-norm error analysis}
We define the global matrix parametrix $M(z)$ piecewise on $\C$:
\begin{equation}
    M(z) = \begin{cases}
        M_p(z), & z \in \D(p, \varepsilon) \quad \text{for } p \in \{\alpha, \beta, \beta^{-1}, \alpha^{-1}\}, \\[6pt]
        \frac{1}{2} \begin{pmatrix} k_s(z) + \ell_a(z) \\ k_s(z)- \ell_a(z) \end{pmatrix}, & z \in \C \setminus \bigcup_{p} \D(p, \varepsilon),
    \end{cases}
\end{equation}
where $M_p(z)$ are the local Bessel parametrices constructed in Section \ref{subsec:one_cut_local_par} (such as $M_{\alpha^{-1}}(z)$ in \eqref{eq:local_explicit}), and $k_s(z), \ell_a(z)$ are the symmetric and antisymmetric vector solutions to the model RHP \ref{OuterRHP_1cut} with parameters $\ta \equiv \alpha$, $\tb \equiv \beta$, and phase $\Delta\equiv \Delta_{c,n} = \frac{1}{2\pi i}(\ln(N)\zeta_c(\eta,\tau) + \omega_c(\eta,\tau))$. Note that $(1, 1) M(z) = k_s(z)$ in the outer region.

Following the Deift--Zhou non-linear steepest descent method \cite{Deift1993}, we introduce the error vector:
\begin{equation}\label{eq:rv}
    r(z) = \frac{1}{v_1(0)}v(z) M^{-1}(z), \qquad z \in \C \setminus \Sigma_r,
\end{equation}
(note the prefactor guarantees that $r_1(0) = 1$) where $\Sigma_r$ consists of the disk boundaries $\partial \D(p, \varepsilon)$ and the segments of the lenses and real axis outside the disks. By construction, the jump matrix $J_r(z) = M_-(z) J_v(z) M_+^{-1}(z)$ satisfies:
\begin{itemize}
    \item On the disk boundaries $\partial \D(p, \varepsilon)$ we have $J_v(z) = I$, and the asymptotic matching of the Bessel parametrix to the outer parametrix yields $J_r(z) = I + \cO\big(\frac{1}{\ln N}\big)$ as $N \to \infty$.
    \item On the lens contours outside the disks, Lemma \ref{lem:whole_regime} guarantees that the off-diagonal entries decay, specifically $J_r(z) = I + \cO(N^{-\delta})$ for some $\delta > 0$.
    \item On the gap $(\beta, \beta^{-1})$ outside the disks, the jump of $M(z)$ exactly matches $J_v(z) = e^{\sigma_3(\ln(N)\zeta_c + \omega_c)}$, so $J_r(z) = I$.
\end{itemize}
Consequently, we have the uniform estimate:
\begin{equation}
    \|J_r - I\|_{L^\infty(\Sigma_r)} = \cO\left( \frac{1}{\ln N} \right), \qquad N \to \infty.
\end{equation}
By the small-norm theory for vector Riemann--Hilbert problems found in Appendix \ref{App:SmallNorm} (see also the proof of Prop.~\ref{Prop:AsymSN} for more details) it follows that
\begin{equation}
    r(z) = (1, 1) + \cO\left( \frac{1}{\ln N} \right)
\end{equation}
uniformly for $z$ bounded away from $\Sigma_r$. 

\subsubsection{Asymptotic expansion of the vector solution near the origin}
We now unwind the transformations to recover the vector $q(z)$ in a neighborhood of the origin $z = 0$. Since the origin lies outside all disks and lenses ($0 < \alpha < \beta < 1$), we have $\wh q(z) = q(z) $. Furthermore, $v(z) = p(z) = \wh q(z) e^{\sigma_3(\ln(N)g(z) + f(z))}$. Therefore from \eqref{eq:rv} it follows that:
\begin{equation}
\label{eq:qApprox}
\begin{split}
    q(z) &= v_1(0)\left[r(z) M(z) + \cO\left( \frac{1}{\ln N} \right)\right]  e^{-\sigma_3(\ln(N)g(z) + f(z))}
    \\
    &= v_1(0)\left[k_s(z)  + \cO\left( \frac{1}{\ln N} \right)\right]e^{-\sigma_3(\ln(N)g(z) + f(z))}.
\end{split}
\end{equation}

From RHP \ref{RHP:q}, the exact vector $q(z)$ satisfies the normalization condition $q_1(0)q_2(0) = 1$ with $q_1(0) > 0$. Additionally, we have by construction $k_{s,1}(0; \Delta_{c,n}) = 1$. It now follows that $v_1^{2}(0)[k_{s,2}(0; \Delta_{c,n}) + \cO(1/\ln N)] = 1$. Solving for $v_1(0)$ and substituting this in above we obtain 
\begin{align}\label{eq:m1_asymptotic_expansion}
    q_1(z)&= \frac{1}{\sqrt{k_{s,2}(0;\Delta_{c,n})}} k_{s,1}(z;\Delta_{c,n}) e^{-\ln(N)g(z) - f(z)} \left( 1 + \cO\left( \frac{1}{\ln N} \right) \right)\,.
\end{align}

We now express $k_{s,2}(0;\Delta_{c,n})$ explicitly in terms of the Jacobi theta function $\theta_3$. Recall from Proposition \ref{prop:model_1cut_solutions} that:
\begin{equation}
    k_s(z) = \gamma(z)^{-1} \Big( E_s(\cA(z)), \, E_s(-\cA(z)) \Big), \qquad E_s(A) = c'(\Delta_{c,n}) \frac{\theta_3(2A - \Delta_{c,n} + \frac{1}{2} \mid 2B)}{\theta_3(2A + \frac{1}{2} \mid 2B)}.
\end{equation}
Let $A_0 \coloneqq \cA(0^{(1)}) \in (1/4, 1/2)$ be the Abel map of the origin on the upper sheet. Since $\lim_{z\to0} \gamma(z) = 1$, the normalization $k_{s,1}(0) = E_s(A_0) = 1$ uniquely determines the constant $c'(\Delta_{c,n})$:
\begin{equation}
    c'(\Delta_{c,n}) = \frac{\theta_3(2A_0 + \frac{1}{2} \mid 2B)}{\theta_3(2A_0 - \Delta_{c,n} + \frac{1}{2} \mid 2B)}.
\end{equation}
Evaluating the second component at $z = 0$, where $\cA(0^{(2)}) = -A_0$, gives:
\begin{equation}
    k_{s,2}(0;\Delta_{c,n}) = E_s(-A_0) = \frac{\theta_3(2A_0 + \frac{1}{2} \mid 2B)}{\theta_3(2A_0 - \Delta_{c,n} + \frac{1}{2} \mid 2B)} \frac{\theta_3(-2A_0 - \Delta_{c,n} + \frac{1}{2} \mid 2B)}{\theta_3(-2A_0 + \frac{1}{2} \mid 2B)}.
\end{equation}
Using the parity and periodicity properties of the Jacobi theta function \cite[20.4]{dlmf} we have:
\begin{align}
    &\theta_3\left(-2A_0 + \frac{1}{2} \;\middle|\; 2B\right) = \theta_3\left(2A_0 + \frac{1}{2} \;\middle|\; 2B\right), \\
    &\theta_3\left(-2A_0 - \Delta_{c,n} + \frac{1}{2} \;\middle|\; 2B\right)  = \theta_3\left(2A_0 + \Delta_{c,n} + \frac{1}{2} \;\middle|\; 2B\right).
\end{align}
Consequently $k_{s,2}(0;\Delta_{c,n})$ reads:
\begin{equation}
\label{eq:ks2_explicit_theta}
    k_{s,2}(0;\Delta_{c,n}) = \frac{\theta_3(2A_0 + \frac{1}{2} + \Delta_{c,n} \mid 2B)}{\theta_3(2A_0 + \frac{1}{2} - \Delta_{c,n} \mid 2B)}.
\end{equation}

\subsubsection{Asymptotics of $a^{(q)}_n(t)$ and $b^{(q)}_n(t)$ in the complete one-cut region}
Using the results of the previous section, we can now compute the asymptotic behavior of $a^{(q)}_n(t),b^{(q)}_n(t)$ in the complete one cut regime.

\paragraph{Asymptotics of $a^{(q)}_n(t)$.}
Evaluating \eqref{eq:m1_asymptotic_expansion} at $z = 0$ gives the factor $A^{(q)}(n,t)$:
\begin{equation}
\label{eq:A_nt_asymptotic}
    A^{(q)}(n,t) = q_1(0) = \sqrt{\frac{\theta_3(2A_0 + \frac{1}{2} - \Delta_{c,n} \mid 2B)}{\theta_3(2A_0 + \frac{1}{2} + \Delta_{c,n} \mid 2B)}} \exp\Big( \ln(N)g_\infty(\eta,\tau) + f_\infty \Big) \left( 1 + \cO\left( \frac{1}{\ln N} \right) \right)\,.
\end{equation}
 From Lemma \ref{lem:recovery_q}, the Toda coefficient $a^{(q)}_n(t)$ is recovered via $a^{(q)}_n(t) = \frac{1}{2} \frac{A^{(q)}(n,t)}{A^{(q)}(n+1,t)}$. Under the discrete shift $n \mapsto n+1$, the slow variable shifts as $\eta \mapsto \eta + \frac{1}{\ln N}$. Expanding the phase gives:
\begin{equation}
    \ln(N) g_\infty\left(\eta + \frac{1}{\ln N}, \tau\right) - \ln(N) g_\infty(\eta, \tau) = \partial_\eta g_\infty(\eta, \tau) + \cO\left( \frac{1}{\ln N} \right)\,.
\end{equation}
 Taking the ratio of \eqref{eq:A_nt_asymptotic} at $n$ and $n+1$, we obtain:
\begin{equation}
\label{eq:an_final_formula}
    a^{(q)}_n(t) = a_{\mathrm{bg}}(\eta,\tau) \sqrt{ \frac{\theta_3(2A_0 + \frac{1}{2} + \Delta_{c,n+1} \mid 2B) \theta_3(2A_0 + \frac{1}{2} - \Delta_{c,n} \mid 2B)}{\theta_3(2A_0 + \frac{1}{2} - \Delta_{c,n+1} \mid 2B) \theta_3(2A_0 + \frac{1}{2} + \Delta_{c,n} \mid 2B)} } \left( 1 + \cO\left( \frac{1}{\ln N} \right) \right).
\end{equation}
where we defined the background value $a_{\mathrm{bg}}(\eta,\tau)  = \frac{1}{2}\exp(-\partial_\eta g_\infty(\eta,\tau))$. Straightforward calculations show that
\begin{equation}
\label{eq:a_bg_explicit}
    a_{\mathrm{bg}}(\eta,\tau) = \frac{1}{2\alpha} \exp\left( - \int_0^\alpha \left( \frac{\frac{s+s^{-1}}{2} - \frac{I_1}{I_0}}{s R_1(s)} - \frac{1}{s} \right) \di s \right).
\end{equation}

\paragraph{Asymptotics of $b^{(q)}_n(t)$.}
To extract $b^{(q)}_n(t)$, we evaluate the $\cO(z)$ term of $q_1(z)$. From Lemma \ref{lem:recovery_q}, $q_1(z) = A^{(q)}(n,t)(1 - 2B^{(q)}(n-1,t)z + \cO(z^2))$, which implies:
\begin{equation}
    -2B^{(q)}(n-1,t) = \left. \frac{\di}{\di z} \ln q_1(z; n, t) \right|_{z=0}.
\end{equation}
Differentiating the logarithm of \eqref{eq:m1_asymptotic_expansion} at $z = 0$ gives:
\begin{equation}
\label{eq:log_deriv_m1}
    -2B^{(q)}(n-1,t) = \left. \frac{\di}{\di z} \ln k_{s,1}(z; \Delta_{c,n}) \right|_{z=0} - \ln(N) g'(0;\eta,\tau) - f'(0) + \cO\left( \frac{1}{\ln N} \right).
\end{equation}
Since $k_{s,1}(z) = \gamma(z)^{-1} E_s(\cA(z))$, we have $\ln k_{s,1}(z) = -\ln \gamma(z) + \ln E_s(\cA(z))$.
Using the value $\left. \frac{\di}{\di z} \ln \gamma(z) \right|_{z=0} = -\frac{1}{2}(\wt\alpha - \wt\beta)$, and differentiating $\ln E_s(A)$ with respect to $A$ via the chain rule with $\cA'(0) = \left.\frac{\di}{\di z} \cA(z)\right|_{z=0} = \frac{1}{2\tg K(k)}$, we find:
\begin{equation}
    \left. \frac{\di}{\di z} \ln k_{s,1}(z; \Delta_{c,n}) \right|_{z=0} = \frac{1}{2}(\wt\alpha - \wt\beta) + 2 \cA'(0) \frac{\theta_3'(2A_0 - \Delta_{c,n} + \frac{1}{2} \mid 2B)}{\theta_3(2A_0 - \Delta_{c,n} + \frac{1}{2} \mid 2B)} - 2 \cA'(0) \frac{\theta_3'(2A_0 + \frac{1}{2} \mid 2B)}{\theta_3(2A_0 + \frac{1}{2} \mid 2B)}.
\end{equation}
From Lemma \ref{lem:recovery_q}, 

\begin{equation}
    b^{(q)}_n(t) = B^{(q)}(n,t) - B^{(q)}(n-1,t) = \frac{1}{2} \left[ \left. \frac{\di}{\di z} \ln q_1(z; n, t) \right|_{z=0} - \left. \frac{\di}{\di z} \ln q_1(z; n+1, t) \right|_{z=0} \right]\,.
\end{equation}
Subtracting the expressions at $n$ and $n+1$, and defining the background value $b_{\mathrm{bg}}(\eta,\tau) = \frac{1}{2} \partial_\eta \big(g'(0;\eta,\tau)\big) $, we arrive at:
\begin{equation}
\label{eq:bn_final_formula}
    b^{(q)}_n(t) = b_{\mathrm{bg}}(\eta,\tau) + \cA'(0) \left[ \frac{\theta_3'(2A_0 + \frac{1}{2} - \Delta_{c,n} \mid 2B)}{\theta_3(2A_0 + \frac{1}{2} - \Delta_{c,n} \mid 2B)} - \frac{\theta_3'(2A_0 + \frac{1}{2} - \Delta_{c,n+1} \mid 2B)}{\theta_3(2A_0 + \frac{1}{2} - \Delta_{c,n+1} \mid 2B)} \right] + \cO\left( \frac{1}{\ln N} \right)\,,
\end{equation}
where $\theta_3'(v\mid B):=\frac{d}{d v}\theta_3(v\mid B)$. Simple algebraic manipulations show that

\begin{equation}
\label{eq:b_bg_explicit}
    b_{\mathrm{bg}}(\eta,\tau) = \frac{\wt\alpha + \wt\beta}{2} - \frac{I_1}{I_0}.
\end{equation}
We summarize our findings in the following theorem:

\begin{theorem}[Complete One-Cut Asymptotics]
\label{thm:complete_one_cut_asymptotics}
Let the macroscopic phase parameters $(\eta, \tau)$ lie strictly in the interior of the complete one-cut regime $\cU_c$ (Lemma \ref{lem:whole_regime}). As $N \to \infty$, the Toda lattice variables $a^{(q)}_n(t)$ and $b^{(q)}_n(t)$ with $n = \eta \ln N$ and $t = \tau \ln N$ are asymptotically given by a modulated genus-1 finite-gap elliptic wave:
\begin{align}
    a^{(q)}_n(t) &= a_{\mathrm{bg}}(\eta,\tau) \sqrt{ \frac{\theta_3(2A_0 + \frac{1}{2} + \Delta_{c,n+1} \mid 2B) \theta_3(2A_0 + \frac{1}{2} - \Delta_{c,n} \mid 2B)}{\theta_3(2A_0 + \frac{1}{2} - \Delta_{c,n+1} \mid 2B) \theta_3(2A_0 + \frac{1}{2} + \Delta_{c,n} \mid 2B)} } \left( 1 + \cO\left( \frac{1}{\ln N} \right) \right), \label{eq:thm_an_asymp} \\
    b^{(q)}_n(t) &= b_{\mathrm{bg}}(\eta,\tau) + \cA'(0) \left[ \frac{\theta_3'(2A_0 + \frac{1}{2} - \Delta_{c,n} \mid 2B)}{\theta_3(2A_0 + \frac{1}{2} - \Delta_{c,n} \mid 2B)} - \frac{\theta_3'(2A_0 + \frac{1}{2} - \Delta_{c,n+1} \mid 2B)}{\theta_3(2A_0 + \frac{1}{2} - \Delta_{c,n+1} \mid 2B)} \right] + \cO\left( \frac{1}{\ln N} \right), \label{eq:thm_bn_asymp}
\end{align}
where:
\begin{itemize}
    \item $\Delta_{c,n} = \Delta_{n,c}(n,t) = \frac{1}{2\pi i} (\ln(N)\zeta_c(\eta,\tau) + \omega_c)$ is the fast phase, with $\zeta_c(\eta,\tau)$ and $\omega_c$ defined in \eqref{eq:zeta_c} and \eqref{eq:omega_whole};
    \item $A_0 = \cA(0^{(1)}) \in (1/4, 1/2)$ is the Abel map of the origin on the first sheet of $\cR(\alpha,\beta)$, and $B \in i\R_+$ is the $\mathbf{b}$-cycle period \eqref{eq:B_period_g1};
    \item $\cA'(0) = \Gamma = \frac{1}{2\tg K(k)} = -\frac{1}{2 I_0}$ is the Abel map derivative at the origin, see Remark \ref{rem:elliptic_1};
    \item The background amplitudes $a_{\mathrm{bg}}(\eta,\tau)$ and $b_{\mathrm{bg}}(\eta,\tau)$ are given explicitly by:
    \begin{align}
        a_{\mathrm{bg}}(\eta,\tau) &= \frac{1}{2\alpha} \exp\left( - \int_0^\alpha \left( \frac{\frac{s+s^{-1}}{2} - \frac{I_1}{I_0}}{s R_1(s)} - \frac{1}{s} \right) \di s \right), \label{eq:thm_abg_explicit} \\
        b_{\mathrm{bg}}(\eta,\tau) &= \frac{\wt\alpha + \wt\beta}{2} - \frac{I_1}{I_0}. \label{eq:thm_bbg_explicit}
    \end{align}
\end{itemize}
\end{theorem}
An example of the complete one cut solution is displayed in Figure \ref{fig:evolution1}-\ref{fig:evolution2}.

\section{The two-cuts modulated region}
\label{sec:two_cuts}
In this section, we start investigating what happens when the construction in Section \ref{sec:complete_one_cut} fails. In particular, we describe the transition from the complete one-cut to a modulated region, where the $g$-function is supported on $4$ intervals with modulated extremes, which, creatively, we call \textit{two-cuts modulated region}.
In particular, we consider the following set in the $(\eta,\tau)$-plane 

\begin{equation}\label{eq:DefV}
    \left\{(\eta,\tau)\;\vert \; \tau_\beta<\tau<\tau_\alpha,\; \eta>\eta_{1} \right\}\cap \cU_{sn}^c\,,
\end{equation}\
where $\eta_{1}$ is defined in Remark \ref{Rmk:eta1} and $\cU_{sn}$ is defined in \eqref{eq:Usn}. We will not be able to fully describe this region, yet, but only part of it.
With the help of Figure \ref{fig:phase_diagram}, the reader might notice that we are considering the central stripe between $\tau_\beta,\tau_\alpha$ where the boundary of $\cU_c$ is exactly the ellipse $Q(x_v(\eta,\tau),\eta,\tau)=0$. 

 We proceed as in Section \ref{sec:complete_one_cut}, but we require that both the $g,f$-functions satisfy a slightly different RHP. We notice that we use again $g(z),f(z)$ to label the functions, hopefully without confusion.  As before, we consider the transformation \eqref{eq:def_transformation_1} and we require that $g(z),f(z)$ to satisfy the following RHPs

 \begin{rhp}
      \label{rhp:g_function_two_cuts}
      We seek for a scalar function $g(z)$ such that
      \begin{itemize}
            \item $g(z)$ is analytic in $\C\setminus \left(\alpha,\alpha^{-1}\right)$
            \item $g_+(z)+g_-(z)+1 + \tau\Phi(z,\eta,\tau) = 0$ for $z\in (\alpha,b)\cup(a,\beta)$
            \item $g_+(z)+g_-(z) - 1 + \tau\Phi(z,\eta,\tau) = 0$ for $z\in \left(\frac{1}{b},\frac{1}{\alpha}\right)\cup\left(\frac{1}{\beta},\frac{1}{a}\right)$
            \item $g_+(z) - g_-(z) = \zeta_{2,e}(\eta,\tau)$ for $z\in (b,a)\cup(a^{-1},b^{-1})$
            \item $g_+(z) - g_-(z) = \zeta_{2,c}(\eta,\tau)$ for $z\in (\beta,\beta^{-1})$
            \item $g(z^{-1}) = -g(z)$
            \item $g(z) = g_\infty + \cO(\frac{1}{z})$ as $z\to\infty$
            \item $g(z) + \frac{ 1+ \tau\Phi(z,\eta,\tau)}{2} = \cO((z-\alpha)^{1/2})$ as $z\to \alpha$ and $g(z) + \frac{ 1+ \tau\Phi(z,\eta,\tau)}{2} = \pm \frac{1}{2} \zeta_{2,c}+ \cO((z-\beta)^{1/2})$ as $z\to \beta$, where $\pm$ indicates that $z \in \mathbb{C}_\pm$
            \item $g(z) + \frac{ 1+ \tau\Phi(z,\eta,\tau)}{2} = \pm\frac{1}{2}\zeta_{2,e}+\cO((z-a)^{3/2}) $ as $z\to a$ and $g(z) + \frac{ 1+ \tau\Phi(z,\eta,\tau)}{2} = \pm\frac{1}{2}\zeta_{2,e}+\cO((z-b)^{3/2})$ as $z\to b$,  where $\pm$ indicates that $z \in \mathbb{C}_\pm$
\end{itemize}
here $0<\alpha\leq b \leq a \leq \beta<1$.
\end{rhp}

\begin{rhp}
    \label{rhp:f_function_two_cuts}
    We seek for a scalar function $f(z)$ such that

    \begin{itemize}
        \item $f(z)$ is analytic in $\C\setminus \left(\alpha,\alpha^{-1} \right)$
        \item $f_+(z)+f_-(z) + \ln(2\pi ) + \ln(z) + \ln(\mu_1\rho(z)) = 0$ for $z\in(\alpha,b)\cup (a,\beta)$
        \item $f_+(z)+f_-(z)  -\ln(2\pi ) + \ln(z) - \ln(\mu_1\rho(z^{-1})) = 0$ for $z\in(\beta^{-1},a^{-1})\cup (b^{-1},\alpha^{-1})$
        \item $f_+(z)-f_-(z) = \omega_{2,c}$ for $z\in(\beta,\beta^{-1})$
        \item $f_+(z) - f_-(z) = \omega_{2,e}$ for $z\in(b,a)\cup(a^{-1},b^{-1})\,. $
        \item $f(z^{-1}) = -f(z)$
        \item $f(z) = f_\infty + \cO(\frac{1}{z})$ as $z\to\infty$
    \end{itemize}
\end{rhp}

The following proposition gives an explicit expression of the $g$-function and exactly characterize the set $\cU_2$ in the $(\eta,\tau)$-plane where the analysis of this section is valid.

\begin{proposition}
\label{prop:g_two_cut}
Define the triangular domain $\wt\cU_2 = \{ (\wt b,\wt a) \mid 1 < \wt\beta \le \wt a \le \wt b \le \wt\alpha \}$ (where $\wt b = \frac{1}{2}(b + b^{-1})$ etc.) and the functions $g(z)$ and $\zeta_{2,c}(\eta,\tau),\zeta_{2,e}(\eta,\tau)$ as
\begin{align}
    &g(z) = \frac{1-\tau\Phi(z,\eta, \tau) }{2} + \int_{\alpha^{-1}}^z \Omega_2(s,\eta,\tau) \di s \,, \qquad \Omega_2(s,\eta,\tau) =\frac{R_2(s)\left( \frac{\tau}{2}(s +s^{-1}) + \gamma_2(\eta,\tau) \right)}{s} \,,\\
    &\zeta_{2,e}(\eta,\tau) = 2\int_{\alpha^{-1}}^{b^{-1}} \Omega_{2,+}(s,\eta,\tau)\di s\,, \qquad \zeta_{2,c}(\eta,\tau)=\zeta_{2,e}(\eta,\tau) + 2\int_{a^{-1}}^{\beta^{-1}} \Omega_{2,+}(s,\eta,\tau)\di s\,,
\end{align}
where
\begin{equation}
      \label{eq:def_R2_gamma2}
      R_2(z) = \frac{\left(z + \frac{1}{z} - b - \frac{1}{b}\right)^{1/2}\left(z + \frac{1}{z} - a - \frac{1}{a}\right)^{1/2}}{\left(z + \frac{1}{z} - \alpha - \frac{1}{\alpha}\right)^{1/2}\left(z + \frac{1}{z} - \beta - \frac{1}{\beta}\right)^{1/2}}, \qquad \gamma_2(\eta,\tau) =  \eta + \frac{\tau}{4}\left(a+b +\frac{1}{a} + \frac{1}{b} -\alpha-\beta - \frac{1}{\alpha} - \frac{1}{\beta}\right)\,.
\end{equation}
Then the map $T_2: \wt\cU_2 \to \R^2$, defined by assigning to each couple $(\wt b, \wt a)$ the unique pair $(\eta,\tau)$ that satisfies the linear system:
\begin{enumerate}

    \item \label{eq:two_cuts_gap_condition} $
    \begin{cases}
        \text{either }    \int_b^a \Omega_2(s,\eta,\tau) \di s = 0 \text{ for } b < a, 
        \\
       \text{or } 2\tau x_v + \eta - \frac{\tau}{4}\left(\alpha+\beta +\frac{1}{\alpha} + \frac{1}{\beta}\right) = 0 \text{ for } \ x_v = \wt a = \wt b,
    \end{cases}$  
    \item \label{eq:two_cuts_normalization} $  \int_\beta^1 \Omega_2(s,\eta,\tau) \di s = \frac{1}{2}$
\end{enumerate}
is well-defined and acts as a global homeomorphism onto its image $\cU_2 \coloneqq T_2(\wt\cU_2)$ (we recall that $x_v$ is the vertex of $Q(x)$ \eqref{eq:Q_poly}). Consequently, for any $(\eta,\tau) \in \cU_2$, there exists a unique couple $(\wt b(\eta,\tau),\wt a(\eta,\tau)) = T_2^{-1}(\eta,\tau) \in \wt\cU_2$ solving the above equations (which equivalently defines $a(\eta,\tau)$ and $b(\eta,\tau)$ with $0 < \alpha \le b \le a \le \beta < 1$), in which case $g(z)$ solves the RHP \ref{rhp:g_function_two_cuts}.
\end{proposition}
\begin{remark}
\label{rem:R_two_cuts}
 Notice that $R_2(z)$ is analytic in $\C\setminus\left\{(\alpha,b)\cup(a,\beta)\cup (\beta^{-1},a^{-1})\cup(b^{-1},\alpha^{-1})\right\}$, and we chose the branch-cut so that $R_2(z)\to 1$ as $z\to\infty$, this determines $R_2(z)$ uniquely.
    Furthermore, $R_{2,\pm}(z)\in i\R_\pm$ for $z\in(\alpha,b)\cup(\beta^{-1},a^{-1})$,  $R_{2,\pm}(z)\in i\R_\mp$ for $z\in(a,\beta)\cup(b^{-1},\alpha^{-1})$, and $R_{2}(z) > 0$ for $z \in (b, a) \cup (\beta, \beta^{-1}) \cup (a^{-1}, b^{-1})$.
\end{remark}
\begin{proof}
    We perform the analysis in the Joukowsky variables $x = \frac{1}{2}(s+s^{-1})$. The map is monotonically decreasing for $s\in(0,1)$, meaning the physical branch points map to the ordering $1 < \wt \beta \leq \wt a \leq \wt b \leq \wt \alpha$. Then conditions \ref{eq:two_cuts_gap_condition}, \ref{eq:two_cuts_normalization} above can be rewritten as:
    \begin{align}
    \label{eq:2cuts_sys}
        F_1(\wt a, \wt b; \eta, \tau) &=\begin{cases}
            \int_{\wt a}^{\wt b}  L(x,\eta,\tau)\di\mu_1(x) = 0, \quad &\text{ for } \wt a < \wt b
            \\
            L(x_v, \eta, \tau) = 0, \quad &\text{ for } x_v = \wt a = \wt b \
        \end{cases}\,,
        \\\label{eq:2cuts_sys2}
        F_2(\wt a, \wt b; \eta, \tau) &= \int_1^{\wt \beta} L(x,\eta,\tau) \di \mu_2(x) - \frac{1}{2} = 0\,,
    \end{align}
    where we defined the positive measure

    \begin{align}
    \di \mu_1(x) &= \frac{\sqrt{ x-\wt a}\sqrt{\wt b - x}}{\sqrt{\wt \alpha - x}\sqrt{x-\wt\beta}\sqrt{x^2-1}}\di x\, , \qquad x \in (\wt a, \wt b)\,,
    \\
    \di\mu_2(x)&= \frac{\sqrt{\wt a - x}\sqrt{\wt b - x}}{\sqrt{\wt \alpha - x}\sqrt{\wt\beta-x}\sqrt{x^2-1}}\di x\,, \qquad x \in (1, \wt \beta),
    \end{align}
    and the linear integrand
    \begin{equation}
           \label{eq:linear_2cuts}
        L(x,\eta,\tau) = \tau x + \eta + \frac{\tau}{2}\left(\wt a +\wt b - \wt\alpha - \wt\beta \right)\,.
    \end{equation}
First we need to show that $T_2 \colon \wt \cU_2 \to \cU_2$ is well-defined, i.e.~for each $(\wt b, \wt a) \in \wt U_2$ there is a unique solution $(\eta, \tau)$ of \eqref{eq:2cuts_sys}. Away from the boundary $\wt b=\wt a$, the nonlinear system of equations \eqref{eq:2cuts_sys}, can be rewritten in matrix form as follows

\begin{equation}\label{eq:MatrixMu}
    \pmtwo{\int_{\wt a}^{\wt b} \left(x + \frac{\wta +\wt b -\wt\alpha -\wt\beta}{2}\right)\di\mu_1}{\int_{\wt a}^{\wt b} \di\mu_1}{\int_{1}^{\wt\beta} \left(x + \frac{\wta +\wt b -\wt\alpha -\wt\beta}{2}\right)\di\mu_2}{\int_{1}^{\wt \beta}\di\mu_2} \vector{\tau}{\eta}= \vector{0}{\frac{1}{2}}\,.
\end{equation}
Therefore, we must show that the matrix on the left-hand side has nonzero determinant

\begin{equation}
    \det \pmtwo{\int_{\wt a}^{\wt b} \left(x + \frac{\wta +\wt b -\wt\alpha -\wt\beta}{2}\right)\di\mu_1}{\int_{\wt a}^{\wt b} \di\mu_1}{\int_{1}^{\wt\beta} \left(x + \frac{\wta +\wt b -\wt\alpha -\wt\beta}{2}\right)\di\mu_2}{\int_{1}^{\wt \beta}\di\mu_2} \neq 0\,,
\end{equation}
for $\wt a < \wt b$. Straightforward calculations show that
\begin{equation}
\label{eq:determinant}
\begin{split}
        \det \pmtwo{\int_{\wt a}^{\wt b} \left(x + \frac{\wta +\wt b -\wt\alpha -\wt\beta}{2}\right)\di\mu_1}{\int_{\wt a}^{\wt b} \di\mu_1}{\int_{1}^{\wt\beta} \left(x + \frac{\wta +\wt b -\wt\alpha -\wt\beta}{2}\right)\di\mu_2}{\int_{1}^{\wt \beta}\di\mu_2}& = \int_{1}^{\wt \beta}\di \mu_2(x)\int_{\wt a}^{\wt b}\di \mu_1(x)\left(\frac{\int_{\wt a}^{\wt b} x\di\mu_1(x)}{\int_{\wt a}^{\wt b} \di\mu_1(x)} -  \frac{\int_{1}^{\wt\beta} x\di\mu_2(x)}{\int_{1}^{\wt \beta} \di\mu_2(x)}\right) \\ &> \int_{1}^{\wt \beta}\di \mu_2(x)\int_{\wt a}^{\wt b}\di \mu_1(x)(\wt a -\wt\beta) \geq 0\,.
\end{split}
\end{equation}
Therefore, we proved that the map $T_2$ is well defined on $\wt\cU_2 \setminus \lbrace \wt a = \wt b \rbrace$. The case $\wt a = \wt b$ can be done similarly by dividing the first row by $\int_{\wt a}^{\wt b}\di \mu_1$ and then taking the limit as $\wt a -\wt b\to0$. This resolves the singularity and has the effect of substituting $\partial_\tau L(x_v, \eta, \tau)$ and $\partial_\eta L(x_v, \eta, \tau)$ (with $x_v = \wt a = \wt b$) into the first row of the matrix in \eqref{eq:MatrixMu}. 

To show continuity, observe that the entries of the matrix in \eqref{eq:MatrixMu} are continuous in $\wt \cU_2 \setminus \lbrace \wt a = \wt b \rbrace$, which together with the uniform bound on the determinant \eqref{eq:determinant} implies that $T_2|_{\wt \cU_2 \setminus \lbrace \wt a = \wt b \rbrace}$ is continuous. To see that $T_2$ extends to a continuous mapping on all of $\wt \cU_2$, one again uses the homogeneity of the first row to divide by $\int_{\wt a}^{\wt b} d\mu_1$ without changing the equation, in which case both entries can be \emph{continuously} extended to $\partial_\tau L(x_v, \eta, \tau)$ and $\partial_\eta L(x_v, \eta, \tau) =1$ for $\wt a \uparrow x_v$ and $\wt b \downarrow x_v$

Next, we want to show that $T_2$ is an homeomorphism. Note that since $T_2 \colon \wt \cU_2 \to T_2(\wt \cU_2)$ is by definition surjective, it suffices to show that $T_2$ is injective and continuous. In fact, as $\wt \cU_2$ is compact and $T_2(\wt \cU_2) \subset \R^2$ is Hausdorff, a standard result in topology (see for instance \cite[Section 26]{munkres2000topology}) tells us that any continuous bijection is already a homeomorphism.

Let us take an arbitrary $(\wt a_*, \wt b_*)$ in the interior of $\wt \cU_2$ and let $(\eta_*, \tau_*) = T_2(\wt a_*, \wt b_*)$. Define $C = \lbrace (\wt a, \wt b) \in \wt \cU_2 \colon F_1(\wt a, \wt b; \eta_*, \tau_*) = 0\rbrace$. Obviously $(\wt a_*, \wt b_*) \in C$. To prove injectivity of $T_2$ we need to show that on $C$ the function $F_2( \, \cdot \, , \, \cdot \ ; \eta_*, \tau_*)$ has only one zero (which necessary will be $(\wt a_*, \wt b_*)$). 

To this end, observe that the condition $F_1(\wt a, \wt b; \eta_*, \tau_*) = 0$ implies that $L(x; \eta_*, \tau_*)$ must change sign in $(\wt a, \wt b)$ (as $L$ is linear, this will be its only zero in $x$). Meanwhile, as $F_2 = 0$ is equivalent to 
\eqref{eq:2cuts_sys2}, it follows that at simultaneous solutions $(\wt a, \wt b)$ of $F_1|_{\eta_*, \tau_*} = F_2|_{\eta_*, \tau_*} = 0$, the integrand $L(x; \eta_*, \tau_*)$ must be strictly positive for $x < \wt \beta$, as we already localized its only zero to $(\wt a, \wt b)$. As at least one such simultaneous solution $(\wt a_*, \wt b_*)$ exists by assumption, it now follows that $L$ must be decreasing in $x$, or equivalently $\tau_* < 0$, which importantly will hold \emph{for all} $\wt a, \wt b$, as it is only a condition on $\tau_*$. Therefore whenever $F_1(\wt a, \wt b; \eta_*, \tau_*) = 0$, it follows that $L(\wt a,\eta_*,\tau_*) > 0$ and $L(\wt b,\eta_*,\tau_*) < 0$, independently of the sign of $F_2(\wt a, \wt b; \eta_*, \tau_*)$ (a corollary of this observation is that $T_2(\wt \cU_2)$ is fully contained in the half-plane $\lbrace (\eta, \tau) \colon \tau < 0 \rbrace$). 

The next key observation are the inequalities \begin{align}\label{Eq:PartFb}
    \partial_{\wt b} F_1 &= \frac{L(\wt b,\eta_*,\tau_*)}{2} \int_{\wt a}^{\wt b} \frac{\di \mu_1(x)}{\wt b-x} < 0\,, \qquad \partial_{\wt b} F_2 = \frac{L(\wt b,\eta_*,\tau_*)}{2} \int_{1}^{\wt \beta} \frac{\di \mu_2(y)}{\wt b-y} < 0\,,
\end{align}
and 
\begin{align} \label{Eq:PartFa}
    \partial_{\wt a} F_1 &= -\frac{L(\wt a,\eta_*,\tau_*)}{2} \int_{\wt a}^{\wt b} \frac{\di \mu_1(x)}{x-\wt a} < 0\,, \qquad \partial_{\wt a} F_2 = \frac{L(\wt a,\eta_*,\tau_*)}{2} \int_{1}^{\wt \beta} \frac{\di \mu_2(y)}{\wt a-y} > 0\,.
\end{align}
These formulas follow from the identities $2(x-\wt a)\partial_{\wt a} L(x; \eta, \tau) - L(x; \eta, \tau) =- L(\wt a; \eta, \tau)$ and $2(x-\wt b)\partial_{\wt b} L(x; \eta, \tau)- L(x; \eta, \tau) = -L(\wt b; \eta, \tau)$, while the inequalities follow from the signs $L(\wt a,\eta_*,\tau_*) > 0$ and $L(\wt b,\eta_*,\tau_*) < 0$ we just established.

By the implicit function theorem, it follows that the set $C$ can be locally parametrized by $\wt a \mapsto (\wt{b}(\wt a), \wt a)$ with $\partial_{\wt a} \wt b < 0$, i.e.~locally $C$ is a curve moving from lower-right to upper-left in the $(\wt b, \wt a)$-plane, with the sign of $F_1|_{\eta_*, \tau_*}$ being as indicated in Fig.~\ref{Fig:SignF}. As the equality $F_1(\wt a, \wt b(\wt a); \eta, \tau) = 0$ remains true on $C$, this curve can be extended until it connects the boundary $\lbrace (\wt b, \wt a) \colon \wt \beta < \wt a = \wt b < \wt \alpha \rbrace$ to $\lbrace (\wt \alpha, \wt a) \colon \wt a \in [\wt \beta, \wt \alpha)\rbrace \cup \lbrace (\wt b, \wt \beta) \colon \wt b \in  (\wt \beta, \wt \alpha]\rbrace$. Due to the local zero structure of $F_1|_{\eta_*, \tau_*}$ which holds near any point of $C$, it follows that the global sign structure of $F_1|_{\eta_*, \tau_*}$ must be as indicated in Fig.~\ref{Fig:SignF} (see the caption for the right image). In particular, the image of the parametrized curve $\wt a \mapsto (\wt b(\wt a), \wt a)$ coincides with $C$ (and is not merely a subset of it). 
\begin{figure}
\begin{center}
\begin{tikzpicture}[scale=1.5]

  % --- Left Diagram ---
  \begin{scope}[xshift=-4cm, yshift=1.5cm]
    % Dashed region boundary
    \draw[dashed, thick] (0,0) circle (1.8cm);
    
    % Solid splitting curve
    \draw[thick] (-1.2,1.5) .. controls (-0.8,-0.2) and (0.2,-0.2) .. (1.3,-1.5);
    
    % Point on the curve
    \coordinate (P1) at (-0.25, -0.12);
    \fill (P1) circle (1.2pt) node[right] {$(\widetilde{b}_*, \widetilde{a}_*)$};
    
    % Region labels
    \node at (0.2, 0.7) {$F_1 < 0$};
    \node at (-0.8, -0.7) {$F_1 > 0$};
  \end{scope}

  % --- Right Diagram ---
  \begin{scope}
    % Axes
    \draw[thick] (0,-0.5) -- (0,3.5) node[above] {$\widetilde{a}$};
    \draw[thick] (-0.8,0) -- (3.5,0) node[right] {$\widetilde{b}$};
    
    % Origin label
    \node[below left] at (0,0) {$(\widetilde{\beta}, \widetilde{\beta})$};
    
    % Axis ticks/labels
    \node[left] at (0,3) {$\widetilde{\alpha}$};
    \node[below] at (3,0) {$\widetilde{\alpha}$};
    
    % Dashed horizontal line to triangle tip
    \draw[dashed, thick] (0,3) -- (3,3);
    
    % Main Triangle: (0,0) to (3,3) to (3,0)
    \draw[thick] (0,0) -- (3,3) -- (3,0);
    
    % Internal splitting curve
    \draw[thick] (1.1, 1.1) .. controls (1.5, 0.8) and (2.0, 0.5) .. (2.5, 0);
    
    % Point on internal curve
    \coordinate (P2) at (1.7, 0.65);
    \fill (P2) circle (1.2pt) node[right, xshift=2pt] {$(\widetilde{b}_*, \widetilde{a}_*)$};
    
    % Dashed arrow from point to the upper region
    \draw[->, dashed, thick] (P2) .. controls (2.0, 1.5) .. (2.4, 2.0);
    
    % Signs (+ and -)
    \node[text=blue] at (1.4, 0.7) {$+$};
    \node[text=red] at (1.5, 1.0) {$-$};
    \node[text=red] at (2.6, 2.1) {$-$};
  \end{scope}

\end{tikzpicture}
\end{center}
\caption{\textbf{Left}: Local structure of the set $C$ is a neighborhood of $(\wt b_*, \wt a_*)$ (or in fact any of its interior points). \textbf{Right}: The set $C$ embedded in $\wt \cU_2$. Note that $F_1|_{\eta_*, \tau_*}$ cannot have a zero on the dashed path, as otherwise this would contradict the local picture on the left. It follows that $F_1 < 0$ in the upper-right part of $\wt \cU_2$ and analogously $F_1 > 0$ in the lower-left part. In particular, the embedded curved pictured coincides with the set $C$.\label{Fig:SignF}}
\end{figure}
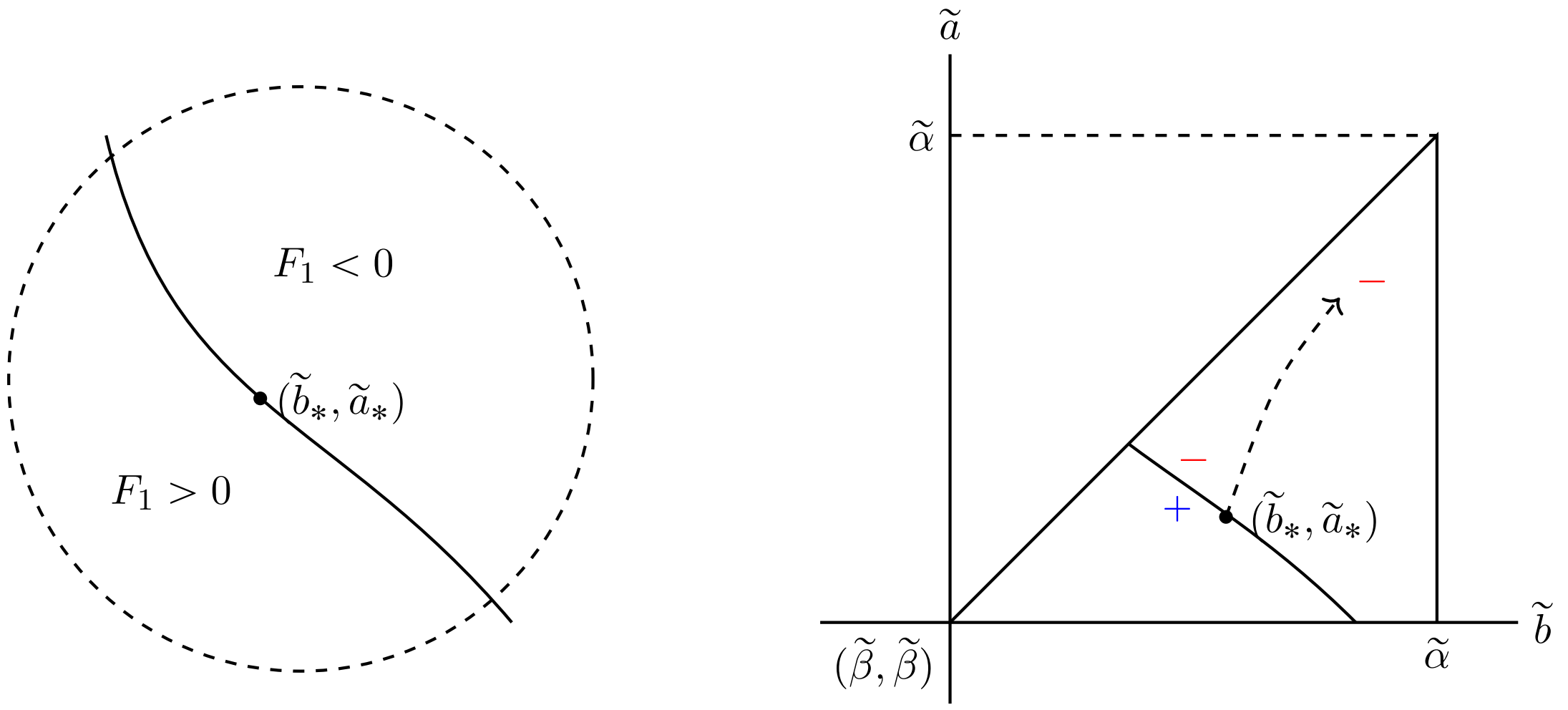

It remains to prove monotonicty of $F_2|_{\eta_*, \tau_*}$ on $C$. We obtain 
\begin{align*}
    \frac{d}{d\wt a} F_2(\wt a, \wt b(\wt a); \eta_*, \tau_*) = \underbrace{\partial_{\wt a}F_2}_{> \, 0} + \underbrace{\partial_{\wt b} F_2}_{< \, 0} \cdot \underbrace{\partial_{\wt a} \wt b}_{< \, 0} > 0,
\end{align*}
showing that indeed $(\wt b_*, \wt a_*)$ is the only solution of $F_1(\wt a, \wt b; \eta_*, \tau_*) = F_2(\wt a, \wt b; \eta_*, \tau_*) = 0$, thereby proving injectivity in the interior of $\wt \cU_2$. With few modifications (like just taking one sided derivatives in \eqref{Eq:PartFb}, \eqref{Eq:PartFa}) this proof can be extended to all of $\wt \cU_2$.
\end{proof}
The modulated two-cut region is strictly speaking only given by the interior $\mathring{\cU_2}$. However, the construction of the $g$-function on the boundary $\partial \cU_2$ will allow us to localize $\cU_2$ in the $(\eta,\tau)$-plane in relation to the other regions. In fact, the transition to the complete-one cut region is the topic of the next subsection 

\subsection{Transition between the complete one-cut region and the two-cuts modulated region}
 Next  we analyze the transition from the complete one-cut regime, occurring at the critical phase parameter $\eta = \eta_{\tau}$, see the proof of Lemma \ref{lem:building_Uc}. More precisely we consider the curve
 \begin{align*}
     \sigma_{x_v} = \left\{(\eta_1(\tau),\tau)\;\vert\; \tau_\beta<\tau<\tau_\alpha\right\}\,.
 \end{align*}
On this curve the 
vertex $x_v(\eta_1,\tau)$ of the complete one-cut polynomial $Q(x,\eta_1,\tau)$ touches zero in $(\wt \beta, \wt \alpha)$. In fact, choosing any $x_v \in (\wt \beta, \wt \alpha)$, consider the solution $(\eta_v, \tau_v)$ of \eqref{eq:2cuts_sys}, \eqref{eq:2cuts_sys2} for $\wt a = \wt b = x_v$:
\begin{align*}
 F_1(x_v, x_v; \eta_v, \tau_v) &= L(x_v, \eta_v, \tau_v) = 0\,,
 \\
    F_2(x_v, x_v; \eta_v, \tau_v) &= \int_1^{\wt \beta} \frac{(x_v-x) L(x, \eta, \tau)}{\sqrt{\wt \alpha - x}\sqrt{\wt \beta -x}\sqrt{x^2-1}} - \frac{1}{2} = 0.
\end{align*}
Comparing with \eqref{eq:IntQ}, we see that necessarily $(x-x_v)L(x,\eta_v,\tau_v) = Q(x, \eta_1(\tau_v), \tau_v)$. This implies that the curve $\sigma_{x_v}$ is equal to the boundary of the modulated two-cut region given by the vanishing of the gap $(\wt a,\wt b)$, i.e,~$\wt a = \wt b$. Alternatively, viewed from the inside of the complete cut region, the curve $\sigma_{x_v}$ corresponds to the birth of a gap. 

Next we want to study how the gap develops if starting on $\sigma_{x_v}$ we increase $\eta > \eta_1(\tau)$ (for a fixed $\tau$). Unfortunately, the first constraint \eqref{eq:2cuts_sys} has different forms depending on whether $\wt a = \wt b$ or $\wt a < \wt b$. To track the solution continuously, we introduce regularized variables: the center $\wt c = \frac{1}{2}(\wt a + \wt b)$ and the half-width $\varepsilon = \frac{1}{2}(\wt b-\wt a)$. The constraint $F_1$ for $\varepsilon > 0$ is rewritten using the substitution $x = \wt c + \varepsilon u$ as 
    \begin{equation}
        F_1(\wt c,\varepsilon; \eta, \tau) = \varepsilon^2\int_{-1}^1 \frac{\sqrt{1-u^2}}{W_1(\wt c+\varepsilon u)} L(\wt c+\varepsilon u,\eta,\tau) \di u\,\,, \quad W_1(x) = \sqrt{\wt \alpha - x}\sqrt{x-\wt\beta}\sqrt{x^2-1}\,.
    \end{equation}
    We define the regularized gap constraint $\wt F_1(\wt c, \varepsilon; \eta, \tau) = \varepsilon^{-2} F_1(c, \varepsilon; \eta, \tau)$:
    \begin{equation}
        \wt F_1(\wt c, \varepsilon; \eta, \tau) = \int_{-1}^1 \frac{\sqrt{1-u^2}}{W_1(\wt c+\varepsilon u)} L(\wt c+\varepsilon u, \eta, \tau) \di u\,.
    \end{equation}
    As a sanity check, one can verify that, after applying the dominated convergence Theorem,

    \begin{equation}
        \lim_{\varepsilon\to 0^+} =  \wt F_1(\wt c, \varepsilon; \eta, \tau) = \frac{\pi}{2 W_1(\wt c)} \left( \tau \wt c + \eta - \frac{\tau}{2}(\wt \alpha + \wt \beta) \right) \ \propto \  L(\wt c, \eta, \tau),
    \end{equation}
    consistent with \eqref{eq:2cuts_sys}. In particular, as $L(x_v, \eta_1, \tau) = 0$, it follows that $\partial_{\wt c} \wt F_1(\wt c, 0; \eta_1, \tau)|_{\wt c = x_v} = \frac{\tau \pi}{2 W_1(x_v)} < 0$ (recall that $\cU_2 \subset \lbrace (\eta, \tau) \colon \tau < 0\rbrace$).
    
    Next we introduce the squared half-width $\delta = \varepsilon^2$, and then write the constraint $F_2$ in these regularized coordinates. The roots in the numerator become $\sqrt{(\wt a - x)(\wt b - x)} = \sqrt{(\wt c-x)^2 - \delta}$. Thus, the second condition reads:
\begin{equation}
    F_2(\wt c, \delta; \eta, \tau) = \int_1^{\wt \beta} \frac{\sqrt{(\wt c-x)^2 - \delta}}{W_2(x)} L(x, \wt c; \eta, \tau) \di x - \frac{1}{2} = 0\,,
\end{equation}
where $W_2(x) = \sqrt{\wt \alpha - x}\sqrt{\wt \beta - x}\sqrt{x^2-1}$ and we introduced $L(x, \wt c; \eta, \tau) = \tau x + \eta + \tau\left(\wt c - \frac{\wt\alpha + \wt\beta}{2} \right)$ to emphasize the $\wt c$-dependence of $L$. We then compute $\partial_{\wt c} F_2(\wt c, 0; \eta_{\tau}, \tau)|_{\wt c = x_v(\eta_1, \tau)}$ obtaining that
(abbreviating $x_v(\eta_1, \tau)$ as $x_v$ here and in the following)
\begin{equation}
    \partial_{\wt c} F_2(\wt c, 0; \eta_{1}, \tau)|_{\wt c = x_v} = \int_1^{\wt \beta} \frac{\overbrace{L(x, x_v; \eta_1, \tau) + (x_v-x)\tau}^{L(x_v, x_v; \eta_1, \tau) = 0}}{W_2(x)} \di x = 0\,.
\end{equation}
Differentiating $F_2$ with respect to the squared width $\delta$ yields (using again that $\tau < 0$):
\begin{equation}
    \partial_\delta F_2(x_v, \delta; \eta_{\tau}, \tau)|_{\delta = 0} = \int_1^{\wt \beta} \frac{-1/2}{x_v-x} \frac{\overbrace{L(x, x_v; \eta_1, \tau)}^{\tau(x-x_v)}}{W_2(x)} \di x = \frac{\tau}{2} \int_1^{\wt \beta} \frac{1}{W_2(x)} \di x < 0\,.
\end{equation}
Additionally, we have $\partial_\eta F_2(x_v, 0; \eta, \tau)|_{\eta= \eta_1} = \int_1^{\wt \beta} \frac{x_v-x}{W_2(x)} \di x > 0$. Let us now denote by $\wt c_\tau(\eta)$, $\delta_\tau(\eta)$ the unique solution of $F_1(\wt c, \delta; \eta, \tau) = F_2(\wt c, \delta; \eta, \tau) = 0$ with $(\eta, \tau) \in \cU_2$ (existence and uniqueness in the equivalent $(\wt b, \wt a)$-coordinates has been established in Prop.~\ref{prop:g_two_cut}). In fact, as the determinant of the regularized Jacobian $\det J_{\text{reg}} = \partial_{\wt c} \wt F_1 \cdot \partial_\delta F_2 - \partial_\delta \wt F_1 \cdot \partial_{\wt c} F_2$ satisfies $\det J_{\text{reg}}|_{x_v, 0, \eta_1, \tau} = (\partial_{\wt c} \wt F_1)(\partial_\delta F_2)|_{x_v, 0, \eta_1, \tau} > 0$, it follows that $\wt c_\tau, \delta_\tau$ are in fact differentiable in $\eta$ sufficiently close to $\eta_1$.
As observed earlier, we have $\delta_\tau(\eta_1) = 0$ and $\wt c_\tau(\eta_1) = x_v(\eta_1, \tau)$. 
From
\begin{align*}
    0 = \frac{d}{d\eta} F_2(\wt c_\tau(\eta), \delta_\tau(\eta); \eta, \tau) = (\partial_{\wt c} F_2) \wt c'_\tau(\eta) + (\partial_{\delta} F_2) \delta_\tau'(\eta) + \partial_\eta F_2
\end{align*}
it follows that $\delta_\tau'(\eta)|_{\eta = \eta_1} = - \frac{\partial_\eta F_2}{\partial_{\delta} F_2}|_{\eta=\eta_1} > 0$. In particular, as we always have $\delta \geq 0$, we see that the region $\cU_2$ lies to the right of the complete one-cut region $\cV_{x_v}$, and that at least close to the boundary the gap width strictly opens ($\delta_\tau$ increases) as $\eta$ moves beyond the critical threshold $\eta_{\tau}$. More precisely we will have $\wt b- \wt a = 2\sqrt{\delta_\tau(\eta)} \, \propto \,  \sqrt{\eta - \eta_1}$.

Next we will analyze the two remaining boundary arcs of $\partial \cU_2$. Consider the boundary of the triangle $\wt \cU_2$ contained in the line $\wt a =\wt \beta$; along this curve the map $T$ is well defined and we can view the corresponding $\eta$, $\tau$ as functions of $\wt b$. By implicit differentiation we get
\begin{equation}
    J \vector{\tau'(\wt b)}{\eta'(\wt b)} = \pmtwo{\partial_\tau F_1}{\partial_\eta F_1}{\partial_\tau F_2}{\partial_\eta F_2} \vector{\tau'(\wt b)}{\eta'(\wt b)} = -\vector{\partial_{\wt b} F_1}{\partial_{\wt b}F_2}\,.
\end{equation}
As we already observed in the proof of Proposition \ref{prop:g_two_cut}, the gap constraint $F_1 = 0$ implies that $L(x,\eta,\tau)$ must change sign inside $(\wt a, \wt b)$, and in fact $L(\wt a,\eta,\tau) > 0$ and $L(\wt b,\eta,\tau) < 0$. Furthermore, recall that the partial derivatives of $F_1$ with respect to $\wt b$ evaluate to (cf.~\ref{Eq:PartFb}):
\begin{equation}
    \partial_{\wt b} F_1 = \frac{L(\wt b,\eta,\tau)}{2} \int_{\wt\beta}^{\wt b} \frac{\di \mu_1(x)}{\wt b-x} < 0\,, \qquad \partial_{\wt b} F_2 = \frac{L(\wt b,\eta,\tau)}{2} \int_{1}^{\wt \beta} \frac{\di \mu_2(y)}{\wt b-y} < 0\,.
\end{equation}
Using Cramer's rule and expanding the numerator as a double integral, we isolate $\tau'(\wt b)$:
\begin{equation}
\begin{split}
    \tau'(\wt b) &= \frac{-\partial_{\wt b} F_1 \partial_\eta F_2 + \partial_{\wt b} F_2 \partial_\eta F_1}{\det J} \\
    &= \frac{L(\wt b,\eta,\tau)}{2 \det J} \int_{\wt\beta}^{\wt b} \di \mu_1(x) \int_{1}^{\wt\beta} \di \mu_2(y) \left( -\frac{1}{\wt b-x} + \frac{1}{\wt b-y} \right) \\
    &= \frac{-L(\wt b,\eta,\tau)}{2 \det J} \int_{\wt\beta}^{\wt b}\int_{1}^{\wt\beta} \frac{x-y}{(\wt b-x)(\wt b-y)} \di \mu_1(x) \di \mu_2(y) > 0\,,
\end{split}
\end{equation}
where $\det J > 0$ was established in \eqref{eq:determinant}. Because $x \in (\wt\beta, \wt b)$ and $y \in (1, \wt\beta)$, we strictly have $x > y$. Thus, the double integral evaluates to a strictly positive quantity, and since $-L(\wt b,\eta,\tau) > 0$, we conclude that $\tau'(\wt b)$ is strictly positive.

An analogous calculation applies to the boundary $\wt b = \wt \alpha$. Along this curve, implicit differentiation yields $J (\tau'(\wt a), \eta'(\wt a))^\mathsf{T} = -(\partial_{\wt a} F_1, \partial_{\wt a} F_2)^\mathsf{T}$. We recall that the partial derivatives with respect to $\wt a$ are (cf.~\ref{Eq:PartFb}):
\begin{equation}
    \partial_{\wt a} F_1 = -\frac{L(\wt a,\eta,\tau)}{2} \int_{\wt a}^{\wt \alpha} \frac{\di \mu_1(x)}{x-\wt a} < 0\,, \qquad \partial_{\wt a} F_2 = \frac{L(\wt a,\eta,\tau)}{2} \int_{1}^{\wt \beta} \frac{\di \mu_2(y)}{\wt a-y} > 0\,.
\end{equation}
Using Cramer's rule to isolate $\tau'(\wt a)$ yields:
\begin{equation}
\begin{split}
    \tau'(\wt a) &= \frac{-\partial_{\wt a} F_1 \partial_\eta F_2 + \partial_{\wt a} F_2 \partial_\eta F_1}{\det J} \\
    &= \frac{L(\wt a,\eta,\tau)}{2 \det J} \int_{\wt a}^{\wt \alpha} \di \mu_1(x) \int_{1}^{\wt \beta} \di \mu_2(y) \left( \frac{1}{x-\wt a} + \frac{1}{\wt a-y} \right) \\
    &= \frac{L(\wt a,\eta,\tau)}{2 \det J} \int_{\wt a}^{\wt \alpha}\int_{1}^{\wt \beta} \frac{x-y}{(x-\wt a)(\wt a-y)} \di \mu_1(x) \di \mu_2(y) > 0\,.
\end{split}
\end{equation}
Again, since $x \in (\wt a, \wt \alpha)$ and $y \in (1, \wt \beta)$, we strictly have $x > y$, making the double integral strictly positive. Since $L(\wt a,\eta,\tau) > 0$, the entire expression evaluates to a strictly positive quantity.

Because $\tau'(\wt b) > 0$ on the first boundary and $\tau'(\wt a) > 0$ on the second boundary, both boundary curves are strictly monotonic in the $\tau$-coordinate. In particular, on the boundary given by $\wt a = \wt \beta$, we can invert $\tau(\wt b)$, and write instead $\wt b(\tau)$, with $\tau \in (\tau_\beta, \tau_{\text{sn}})$ (we justify the appearance of $\tau_{\text{sn}}$ further below). This gives us a parametrization
\begin{align*}
    \tau \in (\tau_\beta, \tau_{\text{sn}}] \mapsto (\eta(\wt b(\tau)), \tau). 
\end{align*}
Similarly we can obtain a parametrization of the second boundary via 
\begin{align*}
    \tau \in [\tau_{\text{sn}}, \tau_\alpha) \mapsto (\eta(\wt a(\tau)), \tau).
\end{align*}
 We now introduce the notation 
 \begin{align} \label{eq:eta2}
   \eta_2(\tau) =   \begin{cases}
         \eta(\wt b(\tau)) \ \ \text{for} \ \  \tau \in (\tau_\beta, \tau_{\text{sn}}],
         \\
         \eta(\wt a(\tau)) \ \ \text{for} \ \  \tau \in [\tau_{\text{sn}}, \tau_a).
     \end{cases}
 \end{align}
It follows that the two-cut region can be characterized by
 \begin{align*}
     \cU_2 = \lbrace (\eta, \tau) \colon \eta_1(\tau) < \eta < \eta_2(\tau) \ \text{with}\ \tau \in (\tau_\beta, \tau_\alpha) \rbrace.
 \end{align*}
 Note that we necessarily have $\eta(\wt b(\tau_{\text{sn}})) = \eta(\wt a(\tau_{\text{sn}}))$, as both $(\eta, \tau)$ coordinates lead to the same (uniquely solvable) conditions $F_1 = F_2 = 0$ corresponding to $\wt a = \wt \beta$ and $\wt b = \wt \alpha$. Writing out these conditions explicitly (note that $d\mu_1(x)$, $d\mu_2(x)$ simplify and the integral becomes explicitly solvable), we obtain 
 \begin{align*}
     0 &= F_1(\wt \beta, \wt \alpha; \eta, \tau) = \int_{\wt \beta}^{\wt \alpha} L(x, \eta, \tau) \frac{\di x }{\sqrt{x^2-1}} = -\frac{1}{2}\tau \wt \Phi(\wt \alpha, \eta, \tau) + \frac{1}{2}\tau \wt \Phi(\wt \beta, \eta, \tau)
     \\
     0 &= F_2(\wt \beta, \wt \alpha; \eta, \tau) = \int_{1}^{\wt \beta} L(x,  \eta, \tau) \frac{\di x }{\sqrt{x^2-1}} - \frac{1}{2} = -\frac{1}{2}\tau \wt \Phi(\wt \beta, \eta, \tau) - \frac{1}{2},
 \end{align*}
where we introduced $\wt \Phi(x, \eta, \tau) = -2\sqrt{x^2-1} + 2\frac{\eta}{\tau}\ln(x - \sqrt{x^2-1}) = \Phi(s, \eta, \tau)$, where $s = x - \sqrt{x^2-1}$, $x \in (1, \infty)$. Comparing to \eqref{eq:Dsn} we see that in fact $(\eta_{\text{sn}}, \tau_{\text{sn}}) = D_{\text{sn}}$ solves the above system of equations.

 See Figure \ref{fig:phase_diagram} for a few illustrative examples. We summarize our findings in the following remark:
 \begin{remark}
\label{rem:big_remark}
    The previous proof establishes the following properties of the phase space:
    \begin{enumerate}
        \item \textbf{Corner Mappings:} The vertices of the triangular domain map precisely to the critical transition points in the physical plane: $T_2(\wt\alpha,\wt\beta) = D_{\text{sn}}$, $T_2(\wt\alpha,\wt\alpha)=D_{\alpha}$, and $T_2(\wt\beta,\wt\beta)=D_{\beta}$.
        \item \textbf{Boundary Curves:} There exist two continuous boundary curves,
        \begin{align*}
            \wt a &\mapsto (\eta(\wt a), \tau(\wt a)), \qquad \wt a \in (\wt \beta, \wt \alpha), 
            \\
            \wt b &\mapsto (\eta(\wt b), \tau(\wt b)), \qquad b \in (\wt \beta, \wt \alpha)
        \end{align*}
        connecting $D_{\alpha}$ to $D_{\text{sn}}$ and $D_{\beta}$ to $D_{\text{sn}}$, respectively. On the first curve, $\wt b= \wt \alpha$ while on the second, $\wt a=\wt \beta$. Furthermore, $\tau(\wt a)$ strictly increases on the path from $D_{\text{sn}}$ to $D_\beta$, while $\tau(\wt b)$ strictly decreases on the path from  $D_{\text{sn}}$ to $D_\alpha$. 
        \item \textbf{Generalized Hodograph:} The technique utilized to invert this system is a direct application of the Tsarev Generalized Hodograph Method \cite{tsarev1985poisson}, which diagonalizes the Whitham modulation equations by interchanging the physical variables $(\eta, \tau)$ with the two modulated extremes $(\wt a, \wt b)$.
    \end{enumerate}
\end{remark}
\begin{remark}
    We notice that the previous proof implies that at the critical point $D_{\text{sn}}$ there is a transition from the 2-cut regime to the small norm regime; to the best of our knowledge, this is the first time this situation is encountered.
\end{remark}

\begin{figure}
    \centering
    \includegraphics[scale=0.6]{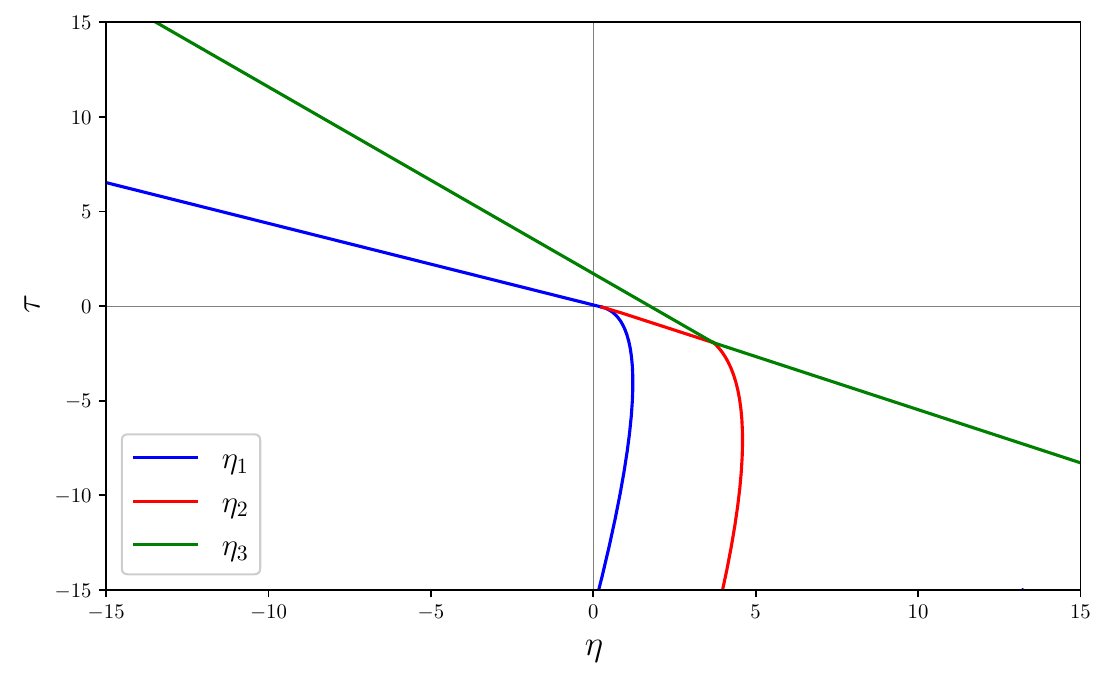}
    \caption{Example of the boundary curve $\eta_1,\eta_2,\eta_3$ for $\alpha=0.14,\, \beta=0.75$. The red boundary will eventually hit the blue boundary at the point $D_{\beta}=(\eta_\beta,\tau_\beta)$, and terminate there.}
    \label{fig:boundary}
\end{figure}

\begin{remark}\label{Rmk:eta2}
    With the introduction of the map $[\tau_\beta, \tau_\alpha] \ni \tau \mapsto \eta_2(\tau)$, we have now completed the collection of functions $\tau \mapsto \eta_j(\tau)$, $j = 1,2,3$ (cf.~Remark.~\ref{Rmk:eta3}, \ref{Rmk:eta1}), that fully describe the boundaries of the complete-cut, two-cut, one-cut modulated and small norm regions (the one-cut modulated region will be studied in Sect.~\ref{sec:one_cut_modulated_regions}). We have $\eta_1(\tau) < \eta_3(\tau)$ for $\tau \in \mathbb R$, and additionally $\eta_1(\tau) < \eta_2(\tau) < \eta_3(\tau)$ for $\tau \in (\tau_\beta, \tau_\alpha)$, explaining the order of numbering. See Figure \ref{fig:boundary} for a pictorial representation of the boundaries.
\end{remark}

Regarding the function $f(z)$ defined in RHP \ref{rhp:f_function_two_cuts}, we can prove the following 
\begin{proposition}
\label{prop:sol_f_two_cuts}
Define the cuts as $\Sigma_2 = (\alpha, b) \cup (a, \beta)$, and $\Sigma_2^{-1} = (\beta^{-1},a^{-1})\cup(b^{-1},\alpha^{-1})$, and define the boundary function:
\begin{equation}
    H(s) =
        \ln(2\pi \mu_1\rho(s)) \,,
\end{equation}
and the square-root function

\begin{equation}
    R_4(z) = \left(z+\frac{1}{z} - \alpha - \frac{1}{\alpha}\right)^{1/2}\left(z+\frac{1}{z} - \beta - \frac{1}{\beta}\right)^{1/2}\left(z+\frac{1}{z} - a - \frac{1}{a}\right)^{1/2}\left(z+\frac{1}{z} - b - \frac{1}{b}\right)^{1/2}
\end{equation}
Then, the explicit solution to RHP \ref{rhp:f_function_two_cuts} is given by: 
\begin{equation}
\begin{split}
       f(z) & = \frac{R_4(z)}{2\pi i} \left( \int_{\Sigma_2^{-1}} \frac{H(s^{-1})(s+z)}{R_{4,+}(s)2s(s-z)} \di s -\int_{\Sigma_2} \frac{H(s)(s+z)}{R_{4,+}(s)2s(s-z)} \di s - \int_{\Sigma_2\cup \Sigma_2^{-1}} \frac{\ln(s)(s+z)}{R_{4,+}(s)2s(s-z)} \di s \right) \\
       & + \frac{R_4(z)\omega_{2,c}}{2\pi i} \int_{\beta}^{\beta^{-1}} \frac{s+z}{2s(s-z)R_{4}}(s)\, \di s\\
       & + \frac{R_4(z)\omega_{2,e}}{2\pi i}\left( \int_{b}^a \frac{s+z}{2s(s-z)R_{4}}(s)\, \di s +  \int_{a^{-1}}^{b^{-1}} \frac{s+z}{2s(s-z)R_{4}(s)}\, \di s\right)\,.
\end{split}
\end{equation}
Where, defining 

\begin{align}
    & \tm_1 = -\int_{\Sigma_2^{-1}} \frac{H(s^{-1})}{2R_{4,+}(s)s} \di s + \int_{\Sigma_2} \frac{H(s)}{2R_{4,+}(s)s} \di s + \int_{\Sigma_2\cup \Sigma_2^{-1}} \frac{\ln(s)}{2R_{4,+}(s)s} \di s\,, \\[4pt]
   &\tm_2 = -\int_{\Sigma_2^{-1}} \frac{H(s^{-1})}{R_{4,+}(s)} \di s + \int_{\Sigma_2} \frac{H(s)}{R_{4,+}(s)} \di s + \int_{\Sigma_2\cup \Sigma_2^{-1}} \frac{\ln(s)}{R_{4,+}(s)} \di s\,,
\end{align}
 $\omega_{2,c},\omega_{2,e}$ are such that they solve the following system of equations

\begin{equation}
\label{eq:system_omega}
    \begin{cases}
        \omega_{2,c} \int_{\beta}^{\beta^{-1}} \frac{1}{2sR_{4}(s)}\, \di s + \omega_{2,e} \left( \int_{b}^a \frac{1}{2sR_{4}(s)}\, \di s +  \int_{a^{-1}}^{b^{-1}} \frac{1}{2sR_{4}(s)}\, \di s\right)  = \tm_1\,, \\
                \omega_{2,c} \int_{\beta}^{\beta^{-1}} \frac{1}{R_{4}(s)}\, \di s + \omega_{2,e} \left( \int_{b}^a \frac{1}{R_{4}(s)}\, \di s +  \int_{a^{-1}}^{b^{-1}} \frac{1}{R_{4}(s)}\, \di s\right)  = \tm_2\,. \\
    \end{cases}
\end{equation}

\end{proposition}
\begin{remark}
    We notice that the solution to \eqref{eq:system_omega} always exists and is unique. In fact using that $R_4(s^{-1}) = R_4(s)$, we obtain that $\int_\beta^{\beta^{-1}} \frac{1}{2sR_4(s)} \di s = \int_\beta^1 \frac{1}{sR_4(s)} \di s$ and $\int_\beta^{\beta^{-1}} \frac{1}{R_4(s)} \di s = \int_\beta^1 \frac{s+s^{-1}}{sR_4(s)} \di s$. It follows that \eqref{eq:system_omega} is equivalent to 
\begin{align*}
     \begin{cases}
        \omega_{2,c} \int_{\beta}^{1} \frac{1}{sR_{4}(s)}\, \di s + \omega_{2,e}  \int_{b}^a \frac{1}{sR_{4}(s)}\, \di s  = \tm_1\,, \\
                \omega_{2,c} \int_{\beta}^{1} \frac{s+s^{-1}}{sR_{4}(s)}\, \di s + \omega_{2,e}  \int_{b}^a \frac{s+s^{-1}}{sR_{4}(s)} = \tm_2\,. \\
    \end{cases}
\end{align*}
This system is invertible if and only if
\begin{align*}
    \frac{\int_{\beta}^{1} \frac{s+s^{-1}}{2sR_{4}(s)}\, \di s}{\int_{\beta}^{1} \frac{1}{sR_{4}(s)}\, \di s} \not = \frac{ \int_{b}^a \frac{s+s^{-1}}{2sR_{4}(s)}\, \di s}{ \int_{b}^a \frac{1}{sR_{4}(s)}\, \di s}.
\end{align*}
The left-hand side can be viewed as an average of $\wt s = (s+s^{-1})/2$ and therefore lies in the interval $(\wt \beta, 1)$, while the right-hand side must analogously lie in the interval $(\wt a, \wt b)$. Both intervals are disjoint.
\end{remark}

\begin{proof}
The proof proceeds by verifying that the explicit integral representation satisfies all the conditions of the Riemann-Hilbert Problem \ref{rhp:f_function_two_cuts}. We utilize the symmetric Cauchy kernel $\omega(s,z)\di s = \frac{s+z}{2s(s-z)}\di s$, which possesses the inversion symmetry $\omega(s^{-1}, z^{-1})d(s^{-1}) = \omega(s, z)\di s$.

By construction, $f(z)$ is the Cauchy transform of bounded functions integrated over the segments $\Sigma_2 \cup \Sigma_2^{-1}$ and the gaps, multiplied by the root function $R_4(z)$. Consequently, $f(z)$ is analytic in $\C \setminus [\beta^{-1}, \beta]$. 

To verify the anti-symmetry $f(z^{-1}) = -f(z)$, we leverage the properties $R_4(z^{-1}) = R_4(z)$ and $R_{4,+}(s^{-1}) = R_{4,-}(s) = -R_{4,+}(s)$ for $s$ on the cuts. Applying the substitution $s = u^{-1}$ (which maps $\Sigma_2^{-1} \to \Sigma_2$ and reverses the contour orientation, yielding an additional minus sign), the integral over $\Sigma_2^{-1}$ evaluated at $z^{-1}$ exactly maps to the integral over $\Sigma_2$ evaluated at $z$. The same mechanism applies to the gap integrals. The combination of these domains guarantees that $f(z^{-1}) = -f(z)$ globally.

To verify the jump relations, one has just to use the Sokhotski-Plemelj formula for the symmetric kernel.
Finally, we require $f(z) = f_\infty + \mathcal{O}(z^{-1})$ as $z \to \infty$. The root function expands as $R_4(z) = \mathcal{O}(z^2)$. Therefore, the integral bracket must be $\mathcal{O}(z^{-2})$ to suppress the growth. 
We Taylor expand the Cauchy kernel as $z \to \infty$:
\begin{equation}
    \frac{s+z}{2s(s-z)} = -\frac{1}{2s}\left(1 + \frac{2s}{z} + \frac{2s^2}{z^2} + \dots \right) = -\frac{1}{2s} - \frac{1}{z} + \mathcal{O}(z^{-2})\,.
\end{equation}
Letting $\phi(s)$ denote the complete weighted measure from all integral terms in $f(z)$, the expansion of $f(z)$ reads:
\begin{equation}
    f(z) = \frac{R_4(z)}{2\pi i} \left[ -\left(\int \frac{\phi(s)}{2s} ds\right) - \frac{1}{z}\left(\int \phi(s) ds\right) + \mathcal{O}(z^{-2}) \right]\,.
\end{equation}
For $f(z)$ to be bounded at infinity, the $\mathcal{O}(z^0)$ and $\mathcal{O}(z^{-1})$ coefficients inside the bracket must vanish. Setting $\int \frac{\phi(s)}{2s} ds = 0$ directly generates the first equation of \eqref{eq:system_omega} (equating the $\omega$ terms to $\tm_1$). Setting $\int \phi(s) ds = 0$ directly generates the second equation (equating the $\omega$ terms to $\tm_2$). 

Thus, the condition that $f(z) = \mathcal{O}(1)$ uniquely fixes the gap constants through the non-degenerate linear system \eqref{eq:system_omega}. 
\end{proof}
We can now proceed and explicitly apply the two previous propositions to the jump matrix $J_p(z)$ \eqref{eq:jump_p}, obtaining the following jump matrix

            \begin{equation}
                  J_{p}(z) =\begin{cases}
                   \begin{pmatrix}
                      e^{2\ln(N)\int_\alpha^z\Omega_{2,+}(s)\di s + f_+-f_-} & 0 \\
                      i & e^{-2\ln(N)\int_\alpha^z\Omega_{2,+}(s)\di s -f_+ + f_-} 
                  \end{pmatrix} &\quad z\in(\alpha,b)\\\\
                  \begin{pmatrix}
                      e^{\ln(N)\zeta_{2,e}(\eta,\tau) + \omega_{2,e} }& 0\\
                      \Xi_1(z)e^{2\ln(N)\int_a^z\Omega_2(s,\eta,\tau)\di s} &  e^{-\ln(N)\zeta_{2,e}(\eta,\tau) - \omega_{2,e}}
                  \end{pmatrix} &\quad z\in (b,a)\\\\
                  \begin{pmatrix}
                      e^{\ln(N)\left(2\int_\beta^z \Omega_{2,+}(s)\di s + \zeta_{2,c}(\eta,\tau)\right) + f_+-f_-} & 0 \\
                      i & e^{-\ln(N)\left(2\int_\beta^z \Omega_{2,+}(s)\di s + \zeta_{2,c}(\eta,\tau)\right) - f_++f_-}
                  \end{pmatrix}&\quad z\in(a,\beta)\\\\
                        \begin{pmatrix}
                            e^{\ln(N)\zeta_{2,c}(\eta,\tau) + \omega_{2,c}} & 0\\
                            0 &  e^{-\ln(N)\zeta_{2,c}(\eta,\tau) - \omega_{2,c} }
                        \end{pmatrix} &\quad z\in \left(\beta,\beta^{-1}\right) \\\\
                        \begin{pmatrix}
                            e^{\ln(N)\left(\zeta_{2,e}(\eta,\tau) + 2 \int_{a^{-1}}^z\Omega_{2,+}(s)\di s\right) + f_+-f_-} & -i\\
                            0 & e^{-\ln(N)\left(\zeta_{2,e}(\eta,\tau) + 2 \int_{a^{-1}}^z\Omega_{2,+}(s)\di s\right) - f_+ + f_-} 
                        \end{pmatrix} &\quad z\in\left(\beta^{-1},a^{-1}\right)\\\\[5pt]
                        \begin{pmatrix}
                            e^{\ln(N) \zeta_{2,e}(\eta,\tau) + \omega_{2,e}} & \Xi_2(z) e^{- 2\ln(N)\int_{b^{-1}}^z\Omega_2(s)\di s} \\
                            0  & e^{-\ln(N) \zeta_{2,e}(\eta,\tau) - \omega_{2,e}}
                        \end{pmatrix}&\quad  z\in \left(a^{-1},b^{-1}\right)\\\\[5pt]
                         \begin{pmatrix} 
                              e^{f_+ - f_- + 2\ln(N)\int_{\alpha^{-1}}^z \Omega_{2,+}(s)\di s} & -i \\ 0 & e^{-(f_+ - f_-) -  2\ln(N)\int_{\alpha^{-1}}^z \Omega_{2,+}(s)\di s}
                        \end{pmatrix}  &\quad z \in \left(b^{-1},\alpha^{-1}\right)
                  \end{cases}
            \end{equation}
             where we defined $\Xi_1(z) = 2\pi i z\mu_1\rho(z)e^{f_+(z)+f_-(z)}, \Xi_2(z) = -2\pi i \frac{\mu_1\rho(z^{-1})}{z}e^{-f_+(z)-f_-(z)}$, and we suppressed the dependence of $\Omega_2(s,\eta,\tau)$ from $\eta,\tau$ to shorten the notation.

\begin{remark}
\label{rem:sym_off_diag}
    Notice that, in view of Proposition \ref{prop:g_two_cut} 

    \begin{align}
        &\int_a^z\Omega_2(s,\eta,\tau)\di s = - \int_{z}^b\Omega_2(s,\eta,\tau)\di s \qquad z\in (b,a)\,,\\
        & \int_{z}^{a^{-1}} \Omega_2(s,\eta,\tau)\di s = - \int_{b^{-1}}^z\Omega_2(s,\eta,\tau)\di s \qquad z\in (a^{-1},b^{-1})\,.
    \end{align}
\end{remark}

\subsection{Lenses opening}
\label{subsec:Lenses_two_cuts}
We are now in position to open the lenses around the four cuts. Following the factorizations introduced previously, we factor the jump matrix $J_{p}(z)$ across the four segments $\Sigma_2\cup\Sigma_2^{-1}$.

Using the ``UTU'' factorization \eqref{eq:UTU} on the left-side cuts $(\alpha, b)$ and $(a, \beta)$, we deduce:
\begin{equation}
\begin{split}
    &\begin{pmatrix}
        e^{2\ln(N)\int_{\alpha}^z \Omega_{2,+}(s)\di s + f_+(z) -f_-(z)}& 0 \\
        i& e^{-2\ln(N)\int_{\alpha}^z \Omega_{2,+}(s)\di s - f_+(z) + f_-(z)}
    \end{pmatrix} = \\ 
    &\begin{pmatrix}
        1 &  -ie^{-2 f_-(z) - \ln(2 \pi z \mu_1 \rho(z))  -2\ln(N)\int_{\alpha}^z \Omega_{2,-}(s)\di s} \\
        0 & 1
    \end{pmatrix}\\
    &\times
    \begin{pmatrix} 0 & i \\ i & 0 \end{pmatrix} 
    \begin{pmatrix}
        1 &  -ie^{-2 f_+(z) - \ln(2 \pi z \mu_1 \rho(z))  -2\ln(N)\int_{\alpha}^z \Omega_{2,+}(s)\di s} \\
        0 & 1
    \end{pmatrix},
\end{split}
\end{equation}
and similarly for $z \in (a, \beta)$, where the exponential phase becomes $\ln(N)\left(\zeta_{2,c} + 2\int_{\beta}^z \Omega_{2,\pm}(s)\di s\right)$.

By enforcing the ``LTL'' factorization \eqref{eq:LTL} on the right-side cuts $(\beta^{-1}, a^{-1})$ and $(b^{-1}, \alpha^{-1})$, we obtain:
\begin{equation}
\begin{split}
    &\begin{pmatrix}
        e^{2\ln(N)\int_{\alpha^{-1}}^z \Omega_{2,+}(s)\di s + f_+(z) -f_-(z)} &-i\\
        0 & e^{-2\ln(N)\int_{\alpha^{-1}}^z \Omega_{2,+}(s)\di s - f_+(z) + f_-(z)}
    \end{pmatrix} = \\
    &\begin{pmatrix}
        1 & 0 \\
        ie^{2\ln(N)\int_{\alpha^{-1}}^z \Omega_{2,-}(s)\di s + 2f_-(z) - \ln(2 \pi \mu_1 \rho(z^{-1})/z)} & 1
    \end{pmatrix}\\ 
    &\times  
    \begin{pmatrix} 0 & -i \\ -i & 0 \end{pmatrix} 
    \begin{pmatrix}
        1 & 0 \\
        ie^{2\ln(N)\int_{\alpha^{-1}}^z \Omega_{2,+}(s)\di s + 2f_+(z) - \ln(2 \pi \mu_1 \rho(z^{-1})/z)} & 1
    \end{pmatrix},
\end{split}
\end{equation}
and similarly for $z \in (\beta^{-1}, a^{-1})$ with the phase is adjusted to include $\zeta_{2,e}$.

We define the new piecewise vector function $v(z)$ as: 
\begin{equation}
\label{eq:v_two_cuts}
    v(z) = \begin{cases}
        p(z)\pmtwo{1}{-ie^{-2 f(z) - \ln(2 \pi z \mu_1 \rho(z))  -2\ln(N)\int_{\alpha}^z \Omega_{2}(s)\di s}}{0}{1} \quad &z \text{ in lower lens below } (\alpha,b) \\\\
        p(z)\pmtwo{1}{ie^{-2 f(z) - \ln(2 \pi z \mu_1 \rho(z))  -2\ln(N)\int_{\alpha}^z \Omega_{2}(s)\di s}}{0}{1}  \quad &z \text{ in upper lens above } (\alpha,b) \\\\
        p(z)\pmtwo{1}{-ie^{-2 f(z) - \ln(2 \pi z \mu_1 \rho(z))  -\ln(N)\left(\zeta_{2,c} + 2\int_{\beta}^z \Omega_{2}(s)\di s\right)}}{0}{1} \quad &z \text{ in lower lens below } (a,\beta) \\\\
        p(z)\pmtwo{1}{ie^{-2 f(z) - \ln(2 \pi z \mu_1 \rho(z))  -\ln(N)\left(\zeta_{2,c} + 2\int_{\beta}^z \Omega_{2}(s)\di s\right)}}{0}{1}  \quad &z \text{ in upper lens above } (a,\beta) \\\\
        p(z) \pmtwo{1}{0}{ie^{\ln(N)\left(\zeta_{2,e} + 2\int_{a^{-1}}^z \Omega_{2}(s)\di s\right) + 2f(z) - \ln(2 \pi \mu_1 \rho(z^{-1})/z)}}{1} \quad &z \text{ in lower lens below } (\beta^{-1},a^{-1}) \\\\
        p(z)\pmtwo{1}{0}{-ie^{\ln(N)\left(\zeta_{2,e} + 2\int_{a^{-1}}^z \Omega_{2}(s)\di s\right) + 2f(z) - \ln(2 \pi \mu_1 \rho(z^{-1})/z)}}{1}  \quad &z \text{ in upper lens above } (\beta^{-1},a^{-1}) \\\\
        p(z) \pmtwo{1}{0}{ie^{2\ln(N)\int_{\alpha^{-1}}^z \Omega_{2}(s)\di s + 2f(z) - \ln(2 \pi \mu_1 \rho(z^{-1})/z)}}{1} \quad &z \text{ in lower lens below } (b^{-1},\alpha^{-1}) \\\\
        p(z)\pmtwo{1}{0}{-ie^{2\ln(N)\int_{\alpha^{-1}}^z \Omega_{2}(s)\di s + 2f(z) - \ln(2 \pi \mu_1 \rho(z^{-1})/z)}}{1}  \quad &z \text{ in upper lens above } (b^{-1},\alpha^{-1}) \\\\
        p(z) \quad & \text{otherwise}
    \end{cases}\,.
\end{equation}

\begin{figure}[p]
    \centering
    \rotatebox{90}{%
    \resizebox{0.95\textheight}{!}{%
    \begin{tikzpicture}[
        decoration={markings, mark=at position 0.55 with {\arrow{Latex}}},
        scale=1.2
    ]

        % Draw the real axis
        \draw[dashed,thin, black, ->] (-11, 0) -- (11, 0) node[right] {$\mathbb{R}$};

        % Coordinates for the branch points
        \coordinate (alpha) at (-9.5, 0);
        \coordinate (b) at (-6.5, 0);
        \coordinate (a) at (-3.5, 0);
        \coordinate (beta) at (-1.2, 0);
        \coordinate (beta_inv) at (1.2, 0);
        \coordinate (a_inv) at (3.5, 0);
        \coordinate (b_inv) at (6.5, 0);
        \coordinate (alpha_inv) at (9.5, 0);

        % ==========================================
        % LEFT CUT 1: (\alpha, b)
        % ==========================================
        \draw[thick, red, postaction={decorate}] (alpha) -- (b) node[midway, black, fill=white, inner sep=1pt] {$\pmtwo{0}{i}{i}{0}$};
        \draw[thick, blue, postaction={decorate}] (alpha) to[out=60, in=120, looseness=1.4] (b);
        \draw[thick, blue, postaction={decorate}] (alpha) to[out=-60, in=-120, looseness=1.4] (b);

        \node at (-8.0, 1.8) {{\small $\pmtwo{1}{{\color{gray}-ie^{-2 f - \ln(\dots)  -2\ln(N)\int_{\alpha}^z \Omega_{2}\di s}}}{0}{1}$}};
        \node at (-8.0, -1.8) {{\small $\pmtwo{1}{{\color{gray}-ie^{-2 f - \ln(\dots)  -2\ln(N)\int_{\alpha}^z \Omega_{2}\di s}}}{0}{1}$}};

        % ==========================================
        % LEFT GAP: (b, a)
        % ==========================================
        \draw[thick, green!70!black, postaction={decorate}] (b) -- (a);
        \node (gap1_mat) at (-5.0, 3.5) {{\small $\pmtwo{e^{\ln(N)\zeta_{2,e} + \omega_{2,e}}}{0}{{\color{gray}\Xi_1(z)e^{2\ln(N)\int_a^z \Omega_2\di s}}}{e^{-\ln(N)\zeta_{2,e} - \omega_{2,e}}}$}};
        \draw[->,dashed, thin, shorten >= 3pt] (gap1_mat) -- (-5.0, 0);

        % ==========================================
        % LEFT CUT 2: (a, \beta)
        % ==========================================
        \draw[thick, red, postaction={decorate}] (a) -- (beta) node[midway, black, fill=white, inner sep=1pt] {$\pmtwo{0}{i}{i}{0}$};
        \draw[thick, blue, postaction={decorate}] (a) to[out=60, in=120, looseness=1.5] (beta);
        \draw[thick, blue, postaction={decorate}] (a) to[out=-60, in=-120, looseness=1.5] (beta);

        \node at (-2.35, 1.8) {{\small $\pmtwo{1}{{\color{gray}-ie^{-2 f - \dots -2\ln(N)\int_{\beta}^z \Omega_{2}\di s}}}{0}{1}$}};
        \node at (-2.35, -1.8) {{\small $\pmtwo{1}{{\color{gray}-ie^{-2 f - \dots -2\ln(N)\int_{\beta}^z \Omega_{2}\di s}}}{0}{1}$}};

        % ==========================================
        % CENTRAL GAP: (\beta, \beta^{-1})
        % ==========================================
        \draw[thick, green!70!black, postaction={decorate}] (beta) -- (beta_inv) node[midway, above=3pt, black] {$e^{\sigma_3\left(\ln(N)\zeta_{2,c} + \omega_{2,c}\right)}$};
        \filldraw[black] (0,0) circle (1.5pt) node[below=2pt] {$1$};

        % ==========================================
        % RIGHT CUT 1: (\beta^{-1}, a^{-1})
        % ==========================================
        \draw[thick, red, postaction={decorate}] (beta_inv) -- (a_inv) node[midway, black, fill=white, inner sep=1pt] {$\pmtwo{0}{-i}{-i}{0}$};
        \draw[thick, blue, postaction={decorate}] (beta_inv) to[out=60, in=120, looseness=1.5] (a_inv);
        \draw[thick, blue, postaction={decorate}] (beta_inv) to[out=-60, in=-120, looseness=1.5] (a_inv);

        \node at (2.35, 1.8) {{\small $\pmtwo{1}{0}{{\color{gray}ie^{\dots + 2f +2\ln(N)\int_{a^{-1}}^z \Omega_{2}\di s}}}{1}$}};
        \node at (2.35, -1.8) {{\small $\pmtwo{1}{0}{{\color{gray}ie^{\dots + 2f + 2\ln(N)\int_{a^{-1}}^z \Omega_{2}\di s}}}{1}$}};

        % ==========================================
        % RIGHT GAP: (a^{-1}, b^{-1})
        % ==========================================
        \draw[thick, green!70!black, postaction={decorate}] (a_inv) -- (b_inv);
        \node (gap2_mat) at (5.0, -3.5) {{\small $\pmtwo{e^{\ln(N)\zeta_{2,e} + \omega_{2,e}}}{{\color{gray}\Xi_2(z)e^{-2\ln(N)\int_{b^{-1}}^z \Omega_2\di s}}}{0}{e^{-\ln(N)\zeta_{2,e} - \omega_{2,e}}}$}};
        \draw[->,dashed, thin, shorten >= 3pt] (gap2_mat) -- (5.0, 0);

        % ==========================================
        % RIGHT CUT 2: (b^{-1}, \alpha^{-1})
        % ==========================================
        \draw[thick, red, postaction={decorate}] (b_inv) -- (alpha_inv) node[midway, black, fill=white, inner sep=1pt] {$\pmtwo{0}{-i}{-i}{0}$};
        \draw[thick, blue, postaction={decorate}] (b_inv) to[out=60, in=120, looseness=1.4] (alpha_inv);
        \draw[thick, blue, postaction={decorate}] (b_inv) to[out=-60, in=-120, looseness=1.4] (alpha_inv);

        \node at (8.0, 1.8) {{\small $\pmtwo{1}{0}{{\color{gray}ie^{2\ln(N)\int_{\alpha^{-1}}^z \Omega_{2}\di s + 2f - \dots}}}{1}$}};
        \node at (8.0, -1.8) {{\small $\pmtwo{1}{0}{{\color{gray}ie^{2\ln(N)\int_{\alpha^{-1}}^z \Omega_{2}\di s + 2f - \dots}}}{1}$}};

        % Branch points
        \filldraw[black] (alpha) circle (1.5pt) node[below left] {$\alpha$};
        \filldraw[black] (b) circle (1.5pt) node[below right] {$b$};
        \filldraw[black] (a) circle (1.5pt) node[below left] {$a$};
        \filldraw[black] (beta) circle (1.5pt) node[below right] {$\beta$};
        \filldraw[black] (beta_inv) circle (1.5pt) node[below left] {$\beta^{-1}$};
        \filldraw[black] (a_inv) circle (1.5pt) node[below right] {$a^{-1}$};
        \filldraw[black] (b_inv) circle (1.5pt) node[below left] {$b^{-1}$};
        \filldraw[black] (alpha_inv) circle (1.5pt) node[below right] {$\alpha^{-1}$};

    \end{tikzpicture}%
    }%
    }
    \caption{RHP for $v(z)$ in the two-cuts modulated regime. Lenses Opening: Notice that, in the appropriate regime, the off-diagonal terms in gray converge to zero as $N$ grows.}
    \label{fig:jump_v_two_cuts}
\end{figure}
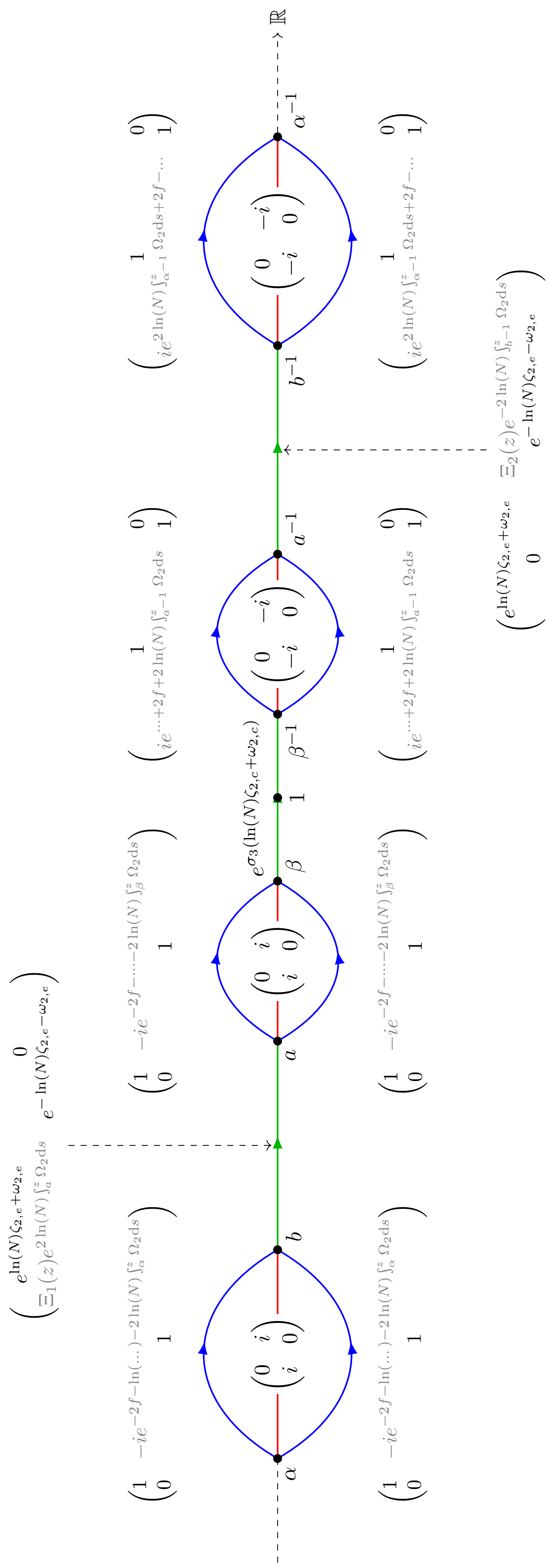

The vector $v(z)$ satisfies the jump conditions $v_+(z) = v_-(z)J_v(z)$, with the symmetry $v(z^{-1}) = v(z)\sigma_1$ preserved, and $\lim_{z\to 0} v_1(z)v_2(z)=1$. The resulting jump matrices for $v(z)$ are illustrated in Figure \ref{fig:jump_v_two_cuts}. 

To show that the RHP for $v(z)$ reduces, in the small norm setting, to the model problem \ref{OuterRHP_2cuts}; we prove the following two lemmas showing the decay of the off-diagonal terms on the gaps and the decay on the lips of the lenses.

\begin{lemma}
\label{lem:off_decay_two_cuts_gap}
    Let $(\eta,\tau)\in \cU_2$ and let $a \equiv a(\eta,\tau)$ and $b \equiv b(\eta,\tau)$ be defined as in Proposition \ref{prop:g_two_cut}. Then, the off-diagonal entries of the jump matrix $J_p(z)$ for $z\in(b,a)\cup(a^{-1},b^{-1})$ converge to zero as $N\to \infty$.
\end{lemma}

\begin{proof}
    We focus on the gap interval $z \in (b, a)$. The decay of the off-diagonal entry is governed by the exponent $  2\ln(N)\int_a^z \Omega_2(s,\eta,\tau)\di s$. To ensure exponential decay, we must show that the integral $\nu(z) = \int_a^z \Omega_2(s,\eta,\tau)\di s$ is strictly negative for all $z \in (b, a)$. 
    
    We analyze this using the Joukowsky variable $x = \frac{1}{2}(s+s^{-1})$. As established in the proof of Proposition \ref{prop:g_two_cut}, the linear factor $L(x,\eta,\tau)$ \eqref{eq:linear_2cuts} is strictly decreasing because $\tau < 0$. Furthermore, to satisfy the Condition \ref{eq:two_cuts_gap_condition} of Proposition \ref{prop:g_two_cut}, $L(x,\eta,\tau)$ must change sign inside $(\wt a, \wt b)$, which implies $L(\wt a,\eta,\tau) > 0$ and $L(\wt b,\eta,\tau) < 0$. 
    
    Because $L(x,\eta,\tau)$ is linear, it crosses zero exactly once in this interval, which implies that the function $\nu(z)$ has a unique minimum in the interval. Furthermore,  $\nu(a) = 0$ by definition, and $\nu(b) = \int_a^b \Omega_2(s)\di s = 0$ due to Condition \ref{eq:two_cuts_gap_condition} in Proposition \ref{prop:g_two_cut}. Therefore, $\nu(z)$ must decrease from zero to a strictly negative minimum before increasing back to zero at $z = b$. So, we conclude that $\nu(z) < 0$ for all $z \in (b, a)$, which guarantees the decay of the off-diagonal term of $J_p(z)$ for $z\in(b,a)$. By Remark \ref{rem:sym_off_diag}, we conclude the same for $z\in(a^{-1}, b^{-1})$.
\end{proof}

\begin{lemma}
    \label{lem:off_decay_two_cuts_lens}
    Let $(\eta,\tau)\in \cU_2$ and let $a, b$ be defined as in Proposition \ref{prop:g_two_cut}. There exist neighborhoods $V_1, V_2, V_3, V_4$ of the cuts $(\alpha,b)$, $(a,\beta)$, $(\beta^{-1},a^{-1})$, and $(b^{-1},\alpha^{-1})$ respectively, such that
    \begin{align}
        &\Re\left(\int_\alpha^z \Omega_2(s,\eta,\tau)\di s\right) \ge 0 \quad \forall\, z \in V_1\,, \qquad \Re\left(\int_\beta^z \Omega_2(s,\eta,\tau)\di s\right) \ge 0 \quad \forall\, z \in V_2\,,\\
        &\Re\left(\int_{a^{-1}}^z \Omega_2(s,\eta,\tau)\di s\right) \le 0 \quad \forall\, z \in V_3\,, \qquad \Re\left(\int_{\alpha^{-1}}^z \Omega_2(s,\eta,\tau)\di s\right) \le 0 \quad \forall\, z \in V_4\,.
    \end{align}
    Which implies that the jump matrix $J_v(z)\to I$ as $N\to\infty$ on the lips of the lenses, see Figure \ref{fig:jump_v_two_cuts}.
\end{lemma}

\begin{proof}
    We begin by analyzing the upper lens of the cut $(\alpha, b)$. On the real axis, the integral $\int_\alpha^z \Omega_{2,+}(s,\eta,\tau)\di s$ is purely imaginary. By the Cauchy-Riemann equations, we can ensure the required positivity in the upper lens by proving that the derivative $\Omega_{2,+}(z,\eta,\tau)$ has a strictly negative imaginary part for $z\in (\alpha, b)$.
    
    Mapping to the Joukowsky variable $x = \frac{1}{2}(z+z^{-1})$, the cut $z \in (\alpha, b)$ corresponds to $x \in (\tilde{b}, \tilde{\alpha})$. From Lemma \ref{lem:off_decay_two_cuts_gap} and Proposition \ref{prop:g_two_cut}, we know that  $L(\wt b,\eta,\tau) < 0$ \eqref{eq:linear_2cuts}. It follows that
    $$L(x,\eta,\tau) < L(\wt b,\eta,\tau) < 0 \quad x \in (\wt b, \wt \alpha)\,.$$
    Since the sign $\Im(\Omega_{2,+}(s))$ is the same as $L(x,\eta,\tau)$, see Remark \ref{rem:R_two_cuts}, we showed that $\Im(\Omega_{2,+}(s))<0$. This ensures $\Re\left(\int_\alpha^z \Omega_{2,+}(s)\di s\right) > 0$ as $z$ moves into the upper lens.

   An analogous argument applies to all the other gaps.
\end{proof}

\begin{remark}
    We notice that the previous lemmas imply that all the terms in gray in Figure \ref{fig:jump_v_two_cuts} converge to zero as $N\to\infty$.
\end{remark}

Applying the previous lemmas, we are led to consider as the outer parametrix problem the model problem RHP \ref{OuterRHP_2cuts} in its symmetric version, where $\ta \equiv \alpha,\, \tb\equiv a,\, \tc\equiv b,\, \td \equiv \beta$ and $\Delta_1 = \frac{1}{2\pi i}(\ln(N)\zeta_{2,e}(\eta,\tau) + \omega_{2,e}),\, \Delta_2 = \frac{1}{2\pi i}\left(\ln(N)\zeta_{2,c}(\eta,\tau)+\omega_{2,c}\right)$ and 

\begin{equation}
\label{eq:Delta2n}
    \vec{\Delta}_{2,n} = (\Delta_1,\Delta_2)\,.
\end{equation}
 To approximate $v(z)$, we need to construct a local approximation near the endpoints of the intervals $(\alpha,b),\,(a,\beta),\,(\beta^{-1},a^{-1}),\,(b^{-1},\alpha^{-1})$. 

\subsection{Local parametrices}
\label{subsec:airy_local_par}

In this subsection, we construct the local parametrices for the approximate solution to the two-cuts case. Since the construction around the hard edges $\alpha,\beta, \alpha^{-1},\beta^{-1}$ is completely analogous to the Bessel parametrix detailed in Section \ref{subsec:one_cut_local_par}, and the treatment for the other soft branch points is standard, we only construct the local parametrix explicitly around the moving soft edge $b^{-1}$. All other soft edges can be constructed using analogous symmetry and orientation transformations; furthermore, they do not contribute to the leading-order asymptotic of the solution.

In particular, we focus on a small disk $\mathbb{D}(b^{-1},\varepsilon)$ centered at $b^{-1}$ with radius $\varepsilon>0$. Our aim is to find a matrix valued function $M_{b^{-1}}(z)$ that solves the following RHP:

\begin{rhp}
\label{rhp:local_b_inv}
    We seek a matrix valued function $M_{b^{-1}}(z)$ such that
    \begin{enumerate}
        \item $M_{b^{-1}}(z)$ is analytic in $\mathbb{D}(b^{-1},\varepsilon)\setminus\{(a^{-1}, \alpha^{-1})\cup \gamma_1\cup\gamma_2\}$
        \item $M_{b^{-1}}(z)_+=M_{b^{-1}}(z)_- J_{b^{-1}}(z)$ where 

        \begin{equation}
            J_{b^{-1}}(z) = \begin{cases}
                 \pmtwo{1}{0}{ie^{\ln(N)(2\int_{b^{-1}}^z \Omega_{2}(s)\di s +\zeta_{e,2}) + 2f(z) - \ln(2 \pi \mu_1 \rho(z^{-1})/z)}}{1}\qquad &z\in \gamma_1\cap\mathbb{D}(b^{-1},\varepsilon)\\\\
                 \pmtwo{1}{0}{ie^{\ln(N)(2\int_{b^{-1}}^z \Omega_{2}(s)\di s -\zeta_{e,2}) + 2f(z) - \ln(2 \pi \mu_1 \rho(z^{-1})/z)}}{1}\qquad &z\in \gamma_2\cap\mathbb{D}(b^{-1},\varepsilon)\\\\
    
    \begin{pmatrix}
        0 & -i
        \\
        -i & 0
    \end{pmatrix} \qquad &z\in (b^{-1},\alpha^{-1})\cap\mathbb{D}(b^{-1},\varepsilon)\\\\
    
    \begin{pmatrix}
        e^{\ln(N)\zeta_{2,e}(\eta,\tau) + \omega_{2,e}} & \Xi_2(z)e^{-2\ln(N)\int_{b^{-1}}^z \Omega_2(s)\di s}
        \\
        0 & e^{-\ln(N)\zeta_{2,e}(\eta,\tau) - \omega_{2,e}}
    \end{pmatrix} \qquad &z\in (a^{-1},b^{-1})\cap\mathbb{D}(b^{-1},\varepsilon)
            \end{cases}
        \end{equation}
        where $\Xi_2(z) = -2\pi i \frac{\mu_1\rho(z^{-1})}{z}e^{-f_+(z)-f_-(z)}$.
    \end{enumerate}
\end{rhp}

\begin{figure}[htbp]
    \centering
    \begin{tikzpicture}[
        decoration={markings, mark=at position 0.65 with {\arrow{Latex}}},
        scale=1.5
    ]
        % Draw the real axis
        \draw[dashed,thin, black, ->] (-2, 0) -- (4, 0) node[right] {$\mathbb{R}$};

        % Coordinates
        \coordinate (a_inv) at (-1.5, 0);
        \coordinate (b_inv) at (0, 0);
        \coordinate (alpha_inv) at (2, 0);

        % Gap
        \draw[thick, green!70!black, postaction={decorate}] (-2, 0) -- (b_inv);
        
        % Cut
        \draw[thick, red, postaction={decorate}] (b_inv) -- (alpha_inv);
        
        % Lenses
        \draw[thick, blue, postaction={decorate}] (b_inv) to[out=60, in=100, looseness=2] (alpha_inv);
        \draw[thick, blue, postaction={decorate}] (b_inv) to[out=-60, in=-100, looseness=2] (alpha_inv);

        % Labels
        \filldraw[black] (a_inv) circle (1.5pt) node[below=3pt] {$a^{-1}$};
        \filldraw[black] (b_inv) circle (1.5pt) node[below left=2pt] {$b^{-1}$};
        \filldraw[black] (alpha_inv) circle (1.5pt) node[above right=3pt] {$\alpha^{-1}$};

        % Disk
        \draw[dashed, thick, black] (b_inv) circle (1.0);
        \node at (-0.3, 0.4) {$\mathbb{D}(b^{-1},\varepsilon)$};
        
        \node[above] at (1.6, 0.45) {$\gamma_1$};
        \node[below] at (1.6, -0.45) {$\gamma_2$};

    \end{tikzpicture}
    \caption{The contour for the local parametrix construction around the soft edge $b^{-1}$ in the two-cuts case.}
    \label{fig:local_zoom_two_cuts}
\end{figure}
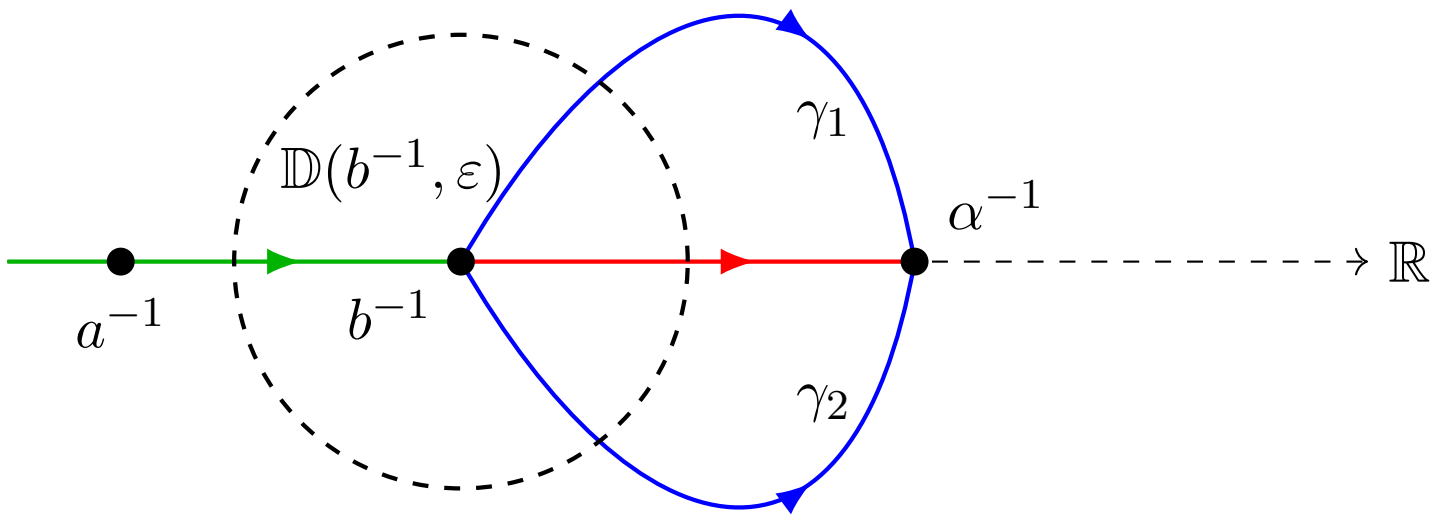

Here, $\gamma_1, \gamma_2$ are upper and lower lenses connecting $b^{-1}$ with $\alpha^{-1}$, with the geometric restriction that $\gamma_1 \subset \mathbb{C}_+$ and $\gamma_2 \subset \mathbb{C}_-$. A sketch of the contours is shown in Figure \ref{fig:local_zoom_two_cuts}. We do not impose any asymptotic condition at the boundary of the domain, as this will be fixed through matching with the global parametrix.

To neutralize the phase and isolate the local fractional powers, we define a local transformation inside $\mathbb{D}(b^{-1},\varepsilon)$ as:

\begin{equation}
    N_{b^{-1}}(z) = \begin{cases}
        M_{b^{-1}}(z) \exp\left(-\frac{\sigma_3}{2}\left(\ln(N)\zeta_{2,e}  +2f(z) - \ln(2\pi \mu_1\rho(z^{-1})/z)\right) \right)\, \quad &\Im(z) >0 \\
        M_{b^{-1}}(z) \exp\left(-\frac{\sigma_3}{2}\left(-\ln(N)\zeta_{2,e} + 2f(z) - \ln(2\pi \mu_1\rho(z^{-1})/z)\right) \right)\, \quad &\Im(z) <0
    \end{cases}\,.
\end{equation}

As a result, tracking the algebraic simplifications evaluated along the respective branch contours, $N_{b^{-1}}(z)$ solves the exact following reduced RHP:

\begin{rhp}
        We seek a matrix valued function $N_{b^{-1}}(z)$ such that
    \begin{enumerate}
        \item $N_{b^{-1}}(z)$ is analytic in $\mathbb{D}(b^{-1},\varepsilon)\setminus ((a^{-1}, \alpha^{-1}) \cup \gamma_1 \cup \gamma_2)$
        \item $N_{b^{-1}}(z)_+=N_{b^{-1}}(z)_- \wt J_{b^{-1}}$ where 

        \begin{equation}
            \wt J_{b^{-1}}(z) = \begin{cases}
                \begin{pmatrix}
        1 &  0 \\
        ie^{ 2\ln(N)\int_{b^{-1}}^z \Omega_{2}(s)\di s} & 1
    \end{pmatrix}\qquad &z\in \gamma_1\cap\mathbb{D}(b^{-1},\varepsilon)\\\\
    
    \begin{pmatrix}
        1 &  0 \\
        ie^{ 2\ln(N)\int_{b^{-1}}^z \Omega_{2}(s)\di s} & 1
    \end{pmatrix} \qquad &z\in \gamma_2\cap\mathbb{D}(b^{-1},\varepsilon)\\\\
    \pmtwo{0}{-i}{-i}{0} \qquad &z\in (b^{-1},\alpha^{-1})\cap\mathbb{D}(b^{-1},\varepsilon)\\\\
    \pmtwo{1}{-ie^{-2\ln(N)\int_{b^{-1}}^z \Omega_2(s)\di s}}{0}{1} \qquad &z\in (a^{-1},b^{-1})\cap\mathbb{D}(b^{-1},\varepsilon)
            \end{cases}
        \end{equation}
    \end{enumerate}
\end{rhp}

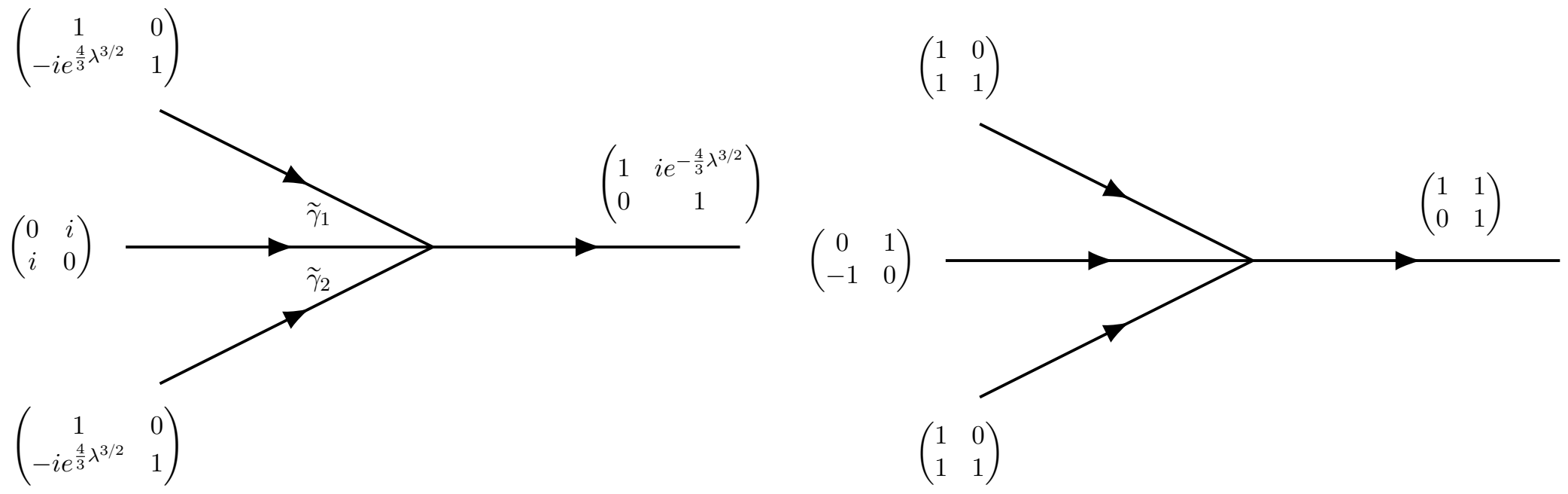
\begin{figure}[htbp]
    \centering
    \scalebox{0.6}{
    \begin{tikzpicture}[
    >={Latex[scale=1.3]},
    decoration={markings, mark=at position 0.55 with {\arrow{>}}},
    ray/.style={very thick, postaction={decorate}}
]
    \coordinate (O) at (0,0);

    % Right ray (gap)
    \draw[ray] (O) -- (4.5, 0) 
        node[midway, above right, yshift=2mm] {$\begin{pmatrix} 1 & ie^{-\frac{4}{3}\lambda^{3/2}} \\ 0 & 1 \end{pmatrix}$};

    % Left ray (cut)
    \draw[ray] (-4.5, 0) -- (O);
    \node[left] at (-4.8, 0) {$\begin{pmatrix} 0 & i \\ i & 0 \end{pmatrix}$};

    % Top left ray (lens)
    \draw[ray] (-4, 2) -- (O) node[midway, below right, yshift=-2mm] {$\wt \gamma_1$};
    \node[above left] at (-3.5, 2.2) {$\begin{pmatrix} 1 & 0 \\ -ie^{\frac{4}{3}\lambda^{3/2}} & 1 \end{pmatrix}$};

    % Bottom left ray (lens)
    \draw[ray] (-4, -2) -- (O) node[midway, above right, yshift=2mm] {$\wt \gamma_2$} ;
    \node[below left] at (-3.5, -2.2) {$\begin{pmatrix} 1 & 0 \\ -ie^{\frac{4}{3}\lambda^{3/2}} & 1 \end{pmatrix}$};
\end{tikzpicture}}
\scalebox{0.6}{
    \begin{tikzpicture}[
    >={Latex[scale=1.3]},
    decoration={markings, mark=at position 0.55 with {\arrow{>}}},
    ray/.style={very thick, postaction={decorate}}
]
    \coordinate (O) at (0,0);

    % Right ray
    \draw[ray] (O) -- (4.5, 0) 
        node[midway, above right, yshift=2mm] {$\begin{pmatrix} 1 & 1 \\ 0 & 1 \end{pmatrix}$};

    % Left ray
    \draw[ray] (-4.5, 0) -- (O);
    \node[left] at (-4.8, 0) {$\begin{pmatrix} 0 & 1 \\ -1 & 0 \end{pmatrix}$};

    % Top left ray
    \draw[ray] (-4, 2) -- (O) ;
    \node[above left] at (-3.5, 2.2) {$\begin{pmatrix} 1 & 0 \\ 1 & 1 \end{pmatrix}$};

    % Bottom left ray
    \draw[ray] (-4, -2) -- (O) ;
    \node[below left] at (-3.5, -2.2) {$\begin{pmatrix} 1 & 0\\ 1 & 1 \end{pmatrix}$};
\end{tikzpicture}}
    \caption{\textbf{Left Panel:} Jump matrices for $\fQ(\lambda)$ centered at $b^{-1}$, notice that we changed orientation compared to the jump for $N_{b^{-1}}$. \textbf{Right Panel:} Jump matrices for the universal standard  $\Psi_{\mathrm{Ai}} (\lambda)$.}
    \label{fig:local_airy_two_cuts}
\end{figure}

Because $\int_{b^{-1}}^z \Omega_2(s)\di s = \mathcal{O}((z-b^{-1})^{3/2})$ as $z\to b^{-1}$ and $\int_{b^{-1}}^z \Omega_{2,+}\di s = - \int_{b^{-1}}^z \Omega_{2,-}\di s$ for $z \in (b^{-1}, \alpha^{-1})$, we can define the local conformal map:
\begin{equation}
    \frac{4}{3}\lambda(z)^{3/2} = 2\ln(N)\int_{b^{-1}}^z \Omega_2(s,\eta,\tau)\di s\,.
\end{equation}
As $\int_{b^{-1}}^z \Omega_2 < 0$ for $z \in (a^{-1}, b^{-1})$ by Lemma \ref{lem:off_decay_two_cuts_gap} and Proposition \ref{prop:g_two_cut}, it follows that $\lambda'(b^{-1}) < 0$.
We fix $\gamma_1, \gamma_2$ such that $\arg(\lambda(\gamma_1\cap \mathbb{D}(b^{-1},\varepsilon))) \equiv -\frac{2\pi}{3}$ and $\arg(\lambda(\gamma_2\cap \mathbb{D}(b^{-1},\varepsilon))) \equiv \frac{2\pi}{3}$. We aim to solve the corresponding RHP in the mapped $\lambda$-plane for the function $\fQ(\lambda)$:
\begin{rhp}
        We seek a matrix valued function $\fQ(\lambda)$ such that
    \begin{enumerate}
        \item $\fQ(\lambda)$ is analytic in $\mathbb{C}\setminus\{\mathbb{R}\cup\wt\gamma_1\cup\wt\gamma_2\}$
        \item $\fQ_+(\lambda)=\fQ_-(\lambda) J_\fQ(\lambda)$ where 

        \begin{equation}
             J_\fQ(\lambda) = \begin{cases}
                \begin{pmatrix}
        1 &  0 \\
        -ie^{ \frac{4}{3}\lambda^{3/2}} & 1
    \end{pmatrix}\qquad &\lambda\in \wt\gamma_1\\\\
    \begin{pmatrix}
        1 &  0 \\
     -ie^{ \frac{4}{3}\lambda^{3/2}} & 1
    \end{pmatrix}\qquad &\lambda\in \wt\gamma_2\\\\
    
    \begin{pmatrix}
        0 & i
        \\
        i & 0
    \end{pmatrix} \qquad &\lambda\in (-\infty,0)\\\\
    \pmtwo{1}{ie^{-\frac{4}{3}\lambda^{3/2}}}{0}{1} \qquad & \lambda \in(0,+\infty)
            \end{cases}
        \end{equation}
    \end{enumerate}
\end{rhp}
where the two contours $\wt\gamma_1, \wt\gamma_2$ are precisely:
\begin{equation}
    \wt \gamma_1 = \left\{\lambda \in \mathbb{C}_+ \; \vert\; \arg(\lambda)=\frac{2\pi}{3}\right\}\,,\qquad \wt \gamma_2 = \left\{\lambda \in \mathbb{C}_- \; \vert\; \arg(\lambda)=-\frac{2\pi}{3}\right\}\,.
\end{equation}
and the orientations flow exactly as indicated in the left panel of Figure \ref{fig:local_airy_two_cuts}.

To reduce $\fQ(\lambda)$ to the  Airy parametric  we perform the following transformation:
\begin{equation}
   \fQ(\lambda) =  \Psi_{\mathrm{Ai}}(\lambda)e^{\left(\frac{2}{3}\lambda^{3/2} - i\frac{\pi}{4}\right)\sigma_3}\,.
\end{equation}
 The function $\Psi_{\mathrm{Ai}}(\lambda)$ solves:

\begin{rhp}
    We seek a matrix valued function $\Psi_{\mathrm{Ai}}(\lambda)$ such that 
    \begin{enumerate}
        \item $\Psi_{\mathrm{Ai}}(\lambda)$ is analytic in $\mathbb{C}\setminus(\mathbb{R}\cup\wt\gamma_1\cup\wt\gamma_2)$
        \item $\Psi_{\mathrm{Ai},+}(\lambda)=\Psi_{\mathrm{Ai},-}(\lambda) J_\Psi(\lambda)$ where 

        \begin{equation}
             J_\Psi(\lambda) = \begin{cases}
                \begin{pmatrix}
        1 &  0
        \\
        1 & 1
    \end{pmatrix}\qquad &\lambda\in \wt\gamma_1\cup\wt\gamma_2\\\\
    
    \begin{pmatrix}
        0 & 1
        \\
        -1 & 0
    \end{pmatrix} \qquad &\lambda\in (-\infty,0)\\\\
    \pmtwo{1}{1}{0}{1} \qquad & \lambda \in(0,+\infty)
            \end{cases}
        \end{equation}
    \end{enumerate}
\end{rhp}
This RHP is solved in \cite{GGJM} using the Airy function $\Ai(\lambda)$ \cite[Ch.~9]{dlmf}:
\begin{equation}
\Psi_{\Ai}(\lambda)=\begin{cases}
    \sqrt{2\pi}\begin{pmatrix}   \Ai( \lambda)&  - \fw^2 \Ai ( \fw^2 \lambda) \\
  -i\Ai'( \lambda) &  i \fw \Ai'( \fw^2 \lambda)
\end{pmatrix} &   0< \arg  \lambda < \frac{2\pi }{3} \\
\sqrt{2\pi} \begin{pmatrix}  - \fw \Ai( \fw \lambda)&  - \fw^2 \Ai ( \fw^2 \lambda) \\
  i \fw^2 \Ai'( \lambda) &   i \fw \Ai'( \fw^2 \lambda)
\end{pmatrix}  &     \frac{2\pi }{3} < \arg  \lambda < \pi \\
\sqrt{2\pi} \begin{pmatrix}  - \fw^2 \Ai ( \fw^2 \lambda) &    \fw \Ai( \fw \lambda)\\
   i \fw \Ai'( \fw^2 \lambda) &   -i \fw^2 \Ai'( \lambda) 
\end{pmatrix}  &    -\pi < \arg  \lambda< - \frac{2\pi }{3} \\
\sqrt{2\pi} \begin{pmatrix} \Ai ( \lambda) &    \fw \Ai( \fw \lambda)\\
   - i \Ai'( \lambda) &  -i \fw^2 \Ai'( \lambda) 
\end{pmatrix}  &    - \frac{2\pi }{3}< \arg  \lambda< 0 \ ,
\end{cases}
\end{equation}
where $\fw = e^{\frac{2}{3}\pi i}$.

Therefore, by unwinding the transformations, we solve the exact local parametrix problem $M_{b^{-1}}(z)$ as:
\begin{equation}
   M_{b^{-1}}(z) = \begin{cases}
        C_{b^{-1}}(z)\Psi_{\mathrm{Ai}}(\lambda(z)) e^{\left(\frac{2}{3}\lambda(z)^{3/2} - i\frac{\pi}{4}+  \frac{\ln(N)}{2}\zeta_{2,e}(\eta,\tau) + f(z) - \frac{\ln(2\pi \mu_1\rho(z^{-1})/z)}{2}\right)\sigma_3}\,, \ \Im(z) > 0 \\
        C_{b^{-1}}(z)\Psi_{\mathrm{Ai}}(\lambda(z)) e^{\left(\frac{2}{3}\lambda(z)^{3/2}- i\frac{\pi}{4} - \frac{\ln(N)}{2}\zeta_{2,e}(\eta,\tau) + f(z) - \frac{\ln(2\pi \mu_1\rho(z^{-1})/z)}{2}\right)\sigma_3}\,, \ \Im(z) < 0
    \end{cases}
\end{equation}
where $C_{b^{-1}}(z)$ is analytic strictly within $\mathbb{D}(b^{-1},\varepsilon)$. Its value is fixed entirely by enforcing that the jump between the local parametrix and the global outer parametrix $M(z)$ uniformly approaches the identity matrix $\mathcal{O}(\ln(N)^{-1})$ on the disk's boundary. 

Matching $M_{b^{-1}}(z)$ to $M(z)$ using the known uniform expansions of the Airy functions as $\lambda \to \infty$:
\begin{equation}
    \begin{split}
        & \Psi_{\mathrm{Ai}}(\lambda) = \lambda^{-\sigma_3/4} \frac{1}{\sqrt{2}} \begin{pmatrix} 1 & i \\ i & 1 \end{pmatrix} \mathcal{E}_{\mathrm{Ai}}(\lambda) e^{-\frac{2}{3}\lambda^{3/2}\sigma_3}, \\
        & \mathcal{E}_{\mathrm{Ai}}(\lambda) = I + \mathcal{O}\left(\lambda^{-\frac{3}{2}}\right), \quad \lambda \to \infty,
    \end{split}
\end{equation}
we define the regularizer:
\begin{align}\label{eq:AiryAsym}
   C_{b^{-1}}(z) =
   \frac{1}{\sqrt{2}}M(z)\begin{cases}
        \exp\left(\sigma_3 \left(i\frac{\pi}{4} -\frac{\ln(N)}{2}\zeta_{2,e}(\eta,\tau) - f(z) +  \frac{\ln(2\pi \mu_1\rho(z^{-1})/z)}{2}\right)\right)\mathfrak{L}\lambda(z)^{\frac{\sigma_3}{4}}\; &\Im(z)>0 \\\\
        \exp\left(\sigma_3 \left( i\frac{\pi}{4}+\frac{\ln(N)}{2}\zeta_{2,e}(\eta,\tau) - f(z) +  \frac{\ln(2\pi \mu_1\rho(z^{-1})/z)}{2}\right)\right)\mathfrak{L}\lambda(z)^{\frac{\sigma_3}{4}}; &\Im(z)<0
    \end{cases}
\end{align}
where $\mathfrak{L}=\pmtwo{1}{-i}{-i}{1}$.
This correctly compensates the local branch cuts, so  that $C_{b^{-1}}(z)$ remains analytic everywhere inside $\mathbb{D}(b^{-1},\varepsilon)$. With this choice one can now check using \eqref{eq:AiryAsym} that on $\partial \mathbb{D}(b^{-1}, \varepsilon)$ the following matching asymptotics hold:
\begin{align*}
    M_{b^{-1}}(z) = M(z)\left(I +  \cO(\lambda^{-3/2}(z))\right) = M(z)\left(I + \cO\left(\frac{1}{\ln(N)}\right)\right)\,.
\end{align*}

\subsection{Genus $2$ reduction of the symmetric vector solution}\label{Sect:genus2Red}
As in the previous cases, the asymptotics of solutions is given through the \emph{symmetric} vector solution. It turns out that due to its symmetry $k_s^{(2)}(z^{-1}) = k_s^{(2)}(z)\sigma_1$, we can rewrite the symmetric solution in terms of $\Theta$-functions on a Riemann surface $\widehat \cR_2$ which has $\mathcal \cR_2$ as a double cover. As the construction is quite similar to the one found in Section \ref{sec:model_problems} we refer to this section for more details. 

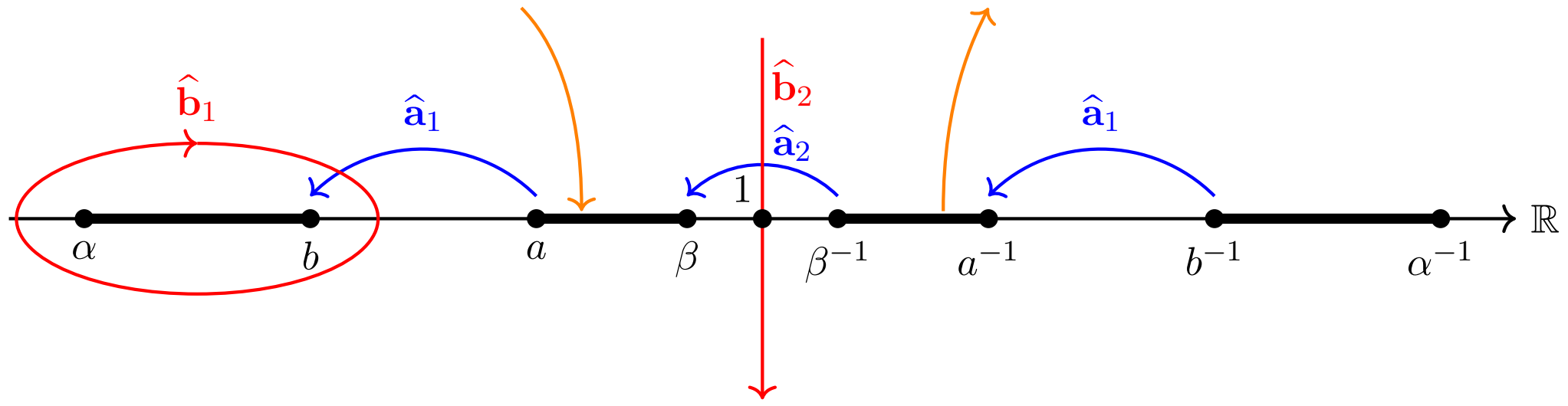
\begin{figure}[htbp]
    \centering
    \begin{tikzpicture}[scale=1.3, every node/.style={scale=0.9}]
        % Real axis
        \draw[thick, ->] (-5, 0) -- (5, 0) node[right] {$\mathbb{R}$};
        
        % Cuts
        \draw[line width=2.5pt, black] (-4.5, 0) -- (-3, 0); %node[midway, below=3pt, black] {$C_1$};
        \draw[line width=2.5pt, black] (-1.5, 0) -- (-0.5, 0); %node[midway, below=3pt, black] {$C_2$};
        \draw[line width=2.5pt, black] (0.5, 0) -- (1.5, 0); %node[midway, below=3pt, black] {$C_3$};
        \draw[line width=2.5pt, black] (3, 0) -- (4.5, 0); %node[midway, below=3pt, black] {$C_4$};

        % A-cycles
        \draw[->, thick, blue] (-1.5, 0.15) to[out=135, in=45] (-3, 0.15); \node[blue] at (-2.25, 0.7) {$\wh{\mathbf{a}}_1$};
        \draw[->, thick, blue] (0.5, 0.15) to[out=135, in=45] (-0.5, 0.15); \node[blue] at (0.2, 0.5) {$\wh{\mathbf{a}}_2$};
        \draw[->, thick, blue] (3, 0.15) to[out=135, in=45] (1.5, 0.15); \node[blue] at (2.25, 0.7) {$\wh{\mathbf{a}}_1$};

        % B-cycles (Fully visible, counterclockwise)
        \draw[<-, thick, red] (-3.75, 0.5) arc (90:450:1.2 and 0.5); \node[red] at (-3.75, 0.8) {$\wh{\mathbf{b}}_1$};
        %\draw[->, thick, red] (3.75, 0.5) arc (90:450:1.2 and 0.5); \node[red] at (3.75, 0.8) {$\mathbf{b}_3$};
        \draw[->, thick, red] (0, 1.2) -- (0, -1.2); \node[red] at (0.2, 0.9) {$\wh{\mathbf{b}}_2$};

             % Points
        \foreach \x/\l in {-4.5/\alpha, -3/b, -1.5/a, -0.5/\beta, 0.5/\beta^{-1}, 1.5/a^{-1}, 3/b^{-1}, 4.5/\alpha^{-1}} {
            \filldraw (\x,0) circle (1.5pt) node[below=2pt] {$\l$};
        }
        \filldraw (0,0) circle (1.5pt) node[above=0.5pt] {$1 \quad$};

        % Orange path entering the second cut and leaving the third cut
        \draw[thick, orange, ->]
            (-1.6, 1.4)
            .. controls (-1.4, 1.2) and (-1.2, 0.8) ..
            (-1.2, 0.05);

        \draw[thick, orange, ->]
            (1.2, 0.05)
            .. controls (1.2, 0.8) and (1.4, 1.2) ..
            (1.5, 1.4);
    \end{tikzpicture}
    \caption{The homology basis of the Riemann surface $\wh \cR_2$. The orange path is a continuous path on the Riemann surface and indicates the glueing of the second and third cut.}
    \label{fig:cycles_Rhat}
\end{figure}

To construct $\widehat \cR_2$ observe that $k_{s}^{(2)}(z) = (k_{s,1}^{(2)}(z), \, k_{s,1}^{(2)}(z^{-1}))$, implying that $k_s^{(2)}$ can be expressed in terms of the \emph{scalar} function $k_{s,1}^{(2)}(z)$ only. Let us define $u(z) = \gamma(z)k_{s,1}^{(2)}(z)$ on $\C$ with the cuts removed. Then on the cuts (oriented from left to right as usual), we have $u_\pm(z) = u_{\pm}(z^{-1})$. Hence, $u(z)$ can be extended to a function on the Riemann surface $\wh\cR_2$ obtained by gluing the cut $[\alpha, b]$ with $[b^{-1}, \alpha^{-1}]$ and $[a, \beta]$ with $[\beta^{-1}, a^{-1}]$ such that a path entering say the cut $[\alpha, b]$ from the upper-half plane, will enter the upper-half plane again through the cut $[b^{-1}, \alpha^{-1}]$, see Figure \ref{fig:cycles_Rhat}.

Observe that the mapping $\cR_2 \to \wh \cR_2$ induced by
\begin{align*}
    \begin{cases}
        z^{(1)} \mapsto z
        \\
        z^{(2)} \mapsto z^{-1}
    \end{cases},
\end{align*}
where $z^{(j)}$ denotes the point $z$ on the $j^{th}$ sheet (note that $\wh \cR_2$ has only one sheet) is an unbranched covering mapping of degree $2$. It follows from the Riemann--Hurwitz formula that $\wh \cR_2$ is of genus $2$. With this observation, it is now easy to find a canonical basis of cycles $\lbrace \wh{\bf{a}}_j, \wh{\bf{b}}_j \rbrace_{j = 1,2}$, see Figure \ref{fig:cycles_Rhat}. 
Note that $\wh \cR_2$ is obtained from $\cR_2$ by identifying points $p$ and $\iota(p)$, where $\iota$ is the involution $z^{(1)} \mapsto (z^{-1})^{(2)}$ that we encountered in Section \ref{Sect:ExplSol2}. Consider the two holomorphic differentials on $\cR_2$
\begin{align*}
    \wh \omega_1 &= \omega_1 + \omega_3
    \\
    \wh \omega_2 &= 2\omega_2.
\end{align*}
Since $\iota^* \omega_1 = \omega_3$ and $\iota^* \omega_2 = \omega_2$, it follows that $\iota^* \wh\omega_j = \wh \omega_j$ for $j =1,2$. In particular, $\wh \omega_j$ can be viewed as holomorphic differentials living on $\wh \cR_2$ (we keep the notation $\wh \omega_j$). One can now compute
\begin{align*}
    \int_{\wh{\bf{a}}_1} \wh \omega_1 = \frac{1}{2}\left( \int_{\bf{a}_1} (\omega_1 + \omega_3) + \int_{\bf{a}_3} (\omega_1 + \omega_3)\right) = 1, \qquad \int_{\wh{\bf{a}}_2} \wh \omega_1 = \frac{1}{2} \int_{\bf{a}_2} (\omega_1 +\omega_3) = 0,
    \end{align*}
    and similarly $\int_{\wh{\bf{a}}_1} \wh \omega_2 = 0$, $\int_{\wh{\bf{a}}_2} \wh \omega_2 = 1$. Furthermore, as $\wh{\bf{b}}_j = \bf{b}_j$, $j=1,2$, it follows that 
    \begin{align*}
        \wh B = \begin{pmatrix}
            B_{11} + B_{13} & 2B_{12}
            \\
            B_{21} + B_{23} & 2B_{22}
        \end{pmatrix}, \quad \text{with} \ \wh B_{jk} = \int_{\wh{\bf{b}}_j} \wh{\omega}_k.
    \end{align*}
    As $B_{21} = B_{12} = B_{23}$, it indeed follows that $\wh B^\mathsf{T} = \wh B$. 

    Recall that we have defined the function $u(z) = \gamma(z)k_{s,1}^{(2)}(z)$ on the Riemann surface $\wh \cR_2$. Then $u$ will have jump discontinuities along the gaps $\wh G_1 = [b, a] \cup [a^{-1}, b^{-1}]$ and $\wh G_2 = [\beta, \beta^{-1}]$. Note that topologically both gaps are in fact cycles on the Riemann surface which are homologically equivalent to the $\wh{\bf{a}}$-cycles; we orient them from left to right as usual, i.e.~opposite the orientation of the $\wh{\bf{a}}$-cycles. Then the jump condition of $u(z)$ reads:
    \begin{align*}
        u_+(z) = u_-(z)
            e^{2\pi i \Delta_j}, \qquad \text{for } \ z \in \wh G_j,
    \end{align*}
    with $\Delta_1 = (\ln(N)\zeta_{2,e}(\eta,\tau) + \omega_{2,e})/(2\pi i)$ and $\Delta_2 = (\ln(N)\zeta_{2,c}(\eta, \tau) + \omega_{2,c})/(2\pi i)$. 

    As in the construction of $k_s^{(2)}$, the poles of $u$ are restricted to $b$  (also $=b^{-1}$ on $\wh\cR_2$) and $\beta$  (also $=\beta^{-1}$ on $\wh\cR_2$). If we define the underlying Abel map $\wh \cA$ to have base point $\alpha$, then we have 
    \begin{align*}
        \wh \cA(b) = \frac{1}{2}\begin{pmatrix}
            \wh B_{11}
            \\
            \wh B_{12}
        \end{pmatrix}, \qquad  \wh \cA(\beta) = \frac{1}{2}\begin{pmatrix}
            \wh B_{21} 
            \\
            \wh B_{22}
        \end{pmatrix} - \frac{1}{2}\begin{pmatrix}
            1
            \\
            0
        \end{pmatrix}\,.
    \end{align*}
    Let us define the pole divisor $\wh{D}_\infty = b + \beta$, and $\wh K$ the vector of Riemann constants. It remains to find $\wh{z}_1$, $\wh{z}_2$ solving the Jacobi inversion problem 
    \begin{align}\label{eq:JacInvHat}
    \wh\cA(\wh{D}_0) - \wh \cA(\wh{D}_\infty) = (\Delta_1, \
        \Delta_2)^\mathsf{T}
    \end{align}
    with $\wh{D}_0 = \wh{z}_1 + \wh{z}_2$. As before the mapping $\wh\mu \colon \wh G_1 \times \wh G_2 \to \R^2 / \Z^2$ is a bijection (cf.~Lem.~\ref{lem:mu}), giving us a unique choice of $\wh z_1 \in \wh G_1$, $\wh z_2 \in \wh G_2$ (in particular due to uniqueness all divisors are nonspecial). In fact, expressing $\wh \omega_j$ in terms of the $\omega_k$ on the original Riemann surface $\cR_2$, we see that the above Jacobi inversion problem is completely equivalent to the one studied in \eqref{eq:z_1_z_2}. 
    
    We conclude as in Section \ref{Sect:ExplSol2} that $u(z)$ must have the form
    \begin{align*}
        u(z) = \frac{\Theta(\wh \cA(z) - \wh \cA(\wh D_0) + K \, | \ \wh B \, )\Theta(\wh \cA(0) - \wh \cA(\wh D_\infty) + K \, | \ \wh B \,)}{\Theta(\wh \cA(z) - \wh \cA(\wh D_\infty) + K \, | \ \wh B \,)\Theta(\wh \cA(0) - \wh \cA(\wh D_0) + K \, | \ \wh B \,)}\,,
    \end{align*}
    which via 
    \begin{align}\label{eq:uForumala}
        k_s^{(2)}(z) = \gamma^{-1}(z) \big( u(z), \, u(z^{-1}))
    \end{align}
    demonstrates that $k_s^{(2)}$ can be expressed via genus $2$ Riemann Theta functions. 

\subsection{The modulated two-cuts solution}
\label{subsec:two_cuts_solution}

With the global outer parametrix $k_s^{(2)}(z)$ explicitly constructed in Section \ref{Sect:genus2Red} on the genus $2$ Riemann surface $\widehat{\mathcal{R}}_2 = \widehat{\mathcal{R}}_2(\alpha, b(\eta,\tau), a(\eta,\tau), \beta)$, and the local parametrices handling the branch point singularities as in Section \ref{subsec:airy_local_par}, we are now in the position to extract the leading order asymptotics for the Toda lattice coefficients $a_n(t)$ and $b_n(t)$ in the two-cuts regime.

\subsubsection{Global parametrix and small-norm error analysis}
Unlike the complete one-cut regime, the two internal branch points $b \equiv b(\eta,\tau)$ and $a \equiv a(\eta,\tau)$ are dynamically modulated by the Whitham equations. We define the global matrix parametrix $M_2(z)$ piecewise on $\C$:
\begin{equation}
    M_2(z) = \begin{cases}
        M_p(z),                                                                                                     & z \in \D(p, \varepsilon) \quad \text{for } p \in \{\alpha, b, a, \beta, \beta^{-1}, a^{-1}, b^{-1}, \alpha^{-1}\}, \\[6pt]
        \frac{1}{2} \begin{pmatrix} k_s^{(2)}(z) + \ell_a^{(2)}(z) \\ k_s^{(2)}(z) - \ell_a^{(2)}(z) \end{pmatrix}, & z \in \C \setminus \bigcup_{p} \D(p, \varepsilon),
    \end{cases}
\end{equation}
where $M_p(z)$ are the local parametrices, and $k_s^{(2)}(z), \ell_a^{(2)}(z)$ are the symmetric and antisymmetric vector solutions to the model RHP on the genus 3 Riemann surface, constructed in Section \ref{sec:model_problems} (with $k_s^{(2)}(z)$ being further simplified to a genus $2$ solution in Sect.~\ref{Sect:genus2Red}). Note that $(1, 1) M_2(z) = k_s^{(2)}(z)$ in the outer region.

Following the Deift--Zhou non-linear steepest descent method \cite{Deift1993}, we define the error vector:
\begin{equation}\label{eq:Defr2}
    r^{(2)}(z) = \frac{1}{v_1(0)}v(z) M_2^{-1}(z), \qquad z \in \C \setminus \Sigma_r,
\end{equation}
where $\Sigma_r$ consists of the disk boundaries $\partial \D(p, \varepsilon)$ and the segments of the lenses and real axis outside the disks. As established by the lens opening analysis (see Section \ref{subsec:Lenses_two_cuts}), the jump matrix $J_r(z)$ satisfies:
\begin{itemize}
    \item On the disk boundaries $\partial \D(p, \varepsilon)$, the asymptotic matching yields $J_r(z) = I + \cO\big(\frac{1}{\ln N}\big)$ as $N \to \infty$.
    \item On the lens contours outside the disks, the off-diagonal entries decay algebraically, therefore $J_r(z) = I + \cO(N^{-\delta})$ for some $\delta > 0$.
    \item On the gaps outside the disks, the jump of $M_2(z)$ exactly matches $J_v(z)$, so $J_r(z) = I$.
\end{itemize}
Consequently, we have the uniform estimate:
\begin{equation}
    \|J_r - I\|_{L^\infty(\Sigma_r)} = \cO\left( \frac{1}{\ln N} \right), \qquad N \to \infty.
\end{equation}
By the small-norm theory for vector Riemann--Hilbert problems (see Appendix ~\ref{App:SmallNorm}), the singular integral operator $I - C_{w_r}$ is invertible, which implies that:
\begin{equation}\label{eq:r2Asym}
    r^{(2)}(z) = (1, 1) + \cO\left(\frac{1}{\ln N}\right)
\end{equation}
uniformly for $z$ bounded away from $\Sigma_r$.

\subsubsection{Asymptotic expansion of the vector solution near the origin}
We now unwind the transformations to recover the vector $q(z)$ in a neighborhood of the origin $z = 0$. Since the origin lies outside all the local parametrix' disks and lenses ($0 < \alpha < b < a < \beta < 1$), we have $\wh q(z) = q(z)$. Furthermore, $v(z) = p(z) = \wh q(z) e^{\sigma_3(\ln(N)g(z) + f(z))}$. Therefore from \eqref{eq:Defr2} and \eqref{eq:r2Asym} it follows that:
\begin{align*}
    q(z) &= v_1(0)\left[r^{(2)}(z) M_2(z) + \cO\left( \frac{1}{\ln N} \right)\right]  e^{-\sigma_3(\ln(N)g(z) + f(z))}
    \\
    &= v_1(0)\left[k_s^{(2)}(z)  + \cO\left( \frac{1}{\ln N} \right)\right]e^{-\sigma_3(\ln(N)g(z) + f(z))}.
\end{align*}
From RHP \ref{RHP:q}, the exact vector $q(z)$ satisfies the normalization condition $q_1(0)q_2(0) = 1$ with $q_1(0) > 0$. Additionally, by construction, we have that $k_{s,1}^{(2)}(0; \vec{\Delta}_{2,n}) = 1$ (recall the definition of $\vec{\Delta}_{2,n}$ in \eqref{eq:Delta2n}). It now follows that $v_1^{2}(0)[k_{s,2}^{(2)}(0; \vec{\Delta}_{2,n}) + \cO(1/\ln N)] = 1$. Solving for $v_1(0)$ and substituting this into the expression for the first component, we obtain 
\begin{align}\label{eq:m1_asymptotic_expansion_two_cuts}
    q_1(z)&= \frac{1}{\sqrt{k_{s,2}^{(2)}(0;\vec{\Delta}_{2,n})}} k_{s,1}^{(2)}(z;\vec{\Delta}_{2,n}) e^{-\ln(N)g(z) - f(z)} \left( 1 + \cO\left( \frac{1}{\ln N} \right) \right)\,.
\end{align}

We now express $k_{s,2}^{(2)}(0;\vec{\Delta}_{2,n})$ explicitly in terms of the genus 2 Riemann theta function $\Theta(A \mid \wh B)$ introduced in Section \ref{Sect:genus2Red}. Recall from \eqref{eq:uForumala} that:
\begin{equation}
    k_s^{(2)}(z;\vec{\Delta}_{2,n}) = \gamma(z)^{-1} \big( u(z), \, u(z^{-1}) \big),
\end{equation}
where 
\begin{align*}
    u(z) = \frac{\Theta(\wh \cA(z) - \wh \cA(\wh D_0) + K \mid \wh B)\Theta(\wh \cA(0) - \wh \cA(\wh D_\infty) + K \mid \wh B)}{\Theta(\wh \cA(z) - \wh \cA(\wh D_\infty) + K \mid \wh B)\Theta(\wh \cA(0) - \wh \cA(\wh D_0) + K \mid \wh B)}.
\end{align*}
Here $\wh D_\infty = b + \beta$ and $\wh D_0 = \wh z_1 + \wh z_2$, where $\wh z_1, \wh z_2$ solve \eqref{eq:JacInvHat} (which is equivalent to \eqref{eq:z_1_z_2}). To simplify the notation, we define the following functions: 
\begin{align}
    \Theta_0(A; \vec{\Delta}_{2, n}) &= \Theta(A - \wh \cA(\wh D_0) + K \mid \wh B)\,,\\
    \Theta_\infty(A) &= \Theta(A - \wh \cA(\wh D_\infty) + K \mid \wh B)\,.
\end{align}
Let $\wh A_0 \coloneqq \wh \cA(0)$. Consider the meromorphic function $y(z) = (z+z^{-1}) - (\alpha + \alpha^{-1})$ on $\wh \cR_2$. The divisor of $y$ is given by $2(\alpha) - (0) - (\infty)$. Abel's theorem implies that $\wh \cA(\infty) = 2\wh \cA(\alpha) -\wh \cA(0) = -\wh \cA(0) = -\wh A_0$ (recalling that the base point is $\alpha$). 
Evaluating the second component of $k_s^{(2)}$ at $z = 0$, where $\lim_{z\to 0}\gamma(z) = 1$ and $z^{-1} \to \infty$, we find:
\begin{equation}
\label{eq:ks2_explicit_theta_two_cuts}
    k_{s,2}^{(2)}(0;\vec{\Delta}_{2,n}) = u(\infty) = \frac{\Theta_0(-\wh A_0; \vec{\Delta}_{2,n})\Theta_\infty(\wh A_0)}{\Theta_\infty(-\wh A_0)\Theta_0(\wh A_0; \vec{\Delta}_{2,n})}.
\end{equation}
Notice that the divisor $\wh D_\infty = b(\eta,\tau) + \beta$ depends only on the slow macroscopic parameters $(\eta,\tau)$ and is completely independent of the fast periodic phase $\vec{\Delta}_{2,n}$. Under the discrete lattice shift $n \mapsto n+1$, the slow variable shifts as $\eta \mapsto \eta + \frac{1}{\ln N}$. Since the Whitham branch points and the Abel map depend smoothly on $\eta$, a first-order Taylor expansion gives:
\begin{equation}
    \wh \cA^{(N+1)}\left(\wh D_\infty\left(\eta + \frac{1}{\ln N}, \tau\right)\right) = \wh \cA^{(N)}(\wh D_\infty(\eta, \tau)) + \cO\left(\frac{1}{\ln N}\right).
\end{equation}
In the previous equation, we stressed the dependence of $\wh\cA$ on $N$ since both the Abel map, and the period matrix depend weakly on it. Since the Riemann theta function $\Theta$ is holomorphic and positive on the real torus, the ratio of the $\Theta_\infty$ factors at $n$ and $n+1$ evaluates to $1 + \cO(1/\ln N)$ and is absorbed into the multiplicative error. Similarly, the geometric quantities $\wh A_0$ and the period matrix $\wh B$ depend smoothly on $\eta$, and their shifts under $n \mapsto n+1$ contribute only $\cO(1/\ln N)$ errors.

\subsubsection{Asymptotics of $a^{(q)}_n(t)$ and $b^{(q)}_n(t)$ in the modulated two cuts region}
Using the results of the previous section, we can now compute the asymptotic behavior of $a^{(q)}_n(t),b^{(q)}_n(t)$ in the modulated two cuts regime.

\paragraph{Asymptotics of $a^{(q)}_n(t)$.}
Evaluating \eqref{eq:m1_asymptotic_expansion_two_cuts} at $z = 0$ gives the factor $A^{(q)}(n,t)$:
\begin{equation}
\label{eq:A_nt_asymptotic_two_cuts}
    A^{(q)}(n,t) = q_1(0) = \sqrt{\frac{\Theta_0(\wh A_0; \vec{\Delta}_{2, n})}{\Theta_\infty(\wh A_0)} \frac{\Theta_\infty(-\wh A_0)}{\Theta_0(-\wh A_0; \vec{\Delta}_{2, n})}} \exp\Big( \ln(N)g_\infty(\eta,\tau) + f_\infty \Big) \left( 1 + \cO\left( \frac{1}{\ln N} \right) \right)\,.
\end{equation}
From Lemma \ref{lem:recovery_q}, the Toda coefficient $a^{(q)}_n(t)$ is recovered via $a^{(q)}_n(t) = \frac{1}{2} \frac{A^{(q)}(n,t)}{A^{(q)}(n+1,t)}$. Under the discrete shift $n \mapsto n+1$, the slow variable shifts as $\eta \mapsto \eta + \frac{1}{\ln N}$, leading to the phase difference:
\begin{equation}
    \ln(N)\left(g_\infty(\eta,\tau) - g_\infty\left(\eta + \frac{1}{\ln(N)},\tau\right) \right) = -\partial_\eta g_\infty(\eta,\tau) + \cO\left(\frac{1}{\ln N}\right)\,.
\end{equation}
As established above, the ratio of the $\Theta_\infty$ terms evaluates to $1 + \cO(1/\ln N)$ and is absorbed into the error. Defining the background amplitude $a^{(2)}_{\mathrm{bg}}(\eta,\tau) = \frac{1}{2}\exp(-\partial_\eta g_\infty(\eta,\tau))$, we obtain:
\begin{equation}
    \label{eq:an_final_formula_two_cuts}
    a^{(q)}_n(t) = a^{(2)}_{\mathrm{bg}}(\eta,\tau) \sqrt{ \frac{\Theta_0(-\wh A_0; \vec{\Delta}_{2, n+1}) \Theta_0(\wh A_0; \vec{\Delta}_{2, n})}{\Theta_0(\wh A_0; \vec{\Delta}_{2, n+1}) \Theta_0(-\wh A_0; \vec{\Delta}_{2, n})} } \left( 1 + \cO\left( \frac{1}{\ln N} \right) \right).
\end{equation}

\paragraph{Asymptotics of $b^{(q)}_n(t)$.}
To extract $b^{(q)}_n(t)$, we evaluate the $\cO(z)$ term of $q_1(z)$. From Lemma \ref{lem:recovery_q}, $q_1(z) = A^{(q)}(n,t)(1 - 2B^{(q)}(n-1,t)z + \cO(z^2))$, which implies:
\begin{equation}
    -2B^{(q)}(n-1,t) = \left. \frac{\di}{\di z} \ln q_1(z; n, t) \right|_{z=0}.
\end{equation}
Differentiating the logarithm of \eqref{eq:m1_asymptotic_expansion_two_cuts} at $z = 0$ gives:
\begin{equation}
    -2B^{(q)}(n-1,t) = \left. \frac{\di}{\di z} \ln k_{s,1}^{(2)}(z; \vec{\Delta}_{2,n}) \right|_{z=0} - \ln(N) g'(0;\eta,\tau) - f'(0) + \cO\left( \frac{1}{\ln N} \right).
\end{equation}
From Lemma \ref{lem:recovery_q},
\begin{equation}
    b^{(q)}_n(t) = B^{(q)}(n,t) - B^{(q)}(n-1,t) = \frac{1}{2} \left[ \left. \frac{\di}{\di z} \ln q_1(z; n, t) \right|_{z=0} - \left. \frac{\di}{\di z} \ln q_1(z; n+1, t) \right|_{z=0} \right]\,.
\end{equation}
Subtracting the expressions at $n$ and $n+1$, the contribution of the slow terms involving $\Theta_\infty$ and $\gamma(z)$ is of order $\cO(1/\ln N)$. Defining the background value $b^{(2)}_{\mathrm{bg}}(\eta,\tau) = \frac{1}{2} \partial_\eta \big(g'(0;\eta,\tau)\big)$, we arrive at:
\begin{equation}
    \label{eq:bn_final_formula_two_cuts}
    b^{(q)}_n(t) = b^{(2)}_{\mathrm{bg}}(\eta,\tau) + \frac{1}{2} \left[ \frac{\Theta_0'(\wh A_0; \vec{\Delta}_{2, n})}{\Theta_0(\wh A_0; \vec{\Delta}_{2, n})} - \frac{\Theta_0'(\wh A_0; \vec{\Delta}_{2, n+1})}{\Theta_0(\wh A_0; \vec{\Delta}_{2, n+1})} \right] + \cO\left( \frac{1}{\ln N} \right)\,,
\end{equation}
where $\Theta_0'$ denotes the directional derivative along $\wh\cA'(0)$, i.e.
\begin{equation}
    \Theta_0'(\wh A_0; \vec{\Delta}_{2, n}) = \left. \nabla_{\wh\cA(z)}\Theta_0(\wh\cA(z);\vec\Delta_{2,n}) \cdot \partial_z \wh\cA(z) \right|_{z=0} \,.
\end{equation}
Therefore, we proved the following: 
\begin{theorem}[Modulated Two-Cuts Asymptotics]
\label{thm:modulated_two_cuts_asymptotics}
Let the phase parameters $(\eta, \tau)$ lie strictly within the two-cuts regime $\cU_2$. As $N \to \infty$, the Toda lattice variables $a^{(q)}_n(t)$ and $b^{(q)}_n(t)$ with $n = \eta \ln N$ and $t = \tau \ln N$ are asymptotically given by a modulated genus-$2$ hyperelliptic wave:
\begin{align}
    a^{(q)}_n(t) & = a^{(2)}_{\mathrm{bg}}(\eta,\tau) \sqrt{ \frac{\Theta_0(-\wh A_0; \vec{\Delta}_{2, n+1}) \Theta_0(\wh A_0; \vec{\Delta}_{2, n})}{\Theta_0(\wh A_0; \vec{\Delta}_{2, n+1}) \Theta_0(-\wh A_0; \vec{\Delta}_{2, n})} } \left( 1 + \cO\left( \frac{1}{\ln N} \right) \right), \label{eq:thm_an_asymp_two_cuts}              \\
    b^{(q)}_n(t) & = b^{(2)}_{\mathrm{bg}}(\eta,\tau) + \frac{1}{2} \left[ \frac{\Theta_0'(\wh A_0; \vec{\Delta}_{2, n})}{\Theta_0(\wh A_0; \vec{\Delta}_{2, n})} - \frac{\Theta_0'(\wh A_0; \vec{\Delta}_{2, n+1})}{\Theta_0(\wh A_0; \vec{\Delta}_{2, n+1})} \right] + \cO\left( \frac{1}{\ln N} \right), \label{eq:thm_bn_asymp_two_cuts}
\end{align}
where:
\begin{itemize}
    \item $\vec{\Delta}_{2, n} = \frac{1}{2\pi i} \left( \ln(N)\vec{\zeta}_2(\eta, \tau) + \vec{\omega}_2(\eta, \tau) \right)$ is the fast phase;
    \item $\wh A_0 = \wh\cA(0)$ is the Abel map of $\wh\cR_2$, and $\wh B$ is the $2 \times 2$ period matrix evaluated on $\wh \cR_2$;
    \item $\Theta_0'$ denotes the directional derivative along $\wh\cA'(0)$, the Abel map derivative at the origin;
    \item The function $\Theta_0(A; \vec{\Delta}_{2,n})$ is defined as $\Theta(A - \wh \cA(\wh D_0) + K \mid \wh B)$ with $\wh D_0$ depending on $\vec{\Delta}_{2,n}$;
    \item The background amplitudes $a^{(2)}_{\mathrm{bg}}(\eta,\tau)$ and $b^{(2)}_{\mathrm{bg}}(\eta,\tau)$ are given explicitly by:
          \begin{align}
              a^{(2)}_{\mathrm{bg}}(\eta,\tau) & = \frac{1}{2}\exp(-\partial_\eta g_\infty(\eta,\tau)), \label{eq:thm_abg_explicit_two_cuts}  \\
              b^{(2)}_{\mathrm{bg}}(\eta,\tau) & = \frac{1}{2} \partial_\eta \big(g'(0;\eta,\tau)\big), \label{eq:thm_bbg_explicit_two_cuts}
          \end{align}
          and are evaluated explicitly in Proposition \ref{prop:background_two_cuts}.
\end{itemize}
\end{theorem}

So, to fully describe the Toda lattice behaviour in the two cut region, we have to compute the two background coefficients $a^{(2)}_{\mathrm{bg}}(\eta,\tau), \, b^{(2)}_{\mathrm{bg}}(\eta,\tau) $.
\subsubsection{Explicit evaluation of the background coefficients in the two-cut region}

To compute the background coefficients $a^{(2)}_{\mathrm{bg}}(\eta,\tau)$ and $b^{(2)}_{\mathrm{bg}}(\eta,\tau)$ in the interior of the two-cuts region, one has to differentiate the $g$-function with respect to the parameter $\eta$. This would lead to expressions involving $\partial_\eta \wt a$ and $\partial_\eta \wt b$, which are implicitly defined by the modulation equations. To avoid this, we use the following technical Lemma that bypasses the endpoint derivatives explicit calculation.

\begin{lemma}
\label{lem:P_eta_two_cut}
Fix $0 < \alpha < b < a < \beta < 1$ and let $H(x) = (x-\wt\alpha)^{1/2}(x-\wt\beta)^{1/2}(x-\wt a)^{1/2}(x-\wt b)^{1/2}$, such that $H(x) \sim x^2$ as $x \to \infty$. 

Define the gap integrals
\begin{align}
    M_j &= \int_{\wt a}^{\wt b} \frac{x^j}{\sqrt{(\wt\alpha-x)(x-\wt\beta)(x-\wt a)(\wt b-x)(x^2-1)}} \, \di x\,,\\
    N_j &= \int_1^{\wt\beta} \frac{x^j}{\sqrt{(\wt\alpha-x)(\wt\beta-x)(\wt a-x)(\wt b-x)(x^2-1)}} \, \di x\,,
\end{align}
for $j=0,1,2$, where the square roots are taken to be positive. There exists a unique monic polynomial of degree two, $P_\eta(x) = x^2 + c_1 x + c_0$, whose coefficients are given explicitly by
\begin{equation}
    c_1 = \frac{M_0 N_2 - M_2 N_0}{M_1 N_0 - M_0 N_1}\,, \qquad c_0 = \frac{M_2 N_1 - M_1 N_2}{M_1 N_0 - M_0 N_1}\,,
\end{equation}
such that the vanishing conditions 
\begin{equation}
\label{eq:vanishing_conditions_derivative}
    \int_{\wt a}^{\wt b} \frac{P_\eta(x)}{H_+(x)\sqrt{x^2-1}} \di x = 0\, \qquad \int_1^{\wt\beta} \frac{P_\eta(x)}{H_+(x)\sqrt{x^2-1}} \di x = 0
\end{equation} hold.
\end{lemma}
\begin{remark}
\label{rem:equivalence_expression}
    we notice that the conditions \eqref{eq:vanishing_conditions_derivative} have the exact same structure as the derivative with respect to $\eta$ of \eqref{eq:two_cuts_gap_condition}-\eqref{eq:two_cuts_normalization}. Indeed, we will use the previous Lemma to simplify those expression, i.e. express $\partial_\eta\wt b, \partial_\eta \wt a$ in terms of known quantities.
\end{remark}
\begin{proof}
The  conditions \eqref{eq:vanishing_conditions_derivative} can be rewritten as the linear system 

\begin{equation}
    \begin{cases}
        M_2 + c_1 M_1 + c_0 M_0 = 0\\[4pt] N_2 + c_1 N_1 + c_0 N_0 = 0\,.
    \end{cases}
\end{equation} 
The determinant of this system is $D = M_1 N_0 - M_0 N_1$. Since $x \in (\wt a, \wt b)$ in the first integral and $x \in (1, \wt\beta)$ in the second, we deduce that $1 < N_1/N_0 < \wt\beta < \wt a < M_1/M_0 < \wt b$, ensuring $D > 0$. Cramer's rule immediately yields the explicit formulas for $c_1$ and $c_0$.
\end{proof}

Given the previous Lemma, we can now express $\partial_\eta g(z)$ in terms of the polynomial $P_\eta(x)$ as follows

\begin{proposition}
\label{prop:g_derivs}
For $z$ in the exterior of the cuts $[\alpha,b]\cup[a,\beta]\cup[\beta^{-1},a^{-1}]\cup[b^{-1},\alpha^{-1}]$, the parameter derivative $\partial_\eta g(z)$ satisfies
\begin{equation}
\label{eq:partial_der_g_eta}
    \partial_\eta g(z)= -\ln(z) + \int_{\alpha^{-1}}^z\frac{P_\eta(x(s))}{s H(x(s))}\di s\,,
\end{equation}
where $P_\eta(x)$ is the polynomial defined in Lemma \ref{lem:P_eta_two_cut}.
\end{proposition}
\begin{proof}
Recall the representation of the $g$-function derivative from Proposition \ref{prop:g_two_cut}, where the integrand is governed by $\Omega_2(z) = \frac{R_2(z) L(x(z))}{z}$ with $L(x) = \tau x + \gamma_2$ and $R_2(z) = \frac{(x(z)-\wt a)(x(z)-\wt b)}{H(x(z))}$. The main difficulty is the explicit differentiation of $\Omega_2(z)$ with respect to $\eta$, indeed one must track the derivative of the moving endpoints $\wt a(\eta,\tau)$ and $\wt b(\eta,\tau)$. To do this, we first compute the logarithmic derivative of $R_2$ 
\begin{equation}
    \frac{\partial_\eta R_2}{R_2} = -\frac{\partial_\eta{\wt a}}{2(x-\wt a)} - \frac{\partial_\eta{\wt b}}{2(x-\wt b)}\,.
\end{equation}
Moreover, from the definition of $\gamma_2(\eta,\tau)$, we have $\partial_\eta L(x) = 1 + \frac{\tau}{2}(\partial_\eta{\wt a} + \partial_\eta{\wt b})$. Combining these, the $\eta$-derivative of the numerator of $\Omega_2(z)$ reads
\begin{equation}
    \partial_\eta(R_2 L) = R_2 \left[ 1 + \frac{\tau}{2}(\partial_\eta{\wt a} + \partial_\eta{\wt b}) - \frac{L(x)\partial_\eta{\wt a}}{2(x-\wt a)} - \frac{L(x)\partial_\eta{\wt b}}{2(x-\wt b)} \right]\,.
\end{equation}
We can absorb the $x$-dependence in the numerator of the fractions by observing that $\tau - \frac{L(x)}{x-e} = -\frac{L(e)}{x-e}$ for $e \in \{\wt a, \wt b\}$. Thus, the expression simplifies to
\begin{equation}
    \partial_\eta(R_2 L) = R_2 \left[ 1 - \frac{L(\wt a)\partial_\eta{\wt a}}{2(x-\wt a)} - \frac{L(\wt b)\partial_\eta{\wt b}}{2(x-\wt b)} \right] = \frac{P_\eta(x)}{H(x)}\,,
\end{equation}
where
\begin{equation}
    P_\eta(x) = (x-\wt a)(x-\wt b) - \frac{\partial_\eta{\wt a} L(\wt a)}{2}(x-\wt b) - \frac{\partial_\eta{\wt b} L(\wt b)}{2}(x-\wt a)\,.
\end{equation}
This explicitly demonstrates that $P_\eta(x)$ is a monic polynomial of degree two in $x$. Since $g(z)$ must satisfy $\partial_\eta \int \Omega_2 \di s = 0$ along the gaps (as the gap conditions are constants), this polynomial $P_\eta(x)$ satisfies the vanishing conditions \eqref{eq:vanishing_conditions_derivative} in Lemma \ref{lem:P_eta_two_cut}, see also Remark \ref{rem:equivalence_expression}. Integration of $\partial_\eta \Omega_2(s)$ from $\alpha^{-1}$ to $z$ yields the claim.
\end{proof}
Given the previous expression for $\partial_\eta g(z)$, we can compute the background coefficients as
\begin{proposition}
\label{prop:background_two_cuts}
In the interior of the two-cut region $\cU_2$, the background coefficients are explicitly given by:
\begin{align}
    b^{(2)}_{\mathrm{bg}}(\eta,\tau) &= \frac{\wt\alpha+\wt\beta+\wt a+\wt b}{2} + c_1\,, \\
    a^{(2)}_{\mathrm{bg}}(\eta,\tau) &= \frac{1}{2\alpha} \exp\left( -\int_{\wt\alpha}^\infty \left( \frac{P_\eta(x)}{H(x)} - 1 \right) \frac{\di x}{\sqrt{x^2-1}} \right)\,.
\end{align}
\end{proposition}
\begin{proof}
Recall the definitions of the background coefficients

\begin{equation}
    b^{(2)}_{\mathrm{bg}} = \frac{1}{2} \partial_\eta g'(0)\,,\qquad a^{(2)}_{\mathrm{bg}} = \frac{1}{2} e^{-\partial_\eta g_\infty}\,.
\end{equation}
Expanding $H(x)$ at large $x$, we find $H(x) = x^2 - \frac{S}{2}x + \mathcal{O}(1)$, where $S = \wt\alpha+\wt\beta+\wt a+\wt b$. Therefore 

\begin{equation}
    \frac{P_\eta(x)}{H(x)} = 1 + \frac{c_1 + S/2}{x} + \mathcal{O}(x^{-2})\,.
\end{equation}
As $z \to 0$, $x(z) = \frac{1}{2z} + \mathcal{O}(z)$, so $x(z)^{-1} = 2z + \mathcal{O}(z^3)$. Taking the $z$ derivative of \eqref{eq:partial_der_g_eta}, and substituting the previous asymptotic into the formula, the $-\frac{1}{z}$ pole given by the derivative of $\log(z)$ cancels with the leading term of the expansion, leaving $\partial_\eta g'(0) = 2c_1 + S$.
This immediately gives the formula for $b^{(2)}_{\mathrm{bg}}$.

For $a^{(2)}_{\mathrm{bg}}$, we can rewrite \eqref{eq:partial_der_g_eta} as

\begin{equation}
    \partial_\eta g(z) = \int_{\alpha^{-1}}^z \frac{1}{s}\left( \frac{P_\eta(x(s))}{H(x(s))} - 1\right) \di s + \ln(\alpha)\,.
\end{equation}
Passing to the Joukowsky coordinates and exponentiation yield to the result.
\end{proof}

\section{The one-cut modulated region}
\label{sec:one_cut_modulated_regions}
In this section, we investigate the remaining parts of the $(\eta,\tau)$-plane. In particular, we describe the region that we call \textit{one-cut modulated}, since the $g$-function will be still supported on two disjoint intervals as in Section \ref{sec:complete_one_cut}, but two of the extremes will be modulated as in Section \ref{sec:two_cuts}. 
Let us define the following two sets in the $(\eta,\tau)$-plane,

\begin{equation}
\label{eq:modulated_regions}
    \cU_{r} = \left\{(\eta,\tau)\in \R^2\,\vert\, \tau\geq\tau_{\text{sn}}\right\}\cap \left(\cU_c \cup \cU_{sn}\cup \cU_2\right)^c\,, \qquad \cU_{\ell} = \left\{(\eta,\tau)\in \R^2\,\vert\, \tau\leq\tau_{\text{sn}}\right\}\cap \left(\cU_c \cup \cU_{sn}\cup \cU_2\right)^c\,,
\end{equation}
where $\tau_{\text{sn}}$ is defined in Remark \ref{Rmk:eta3}. We call the set $\cU_r$ modulated right one-cut region, and $\cU_\ell$ the modulated left one-cut region. In the next subsections, we show how to analyze the asymptotic behavior of the Toda lattice solution in these two sets.

\subsection[Modulated left one-cut]{From the complete one-cut and two-cuts modulated regions to the modulated left one-cut}
In this subsection, we focus on the region $\cU_{\ell}$ \eqref{eq:modulated_regions}; we proceed as in Section \ref{sec:complete_one_cut}, but we require that both the $g,f$-functions satisfy a different RHP. We notice that we use again $g(z),f(z)$ to label the functions, hopefully without confusion.  As before, we consider the transformation \eqref{eq:def_transformation_1} and we require that $g(z),f(z)$ to satisfy the following RHPs

\begin{rhp}
      \label{rhp:g_function_one_cut_left}
      We seek for a scalar function $g(z)$ such that 
      \begin{itemize}
            \item $g(z)$ is analytic in $\C\setminus \left(\alpha,\alpha^{-1}\right)$
            \item $g_+(z)+g_-(z)+1 + \tau\Phi(z,\eta,\tau) = 0$ for $z\in (\alpha,b)$
            \item $g_+(z)+g_-(z) - 1 + \tau\Phi(z,\eta,\tau) = 0$ for $z\in \left(\frac{1}{b},\frac{1}{\alpha}\right)$
            \item $g_+(z) - g_-(z) = \zeta_\ell(\eta,\tau)$ for $z\in (b,b^{-1})$
            \item $g(z^{-1}) = -g(z)$
            \item $g(z) = g_\infty + \cO(\frac{1}{z})$ as $z\to\infty$
            \item $g(z) + \frac{1 + \tau\Phi(z,\eta,\tau)}{2} = \cO((z-\alpha)^{1/2})$ as $z\to \alpha$ and $g(z) + \frac{1 + \tau\Phi(z,\eta,\tau)}{2} = \pm \frac{1}{2}\zeta_{\ell}+\cO((z-b)^{3/2})$ as $z\to b$, where $\pm$ indicates that $z \in \mathbb{C}_\pm$
\end{itemize}
\end{rhp}

 \begin{rhp}
    \label{rhp:f_function_one_cut_left}
    We seek for a scalar function $f(z)$ such that

    \begin{itemize}
        \item $f(z)$ is analytic in $\C\setminus \left(\alpha,\alpha^{-1}\right)$
        \item $f_+(z)+f_-(z) + \ln(2\pi ) + \ln(z) + \ln(\mu_1\rho(z)) = 0$ for $z\in(\alpha,b)$
        \item $f_+(z)+f_-(z)  -\ln(2\pi ) + \ln(z) - \ln(\mu_1\rho(z^{-1})) = 0$ for $z\in(b^{-1},\alpha^{-1})$
        \item $f_+(z)-f_-(z) = \omega_\ell$ for $z\in(b,b^{-1})$
        \item $f(z^{-1}) = -f(z)$
        \item $f(z) = f_\infty + \cO(\frac{1}{z})$ as $z\to\infty$
    \end{itemize}
\end{rhp}

In both RHPs the point $b$ is the same and $b\in[\alpha,\beta]$. In particular, $b$ has to be determined so that the resulting RHP for the vector $p(z)$ has good asymptotic behavior, therefore $b\equiv b(\eta.\tau)$. To determine this point, we use a continuation argument based on the Implicit Function Theorem (IFT) as we did in Section \ref{sec:two_cuts}. Specifically, we can prove the following

\begin{proposition}
\label{prop:g_left_cut}
Let $(\eta, \tau) \in \cU_\ell$. Define the function $g(z)$ as
    \begin{equation}
      g(z) = \frac{1-\tau\Phi(z,\eta, \tau)}{2} + \int_{\alpha^{-1}}^z \Omega_\ell(s,\eta,\tau) \di s \,, \qquad \Omega_\ell(s,\eta,\tau) =\frac{R_\ell(s)\left( \frac{\tau}{2}(s +s^{-1}) + \gamma_\ell(\eta,\tau) \right)}{s} 
\end{equation}
where
\begin{equation}
      \label{eq:def_Rell_gammaell}
      R_\ell(z) = \frac{\left(z + \frac{1}{z} - b - \frac{1}{b}\right)^{1/2}}{\left(z + \frac{1}{z} - \alpha - \frac{1}{\alpha}\right)^{1/2}}, \qquad \gamma_\ell(\eta,\tau) = \eta + \frac{\tau}{4} \left(\frac{1}{b} + b - \frac{1}{\alpha} - \alpha\right)\, \qquad  \zeta_\ell(\eta, \tau) = 2\int_\alpha^b \Omega_{\ell,+}(s, \eta, \tau) \, \di s.
\end{equation}
Then there exists a unique $b \in \left(\alpha,\beta\right)$, such that the integral equation
\begin{equation}
\label{eq:panic_eq_left}
      \int_b^1 \Omega_\ell(s,\eta,\tau) \di s = \frac{1}{2}\,
\end{equation}
holds, in which case $g(z)$ solves RHP \ref{rhp:g_function_one_cut_left}.
\end{proposition}
\begin{remark}
\label{rem:Rell_boundary}
    Notice that $R_\ell(z)$ is analytic in $\C\setminus\left\{(\alpha,b), (b^{-1},\alpha^{-1})\right\}$, and we chose the branch-cut so that $R(z)\to 1$ as $z\to\infty$, this determines $R_\ell(z)$ uniquely. Furthermore, we notice that $R_{\ell,\pm}(z)\in i\R_{\pm}$ for $z\in (\alpha,b)$, $R_{\ell,\pm}(z)\in i\R_{\mp}$ for $z\in (b^{-1},\alpha^{-1})$ and $R_{\ell}(z) \in \R_+$ for $z \in (b, b^{-1})$.
\end{remark}
\begin{proof}
    By standard manipulation, it can be checked that if $g(z)$ is defined as before and \eqref{eq:panic_eq_left} is satisfied, then $g(z)$ solves RHP \ref{rhp:g_function_one_cut_left}; indeed, the condition $\int_b^1 \Omega_\ell(s,\eta,\tau) \di s = 1/2$ guarantees that $\int_{b^{-1}}^b \Omega_\ell(s,\eta,\tau) \di s = -1$, which provides the correct jump $g_+(z)+g_-(z) = -1 - \tau\Phi(z)$ on the cut $(\alpha, b)$. Therefore, we only have to prove the existence of such a point $b$.
    
    We move to the Joukowsky variables by applying the mapping $x(s) = \frac{1}{2}(s + s^{-1})$ \eqref{eq:Joukowsky}. Notice that the map is monotonically decreasing in $s\in(0,1)$, therefore  $1 < \wt \beta < \wt \alpha$, where $\wt \alpha = x(\alpha)$ and $\wt \beta = x(\beta)$.
    
    In these coordinates, using $\frac{\di s}{s} = - \frac{\di x }{\sqrt{x^2-1}}$, the integral equation \eqref{eq:panic_eq_left} is equivalent to finding a unique $\wt b = x(b) \in (\wt \beta, \wt \alpha)$ such that $F_\ell(\wt b, \eta) = 1/2$, where $F_\ell(s,\eta)$ is given by:
    \begin{equation}
        F_\ell(s, \eta) = \int_1^s \frac{(s - x)^{1/2}}{(\wt \alpha - x)^{1/2}\sqrt{x^2 - 1}} \left( \tau x + \eta + \frac{\tau}{2}(s - \wt \alpha) \right) \di x \,, \qquad s \in [1, \wt \alpha], \, \eta \in \mathbb R,
    \end{equation}
    (note that $F_\ell$ also depends on $\tau$, but $\eta$ will be the parameter we will vary). A simple computation gives us
    \begin{align}
        \partial_s F_\ell(s, \eta) &= \int_1^s \frac{1}{(\wt \alpha - x)^{1/2}\sqrt{x^2 - 1}} \left[ \frac{\tau x + \eta + \frac{\tau}{2}(s - \wt \alpha)}{2(s - x)^{1/2}} + \frac{\tau}{2}(s - x)^{1/2} \right] \di x  \nonumber \\\label{eq:DsF}
        &= \frac{3\tau s - \tau \wt \alpha + 2\eta}{4} \underbrace{\int_1^s \frac{1}{\sqrt{s - x}\sqrt{\wt \alpha - x}\sqrt{x^2 - 1}} \di x }_{>0}\,.
    \end{align}
The sign of $\partial_s F_\ell$ depends exclusively on the linear prefactor which vanishes at $s_{\text{max}} =  \frac{\tau \wt \alpha - 2\eta}{3\tau}$. As $\tau < \tau_{\text{sn}} < 0$, we always have that if $s_{\max}$ lies in the domain of $F_\ell$, i.e.~$s_{\max} \in [1, \wt \alpha]$, then it will be a \emph{global} maximum of $F_\ell$. Also, note that we have 
\begin{equation}\label{eq:Feta}
\partial_\eta F_\ell(s, \eta) = \int_1^s \frac{(s - x)^{1/2}}{(\wt \alpha - x)^{1/2}\sqrt{x^2 - 1}} \di x  > 0\,,
\end{equation}
implying that $F_\ell(s_{\text{max}}(\eta), \eta) < F_\ell(s_{\text{max}}(\eta'), \eta')$ whenever $\eta < \eta'$ and $s_{\text{max}}(\eta), s_{\text{max}}(\eta') \in [1, \wt \alpha]$, indeed
\begin{equation}
\begin{split}
        \partial_\eta F_\ell(s_{\max}(\eta),\eta) &  = \left. \partial_sF_\ell(s,\eta)\right\vert_{s=s_{\max}(\eta)}\partial_\eta s_{\max}(\eta) + \left.\partial_\eta F_{\ell}(s,\eta)\right\vert_{s=s_{\max}(\eta)}\\& = \left.\partial_\eta F_{\ell}(s,\eta)\right\vert_{s=s_{\max}(\eta)} >0\,.
\end{split}
\end{equation}

 We now consider two different situations: the transition from the complete one-cut (i.e. when $\tau\leq \tau_\beta$) and the transition from the two-cuts regime (i.e. when $\tau_{\beta}\leq \tau<\tau_{\text{sn}}$).
    First, we analyze the transition from the complete one-cut regime, which occurs precisely on the boundary curve $Q(\wt \beta, \eta, \tau) = 0$. By fixing $\tau < \tau_\beta < 0$ (where $\tau_\beta$ marks the tangency point with the ellipse $Q(x_v(\eta,\tau),\eta,\tau)=0$, see Remark \ref{rem:intersections_one_cut}), following the same notation as Lemma \ref{lem:building_Uc}, we set $\eta_1 = \eta_1(\tau)$ (i.e.~the boundary of the complete-cut region) as the parameter where we expect the transition to happen. 

     We have a natural candidate for the $g$-function at $(\eta_1, \tau)$, given in terms the complete cut $g$-function from Proposition \ref{prop:g_whole} denoted here by $g_{\text{complete cut}}$. We just need to show that $g_{\text{complete cut}}$ has the required form. To this end, we observe that as $Q(\wt \beta, \eta_1, \tau) = 0$, we must have $Q(x, \eta_1, \tau) = (x - \wt \beta)q(x, \eta_1, \tau)$, where $q$ is a linear polynomial. Substituting this expression into the formula for the complete cut $g$-function $g_{\text{complete cut}}$, it reads:
    \begin{align*}
        g_{\text{complete cut}}(z) = \frac{1-\tau \Phi(z, \eta_1, \tau)}{2} + \int_{\alpha^{-1}}^{z} \frac{R_\ell(s)(\frac{\tau}{2}(s+s^{-1}) + c_\ell(\eta_1, \tau))}{s} \di x ,
    \end{align*}
    and it satisfies Riemann--Hilbert problem \ref{rhp:g_function_one_cut_left}.  In particular, due to the normalization condition at infinity (or at $0$), we must have $c_\ell(\eta_1, \tau) = \gamma_{\ell}(\eta_1, \tau)$, and \eqref{eq:IntP} becomes equivalent to \eqref{eq:panic_eq_left} with $\wt b = \wt \beta$. This gives us at least one solution $F_\ell(\wt \beta, \eta_1) = \frac{1}{2}$. To see that it is the unique solution, it is enough to show that $s_\text{max} < \wt \beta$, as then $F_\ell(s, \eta_1)$ will be strictly decreasing for $s \in [\wt \beta, \wt \alpha]$. To this end note that the prefactor in \eqref{eq:DsF} evaluated at $s = \wt \beta$ can be rewritten in terms of the vertex $x_v(\eta, \tau) = -\frac{\eta}{2\tau} + \frac{\wt \alpha + \wt \beta}{4}$:
    \begin{equation}
    \label{eq:prefactor_left_cut}
        3\tau \wt \beta - \tau \wt \alpha + 2\eta_1 = 4\tau(\wt \beta - x_v(\eta_1, \tau))\,.
    \end{equation}
    Because we fixed $\tau < \tau_\beta < 0$ and defined $\eta_1$ such that $Q(\wt \beta,\eta_1,\tau) = 0$, Corollary \ref{cor:techincal_cor} ensures that $x_v(\eta_1, \tau) < \wt \beta$. Since $\tau < 0$ and $\wt \beta - x_v > 0$, the prefactor in \eqref{eq:DsF} is strictly \textit{negative}, implying $\partial_s F_\ell(s, \eta_1)|_{s = \wt \beta} < 0$ and therefore $s_{\text{max}} < \wt \beta$. In particular $\wt b(\eta_1) = \wt \beta$ is the unique solution of $F_\ell(s, \eta_1) = \frac{1}{2}$.

    Now, using \eqref{eq:Feta}, the Implicit Function Theorem (IFT) guarantees that (at least locally) $\wt b(\eta)$ is a monotonically increasing function of $\eta$:
    \begin{equation}
    \label{eq:b_monotone_left}
        \frac{\partial \wt b}{\partial \eta} = - \frac{\partial_\eta F_\ell(\wt b, \eta)}{\partial_s F_\ell(s, \eta)|_{s=\wt b}} > 0\,.
    \end{equation}
Note that the mapping $\eta \mapsto \wt b$ can be continued for $\eta > \eta_1$ using the IFT as long as $\wt b < \wt \alpha$ and $\partial_s F_\ell(s, \eta)|_{s=\wt b} < 0$. If we would have for some $\eta_*$ that $\partial_s F_\ell(s, \eta_*)|_{s=\wt b(\eta_*)} = 0$, then necessarily $\wt b(\eta_*) = s_{\text{max}}$ and therefore $F_\ell(s_{\text{max}}, \eta_*) = \frac{1}{2}$. This is not possible as for $\eta_* > \eta_1$ we have $F_\ell(s_{\text{max}}, \eta_*) > F_\ell(s, \eta_1)$, $s \in [1, \wt \alpha]$ by the monotonicity of $F_\ell$ in $\eta$ \eqref{eq:Feta}. Choosing $s = \wt \beta$ gives the required contradiction $\frac{1}{2} > \frac{1}{2}$. Hence, $\wt b(\eta)$ is a monotonically increasing function of $\eta$ that reaches $\wt b(\eta_{\wt \alpha}) = \wt \alpha$ at some critical $\eta_{\wt \alpha} > \eta_1$, we precisely identify this transition point later in the proof.

Next, we consider the transition from the two-cuts case. This transition occurs on the curve given by $(\eta_2, \tau)$ with $\tau \in (\tau_\beta, \tau_{\text{sn}})$ (cf.~\eqref{eq:eta2}), which in the $(\wt b, \wt a)$-coordinates corresponds to $\wt b(\eta_2, \tau) = \wt b_2 \in (\wt \beta, \wt \alpha)$ and $\wt a = \wt \beta$, see Remark \ref{rem:big_remark}. For the corresponding two-cut $g$-function at $(\eta_2, \tau)$ we have $\wt a = \wt \beta$ and $d\mu_1 = d\mu_2$. Therefore we have (see \eqref{eq:2cuts_sys}, \eqref{eq:2cuts_sys2})
    \begin{align*}
        F_1(\wt \beta, \wt b_2; \eta_2, \tau) + F_2(\wt \beta, \wt b_2; \eta_2, \tau) + \frac{1}{2} = \underbrace{\int_1^{\wt b} \frac{\sqrt{\wt b_2 -x}}{\sqrt{\wt \alpha - x}\sqrt{x^2-1}}\left( \tau x + \eta_2 + \frac{\tau}{2}(\wt b_2 - \wt \alpha) \right) \di x }_{= F_\ell(\wt b_2, \eta_2)} = \frac{1}{2}.
    \end{align*}
Giving us a solution of $F_\ell(\wt b_2, \eta_2) = \frac{1}{2}$ which is consistent with the boundary case of the two-cut region. We want to show as before that $s_{\text{max}} < \wt b_2$. Observe that the linear prefactor in \eqref{eq:DsF} determining the sign of $\partial_sF_\ell(s,\eta)$ becomes

    \begin{equation}
        3\tau \wt b_2 - \tau \wt \alpha + 2\eta_2 = 2L(\wt b_2, \eta_2,\tau)<0\,,
    \end{equation}
    as we proved in Proposition \ref{prop:g_two_cut} (as $\wt b_2 = \wt b(\eta_2, \tau)$). Therefore, also in this case, we can apply the IFT and deduce the existence of an increasing solution $\wt b(\eta)$ for $\eta > \eta_2$ originating from $\wt b_2$ and moving to the right until it hits $\wt \alpha$. 
    
    It remains to find $\eta_{\wt \alpha}$ such that $\wt b(\eta_{\wt \alpha}) = \wt \alpha$. The answer turns out to be the same regardless of $\tau \leq \tau_\beta$ or $\tau \in (\tau_\beta, \tau_\text{sn})$. Evaluating at $\wt \alpha$ gives us
    \begin{equation}
        F_\ell(\wt \alpha, \eta) = \int_1^{\wt \alpha} \frac{\tau x + \eta}{\sqrt{x^2 - 1}} \di x  = -\frac{1}{2}\tau\Phi(\alpha)\,.
    \end{equation}
    We conclude that while the system remains in the left one-cut regime, we are constrained by $1 + \tau\Phi(\alpha) > 0$. By definition of $\eta_3$ we have $F_\ell(\wt \alpha, \eta_3) = \frac{1}{2}$, implying that $\wt b(\eta_3) = \wt \alpha$. This concludes the proof. 
\end{proof}

\begin{figure}
    \centering
    \includegraphics[scale=0.6]{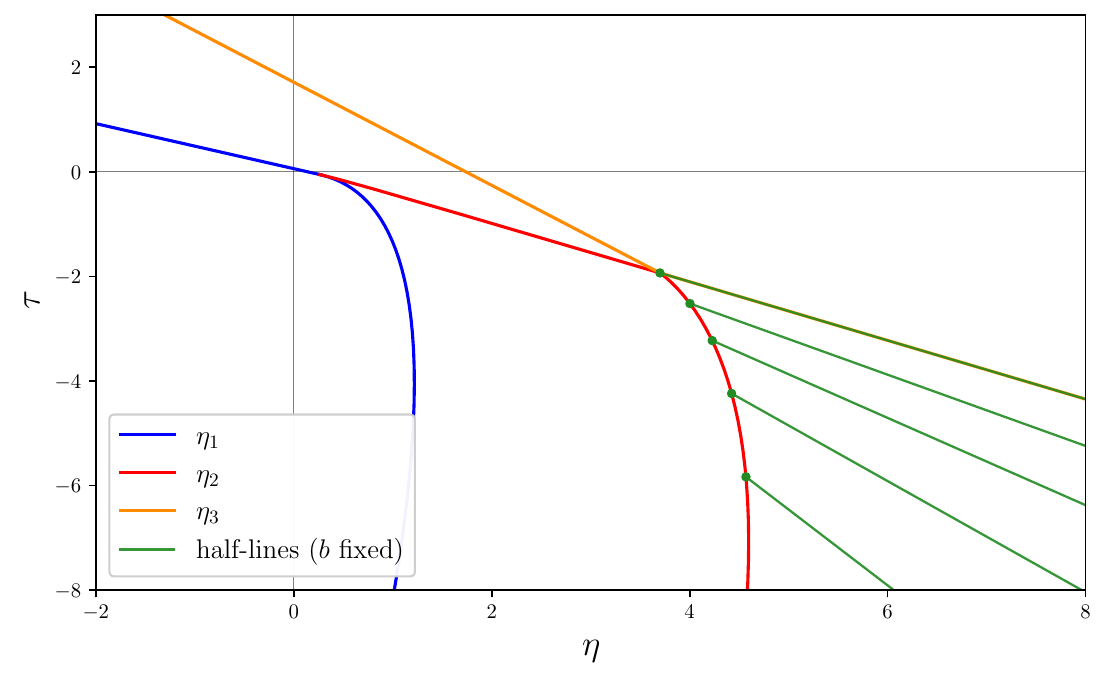}
    \caption{The set $\cU_\ell$ can be covered by affine half-lines corresponding to the lines $b(\eta, \tau)\equiv c\in(\alpha,\beta)$.}
    \label{fig:half_lines_left}
\end{figure}

\begin{remark}
    We notice that condition \eqref{eq:panic_eq_left} for each fixed $b\in (\alpha,\beta)$ represents a line in the $(\eta,\tau)$ plane. This implies that one can think about $\cU_\ell$ as an infinite union of affine half-lines stemming from the boundary of the two-cuts modulated region. See Figure \ref{fig:half_lines_left}.  
\end{remark}

\begin{remark}
    We notice that using \eqref{eq:b_monotone_left} the modulated branch point $\wt b(\eta,\tau)$ is strictly increasing with respect to $\eta$ throughout $\cU_\ell$.
\end{remark}

Regarding $f(z)$, one can prove the following 
\begin{proposition}\label{prop:f_left_anti_symmetric}

Define the function $f(z)$ as

\begin{equation}\label{eq:f_left_symmetric}
f(z) = \frac{\wt R(z)}{2\pi i} \left( \int_{\alpha}^{b} \frac{s+z}{2s(s-z)}\frac{H_1(s)}{\wt R_{+}(s)}\di s + \int_{b^{-1}}^{\alpha^{-1}} \frac{s+z}{2s(s-z)}\frac{H_2(s)}{\wt R_{+}(s)}\di s + \omega_\ell\int_{b}^{b^{-1}} \frac{s+z}{ 2s(s-z)}\frac{1}{\wt R(s)}\di s \right)\,,
\end{equation}
where $b$ is the same as in Proposition \ref{prop:g_left_cut}, $\wt R(z)$ is defined as 

\begin{equation}
    \wt R(s) = \frac{1}{2} \left( s+\frac{1}{s} - b -\frac{1}{b}\right)^\frac{1}{2} \left( s+\frac{1}{s} - \alpha -\frac{1}{\alpha}\right)^\frac{1}{2}\,,
\end{equation}
and the jump functions $H_1(s)$ and $H_2(s)$ are given by \eqref{eq:H_functions}; finally, the constant $\omega_\ell$ is defined as:
 
 \begin{equation}\label{eq:omega_left}
 \omega_\ell =\frac{ \int_{\alpha}^{b} \frac{\ln(2\pi s\rho(s)\mu_1)}{s\wt R_{+}(s)}\di s}{ \int_{1}^{b^{-1}} \frac{1}{s\wt R(s)}\di s },.\end{equation}
 Then $f(z)$ is the unique solution to RHP \ref{rhp:f_function_one_cut_left}.
 \end{proposition}

 The proof follows the same lines as Proposition \ref{prop:sol_f_two_cuts}, therefore we omit it.
\begin{remark}
\label{rem:modulation_eq}
   We notice that defining the elliptic-like integral 

   \begin{equation}
       I_k^{(\ell)} = \int_1^{\wt b} \frac{x^k}{\sqrt{\wt b - x}\sqrt{\wt \alpha - x}\sqrt{x^2 - 1}} \di x \qquad k=0,1,2\,,
   \end{equation}
   differentiating \eqref{eq:panic_eq_left} one (re)obtains the following Whitham modulation equation \cite{Bloch1992,Biondini2024} for the point $\wt b\equiv \wt b(\eta,\tau)$
    \begin{equation}
        \partial_\eta \wt b = 2\frac{\wt b I^{(\ell)}_0 - I^{(\ell)}_1}{(\wt b +\wt \alpha)  I^{(\ell)}_1 + \wt b(\wt b- \wt\alpha)I^{(\ell)}_0 - 2I^{(\ell)}_2 } \partial_\tau \wt b\,,
    \end{equation}
    we notice that the elliptic integrals $I^{(\ell)}_k$ can be computed from Remark \ref{rem:elliptic_1}.
\end{remark}

After applying the previous propositions to RHP \ref{RHP:p}, the jump matrix $J_p(z)$ \eqref{eq:jump_p} becomes

\begin{equation}
\label{eq:J_p_left_dressed}
    	J_{p}(z)=\begin{cases}
  		\begin{pmatrix}
  			e^{\left(2\ln(N)\int_{\alpha}^z \Omega_{\ell,+}(s,\eta,\tau)\di s + f_+(z) -f_-(z)\right)}& 0 \\
  			i& e^{-\left(2\ln(N)\int_{\alpha}^z \Omega_{\ell,+}(s,\eta,\tau)\di s + f_+(z) -f_-(z)\right)}
  		\end{pmatrix} \; &z \in (\alpha,b) \\\\[5pt]
          		\begin{pmatrix}
  			e^{\ln(N)\zeta_\ell(\eta,\tau) + \omega_\ell}& 0 \\
  			\Xi_1(z)e^{2\ln(N)\int_b^z \Omega_{\ell}(s,\eta,\tau)}\di s& e^{-\ln(N)\zeta_\ell(\eta,\tau) - \omega_\ell}
  		\end{pmatrix} \; &z \in (b,\beta) \\\\[5pt]
        \exp\left(\sigma_3\left(\ln(N)\zeta_\ell(\eta,\tau) + \omega_\ell\right)\right)\; & z\in (\beta,\beta^{-1})\\\\[5pt]
          		\begin{pmatrix}
  			e^{\ln(N)\zeta_\ell(\eta,\tau) + \omega_\ell} &\Xi_2(z)e^{2\ln(N)\int_{z}^{b^{-1}} \Omega_\ell(s,\eta,\tau)\di s}\\
  			0 & e^{-\ln(N)\zeta_\ell(\eta,\tau) - \omega_\ell}
  		\end{pmatrix} \; &z\in (\beta^{-1},b^{-1}) \\\\[5pt]
  		\begin{pmatrix}
  			e^{\left(2\ln(N)\int_{\alpha^{-1}}^z\Omega_{\ell,+}\di s + f_+(z) -f_-(z)\right)} &-i\\
  			0 & e^{-\left(2\ln(N)\int_{\alpha^{-1}}^z \Omega_{\ell,+}\di s + f_+(z) -f_-(z)\right)}
  		\end{pmatrix} \; &z\in (b^{-1},\alpha^{-1}) 
  	\end{cases}\,.
  \end{equation}
where we defined

\begin{equation}
    \Xi_1(z) = 2\pi i z\mu_1\rho(z)\exp(f_+(z)+f_-(z))\,, \quad \Xi_2(z)=-2\pi i z^{-1}\mu_1\rho(z^{-1})\exp(-(f_+(z)+f_-(z)))\,.
\end{equation}

\subsubsection{Lenses opening}
\label{subsubsec:Lenses_left_cut}

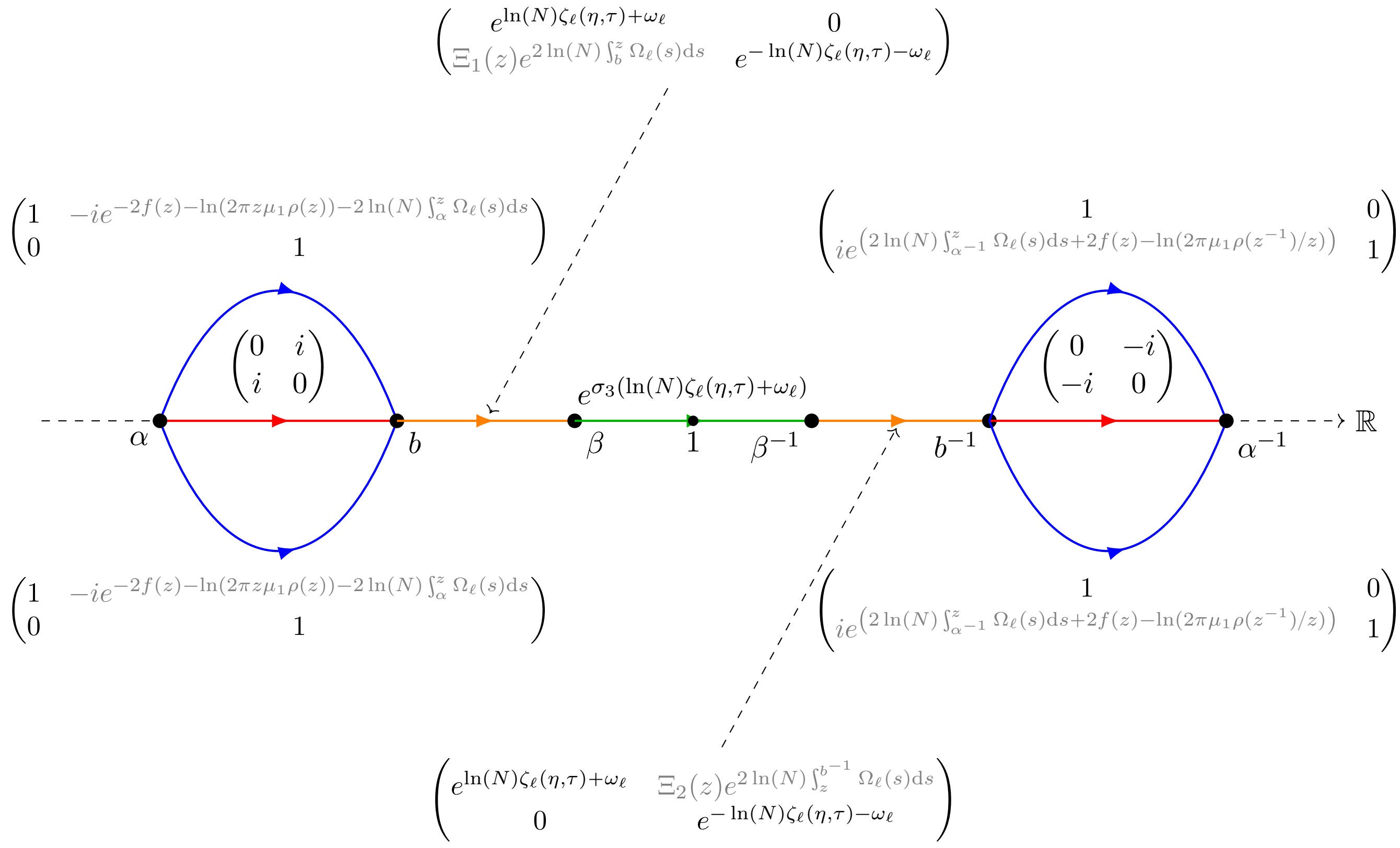
\begin{figure}
    \centering
    \begin{tikzpicture}[
    decoration={markings, mark=at position 0.55 with {\arrow{Latex}}},
    scale=1.5
]

    % Draw the real axis
    \draw[dashed,thin, black, ->] (-5.5, 0) -- (5.5, 0) node[right] {$\mathbb{R}$};

    % Define the coordinates for the branch points
    \coordinate (alpha) at (-4.5, 0);
    \coordinate (b) at (-2.5, 0);
    \coordinate (beta) at (-1, 0);
    \coordinate (beta_inv) at (1, 0);
    \coordinate (b_inv) at (2.5, 0);
    \coordinate (alpha_inv) at (4.5, 0);

    % ==========================================
    % LEFT CUT: (\alpha, b)
    % ==========================================
    
    % Main cut
    \draw[thick, red, postaction={decorate}] (alpha) -- (b) 
        node[midway, above=2pt, black] {$\pmtwo{0}{i}{i}{0}$};
        
    % Upper lens
    \draw[thick, blue, postaction={decorate}] (alpha) to[out=70, in=110, looseness=2] (b);
    \node at (-3.5, 1.6) {{\small $\pmtwo{1}{{\color{gray}-ie^{-2 f(z) - \ln(2 \pi z \mu_1 \rho(z))  -2\ln(N)\int_{\alpha}^z \Omega_{\ell}(s)\di s}}}{0}{1} $}};
    
    % Lower lens
    \draw[thick, blue, postaction={decorate}] (alpha) to[out=-70, in=-110, looseness=2] (b);
    \node at (-3.5, -1.6) {{\small $\pmtwo{1}{{\color{gray}-ie^{-2 f(z) - \ln(2 \pi z \mu_1 \rho(z))  -2\ln(N)\int_{\alpha}^z \Omega_{\ell}(s)\di s}}}{0}{1}$}};

    % Branch points (Left Cut)
    \filldraw[black] (alpha) circle (1.5pt) node[below left] {$\alpha$};
    \filldraw[black] (b) circle (1.5pt) node[below right] {$b$};

    % ==========================================
    % LEFT GAP: (b, \beta)
    % ==========================================
    \draw[thick, orange, postaction={decorate}] (b) -- (beta);
    \filldraw[black] (beta) circle (1.5pt) node[below right] {$\beta$};
    
    % Jump matrix with long arrow (Placed above)
    \node (gap1_mat) at (0, 3.2) {{\small $\pmtwo{e^{\ln(N)\zeta_\ell(\eta,\tau) + \omega_\ell}}{0}{{\color{gray}\Xi_1(z)e^{2\ln(N)\int_b^z \Omega_{\ell}(s)\di s}}}{e^{-\ln(N)\zeta_\ell(\eta,\tau) - \omega_\ell}}$}};
    \draw[->,dashed, thin, shorten >= 3pt] (gap1_mat) -- (-1.75, 0);

    % ==========================================
    % CENTRAL GAP: (\beta, \beta^{-1})
    % ==========================================
    \draw[thick, green!70!black, postaction={decorate}] (beta) -- (beta_inv) 
        node[midway, above=2pt, black] {$e^{\sigma_3\left(\ln(N)\zeta_\ell(\eta,\tau) + \omega_\ell\right)}$};
        
    % Fixed point at z = 1
    \filldraw[black] (0,0) circle (1pt) node[below] {$1$};

    % ==========================================
    % RIGHT GAP: (\beta^{-1}, b^{-1})
    % ==========================================
    \draw[thick, orange, postaction={decorate}] (beta_inv) -- (b_inv);
    \filldraw[black] (beta_inv) circle (1.5pt) node[below left] {$\beta^{-1}$};
    \filldraw[black] (b_inv) circle (1.5pt) node[below left] {$b^{-1}$};

    % Jump matrix with long arrow (Placed below)
    \node (gap2_mat) at (0, -3.2) {{\small $\pmtwo{e^{\ln(N)\zeta_\ell(\eta,\tau) + \omega_\ell}}{{\color{gray}\Xi_2(z)e^{2\ln(N)\int_{z}^{b^{-1}} \Omega_\ell(s)\di s}}}{0}{e^{-\ln(N)\zeta_\ell(\eta,\tau) - \omega_\ell}}$}};
    \draw[->,dashed, thin, shorten >= 3pt] (gap2_mat) -- (1.75, 0);

    % ==========================================
    % RIGHT CUT: (b^{-1}, \alpha^{-1})
    % ==========================================
    
    % Main cut
    \draw[thick, red, postaction={decorate}] (b_inv) -- (alpha_inv) 
        node[midway, above=2pt, black] {$\pmtwo{0}{-i}{-i}{0}$};
        
    % Upper lens
    \draw[thick, blue, postaction={decorate}] (b_inv) to[out=70, in=110, looseness=2] (alpha_inv);
    \node at (3.5, 1.6) {\small {$\pmtwo{1}{0}{{\color{gray}ie^{\left(2\ln(N)\int_{\alpha^{-1}}^z \Omega_{\ell}(s)\di s + 2f(z) - \ln(2 \pi \mu_1 \rho(z^{-1})/z)\right)}}}{1}$}};
    
    % Lower lens
    \draw[thick, blue, postaction={decorate}] (b_inv) to[out=-70, in=-110, looseness=2] (alpha_inv);
    \node at (3.5, -1.6) {{\small $\pmtwo{1}{0}{{\color{gray}ie^{\left(2\ln(N)\int_{\alpha^{-1}}^z \Omega_{\ell}(s)\di s + 2f(z) - \ln(2 \pi \mu_1 \rho(z^{-1})/z)\right)}}}{1}$}};

    % Branch points (Right Cut)
    \filldraw[black] (alpha_inv) circle (1.5pt) node[below right] {$\alpha^{-1}$};

\end{tikzpicture}
    \caption{RHP for $v(z)$ in the modulated left one-cut regime. Lenses Opening: Notice that, in the regime of Lemma \ref{lem:off_decay_left}, the terms in gray go to zero as $N$ grows.}
    \label{fig:jump_v_left}
\end{figure}

We are now in position to open the lenses exactly as we did in the previous sections. We start by noticing that one can factor the jump matrix $J_p(z)$ in the two intervals $(\alpha,b),(b^{-1},\alpha^{-1})$ as

\begin{equation}
\begin{split}
        &\begin{pmatrix}
  			e^{\left(2\ln(N)\int_{\alpha}^z \Omega_{\ell,+}(s,\eta,\tau)\di s + f_+(z) -f_-(z)\right)}& 0 \\
  			i& e^{-\left(2\ln(N)\int_{\alpha}^z \Omega_{\ell,+}(s,\eta,\tau)\di s + f_+(z) -f_-(z)\right)}
  		\end{pmatrix}=\\ &\begin{pmatrix}
        1 &  -ie^{-2 f_-(z) - \ln(2 \pi z \mu_1 \rho(z))  -2\ln(N)\int_{\alpha}^z \Omega_{\ell,-}(s,\eta,\tau)\di s}
        \\
        0 & 1
    \end{pmatrix}\\
    &\times
    \begin{pmatrix}
        0 & i
        \\
        i & 0
    \end{pmatrix} 
    \begin{pmatrix}
        1 &  -ie^{-2 f_+(z) - \ln(2 \pi z \mu_1 \rho(z))  -2\ln(N)\int_{\alpha}^z \Omega_{\ell,+}(s,\eta,\tau)\di s}
        \\
        0 & 1
    \end{pmatrix}\,,
    \end{split}
\end{equation}
and 
\begin{equation}
\begin{split}&
\begin{pmatrix}
  			e^{\left(2\ln(N)\int_{\alpha^{-1}}^z \Omega_{\ell,+}(s,\eta,\tau)\di s + f_+(z) -f_-(z)\right)} &-i\\
  			0 & e^{-\left(2\ln(N)\int_{\alpha^{-1}}^z \Omega_{\ell,+}(s,\eta,\tau)\di s + f_+(z) -f_-(z)\right)}
  		\end{pmatrix} = \\
        &\pmtwo{1}{0}{ie^{\left(2\ln(N)\int_{\alpha^{-1}}^z \Omega_{\ell,-}(s,\eta,\tau)\di s + 2f_-(z) - \ln(2 \pi \mu_1 \rho(z^{-1})/z)\right)}}{1}\\ & \times  \pmtwo{0}{-i}{-i}{0} \pmtwo{1}{0}{ie^{\left(2\ln(N)\int_{\alpha^{-1}}^z \Omega_{\ell,+}(s,\eta,\tau)\di s + 2f_+(z) - \ln(2 \pi \mu_1 \rho(z^{-1})/z)\right)}}{1}
\end{split}\,.
\end{equation}
Then, as we did in Section \ref{sec:complete_one_cut}, we define the following vector $v(z)$ 

\begin{equation}
\label{eq:v_left}
    v(z) = \begin{cases}
        p(z)\pmtwo{1}{-ie^{-2 f(z) - \ln(2 \pi z \mu_1 \rho(z))  -2\ln(N)\int_{\alpha}^z \Omega_{\ell}(s,\eta,\tau)\di s}}{0}{1} \; &z \text{ in the lower lenses below }(\alpha,b) \\\\
        p(z)\pmtwo{1}{ie^{-2 f(z) - \ln(2 \pi z \mu_1 \rho(z))  -2\ln(N)\int_{\alpha}^z \Omega_{\ell}(s,\eta,\tau)\di s}}{0}{1}  \; &z \text{ in the upper lenses above }(\alpha,b) \\\\
        p(z) \pmtwo{1}{0}{ie^{\left(2\ln(N)\int_{\alpha^{-1}}^z \Omega_{\ell}(s,\eta,\tau)\di s + 2f(z) - \ln(2 \pi \mu_1 \rho(z^{-1})/z)\right)}}{1} \; &z \text{ in the lower lenses below }(b^{-1},\alpha^{-1}) \\\\
        p(z)\pmtwo{1}{0}{-ie^{\left(2\ln(N)\int_{\alpha^{-1}}^z \Omega_{\ell}(s,\eta,\tau)\di s + 2f(z) - \ln(2 \pi \mu_1 \rho(z^{-1})/z)\right)}}{1}  \; &z \text{ in the upper lenses above }(b^{-1},\alpha^{-1}) \\\\
        p(z) \; & \text{otherwise}
    \end{cases}\,.
\end{equation}
The vector $v(z)$ satisfies the following conditions
\begin{equation}
\label{eq:v_left_cut}
    v_+(z) = v_-(z)J_v(z)\,, \quad
    \lim_{z\to 0} v_1(z)v_2(z)=1\,, \quad
     v(z^{-1}) = v(z)\sigma_1\,, \quad v_1(0) > 0\,,
\end{equation}
where the jump matrix $J_v(z)$ is as in Figure \ref{fig:jump_v_left}.

We now prove the following lemma that will allow us to reduce our analysis to the model problem RHP \ref{OuterRHP_1cut}.

\begin{lemma}
\label{lem:off_decay_left}
Let $(\eta,\tau)\in \cU_{\ell}$ \eqref{eq:modulated_regions} and $b\equiv b(\eta,\tau)$ be defined as in Proposition \ref{prop:g_left_cut}. Then, the off-diagonal entries of the jump matrix $J_p(z)$ \eqref{eq:J_p_left_dressed} for $z\in(b,\beta)\cup(\beta^{-1},b^{-1})$ converge to zero as $N\to \infty$. 

Furthermore, there exist two neighborhoods $V_1,V_2$ of $(\alpha,b)$ and $(b^{-1},\alpha^{-1})$ respectively such that
\begin{equation}
    \Re\left(\int_\alpha^z \Omega_\ell(s,\eta,\tau)\di s\right) \geq 0\; \forall\,z\in V_1\,, \quad \Re\left(\int_{\alpha^{-1}}^z \Omega_\ell(s,\eta,\tau)\di s\right) \leq 0\; \forall\,z\in V_2\,.
\end{equation}
\end{lemma}

\begin{proof}
We first prove the decay of the off-diagonal terms on the gap $(b, \beta)$. The decay is governed by the exponent $2 \ln(N) \int_b^z \Omega_\ell(s, \eta, \tau) \di s$. We must show that $\nu(z, \tau) = \int_b^z \Omega_\ell(s, \eta, \tau) \di s < 0$ for $z \in (b, \beta)$. 

Fix the branch point $b_0 = b(\eta, \tau) \in (\alpha, \beta)$. The condition $\int_{b_0}^1 \Omega_\ell(s, \eta, \tau) \di s = 1/2$ defines an affine line in the $(\eta, \tau)$-plane. Recalling that $\Omega_\ell(s, \eta, \tau) = \frac{R_\ell(s)}{s} \left( \tau x(s) + \eta + \frac{\tau}{2}(\wt b_0 - \wt{\alpha}) \right)$, where $x(s) = \frac{1}{2}(s+s^{-1})$, we differentiate the modulation equation with respect to $\tau$ along this affine line:
\begin{equation}
    \int_{b_0}^1 \frac{\di \Omega_\ell}{\di \tau} \di s = 0\,, \quad \text{where} \quad \frac{\di \Omega_\ell}{\di \tau} = \frac{R_\ell(s)}{s} \underbrace{\left( x(s) + \frac{\di \eta}{\di \tau} + \frac{1}{2}(\wt{b}_0 - \wt{\alpha}) \right)}_{\equiv M(s)}\,.
\end{equation}
Because $x(s)$ is strictly decreasing for $s \in (0, 1)$, the function $M(s)$ is strictly decreasing. The weight $R_\ell(s)/s$ is strictly positive on $(b_0, 1)$. Since the integral of $M(s)$ over the full interval $(b_0, 1)$ is zero, $M(s)$ must change sign exactly once from positive to negative. Consequently, for any partial interval $(b_0,z) \subsetneq (b_0, 1)$, the integral must be strictly positive:
\begin{equation}
    \frac{\di \nu(z, \tau)}{\di \tau} = \int_{b_0}^z \frac{R_\ell(s)}{s} M(s) \di s > 0\,.
\end{equation}
At the boundary $(\eta_2(\tau_*), \tau_*)$ such that $b(\eta_2(\tau_*), \tau_*) = b_0$ between the two-cut and left one-cut regimes, the point $a$ merges with $\beta$. Here, $\Omega_\ell$ smoothly coincides with $\Omega_2$, and Lemma \ref{lem:off_decay_two_cuts_gap} guarantees that $\nu(z, \tau_*) < 0$. As we move into $\cU_{\ell}$ along the level curve $b(\eta, \tau) = b_0$, the parameter $\tau$ strictly decreases. Since the derivative with respect to $\tau$ is positive, decreasing $\tau$ makes $\nu(z, \tau)$ strictly more negative. Thus, $\nu(z, \tau) < 0$ for all $z \in (b_0, \beta)$, ensuring the wanted decay. By the symmetry established in Remark \ref{rem:sym_off_diag}, the same decay holds on $(\beta^{-1}, b^{-1}_0)$. 

Next, we analyze the decay on the lenses around $(\alpha, b_0)$. On the real axis, $\int_\alpha^z \Omega_{\ell,+}(s, \eta, \tau) \di s$ is purely imaginary. By the Cauchy-Riemann equations, we obtain the required positivity in the upper lens if $\Im(\Omega_{\ell,+}(z, \eta, \tau)) < 0$ for $z \in (\alpha, b_0)$. As before we have $\frac{d\Omega_{\ell,+}}{d\tau} = \frac{R_{\ell, +}(s)}{s} M(s)$. We already established that $M(s)$ is strictly decreasing and positive at $s = b_0$. It follows that $M(s)$ is positive for $s \in (\alpha, b_0)$, and as $\Im(R_{\ell, +}(s)) > 0$ for $s \in (\alpha, b_0)$, it follows that $\frac{d}{d\tau}\Im(\Omega_{\ell, +}) > 0$ there. Starting again at the boundary $(\eta_2(\tau_*), \tau_*)$ such that $b(\eta_2(\tau_*), \tau_*) = b_0$, it follows that $\Im(\Omega_{\ell, +})|_{\eta_2(\tau_*), \tau_*} < 0$ for $s \in (\alpha, b_0)$ by the proof of Lemma \ref{lem:off_decay_two_cuts_lens}. By the aforementioned monotonicity of $\Im(\Omega_{\ell, +})$ in $\tau$, it follows that $\Im(\Omega_{\ell, +})|_{\eta, \tau} < \Im(\Omega_{\ell, +})|_{\eta_2(\tau_*), \tau_*} < 0$ for all $\eta, \tau$ such that $b(\eta, \tau) = b_0$ and $s \in (\alpha, b_0)$ (recall $\tau < \tau_*$). This ensures decay on the upper lens between $\alpha$ and $b_0$. An analogous argument applies to the lower lens and the lenses around $(b^{-1}, \alpha^{-1})$, finishing the proof.
\end{proof}

\begin{remark}
    We notice that the previous Lemma shows that all the entries colored gray in Figure \ref{fig:jump_v_left} decay as $N\to \infty$.
\end{remark}

Applying Lemma \ref{lem:off_decay_left}, we are led to consider as the outer parametrix problem the model problem RHP \ref{OuterRHP_1cut} in its symmetric version, where $\ta \equiv \alpha$,$\tb\equiv b(\eta,\tau)$ and 

$$\Delta \equiv \Delta_{\ell,n}=\frac{1}{2\pi i}(\ln(N)\zeta_\ell(\eta,\tau) + \omega_\ell)\,.$$
To approximate $v(z)$, we need to construct a local approximation near the end points of the $2$ intervals $(\alpha,b),\,(b^{-1},\alpha^{-1})$. Since the construction is analogue to the one done in the previous cases, and that this local approximation do not contribute to the leading order asymptotic, we skip their explicit construction.

\subsubsection{The modulated left one-cut solution}
\label{subsubsec:left_one_cut_solution}

To obtain the asymptotic solution in the left modulated one-cut regime, we follow Section~\ref{subsec:complete_one_cut_solution}. In particular, we use the model solutions $k_s^{(\ell)},k_a^{(\ell)}$ evaluated on the Riemann surface
$\cR_\ell=\cR(\alpha,b(\eta,\tau))$.
Throughout the derivation, we assume that the point $(\eta,\tau),(\eta\pm\ln(N)^{-1},\tau)\in \mathring{\cU}_\ell$.

\paragraph{Global parametrix and expansion near the origin.}
As we did in Section \ref{subsec:complete_one_cut_solution}, we first construct a global matrix approximation $M(z)$ of the solution $v(z)$ as
\begin{equation}
    M(z) = \begin{cases}
        M_p(z), & z \in \D(p, \varepsilon) \quad \text{for } p \in \{\alpha, b(\eta,\tau), b(\eta,\tau)^{-1}, \alpha^{-1}\}, \\[6pt]
        \frac{1}{2} \begin{pmatrix} k_s(z) + \ell_a(z) \\ k_s(z) - \ell_a(z) \end{pmatrix}, & z \in \C \setminus \bigcup_{p} \D(p, \varepsilon),
    \end{cases}
\end{equation}
where $M_{\alpha}(z),M_{\alpha^{-1}}(z)$  are the local Bessel parametrices constructed as in Section \ref{subsec:one_cut_local_par}, $M_{b(\eta,\tau)},M_{b(\eta,\tau)^{-1}}$ are the local Airy parametrices built following Section \ref{subsec:airy_local_par}, and $k_s(z), \ell_a(z)$ are the symmetric and antisymmetric vector solutions to the model RHP \ref{OuterRHP_1cut} with parameters $\ta \equiv \alpha$, $\tb \equiv b(\eta,\tau)$, and phase $\Delta_{\ell,n} = \frac{1}{2\pi i}(\ln(N)\zeta_\ell(\eta,\tau) + \omega_\ell)$. 

Following the same argument as in Section \ref{subsec:complete_one_cut_solution}, we have the uniform estimate:
\begin{equation}
    \|J_r - I\|_{L^\infty(\Sigma_r)} = \cO\left( \frac{1}{\ln N} \right), \qquad N \to \infty.
\end{equation}
By the small-norm theory for vector Riemann--Hilbert problems (see App.~\ref{App:SmallNorm}), it follows that
\begin{equation}
    r(z) = (1, 1) + \cO\left( \frac{1}{\ln N} \right)
\end{equation}
uniformly for $z$ bounded away from $\Sigma_r$. 

We now unwind the transformations to recover the vector $q(z)$ in a neighborhood of the origin $z = 0$. Since the origin lies outside all disks and lenses, we have $\wh q(z) = q(z)$ and $v(z) = p(z) = \wh q(z) e^{\sigma_3(\ln(N)g(z) + f(z))}$. Therefore,
\begin{align*}
    q(z) &= v_1(0)\left[r(z) M(z) + \cO\left( \frac{1}{\ln N} \right)\right]  e^{-\sigma_3(\ln(N)g(z) + f(z))}
    \\
    &= v_1(0)\left[k_s^{(\ell)}(z)  + \cO\left( \frac{1}{\ln N} \right)\right]e^{-\sigma_3(\ln(N)g(z) + f(z))}.
\end{align*}
From RHP \ref{RHP:q}, the exact vector $q(z)$ satisfies the normalization condition $q_1(0)q_2(0) = 1$ with $q_1(0) > 0$. Additionally, we have by construction $k_{s,1}^{(\ell)}(0; \Delta_{\ell,n}) = 1$. It now follows that $v_1^2(0)[k_{s,2}^{(\ell)}(0; \Delta_{\ell,n}) + \cO(1/\ln N)] = 1$. Solving for $v_1(0)$ and substituting this into the expression for the first component, we obtain:
\begin{equation}\label{eq:m1_asymptotic_expansion_left}
 q_1(z)=\frac{1}{\sqrt{k_{s,2}^{(\ell)}(0;\Delta_{\ell,n})}} k_{s,1}^{(\ell)}(z;\Delta_{\ell,n})
 e^{-\ln(N)g(z)-f(z)}
 \left(1+\cO\left(\frac1{\ln N}\right)\right)\,.
\end{equation}
Keeping the notation $A_0^{(\ell)}=\cA_\ell(0^{(1)})$ and
$B^{(\ell)}$ for the Abel-map value and period on $\cR_\ell$,
the same calculation as in \eqref{eq:ks2_explicit_theta} gives:
\begin{equation}
 k_{s,2}^{(\ell)}(0;\Delta_{\ell,n})=
 \frac{\theta_3(2A_0^{(\ell)}+\frac12+\Delta_{\ell,n}\mid2B^{(\ell)})}
 {\theta_3(2A_0^{(\ell)}+\frac12-\Delta_{\ell,n}\mid2B^{(\ell)})}.
\end{equation}

\paragraph{Asymptotics of $a^{(q)}_n(t)$ and $b^{(q)}_n(t)$.}
Using $g(0)=-g_\infty$, $f(0)=-f_\infty$ and Lemma \ref{lem:recovery_q}, evaluating \eqref{eq:m1_asymptotic_expansion_left} at $z = 0$ gives:
\begin{equation}
 A^{(q)}(n,t) = q_1(0) =
 \sqrt{\frac{\theta_3(2A_0^{(\ell)}+\frac12-\Delta_{\ell,n}\mid2B^{(\ell)})}
 {\theta_3(2A_0^{(\ell)}+\frac12+\Delta_{\ell,n}\mid2B^{(\ell)})}}
 e^{\ln(N)g_\infty+f_\infty}
 \left(1+\cO\left(\frac1{\ln N}\right)\right)\,.
\end{equation}
From Lemma \ref{lem:recovery_q}, the Toda coefficient $a^{(q)}_n(t)$ is recovered via $a^{(q)}_n(t) = \frac{1}{2}\frac{A^{(q)}(n,t)}{A^{(q)}(n+1,t)}$. Under the discrete shift $n \mapsto n+1$, the slow variable shifts as $\eta \mapsto \eta + \frac{1}{\ln N}$, leading to the phase difference $\ln(N)[g_\infty(\eta,\tau) - g_\infty(\eta + 1/\ln N, \tau)] = -\partial_\eta g_\infty(\eta,\tau) + \cO(1/\ln N)$. Taking the ratio of $A^{(q)}(n,t)$ at $n$ and $n+1$, we obtain:
\begin{equation}
 a^{(q)}_n(t)=a^{(\ell)}_{\mathrm{bg}}(\eta,\tau)
 \sqrt{\frac{
 \theta_3(2A_0^{(\ell)}+\frac12+\Delta_{\ell,n+1}\mid2B^{(\ell)})
 \theta_3(2A_0^{(\ell)}+\frac12-\Delta_{\ell,n}\mid2B^{(\ell)})}{
 \theta_3(2A_0^{(\ell)}+\frac12-\Delta_{\ell,n+1}\mid2B^{(\ell)})
 \theta_3(2A_0^{(\ell)}+\frac12+\Delta_{\ell,n}\mid2B^{(\ell)})}}
 \left(1+\cO\left(\frac1{\ln N}\right)\right),
\end{equation}
where $a^{(\ell)}_{\mathrm{bg}}(\eta,\tau) = \frac{1}{2}\exp(-\partial_\eta g_\infty)$.

To extract $b^{(q)}_n(t)$, we evaluate the $\cO(z)$ term of $q_1(z)$. From Lemma \ref{lem:recovery_q}, $q_1(z) = A^{(q)}(n,t)(1 - 2B^{(q)}(n-1,t)z + \cO(z^2))$, which implies:
\begin{equation}
 -2B^{(q)}(n-1,t) = \left.\frac{\di}{\di z}\ln q_1(z; n, t)\right|_{z=0} =
 \left.\frac{\di}{\di z}\ln k_{s,1}^{(\ell)}(z;\Delta_{\ell,n})\right|_{z=0}
 -\ln(N)g'(0)-f'(0)+\cO\left(\frac1{\ln N}\right).
\end{equation}
From Lemma \ref{lem:recovery_q},
\begin{equation}
 b^{(q)}_n(t) = B^{(q)}(n,t) - B^{(q)}(n-1,t) = \frac{1}{2} \left[ \left.\frac{\di}{\di z}\ln q_1(z; n, t)\right|_{z=0} - \left.\frac{\di}{\di z}\ln q_1(z; n+1, t)\right|_{z=0} \right].
\end{equation}
Following the same calculations as in Section \ref{subsec:complete_one_cut_solution}, subtracting the expressions at $n$ and $n+1$ yields:
\begin{equation}
 b^{(q)}_n(t)=b^{(\ell)}_{\mathrm{bg}}(\eta,\tau)
 +\frac{1}{2I^{(\ell)}_0}\left[
 \frac{\theta_3'(2A_0^{(\ell)}+\frac12-\Delta_{\ell,n}\mid2B^{(\ell)})}
 {\theta_3(2A_0^{(\ell)}+\frac12-\Delta_{\ell,n}\mid2B^{(\ell)})}
 -\frac{\theta_3'(2A_0^{(\ell)}+\frac12-\Delta_{\ell,n+1}\mid2B^{(\ell)})}
 {\theta_3(2A_0^{(\ell)}+\frac12-\Delta_{\ell,n+1}\mid2B^{(\ell)})}
 \right]+\cO\left(\frac1{\ln N}\right),
\end{equation}
where $b^{(\ell)}_{\mathrm{bg}}(\eta,\tau) = \frac{1}{2}\partial_\eta g'(0)$. 
To explicitly compute this quantities, we first need to compute $\partial_\eta g(z)$ and therefore $\partial_\eta \wt b(\eta,\tau)$. To do so, we use the modulation equation $F_\ell(\wt b,\eta) = 1/2$ in Proposition \ref{prop:g_left_cut}.

Recall that we defined the Joukowsky map as  $x(z)=(z+z^{-1})/2$, $\wt\alpha=x(\alpha)$ and
$\wt b=x(b)$. We use the function $\wt R$ already defined in
Proposition~\ref{prop:f_left_anti_symmetric}:
\[
 \wt R(z)=\frac12
 \left(z+z^{-1}-b-b^{-1}\right)^{1/2}
 \left(z+z^{-1}-\alpha-\alpha^{-1}\right)^{1/2},
 \qquad \wt R(z)\sim\frac z2\quad \text{as }z\to\infty,
\]
with canonical branch-cuts, we deduce that 
\[
 R_\ell(z)=\frac{x(z)-\wt b}{\wt R(z)}.
\]
Furthermore, we recall the integrals $I^{(\ell)}_k$ from Remark~\ref{rem:modulation_eq}:
\begin{equation}\label{eq:left_gap_moments}
 I^{(\ell)}_k=\int_1^{\wt b}
 \frac{x^k}{\sqrt{(x^2-1)(\wt b-x)(\wt\alpha-x)}}\di x>0\,.
 \qquad k=0,1,2\,.
\end{equation}
In the proof of Proposition ~\ref{prop:g_left_cut}  we showed that the equation $F_\ell(\wt b,\eta)=1/2$ reads
\[
 \int_1^{\wt b}
 \frac{\sqrt{\wt b-x}}{\sqrt{\wt\alpha-x}\sqrt{x^2-1}}
 \left(\tau x+\gamma_\ell(\eta,\tau)\right)\di x=\frac12.
\]
This together with equations~\eqref{eq:DsF} and \eqref{eq:Feta} gives
\[
 \left.\partial_\eta F_\ell(s,\eta)\right\vert_{s=\wt b}=\wt bI^{(\ell)}_0-I^{(\ell)}_1,
 \qquad
 \left.\partial_sF_\ell(s,\eta)\right|_{s=\wt b}
 =\frac{\tau\wt b+\gamma_\ell}{2}I^{(\ell)}_0.
\]
Therefore, implicit differentiation yields
\begin{equation}\label{eq:left_endpoint_derivative}
 \partial_\eta\wt b
 =-\frac{2(\wt bI^{(\ell)}_0-I^{(\ell)}_1)}{(\tau\wt b+\gamma_\ell)I^{(\ell)}_0},
 \qquad
 \wt b+\frac{\tau\wt b+\gamma_\ell}{2}\partial_\eta\wt b
 =\frac{I^{(\ell)}_1}{I^{(\ell)}_0}\,.
\end{equation}
Since $\gamma_\ell=\eta+\tau(\wt b-\wt\alpha)/2$, direct differentiation
of $\Omega_\ell$ gives
\begin{align*}
 \partial_\eta\Omega_\ell(z)
 &=\frac{R_\ell(z)}{z}
 \left[1+\frac\tau2\partial_\eta\wt b
 -\frac{\tau x(z)+\gamma_\ell}{2(x(z)-\wt b)}
 \partial_\eta\wt b\right]\\
 &=\frac{x(z)-\wt b
 -\frac12(\tau\wt b+\gamma_\ell)\partial_\eta\wt b}{z\wt R(z)}
 =\frac{x(z)-\frac{I^{(\ell)}_1}{I^{(\ell)}_0}}{z\wt R(z)}.
\end{align*}
Thus the full derivative of the $g$-function is
\begin{equation}\label{eq:left_g_eta}
 \partial_\eta g(z)=-\ln z+
 \int_{\alpha^{-1}}^z\frac{x(s)-\frac{I^{(\ell)}_1}{I^{(\ell)}_0}}{s\wt R(s)}\di s.
\end{equation}
We notice that this derivative has the same structure as the corresponding one in Section \ref{subsec:complete_one_cut_solution} with $\beta$ replaced by $b$ and $I_1/I_0$ by $I^{(\ell)}_1/I^{(\ell)}_0$.
As $z\to0$, we have
\[
 \wt R(z)=\frac1{2z}-\frac{\wt\alpha+\wt b}{2}+\cO(z),
 \qquad
 \frac{x(z)-\frac{I^{(\ell)}_1}{I^{(\ell)}_0}}{\wt R(z)}
 =1+\left(\wt\alpha+\wt b-2\frac{I^{(\ell)}_1}{I^{(\ell)}_0}\right)z+\cO(z^2).
\]
Differentiating \eqref{eq:left_g_eta} in $z$ gives
$\partial_\eta g'(0)=\wt\alpha+\wt b-2I^{(\ell)}_1/I^{(\ell)}_0$.
At infinity, subtracting the logarithmic term as in the derivation of
\eqref{eq:a_bg_explicit} gives
\[
 \partial_\eta g_\infty=\ln\alpha+
 \int_{\wt\alpha}^\infty
 \left(\frac{x-\frac{I^{(\ell)}_1}{I^{(\ell)}_0}}
 {\sqrt{(x-\wt\alpha)(x-\wt b)}}-1\right)
 \frac{\di x}{\sqrt{x^2-1}}.
\]
Changing back to $s\in(0,\alpha)$, the background coefficients are
\begin{align}
 a^{(\ell)}_{\mathrm{bg}}(\eta,\tau)
 &=\frac12e^{-\partial_\eta g_\infty}
 =\frac1{2\alpha}\exp\left\{-\int_0^\alpha
 \left(\frac{\frac{s+s^{-1}}2-\frac{I^{(\ell)}_1}{I^{(\ell)}_0}}{s\wt R(s)}
 -\frac1s\right)\di s\right\},
 \label{eq:thm_abg_left_explicit}\\
 b^{(\ell)}_{\mathrm{bg}}(\eta,\tau)
 &=\frac12\partial_\eta g'(0)
 =\frac{\wt\alpha+\wt b}{2}-\frac{I^{(\ell)}_1}{I^{(\ell)}_0}.
 \label{eq:thm_bbg_left_explicit}
\end{align}
The moments $I^{(\ell)}_k$ can be evaluated using Remark~\ref{rem:elliptic_1}:
\[
 I^{(\ell)}_0=\tg K(k),\qquad
 I^{(\ell)}_1=\tg\left[\wt\alpha K(k)+(\wt b-\wt\alpha)\Pi(n,k)\right],
\]
where $\tg,k,n$ are evaluated with $\wt\beta$ replaced by $\wt b$.
Therefore, we proved the following:
\begin{theorem}[Modulated Left One-Cut Asymptotics]
\label{thm:left_one_cut_asymp}
For $(\eta,\tau)$ in a compact subset of the interior of $\cU_\ell$,
as $N\to\infty$, the Toda lattice variables $a^{(q)}_n(t)$ and $b^{(q)}_n(t)$ with $n = \eta\ln N$ and $t = \tau\ln N$ are asymptotically given by:
\begin{align}
 a^{(q)}_n(t)&=a^{(\ell)}_{\mathrm{bg}}(\eta,\tau)
 \sqrt{\frac{
 \theta_3(2A_0^{(\ell)}+\frac12+\Delta_{\ell,n+1}\mid2B^{(\ell)})
 \theta_3(2A_0^{(\ell)}+\frac12-\Delta_{\ell,n}\mid2B^{(\ell)})}{
 \theta_3(2A_0^{(\ell)}+\frac12-\Delta_{\ell,n+1}\mid2B^{(\ell)})
 \theta_3(2A_0^{(\ell)}+\frac12+\Delta_{\ell,n}\mid2B^{(\ell)})}}
 \nonumber\\
 &\qquad\times\left(1+\cO\left(\frac1{\ln N}\right)\right),
 \label{eq:thm_an_asymp_left}\\
 b^{(q)}_n(t)&=b^{(\ell)}_{\mathrm{bg}}(\eta,\tau)
 +\frac{1}{2I^{(\ell)}_0}\left[
 \frac{\theta_3'(2A_0^{(\ell)}+\frac12-\Delta_{\ell,n}\mid2B^{(\ell)})}
 {\theta_3(2A_0^{(\ell)}+\frac12-\Delta_{\ell,n}\mid2B^{(\ell)})}
 \right.\nonumber\\
 &\qquad\left.
 -\frac{\theta_3'(2A_0^{(\ell)}+\frac12-\Delta_{\ell,n+1}\mid2B^{(\ell)})}
 {\theta_3(2A_0^{(\ell)}+\frac12-\Delta_{\ell,n+1}\mid2B^{(\ell)})}
 \right]+\cO\left(\frac1{\ln N}\right).
 \label{eq:thm_bn_asymp_left}
\end{align}
The background coefficients are $a^{(\ell)}_{\mathrm{bg}},b^{(\ell)}_{\mathrm{bg}}$ given by
\eqref{eq:thm_abg_left_explicit}--\eqref{eq:thm_bbg_left_explicit}.
All quantities in these formulas are evaluated at
$(\eta,\tau)$, while $\Delta_{\ell,n+1}$ is evaluated at
$(\eta+1/\ln N,\tau)$. $\theta'_3(P\mid B)$ denotes differentiation
with respect to the first argument.
\end{theorem}

\begin{remark}
As a sanity check, one verifies that as $b\to\beta$, the background formulas agree with those of the
complete one-cut calculation. Indeed, as $b\to\beta$, they satisfy
$a^{(\ell)}_{\mathrm{bg}}\to a^{(c)}_{\mathrm{bg}}$ and $b^{(\ell)}_{\mathrm{bg}}\to b^{(c)}_{\mathrm{bg}}$.
\end{remark}

\subsection[Modulated right one-cut]{From the complete one-cut and two-cuts modulated regions to the modulated right one-cut}

In this subsection, we focus on the set $\cU_{r}$ \eqref{eq:modulated_regions}; we proceed as in the previous sub-section, but we require that both $g(z),f(z)$-functions satisfy a different RHP. As before, we consider the transformation \eqref{eq:def_transformation_1}, but we seek for $f,g$-functions satisfying the following
\begin{rhp}
      \label{rhp:g_function_one_cut_right}
      We seek for a scalar function $g(z)$ such that
      \begin{itemize}
            \item $g(z)$ is analytic in $\C\setminus \left(a,a^{-1}\right)$
            \item $g_+(z)+g_-(z)+1 + \tau\Phi(z,\eta,\tau) = 0$ for $z\in (a,\beta)$
            \item $g_+(z)+g_-(z) - 1 + \tau\Phi(z,\eta,\tau) = 0$ for $z\in \left(\frac{1}{\beta},\frac{1}{a}\right)$
            \item $g_+(z) - g_-(z) = \zeta_r(\eta,\tau)$ for $z\in (\beta,\beta^{-1})$
            \item $g(z^{-1}) = -g(z)$
            \item $g(z) = g_\infty + \cO(\frac{1}{z})$ as $z\to\infty$
            \item $g(z) + \frac{1 + \tau\Phi(z,\eta,\tau)}{2} = \cO((z-a)^{3/2})$ as $z\to a$ and $g(z) + \frac{1 + \tau\Phi(z,\eta,\tau)}{2} = \pm\frac{1}{2}\zeta_r + \cO((z-\beta)^{1/2})$ as $z\to \beta$, where $\pm$ indicates that $z \in \mathbb{C}_\pm$
\end{itemize}
\end{rhp}

 \begin{rhp}
    \label{rhp:f_function_one_cut_right}
    We seek for a scalar function $f(z)$ such that

    \begin{itemize}
        \item $f(z)$ is analytic in $\C\setminus \left(a,a^{-1}\right)$
        \item $f_+(z)+f_-(z) + \ln(2\pi ) + \ln(z) + \ln(\mu_1\rho(z)) = 0$ for $z\in(a,\beta)$
        \item $f_+(z)+f_-(z)  -\ln(2\pi ) + \ln(z) - \ln(\mu_1\rho(z^{-1})) = 0$ for $z\in(\beta^{-1},a^{-1})$
        \item $f_+(z)-f_-(z) = \omega_r$ for $z\in(\beta,\beta^{-1})$
        \item $f(z^{-1}) = -f(z)$
        \item $f(z) = f_\infty + \cO(\frac{1}{z})$ as $z\to\infty$
    \end{itemize}
\end{rhp}

In both RHP the point $a$ is the same and $a\in[\alpha,\beta]$. In particular, $a$ has to be determined so that the resulting RHP for the vector $p(z)$ has good asymptotic behavior, therefore $a\equiv a(\eta,\tau)$. To determine this point, we use the same kind of continuation argument based that we used in the previous section, but with some important differences.

\begin{proposition}
\label{prop:g_right_cut}
Define the function $g(z)$ as
    \begin{equation}
      g(z) = \frac{1-\tau\Phi(z,\eta, \tau)}{2} + \int_{a^{-1}}^z \Omega_r(s,\eta,\tau) \di s \,, \qquad \Omega_r(s,\eta,\tau) =\frac{R_r(s)\left( \frac{\tau}{2}(s +s^{-1}) + \gamma_r(\eta,\tau) \right)}{s} 
\end{equation}
where
\begin{equation}
      \label{eq:def_Rr_gammar}
      R_r(z) = \frac{\left(z + \frac{1}{z} - a - \frac{1}{a}\right)^{1/2}}{\left(z + \frac{1}{z} - \beta - \frac{1}{\beta}\right)^{1/2}}, \qquad \gamma_r(\eta,\tau) =\eta+ \frac{\tau}{4}   \left(a+\frac{1}{a}-\beta -\frac{1}{\beta }\right)\,, \qquad  \zeta_r(\eta, \tau) = 2\int_a^\beta \Omega_{r,+}(s, \eta, \tau) \, \di s.
\end{equation}
Then there exists a unique $a \in \left(\alpha,\beta\right)$, such that the integral equation
\begin{equation}
\label{eq:panic_eq_right}
      \int_\beta^1 \Omega_r(s,\eta,\tau) ds = \frac{1}{2}\,
\end{equation}
holds, in which case $g(z)$ solves the RHP \ref{rhp:g_function_one_cut_right}.
\end{proposition}
\begin{remark}
\label{rem:Rr_boundary}
    Notice that $R_r(z)$ is analytic in $\C\setminus\left\{(a,\beta)\cup (\beta^{-1},a^{-1})\right\}$, and we chose the branch-cut so that $R_r(z)\to 1$ as $z\to\infty$, this determines $R_r(z)$ uniquely. Furthermore, we notice that $R_{r,\pm}(z)\in i\R_{\mp}$ for $z\in (a,\beta)$, $R_{r,\pm}(z)\in i\R_{\pm}$ for $z\in (\beta^{-1},a^{-1})$, and $R_r(z) > 0$ for $z \in (\beta, \beta^{-1})$.
\end{remark}

\begin{proof}
As before, it is straightforward to show that if $g(z)$ satisfies all the properties above, then it satisfies the RHP \ref{rhp:g_function_one_cut_right}. Therefore, we have just to show that there exists a point $a\equiv a(\eta,\tau)\in(\alpha,\beta)$ such that \eqref{eq:panic_eq_right} is satisfied. To do so, we transition to the Joukowsky variables $x(s) = \frac{1}{2}(s+s^{-1})$ and, for a fixed $\tau>\tau_{\text{sn}}$ \eqref{eq:point_ineq}, we define the function $F_r(s,\eta)$ as follows:
    \begin{equation}
        F_r(s, \eta) = \int_{1}^{\wt \beta} \frac{\sqrt{s-x}}{\sqrt{\tilde{\beta}-x}\sqrt{x^2-1}} \left( \tau x + \eta + \frac{\tau}{2}(s-\wt\beta) \right) \di x, \qquad s \geq \wt \beta, \ \eta \in \R.
    \end{equation}
    Our goal is to prove that there exists a unique root $\wt a \in [\wt \beta, \wt \alpha]$ such that $F_r(\wt a, \eta) = 1/2$. We first analyze the transition from the complete one-cut regime, which occurs precisely on the affine line given by $Q(\wt \alpha, \eta, \tau) = 0$. We notice that in this case, $\tau > \tau_\alpha$, and $\eta_1 = \eta_1(\tau)$ is the $\eta$ value where this transition happens (see Rmk.~\ref{Rmk:eta1}). We then have $Q(x, \eta_1, \tau) = (x - \wt \alpha)q(x, \eta_1, \tau)$ for some linear polynomial $q$, and substituting this into the complete cut $g$-function, we obtain 
    \begin{equation}
      g_{\text{complete cut}}(z) = \frac{1-\tau\Phi(z,\eta_1, \tau)}{2} + \int_{\alpha^{-1}}^z \frac{R_r(s)\left( \frac{\tau}{2}(s +s^{-1}) + c_r(\eta_1, \tau) \right)}{s} ds. 
\end{equation}
Moreover,  $g_{\text{complete cut}}(z)$ will satisfy Riemann--Hilbert problem \ref{rhp:g_function_one_cut_right}.
    In particular, due to the normalization condition at infinity (or at $0$), we must have $c_r(\eta_1, \tau) = \gamma_{r}(\eta_1, \tau)$, and \eqref{eq:IntP} becomes equivalent to \eqref{eq:panic_eq_right} with $\wt a = \wt \alpha$. This gives us at least one solution $F_r(\wt \alpha, \eta_1) = \frac{1}{2}$. 

    To track the root $\wt a(\eta)$ for $\eta > \eta_1$, we use the Implicit Function Theorem (IFT). We compute the partial derivatives of $F_r(s,\eta)$:
    \begin{equation}\label{eq:FrEta}
        \partial_\eta F_r(s, \eta) = \int_1^{\wt \beta} \frac{(s - x)^{1/2}}{(\wt \beta - x)^{1/2}\sqrt{x^2 - 1}} \di x   > 0\,,
    \end{equation}
    \begin{equation}\label{eq:DsFr}
        \partial_s F_r(s, \eta) = \frac{3\tau s - \tau \wt \beta + 2\eta}{4} \underbrace{\int_1^{\wt\beta} \frac{1}{\sqrt{s - x}\sqrt{\wt \beta - x}\sqrt{x^2 - 1}} \di x }_{>0}\,.
    \end{equation}
 The sign of $\partial_s F$ depends exclusively on the linear prefactor. In particular, for $\tau>0$ there is a \textit{minimum} at $s_{\text{crit}}(\eta,\tau) = \frac{\tau \wt\beta - 2\eta}{3\tau}$, while for $\tau<0$ the same point is a \textit{maximum}. At $\tau = 0$ there are no critical points and $F_r(s,\eta)$ will be monotonic in $s$.

At the critical point $(\wt\alpha,\eta_1)$, except in the case $\tau=0$, the linear prefactor can be evaluated in terms of the parabola vertex $x_v(\eta_1,\tau)$ as
 \begin{equation}
     3\tau s - \tau \wt \beta + 2\eta = 4\tau\left(\wt\alpha - x_v(\eta_1,\tau)\right)\,,
 \end{equation}
 this, together with Corollary \ref{cor:techincal_cor}, implies that 
 \begin{equation} \partial_sF_r(s,\eta_1)_{\big\vert_{s=\wt\alpha}} >0\,, \quad \forall \tau>\tau_\alpha\,.
 \end{equation}
 Since $F_r(\wt \alpha, \eta_1) = 1/2$ and $\partial_s F_r(s, \eta_1)|_{s=\wt \alpha} \neq 0$, the IFT guarantees (at least locally) the local existence of a smooth curve $\wt a(\eta)$, $\eta > \eta_1$ originating from $\wt \alpha$ and satisfying
    \begin{equation}
        \frac{\partial \wt a}{\partial \eta} = - \frac{\partial_\eta F_r(\wt a, \eta)}{\partial_s F_r(s, \eta)|_{s = \wt a}} < 0\,.
    \end{equation}
Note that the (strictly decreasing) mapping $\eta \mapsto \wt a$ can be continued for $\eta > \eta_1$ using the IFT as long as $\wt \beta < \wt a$ and $\partial_s F_r(s, \eta)|_{s=\wt a(\eta)} > 0$ (in which case $a(\eta)$ strictly increases, moving away from $\alpha$ towards $\beta$). If we would have for some $\eta_*$ that $\partial_s F_r(s, \eta_*)|_{s=\wt a(\eta_*)} = 0$, then necessarily $\wt a(\eta_*) = s_{\text{crit}}$ and therefore $F_r(s_{\text{crit}}, \eta_*) = \frac{1}{2}$ (in particular $s_{\text{crit}}$ exists and therefore $\tau \not = 0$). In the case that $\tau \in (\tau_\alpha, 0)$, we have that $s_{\text{crit}}$ is a local maximum, and therefore we have $F_r(s_{\text{crit}}, \eta_*) > F_r(s, \eta_1)$, $s \in [\wt \beta, \infty)$ by the monotonicity of $F_r$ in $\eta$ \eqref{eq:FrEta}. Choosing $s = \wt \alpha$ gives the required contradiction $\frac{1}{2} > \frac{1}{2}$.

The case $\tau > 0$ is more subtle as in that case $s_\text{crit}$ is a minimum. If $F_r(s_{\text{crit}}, \eta) = \frac{1}{2}$ for some $\eta > \eta_1$, then upon \emph{decreasing} $\eta$ we would have \emph{two} local branches of $F_r(\wt a(\eta), \eta) = \frac{1}{2}$. One will be moving to the right, i.e.~$\wt a_1(\eta)$ will increase while $\eta$ decreases, until it hits $\wt a_1(\eta_1) = \wt \alpha$; this is the branch we already studied using the IFT. There will be a second branch $\wt a_2(\eta)$ which will move to the left as $\eta$ decreases until it hits $\wt \beta$. As
\begin{equation}
        F_r(\wt\beta, \eta) = \int_1^{\wt\beta} \frac{\tau x + \eta}{\sqrt{x^2-1}}\di x = -\tau \frac{\Phi(\beta,\eta,\tau)}{2},
    \end{equation}
this will happen exactly at $\eta = \eta_3$ (note that so far we have not shown that $\eta_1 < \eta_3$). Moreover we will have then $\partial_s F_r(s, \eta)|_{s = \wt a_2(\eta)} < 0$ for $\eta \in [\eta_3, \eta_*)$, in particular $\partial_s F_r(s, \eta_3)|_{s = \wt a_2(\eta_3)} < 0$ which is via \eqref{eq:DsFr} is equivalent to
\begin{align}\label{eq:LinCond}
    3 \tau \wt \beta - \tau \wt \beta + 2 \eta_3 = \tau (\beta + \beta^{-1}) +2\eta_3 < 0, \qquad \tau > 0.
\end{align}
Recall that $(\eta_3, \tau)$ is characterized (for $\tau \geq \tau_{\text{sn}}$) by the linear condition (see \eqref{eq:Usn})
\begin{align*}
    1+\tau(\beta-\beta^{-1}) + 2\eta_3 \ln(\beta) = 0.
\end{align*}
Substituting this equality into \eqref{eq:LinCond} we obtain 
\begin{align*}
    \tau \left[\beta+\beta^{-1} - \frac{\beta-\beta^{-1}}{\ln (\beta)}\right] - \frac{1}{\ln(\beta)} < 0.
\end{align*}
As $\tau > 0$ by assumption and $\beta \in (0,1)$, it follows that we obtain a contradiction if the term in the square brackets is positive. To show that it is positive, we assume that it is negative which is equivalent to $\ln(\beta)(\beta + \beta^{-1}) > \beta - \beta^{-1}$, or $\mu(\beta) = \ln(\beta)(\beta + \beta^{-1}) - (\beta - \beta^{-1}) > 0$. However, $\mu(1) = 0$ and $\mu'(\beta) = \ln(\beta)(1-\beta^{-2}) > 0$ for $\beta \in (0,1)$, implying that $\mu(\beta) < 0$ for all $\beta \in (0,1)$. Hence, \eqref{eq:LinCond} cannot happen, showing that $\wt a(\eta)$ will decrease from $\wt a(\eta_1) = \wt \alpha$ to $\wt a(\eta_3) = \wt \beta$ as $\eta$ increases from $\eta_1$ to $\eta_3$ (in particular necessarily $\eta_1 < \eta_3$).

It remains to consider the transition from the two-cuts region. This transition occurs on the curve given by $(\eta_2, \tau)$ with $\tau \in (\tau_{\text{sn}}, \tau_{\alpha})$ (cf.~\eqref{eq:eta2}), which in the $(\wt b, \wt a)$-coordinates corresponds to $\wt b = \wt \alpha$ and $\wt a(\eta_2, \tau)  \in (\wt \beta, \wt \alpha)$, see Remark \ref{rem:big_remark}. For the corresponding two-cut $g$-function at $(\eta_2, \tau)$ we have $F_r(\wt a_2, \eta_2) = F_2(\wt a_2, \wt \alpha; \eta_2, \tau) + \frac{1}{2} = \frac{1}{2}$ (see \eqref{eq:2cuts_sys2}). This gives us a solution of $F_r(\wt a_2, \eta_2) = \frac{1}{2}$ which is consistent with the boundary case of the two-cut region. Observe that the linear prefactor in \eqref{eq:DsFr} determining the sign of $\partial_sF_r(s,\eta_2)|_{s = \wt a_2}$ becomes

    \begin{equation}
        3\tau \wt a_2 - \tau \wt \beta + 2\eta_2 = 2L(\wt a_2, \eta_2,\tau)>0\,,
    \end{equation}
    as we proved in Proposition \ref{prop:g_two_cut} (as $\wt a_2 = \wt a(\eta_2, \tau)$). As $s_{\text{crit}}$ can be at most a maximum for $\tau < \tau_{\alpha} < 0$ (if it is in the domain of $F_r$) we can argue as before and apply the IFT to deduce the existence of a decreasing solution $\wt a(\eta)$ for $\eta \in (\eta_2, \eta_3)$ originating from $\wt a_2$ and moving to the left until it hits $\wt \beta$ at $\eta = \eta_3$.

\end{proof}

\begin{figure}
    \centering
    \includegraphics[scale=0.6]{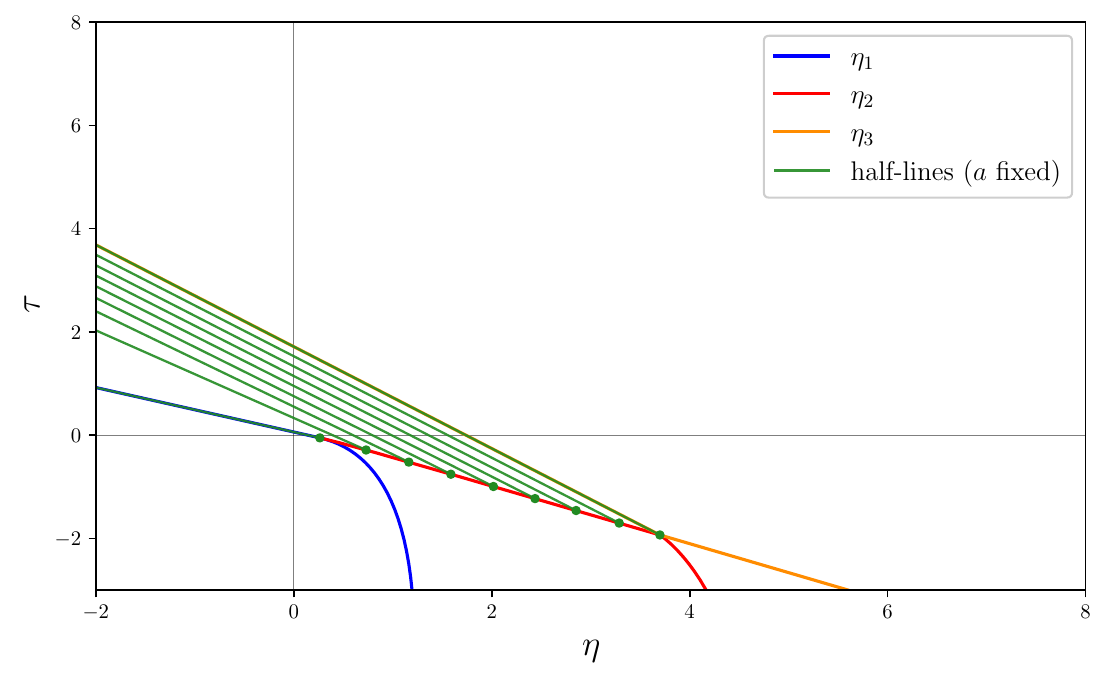}
    \caption{The set $\cU_r$ can be covered by affine half-lines corresponding to the lines $a(\eta,\tau)\equiv c \in(\alpha,\beta)$.}
    \label{fig:half_lines_right}
\end{figure}

\begin{remark}
    As for the one-cut left region, we notice that condition \eqref{eq:panic_eq_right} for each fixed $a\in (\alpha,\beta)$ represents a line in the $(\eta,\tau)$ plane. This implies that one can think about $\cU_r$ as an infinite union of affine half-lines stemming from the boundary of the two-cuts modulated region. See Figure \ref{fig:half_lines_right}.  
\end{remark}

\begin{remark}
\label{rem:modulation_eq_right}
    As we did in the previous section. By defining the elliptic-like integrals

    \begin{equation}
        I^{(r)}_k = \int_1^{\wt \beta} \frac{x^k}{(\wt a - x)^{1/2}(\wt \beta - x)^{1/2}(x^2-1)^{1/2}}\di x\,, \qquad k=0,1,2\,,
    \end{equation}
    and differentiating \eqref{eq:panic_eq_right} one (re)obtains the following Whitham modulation equation \cite{Biondini2024,Bloch1992} for $\wt a(\eta,\tau)\equiv \wt a$

    \begin{equation}
        \partial_\eta \wt a = 2\frac{\wt a I_0^{(r)} - I_1^{(r)}}{(\wt \beta  + \wt a)I_1^{(r)} + \wt a(\wt a - \wt \beta)I^{(r)}_0 - 2 I_2^{(r)}}\partial_\tau \wt a\,.
    \end{equation}
    We notice that $I_k^{(r)}$ can be computed following the same steps as in Remark \ref{rem:elliptic_1}.
\end{remark}

Regarding the function $f(z)$, one can prove the following

\begin{proposition}\label{prop:f_right_one_cut}
Define the function $f(z)$ as
\begin{equation}\label{eq:f_right_symmetric}
f(z) = \frac{\wt R_r(z)}{2\pi i} \left( \int_{a}^{\beta} \frac{s+z}{2s(s-z)}\frac{H_1(s)}{\wt R_{r,+}(s)}\di s + \int_{\beta^{-1}}^{a^{-1}} \frac{s+z}{2s(s-z)}\frac{H_2(s)}{\wt R_{r,+}(s)}\di s + \omega_r\int_{\beta}^{\beta^{-1}} \frac{s+z}{ 2s(s-z)}\frac{1}{\wt R_r(s)}\di s \right)\,,
\end{equation}
where $a$ is the same as in Proposition \ref{prop:g_right_cut}, $\wt R_r(z)$ is defined as 
\begin{equation}
    \wt R_r(s) = \frac{1}{2} \left( s+\frac{1}{s} - a -\frac{1}{a}\right)^\frac{1}{2} \left( s+\frac{1}{s} - \beta -\frac{1}{\beta}\right)^\frac{1}{2}\,,
\end{equation}
and the jump functions $H_1(s)$ and $H_2(s)$ are given by \eqref{eq:H_functions}; finally, the constant $\omega_r$ is defined as:
\begin{equation}\label{eq:omega_right}
 \omega_r = -\frac{ \int_{a}^{\beta} \frac{\ln(2\pi \rho(s)\mu_1)}{s\wt R_{r,+}(s)}\di s}{ \int_{1}^{\beta^{-1}} \frac{1}{s\wt R_r(s)}\di s }\,.
\end{equation}
 Then $f(z)$ is the unique solution to RHP \ref{rhp:f_function_one_cut_right}.
 \end{proposition}
 We omit the proof of the previous proposition since it is analogous to the proof of Proposition \ref{prop:f_whole_anti_symmetric}.

 After applying the previous propositions, the jump matrix $J_p(z)$ \eqref{eq:jump_p} becomes

\begin{equation}
\label{eq:J_p_right_dressed}
    	J_{p}(z)=\begin{cases}
  		\begin{pmatrix}
  			1& 0 \\
  			\Xi_1(z)e^{2\ln(N)\int_a^z \Omega_{r}(s,\eta,\tau)\di s}& 1
  		\end{pmatrix} \; &z \in (\alpha,a) \\\\[5pt]
          		\begin{pmatrix}
  			e^{\left(2\ln(N)\int_a^z \Omega_{r,+}(s,\eta,\tau)\di s + f_+(z) -f_-(z)\right)}& 0 \\
  			i& e^{-\left(2\ln(N)\int_a^z \Omega_{r,+}(s,\eta,\tau)\di s + f_+(z) -f_-(z)\right)}
  		\end{pmatrix} \; &z \in (a,\beta) \\\\[5pt]
        \exp\left(\sigma_3\left(\ln(N)\zeta_r(\eta,\tau) + \omega_r\right)\right)\; & z\in (\beta,\beta^{-1})\\\\[5pt]
          		\begin{pmatrix}
  			e^{\left(2\ln(N)\int_{a^{-1}}^z\Omega_{r,+}\di s + f_+(z) -f_-(z)\right)} &-i\\
  			0 & e^{-\left(2\ln(N)\int_{a^{-1}}^z \Omega_{r,+}\di s + f_+(z) -f_-(z)\right)}
  		\end{pmatrix} \; &z\in (\beta^{-1},a^{-1}) \\\\[5pt]
  		\begin{pmatrix}
  			1 &\Xi_2(z)e^{-2\ln(N)\int_{a^{-1}}^{z} \Omega_r(s,\eta,\tau)\di s}\\
  			0 & 1
  		\end{pmatrix} \; &z\in (a^{-1},\alpha^{-1}) 
  	\end{cases}\,,
  \end{equation}
where we defined

\begin{equation}
    \Xi_1(z) = 2\pi i z\mu_1\rho(z)\exp(f_+(z)+f_-(z))\,, \quad \Xi_2(z)=-2\pi i z^{-1}\mu_1\rho(z^{-1})\exp(-(f_+(z)+f_-(z)))\,.
\end{equation}

\subsubsection{Lenses opening}

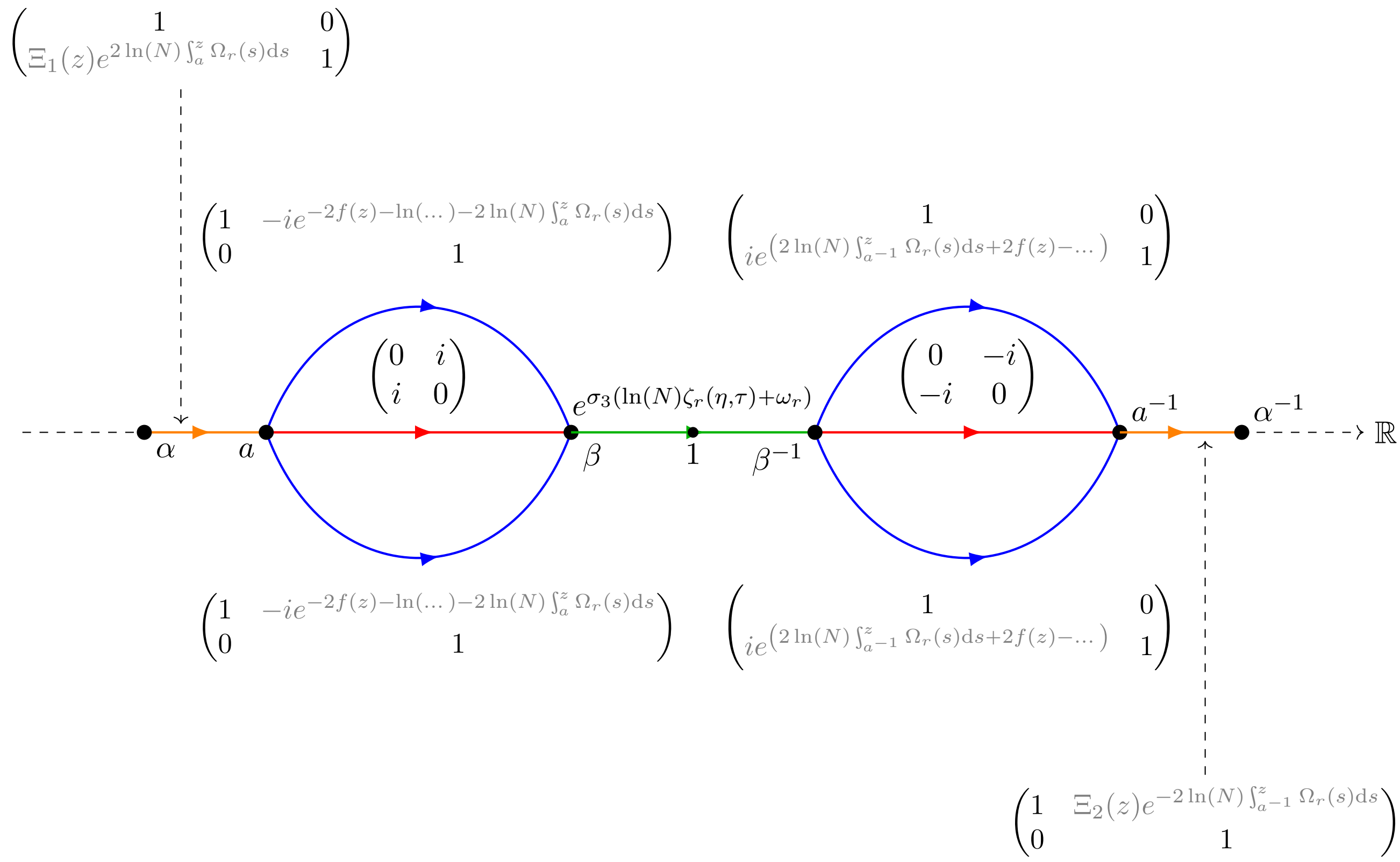
\begin{figure}[ht]
    \centering
    \begin{tikzpicture}[
    decoration={markings, mark=at position 0.55 with {\arrow{Latex}}},
    scale=1.5
]

    % Draw the real axis
    \draw[dashed,thin, black, ->] (-5.5, 0) -- (5.5, 0) node[right] {$\mathbb{R}$};

    % Define the coordinates for the branch points
    \coordinate (alpha) at (-4.5, 0);
    \coordinate (a) at (-3.5, 0);
    \coordinate (beta) at (-1, 0);
    \coordinate (beta_inv) at (1, 0);
    \coordinate (a_inv) at (3.5, 0);
    \coordinate (alpha_inv) at (4.5, 0);

    % ==========================================
    % LEFT GAP: (\alpha, a)
    % ==========================================
    \draw[thick, orange, postaction={decorate}] (alpha) -- (a);
    \filldraw[black] (alpha) circle (1.5pt) node[below right] {$\alpha$};
    
    % Jump matrix with long arrow (Placed above)
    \node (gap1_mat) at (-4.2, 3.2) {{\small $\pmtwo{1}{0}{{\color{gray}\Xi_1(z)e^{2\ln(N)\int_a^z \Omega_{r}(s)\di s}}}{1}$}};
    \draw[->,dashed, thin, shorten >= 3pt] (gap1_mat) -- (-4.2, 0);

    % ==========================================
    % LEFT CUT: (a, \beta)
    % ==========================================
    
    % Main cut
    \draw[thick, red, postaction={decorate}] (a) -- (beta) 
        node[midway, above=2pt, black] {$\pmtwo{0}{i}{i}{0}$};
        
    % Upper lens
    \draw[thick, blue, postaction={decorate}] (a) to[out=70, in=110, looseness=1.5] (beta);
    \node at (-2.1, 1.6) {{\small $\pmtwo{1}{{\color{gray}-ie^{-2 f(z) - \ln(\dots)  -2\ln(N)\int_a^z \Omega_{r}(s)\di s}}}{0}{1} $}};
    
    % Lower lens
    \draw[thick, blue, postaction={decorate}] (a) to[out=-70, in=-110, looseness=1.5] (beta);
    \node at (-2.1, -1.6) {{\small $\pmtwo{1}{{\color{gray}-ie^{-2 f(z) - \ln(\dots)  -2\ln(N)\int_a^z \Omega_{r}(s)\di s}}}{0}{1}$}};

    % Branch points (Left Cut)
    \filldraw[black] (a) circle (1.5pt) node[below left] {$a$};
    \filldraw[black] (beta) circle (1.5pt) node[below right] {$\beta$};

    % ==========================================
    % CENTRAL GAP: (\beta, \beta^{-1})
    % ==========================================
    \draw[thick, green!70!black, postaction={decorate}] (beta) -- (beta_inv) 
        node[midway, above=2pt, black] {$e^{\sigma_3\left(\ln(N)\zeta_r(\eta,\tau) + \omega_r\right)}$};
        
    % Fixed point at z = 1
    \filldraw[black] (0,0) circle (1pt) node[below] {$1$};

    % ==========================================
    % RIGHT CUT: (\beta^{-1}, a^{-1})
    % ==========================================
    
    % Main cut
    \draw[thick, red, postaction={decorate}] (beta_inv) -- (a_inv) 
        node[midway, above=2pt, black] {$\pmtwo{0}{-i}{-i}{0}$};
        
    % Upper lens
    \draw[thick, blue, postaction={decorate}] (beta_inv) to[out=70, in=110, looseness=1.5] (a_inv);
    \node at (2.1, 1.6) {\small {$\pmtwo{1}{0}{{\color{gray}ie^{\left(2\ln(N)\int_{a^{-1}}^z \Omega_{r}(s)\di s + 2f(z) - \dots\right)}}}{1}$}};
    
    % Lower lens
    \draw[thick, blue, postaction={decorate}] (beta_inv) to[out=-70, in=-110, looseness=1.5] (a_inv);
    \node at (2.1, -1.6) {{\small $\pmtwo{1}{0}{{\color{gray}ie^{\left(2\ln(N)\int_{a^{-1}}^z \Omega_{r}(s)\di s + 2f(z) - \dots\right)}}}{1}$}};

    % Branch points (Right Cut)
    \filldraw[black] (beta_inv) circle (1.5pt) node[below left] {$\beta^{-1}$};
    \filldraw[black] (a_inv) circle (1.5pt) node[above right] {$a^{-1}$};

    % ==========================================
    % RIGHT GAP: (a^{-1}, \alpha^{-1})
    % ==========================================
    \draw[thick, orange, postaction={decorate}] (a_inv) -- (alpha_inv);
    \filldraw[black] (alpha_inv) circle (1.5pt) node[above right] {$\alpha^{-1}$};

    % Jump matrix with long arrow (Placed below)
    \node (gap2_mat) at (4.2, -3.2) {{\small $\pmtwo{1}{{\color{gray}\Xi_2(z)e^{-2\ln(N)\int_{a^{-1}}^{z} \Omega_r(s)\di s}}}{0}{1}$}};
    \draw[->,dashed, thin, shorten >= 3pt] (gap2_mat) -- (4.2, 0);

\end{tikzpicture}
    \caption{RHP for $v(z)$ in the modulated right one-cut regime. Lenses Opening: Notice that, in the regime of Lemma \ref{lem:off_decay_right}, the terms in gray go to zero as $N$ grows.}
    \label{fig:jump_v_right}
\end{figure}

We are now in position to open the lenses exactly as we did in the previous sections. We start by noticing that one can factor the jump matrix $J_p(z)$ in the two intervals $(a,\beta),(\beta^{-1},a^{-1})$ as

\begin{equation}
\begin{split}
        &\begin{pmatrix}
  			e^{\left(2\ln(N)\int_a^z \Omega_{r,+}(s,\eta,\tau)\di s + f_+(z) -f_-(z)\right)}& 0 \\
  			i& e^{-\left(2\ln(N)\int_a^z \Omega_{r,+}(s,\eta,\tau)\di s + f_+(z) -f_-(z)\right)}
  		\end{pmatrix}=\\ &\begin{pmatrix}
        1 &  -ie^{-2 f_-(z) - \ln(2 \pi z \mu_1 \rho(z))  -2\ln(N)\int_a^z \Omega_{r,-}(s,\eta,\tau)\di s}
        \\
        0 & 1
    \end{pmatrix}\\
    &\times
    \begin{pmatrix}
        0 & i
        \\
        i & 0
    \end{pmatrix} 
    \begin{pmatrix}
        1 &  -ie^{-2 f_+(z) - \ln(2 \pi z \mu_1 \rho(z))  -2\ln(N)\int_a^z \Omega_{r,+}(s,\eta,\tau)\di s}
        \\
        0 & 1
    \end{pmatrix}\,,
    \end{split}
\end{equation}
and 
\begin{equation}
\begin{split}&
\begin{pmatrix}
  			e^{\left(2\ln(N)\int_{a^{-1}}^z \Omega_{r,+}(s,\eta,\tau)\di s + f_+(z) -f_-(z)\right)} &-i\\
  			0 & e^{-\left(2\ln(N)\int_{a^{-1}}^z \Omega_{r,+}(s,\eta,\tau)\di s + f_+(z) -f_-(z)\right)}
  		\end{pmatrix} = \\
        &\pmtwo{1}{0}{ie^{\left(2\ln(N)\int_{a^{-1}}^z \Omega_{r,-}(s,\eta,\tau)\di s + 2f_-(z) - \ln(2 \pi \mu_1 \rho(z^{-1})/z)\right)}}{1}\\ & \times  \pmtwo{0}{-i}{-i}{0} \pmtwo{1}{0}{ie^{\left(2\ln(N)\int_{a^{-1}}^z \Omega_{r,+}(s,\eta,\tau)\di s + 2f_+(z) - \ln(2 \pi \mu_1 \rho(z^{-1})/z)\right)}}{1}
\end{split}\,.
\end{equation}
Then, as we did in Section \ref{subsubsec:Lenses_left_cut} , we define the following vector $v(z)$ 

\begin{equation}
\label{eq:v_right}
    v(z) = \begin{cases}
        p(z)\pmtwo{1}{-ie^{-2 f(z) - \ln(2 \pi z \mu_1 \rho(z))  -2\ln(N)\int_a^z \Omega_{r}(s,\eta,\tau)\di s}}{0}{1} \; &z \text{ in the lower lenses below }(a,\beta) \\\\
        p(z)\pmtwo{1}{ie^{-2 f(z) - \ln(2 \pi z \mu_1 \rho(z))  -2\ln(N)\int_a^z \Omega_{r}(s,\eta,\tau)\di s}}{0}{1}  \; &z \text{ in the upper lenses above }(a,\beta) \\\\
        p(z) \pmtwo{1}{0}{ie^{\left(2\ln(N)\int_{a^{-1}}^z \Omega_{r}(s,\eta,\tau)\di s + 2f(z) - \ln(2 \pi \mu_1 \rho(z^{-1})/z)\right)}}{1} \; &z \text{ in the lower lenses below }(\beta^{-1},a^{-1}) \\\\
        p(z)\pmtwo{1}{0}{-ie^{\left(2\ln(N)\int_{a^{-1}}^z \Omega_{r}(s,\eta,\tau)\di s + 2f(z) - \ln(2 \pi \mu_1 \rho(z^{-1})/z)\right)}}{1}  \; &z \text{ in the upper lenses above }(\beta^{-1},a^{-1}) \\\\
        p(z) \; & \text{otherwise}
    \end{cases}\,.
\end{equation}
The vector $v(z)$ satisfies the following conditions

\begin{equation}
\label{eq:v_right_cut}
    v_+(z) = v_-(z)J_v(z)\,, \quad
    \lim_{z\to 0} v_1(z)v_2(z)=1\,, \quad
     v(z^{-1}) = v(z)\sigma_1\,, \quad v_1(0) > 1\,,
\end{equation}
where the jump matrix $J_v(z)$ is as in Figure \ref{fig:jump_v_right}.

We now prove the following lemma that will allow us to reduce our analysis to the model problem RHP \ref{OuterRHP_1cut} .

\begin{lemma}
\label{lem:off_decay_right}
Let $(\eta,\tau)\in \cU_{r}$ \eqref{eq:modulated_regions} and $a\equiv a(\eta,\tau)$ be defined as in Proposition \ref{prop:g_right_cut}. Then, the off-diagonal entries of the jump matrix $J_p(z)$ \eqref{eq:J_p_right_dressed} for $z\in(\alpha,a)\cup(a^{-1},\alpha^{-1})$ converge to zero as $N\to \infty$. 

Furthermore, there exist two neighborhoods $V_1,V_2$ of $(a,\beta)$ and $(\beta^{-1},a^{-1})$ respectively such that
\begin{equation}
    \Re\left(\int_a^z \Omega_r(s,\eta,\tau)\di s\right) \geq 0\; \forall\,z\in V_1\,, \quad \Re\left(\int_{a^{-1}}^z \Omega_r(s,\eta,\tau)\di s\right) \leq 0\; \forall\,z\in V_2\,.
\end{equation}
\end{lemma}

\begin{proof}
We prove the decay of the off-diagonal terms on the gap $(\alpha, a)$. The decay is governed by the exponent $2 \ln(N) \int_a^z \Omega_r(s, \eta, \tau) \di s$. We must show that $\nu(z, \tau) = \int_a^z \Omega_r(s, \eta, \tau) \di s < 0$ for $z \in (\alpha, a)$.

Fix the branch point $a_0 = a(\eta, \tau) \in (\alpha, \beta)$. The condition $\int_\beta^1 \Omega_r(s, \eta, \tau) \di s = 1/2$ defines an affine line in the $(\eta, \tau)$-plane. Recalling that $\Omega_r(s, \eta, \tau) = \frac{R_r(s)}{s} \left( \tau x(s) + \eta + \frac{\tau}{2}(\tilde{a}_0 - \tilde{\beta}) \right)$ where $x(s) = \frac{1}{2}(s+s^{-1})$, we differentiate the modulation equation with respect to $\tau$ along this curve:
\begin{equation}
    \int_\beta^1 \frac{\di \Omega_r}{\di \tau} \di s = 0, \quad \text{where} \quad \frac{\di \Omega_r}{\di \tau} = \frac{R_r(s)}{s} \underbrace{\left( x(s) + \frac{\di \eta}{\di \tau} + \frac{1}{2}(\wt a_0 - \tilde{\beta}) \right)}_{\equiv M(s)}\,.
\end{equation}

Because $x(s)$ is strictly decreasing for $s \in (0, 1)$, the function $M(s)$ is strictly decreasing. The weight $R_r(s)/s$ is strictly positive on $(\beta, 1)$ . Since the integral of $M(s)$ against a strictly positive weight over the full interval $(\beta, 1)$ is zero, $M(s)$ must change sign exactly once from positive to negative . Consequently, for $s \le \beta$, $M(s)$ must be strictly positive.

For $z \in (\alpha, a_0)$, we have $z < a_0 < \beta$, meaning $M(s) > 0$ for all $s \in [z, a_0]$. Differentiating $\nu(z, \tau)$ with respect to $\tau$ along the level curve yields:
\begin{equation}
    \frac{\di \nu(z, \tau)}{\di \tau} = \int_{a_0}^z \frac{R_r(s)}{s} M(s) \di s\,.
\end{equation}
Because $z < a_0$ and the integrand is strictly positive on $(z, a_0)$, the integral evaluated from $a_0$ to $z$ is strictly negative, giving $\frac{\di \nu(z, \tau)}{\di \tau} < 0$.

At the boundary $(\eta_2(\tau_*), \tau_*)$ such that $a(\eta_2(\tau_*), \tau_*) = a_0$ separating the two-cut and right one-cut regimes, the left cut collapses ($b$ merges with $\alpha$). Here, $\Omega_r$ coincides with $\Omega_2$, and Lemma \ref{lem:off_decay_two_cuts_gap} guarantees that $\nu(z, \tau_*) < 0$ . As we move into $\cU_{r}$ along the level curve $a(\eta, \tau) = a_0$, the parameter $\tau$ strictly increases (since $\tau > \tau_*$). Because the derivative with respect to $\tau$ is negative, increasing $\tau$ makes $\nu(z, \tau)$ strictly more negative. Thus, $\nu(z, \tau) < 0$ for all $z \in (\alpha, a_0)$, ensuring the wanted decay. By the same argument established in Remark \ref{rem:sym_off_diag}, the identical decay holds on $(a_0^{-1}, \alpha^{-1})$.

Next, we analyze the decay on the lenses around $(a_0, \beta)$. On the real axis, $\int_{a_0}^z \Omega_{r,+}(s, \eta, \tau) \di s$ is purely imaginary. By the Cauchy-Riemann equations, we obtain the required positivity in the upper lens if $\Im(\Omega_{r,+}(z, \eta, \tau)) < 0$ for $z \in (a_0, \beta)$. As before we have $\frac{d\Omega_{r,+}}{d\tau} = \frac{R_{r, +}(s)}{s} M(s)$. We already established that $M(s)$ is strictly decreasing and positive at $s = \beta$. It follows that $M(s)$ is positive for $s \in (\alpha, \beta)$, and as $\Im(R_{r, +}(s)) < 0$ for $s \in (a_0, \beta) \subset (\alpha, \beta)$ (see Rmk.~\ref{rem:Rr_boundary}), it follows that $\frac{d}{d\tau}\Im(\Omega_{r, +}) < 0$ there. Starting again at the boundary $(\eta_2(\tau_*), \tau_*)$ such that $a(\eta_2(\tau_*), \tau_*) = a_0$, it follows that $\Im(\Omega_{r, +})|_{\eta_2(\tau_*), \tau_*} < 0$ for $s \in (a_0, \beta)$ by the proof of Lemma \ref{lem:off_decay_two_cuts_lens}. By the aforementioned monotonicity of $\Im(\Omega_{r, +})$ in $\tau$, it follows that $\Im(\Omega_{r, +})|_{\eta, \tau} < \Im(\Omega_{r, +})|_{\eta_2(\tau_*), \tau_*} < 0$ for all $\eta, \tau$ such that $a(\eta, \tau) = a_0$ and $s \in (a_0, \beta)$ (recall $\tau > \tau_*$). This ensures decay on the upper lens between $a_0$ and $\beta$. An analogous argument applies to the lower lens and the lenses around $(\beta^{-1}, a_0^{-1})$, finishing the proof.
\end{proof}

\begin{remark}
    We notice that the previous Lemma shows that all the entries colored gray in Figure \ref{fig:jump_v_right} decay as $N\to \infty$.
\end{remark}

Applying Lemma \ref{lem:off_decay_right}, we are led to consider as the outer parametrix problem the model problem RHP \ref{OuterRHP_1cut} in its symmetric version, where $\ta \equiv a(\eta,\tau)$, $\tb\equiv \beta$ and $\Delta \equiv \Delta_{r,n}= \frac{1}{2\pi i}(\ln(N)\zeta_r(\eta,\tau) + \omega_r(\eta,\tau))$. To approximate $v(z)$, we need to construct a local approximation near the endpoints of the $2$ intervals $(a,\beta),\,(\beta^{-1},a^{-1})$. Since the construction is analogous to the one done in the previous cases, and since this local approximation does not contribute to the leading-order asymptotics, we skip their explicit construction.

\subsubsection{The modulated right one-cut solution}
\label{subsubsec:right_one_cut_solution}

The construction of  the asymptotic solution in this regime is completely analogous as the one in Section~\ref{subsubsec:left_one_cut_solution}, therefore we only underline the main differences and we state the final result. In particular,
the model solution $k_s^{(r)}, \ell_a^{(r)}$ are evaluated on
$\cR_r=\cR(a(\eta,\tau),\beta)$. Analogously as before, we set $A^{(r)}_0=\cA_r(0^{(1)})$ and the period $B^{(r)}$ on $\cR_r$. Furthermore, we define the two background $a_{\mathrm{bg}}^{(r)}(\eta,\tau),b_{\mathrm{bg}}^{(r)}(\eta,\tau)$ as 
\begin{align}
 a^{(r)}_{\mathrm{bg}}(\eta,\tau)
 &=\frac12e^{-\partial_\eta g_\infty}
 =\frac1{2a}\exp\left\{-\int_0^a
 \left(
 \frac{\frac{s+s^{-1}}2-\frac{I^{(r)}_1}{I^{(r)}_0}}
 {s\wt R_r(s)}-\frac1s
 \right)\di s\right\},
 \label{eq:thm_abg_explicit_right}\\
 b^{(r)}_{\mathrm{bg}}(\eta,\tau)
 &=\frac12\partial_\eta g'(0)
 =\frac{\wt a+\wt\beta}{2}-\frac{I^{(r)}_1}{I^{(r)}_0}\,,
 \label{eq:thm_bbg_explicit_right}
\end{align}
we recall that we defined the integrals $I_k^{(r)}$ in Remark \ref{rem:modulation_eq_right}.
Except for these changes, the derivation of the asymptotic for $q(z)$ and the recovery of $a^{(q)}_n(t), b^{(q)}_n(t)$ via Lemma \ref{lem:recovery_q} follows exactly the same lines as in the previous section, therefore, we just state the final result
\begin{theorem}
\label{thm:right_one_cut_asymp}
For $(\eta,\tau)$ in a compact subset of the interior of $\cU_r$,
as $N\to\infty$, the Toda lattice variables $a^{(q)}_n(t)$ and $b^{(q)}_n(t)$ with $n = \eta\ln N$ and $t = \tau\ln N$ are asymptotically given by:
\begin{align}
 a^{(q)}_n(t)&=a^{(r)}_{\mathrm{bg}}(\eta,\tau)
 \sqrt{\frac{
 \theta_3(2A^{(r)}_0+\frac12+\Delta_{r,n+1}\mid2B^{(r)})
 \theta_3(2A^{(r)}_0+\frac12-\Delta_{r,n}\mid2B^{(r)})}{
 \theta_3(2A^{(r)}_0+\frac12-\Delta_{r,n+1}\mid2B^{(r)})
 \theta_3(2A^{(r)}_0+\frac12+\Delta_{r,n}\mid2B^{(r)})}}
 \nonumber\\
 &\qquad\times\left(1+\cO\left(\frac1{\ln N}\right)\right),
 \label{eq:thm_an_asymp_right}\\
 b^{(q)}_n(t)&=b^{(r)}_{\mathrm{bg}}(\eta,\tau)
 +\frac{1}{2I_0^{(r)}}\left[
 \frac{\theta_3'(2A^{(r)}_0+\frac12-\Delta_{r,n}\mid2B^{(r)})}
 {\theta_3(2A^{(r)}_0+\frac12-\Delta_{r,n}\mid2B^{(r)})}
 \right.\nonumber\\
 &\qquad\left.
 -\frac{\theta_3'(2A^{(r)}_0+\frac12-\Delta_{r,n+1}\mid2B^{(r)})}
 {\theta_3(2A^{(r)}_0+\frac12-\Delta_{r,n+1}\mid2B^{(r)})}
 \right]+\cO\left(\frac1{\ln N}\right).
 \label{eq:thm_bn_asymp_right}
\end{align}
The backgrounds are given by
\eqref{eq:thm_abg_explicit_right}--\eqref{eq:thm_bbg_explicit_right}.
All geometric quantities are evaluated at $(\eta,\tau)$, whereas
$\Delta_{r,n+1}$ is evaluated at $(\eta+1/\ln N,\tau)$.
The prime on $\theta_3$ denotes differentiation in its first argument.
\end{theorem}

\begin{remark}
As a sanity check, one verifies that as $a\to\alpha$, the background formulas agree with those of the
complete one-cut calculation.Indeed, as $a\to\alpha$, they satisfy
$a^{(r)}_{\mathrm{bg}}\to a^{(c)}_{\mathrm{bg}}$ and $b^{(r)}_{\mathrm{bg}}\to b^{(c)}_{\mathrm{bg}}$.
\end{remark}

\section{Probabilistic and small-norm estimates}
\label{sec:prob_estimate}

With the asymptotic analysis in Sections \ref{sec:small_norm}--\ref{sec:one_cut_modulated_regions} complete, we are now in a position to collect the leading-order deterministic variables $\fa(\eta,\tau)$ and $\fb(\eta,\tau)$ introduced in Theorem \ref{thm:main_prob}. Specifically, on each macroscopic region of the phase diagram, $\fa(\eta,\tau)$ and $\fb(\eta,\tau)$ are defined by the leading order asymptotic derived in Corollary \ref{cor:small_norm_sol} for $\cU_{\mathrm{sn}}$, Theorem \ref{thm:complete_one_cut_asymptotics} for $\cU_c$, Theorem \ref{thm:modulated_two_cuts_asymptotics} for $\cU_2$, and Theorems \ref{thm:left_one_cut_asymp}, \ref{thm:right_one_cut_asymp} for $\cU_\ell$ and $\cU_r$. We also define the phase boundary as 
\begin{align*}
    \Sigma_{\mathrm{phase}} = \bigcup_{j=1}^3 \lbrace (\eta_j(\tau), \tau) \mid \tau \in \text{dom}(\eta_j)\rbrace= \partial\cU_{\mathrm{sn}}\cup\partial\cU_{c}\cup\partial\cU_{r}\cup\partial\cU_{\ell},
\end{align*}
see Figure \ref{fig:boundary}. The goal of this final section is to lift these deterministic asymptotics to the full probability space. To this end, we first establish the matrix-valued small-norm approximation for RHP \ref{rhp:m_hat} with initial data in the good set $\fW_N$ \eqref{eq:good_set}, giving us Theorem \ref{thm:m_2_q_approx}. We then prove the moment bounds for the bad set $\fW_N^c$ (Lemma \ref{lem:measure_bad_set}) and conclude the proof of Theorem \ref{thm:main_prob}.

\subsection{Small-norm problem in the good set}\label{Sect:SNGoodSet}

The aim of the present subsection is to prove Theorem \ref{thm:m_2_q_approx}. Existence and uniqueness for the solution of RHP \ref{RHP:q} under Assumption \ref{ass:condensate_assumptions} follows from a slight modification of the existence and uniqueness argument used in \cite[Theorem 2.1]{EPT}. Therefore, from now on, we assume that the initial data  $(\boldsymbol{\zeta},\boldsymbol{\gamma})\in\fW_N$ \eqref{eq:good_set}. To this end, we formulate a \emph{matrix-valued} small norm problem for the RHP obtained after opening lenses, which in turn allows us to use a \emph{vector-valued} small norm problem. 

\begin{proof}[Proof of Theorem \ref{thm:m_2_q_approx}]
Let $M(z)$ denote the matrix-valued global parametrix in the small-norm regime, modulated left/right cut-regime or two-cut regime (the details can be found in the relevant sections). Let $J_r(z) = M_-(z) J_v(z)M_+^{-1}(z)$ be the small-norm jump matrix that satisfies $\Vert J_r - I \Vert_{\infty} = \cO(\frac{1}{\ln N})$. We are looking for a matrix-valued function satisfying
\begin{itemize}
    \item $R_+(z) = R_-(z) J_r(z)$ on the jump contour;
    \item $
    R(0) = \begin{pmatrix}
        1 & *
        \\
        0 & *
    \end{pmatrix}, \qquad R(\infty) = \begin{pmatrix}
        * & 0
        \\
        * & 1
    \end{pmatrix}$.
\end{itemize}
The exact choice of jump contour in question will depend on the region we are considering. In particular, one has that 
\begin{equation}
    (1,\, 1) R(z) = r(z)\,,
\end{equation}
where $r(z)$ is the small-norm vector that we used in the different asymptotic regions, see Subsections \ref{subsec:complete_one_cut_solution}, \ref{subsec:two_cuts_solution}, \ref{subsubsec:left_one_cut_solution}, and \ref{subsubsec:right_one_cut_solution}.
One can check, see also Appendix \ref{App:SmallNorm},  that finding such a solution is equivalent to solving $(1 - C_{w_r})\mu = I$, in which case $\mu = R_-$ and 
\begin{align*}
R(z) = I +\frac{1}{2\pi i} \int_{\Sigma_r} \mu(s) w_r(s)\begin{pmatrix}
        \frac{1}{s-z} - \frac{1}{s} & 0
        \\
        0 & \frac{1}{s-z}
    \end{pmatrix} ds.
\end{align*}
Standard small-norm theory implies that a unique solution $\mu(z)$ exists, which implies that the Riemann--Hilbert solution $R(z)$ exists and is unique. Moreover, due to the symmetries of the jump matrices it follows that $\sigma_1 R(z^{-1}) \sigma_1$ also satisfies the same RHP, hence by uniqueness $R(z) = \sigma_1 R(z^{-1}) \sigma_1$ (note that the global parametrix $M$ satisfies the same symmetry and normalization).

Next let us introduce $V(z) = R(z)M(z)$, which, as it happened for $r(z)$, is related to $v(z)$ by $(1,\,1)V(z) = v(z) v_1(0)^{-1}$. Then $V(z)$ will be a matrix-valued solution to the jump problem $V_+(z) = V_-(z) J_v(z)$ with the same normalization and symmetry properties stated above. In particular, due to the symmetry of $R(z)$, it follows that 

$$R(1) = \begin{pmatrix}
    a & b
    \\
    b & a 
\end{pmatrix}\,,$$
furthermore, in view of the structure of $M(z)$ there exist two constants $d^{(M)}_\pm$ such that

\begin{equation}
    M_\pm(z) = \frac{d^{(M)}_\pm}{z-1} \pmtwo{1}{1}{-1}{-1} + \cO(1) \,,
\end{equation}
which implies that there exist two constants $d^{(V)}_\pm$ such that $V(z)$ has the following residue structure at $z = 1$

\begin{align}\label{eq:V_res}
V_\pm(z) = \frac{d^{(V)}_\pm}{(z-1)} \begin{pmatrix}
    1 & 1
    \\
    -1 & -1
\end{pmatrix} + \cO(1)\,.
\end{align}
We define 
\begin{align*}
    \widehat Q(z) &= \begin{pmatrix}
    e^{\ln (N) g(0) + f(0)} & 0
    \\
    0 & e^{-(\ln(N) g(\infty) + f(\infty))}
\end{pmatrix} V(z) e^{-\sigma_3(\ln N g(z) + f(z))} 
\\
&= e^{\ln N g(0) + f(0)} V(z) e^{-\sigma_3(\ln N g(z) + f(z))}, 
\end{align*}
and finally
\begin{equation}
    Q(z) = \begin{cases}
        \wh Q(z)\begin{pmatrix}
  			1 & 0 \\
  			N\mu_1 e^{t\Phi(z,\eta,\tau)}\int_\alpha^\beta \frac{s\rho(s)}{z-s}\mathrm{d} s& 1
  		\end{pmatrix} \qquad &z\in \mathring{\sigma}\\\\
        \wh Q(z) \begin{pmatrix}
  			1 &-zN\mu_1 e^{-t\Phi(z,\eta,\tau)} \int_{\beta^{-1}}^{\alpha^{-1}} \frac{\rho(s^{-1})}{(z-s)s^2} \mathrm{d} s\\
  			0 & 1
  		\end{pmatrix} \qquad &z^{-1}\in \mathring{\sigma}\\\\
        \wh Q(z) \qquad &\text{otherwise}
    \end{cases}
\end{equation}
Note that $Q(z)$ will be a matrix-valued solution of $Q_+(z) = Q_-(z) J_q(z)$, see \eqref{eq:jump_q}, it satisfies $Q(z) = \sigma_1 Q(z^{-1}) \sigma_1$, it has the residue structure
\begin{align*}
    Q_\pm(z) = \frac{d^{(Q)}_\pm}{(z-1)} \begin{pmatrix}
    1 & 1
    \\
    -1 & -1
\end{pmatrix} e^{-\sigma_3(\ln N g_\pm(1) + f_\pm(1))} + \cO(1)\,,
\end{align*}
and the normalization 
\begin{align}\label{eq:Qres}
     Q(0) = \begin{pmatrix}
        1 & *
        \\
        0 & *
    \end{pmatrix}, \qquad Q(\infty) = \begin{pmatrix}
        * & 0
        \\
        * & 1
    \end{pmatrix}.
\end{align}
Let us now introduce $\dd(z) = \frac{1}{m^{(2)}_1(0)}m^{(2)}(z) Q(z)^{-1}$. Note that as $m^{(2)}(z^{-1}) = m^{(2)}(z) \sigma_1$, it follows that $\dd(z^{-1}) = \dd(z) \sigma_1$. Moreover, $\dd(z)$ is holomorphic around $z = 1$ due to the symmetry of $m^{(2)}(z)$ and the residue structure \eqref{eq:Qres} of $Q(z)$ near $z = 1$.  The jump condition $\dd_+(z) = \dd_-(z)J_{\dd}(z)$ reads as follows: 
\begin{equation}\label{eq:Jdelta}
\begin{split}
    J_{\dd}(z) = R(z)M(z)& \left.
    \begin{cases}
        \begin{pmatrix}
            1 & 0
            \\
            e^{\ln N[\tau \Phi(z) + 2 g(z) + 1] + 2f(z)}\left( \frac{1}{N}\sum_{j=1}^N \frac{\zeta_j\gamma_j}{z-\zeta_j}-\mu_1 \int_\alpha^\beta \frac{s\rho(s)}{z-s}\mathrm{d} s\right) & 1
        \end{pmatrix}, \; z \in \sigma
        \\
        \\
        \begin{pmatrix}
            1 & -e^{-\ln N[\tau \Phi(z) + 2 g(z) - 1] - 2f(z)}z\left(\frac{1}{N}\sum_{j=1}^N \frac{\gamma_j}{z-\zeta^{-1}_j} - \mu_1 \int_{\alpha}^{\beta} \frac{\rho(s)}{z-s^{-1}} \mathrm{d} s\right)
            \\
            0 & 1
        \end{pmatrix}, \; z^{-1} \in \sigma
    \end{cases}\right\}\\ 
    & \times \big(R(z)M(z)\big)^{-1}
    \end{split}
\end{equation}
To see that $J_\dd(z) \to I$ uniformly as $N \to \infty$, we need to study the harmonic functions $\Re(\tau \Phi(z) + 2g(z) \pm 1)$. We will focus on the jump on $z \in \sigma$, surrounding the interval $[\alpha, \beta]$ (the other case is similar). Note that on the support of the $g$-function (i.e.~the cuts) contained in $[\alpha, \beta]$ we have the equality
\begin{align*}
    g_+(z)+g_-(z)+1 + \tau\Phi(z,\eta,\tau) = 0 \quad \text{for} \ z \ \text{on the cuts} \subseteq [\alpha, \beta].
\end{align*}
As $g_+(z) = \overline{g_-(z)}$ everywhere on the real axis, and $\Phi(z,\eta,\tau)$ is purely real for $z > 0$, the above condition is equivalent to $\Re(\tau \Phi(z) + 2g(z) +1) = 0$. Meanwhile, on the gap $[b, a]$ in the two-cut region, or $[b, \beta]$ resp.~$[\alpha, a]$ in the modulated left resp. right cut regions, we have that $e^{\ln N(\tau \Phi + g_+(z) + g_-(z) + 1) + f_+(z) + f_-(z)}$ always appears in the lower off-diagonal entry of the jump matrix $J_v(z)$ ($=J_p(z)$ in the relevant interval). This follows essentially from the definition $p(z) = \wh q(z)e^{\sigma_3(\ln(N)g(z) + f(z)) }$). We have already shown in Lemma \ref{lem:off_decay_two_cuts_gap}, \ref{lem:off_decay_right} and \ref{lem:off_decay_left} that this entry decays exponentially, implying that $\Re(\tau \Phi(z) + 2g(z) +1) < 0$ in the interior of $(b, a)$ or $(b, \beta)$ or $(\alpha, a)$ depending on the regime. 

It now follows that the contour $\sigma$ can always be deformed (if necessary) to enclose $[\alpha, \beta]$ such that $\Re(\tau \Phi(z) + 2g(z) +1) < \varepsilon_\sigma$ for any $\varepsilon_\sigma > 0$ and all $z \in \sigma$. To satisfy the required $\cO(N^{-\delta})$-bound of the theorem, we need set $\varepsilon_\sigma = \varepsilon_\nu - \delta$, which is positive by assumption.  This implies that, if the initial data  $(\boldsymbol{\zeta},\boldsymbol{\gamma})\in\fW_N$ \eqref{eq:good_set}, the lower triangle matrix in \eqref{eq:Jdelta} has the form $I + O(N^{-\varepsilon_\nu+\varepsilon_\sigma})$, uniformly on $\sigma$, and we recall that we defined $\varepsilon_\nu$ above \eqref{eq:good_set}. A similar argument shows that the upper triangular matrix in \eqref{eq:Jdelta} satisfies the same estimate (alternatively one can use the symmetry $J_\dd(z^{-1}) = \sigma_1 J_{\dd}(z) \sigma_1$). As $R(z)$, $M(z)$, and their inverses are uniformly bounded in $N$ away from the intervals $[\alpha, \beta] \cup [\beta^{-1}, \alpha^{-1}]$ (cf.~Remark ~\ref{rmk:UnifBoundPara}), it follows that $\Vert J_\dd - I \Vert_\infty = \cO(N^{-\varepsilon_\nu + \varepsilon_\sigma})$. 
 Therefore we can conclude (recalling the definition of $\dd(z)$)
\begin{align*}
    m^{(2)}(z) &= m_1^{(2)}(0) \dd(z) Q(z) 
    \\
    &= m_1^{(2)}(0)\left[ (1, \, 1) + \cO(N^{-\varepsilon_\nu + \varepsilon_\sigma})\right]Q(z)
    \\
    &= m_1^{(2)}(0) \left[ \frac{1}{q_1(0)}q(z) + \cO(N^{-\varepsilon_\nu + \varepsilon_\sigma})Q(z)\right].
\end{align*}
As in a neighbourhood of $z = 0$ we have $Q(z) = e^{\ln N g(0) + f(0)} V(z) e^{-\sigma_3(\ln N g(z) + f(z))}$, $q(z) = v(z)e^{-\sigma_3(\ln N g(z) + f(z))}$ and $V(0)$ is uniformly bounded, it follows that the above equation can be written as
\begin{align*}
     m^{(2)}(z) =  \left(m_1^{(2)}(0)e^{\ln N g(0) + f(0)}\right)\left[\frac{1}{v_1(0)}v(z) +\cO(N^{-\varepsilon_\nu + \varepsilon_\sigma})\right]e^{-\sigma_3(\ln N g(z) + f(z))}
\end{align*}
This equation has exactly the same form as \eqref{eq:qApprox} for the complete-cut region (and analogously for all other regions except small-norm) with the identification $q(z) \mapsto m^{(2)}(z)$, $r(z) \mapsto \dd(z)$, $M(z) \mapsto Q(z)$, $k_s(z) \mapsto v(z)/v_1(0)$, $\cO(1/\ln \, N) \mapsto \cO(N^{-\varepsilon_\nu + \varepsilon_\sigma})$ (here $k_s(z)$ stands for $k_s^{(2)}(z), k_s^{(\ell)}(z)$ etc.). Essentially, instead of approximating $q(z)$ with the global parametrix $M(z)$ (or $k_s(z)$) as in \eqref{eq:qApprox} and the other regions, we are approximating the exact $N$-soliton solution $m^{(2)}(z)$ with the matrix $Q(z)$ (or $q(z)$). 

The explicit approximation of $m_1^{(2)}(z)$ in a neighborhood of $z = 0$ is now completely analogous to previous sections dealing with the approximation of $q(z)$ and gives us (in a neighborhood of $z = 0$) 
\begin{align*}
    m_1^{(1)}(z) = m_1^{(2)}(z) &= v_1(z)e^{-\ln N g(z) - f(z))}\left( 1 + \cO(N^{-\varepsilon_\nu + \varepsilon_\sigma}) \right)
    \\
    & = q_1(z)\left( 1 + \cO(N^{-\varepsilon_\nu + \varepsilon_\sigma}) \right)
\end{align*}
Here we used repeatedly that $|v_{1,2}(z)|$ stays uniformly away from $0, \infty$, (and therefore $v_1(z) + \cO(N^{-\varepsilon_\nu + \varepsilon_\sigma}) = v_1(z)(1+ \cO(N^{-\varepsilon_\nu + \varepsilon_\sigma}))$ etc.) as $v_1(z) = k_{s,1}(z)/\sqrt{k_{s,2}(0)} + \cO(1/\ln N)$, and we have explicit formulas for $k_{s, j}(z)$.
This finishes the proof.
\end{proof}
\begin{remark}
    We notice that in order to obtain this approximation,  we only need $\sum_{j=1}^N \frac{\zeta_j\gamma_j}{z-\zeta_j}-N\mu_1\int_\alpha^\beta \frac{s\rho(s)}{z-s}\mathrm{d} s = \cO(N^{1-\varepsilon})$ for $z \in \sigma$. This is optimal, as the quantities are themselves of order $\cO(N)$.
\end{remark}
\begin{remark}
    In the above analysis, we have only considered the cases for which we had to construct the $g$-function. For the remaining small-norm regime, the analysis is even simpler. In fact, observe that in the small norm regime we have $1 + \tau \Phi(z) < 0$ for $z \in [\alpha, \beta]$ and $1 - \tau \Phi(z) < 0$ for $z \in [\beta^{-1}, \alpha^{-1}]$, see \eqref{eq:small_norm_req}. Thus choosing $\sigma$ close enough to the interval $[\alpha, \beta]$ we can guarantee that the lower and upper diagonal matrices in \eqref{eq:Jdelta} (with $g \equiv 0$) are of the form $I + \cO(N^{-\varepsilon_\nu})$, implying the stronger estimate $\Vert m^{(2)}(z) - (1, \, 1) \Vert_\infty = \cO(N^{-\varepsilon_\nu})$ away from the jump contour (notice that in this regime $q(z) \equiv (1, \, 1)$).
\end{remark}

\subsection{$L^1$ and almost sure convergence}\label{sec:L1AS}

In this subsection, we conclude the proof of Theorem \ref{thm:main_prob}. First, we recall that from the general theory of the infinite Toda lattice \cite[Theorem 12.6]{Teschl1999}, the following holds:
\begin{lemma}\label{lem:apriori}
For any realization of the spectral data  $(\boldsymbol{\zeta},\boldsymbol{\gamma})\in (\alpha,\beta)^N\times\R_+^N$, and for all spatial indices $n \in \mathbb{Z}$ and times $t \in \R$:
\begin{equation}
    \norm{b_n(t)}_\infty + \norm{a_n(t)}_\infty \le 2\wt \alpha = \alpha + \alpha^{-1}\,.
\end{equation}
\end{lemma}
Next, we prove Lemma \ref{lem:measure_bad_set}, whose statement we report for the reader's convenience.

\begin{lemma}
    Under Assumptions \ref{ass:condensate_assumptions}, for any $\nu > 1$ and $\varepsilon_\nu \in \left(0, \frac{\kappa_\nu}{\nu}\right)$, where $\kappa_\nu = \min\left(\nu-1, \frac{\nu}{2}\right)$, there exists a positive constant $c_{\textrm{bad}}$, depending only on $\sigma,\nu,\mu_\nu$, such that
    \begin{equation}
        \mathbb{P}(\fW_N^c) = |\fW_N^c| \leq c_{\textrm{bad}}N^{-\kappa_\nu+\nu\varepsilon_\nu} = \begin{cases}
            c_{\textrm{bad}} N^{1-\nu+\nu\varepsilon_\nu}\,, & \text{if } 1 < \nu \leq 2\,, \\[4pt]
            c_{\textrm{bad}} N^{-\frac{\nu}{2}+\nu\varepsilon_\nu}\,, & \text{if } \nu > 2\,.
        \end{cases}
    \end{equation}
\end{lemma}
\begin{proof}
Define the following quantities:
\begin{equation}
    f_z(s) = \frac{s}{s-z}\,, \qquad S_N(z) = \frac{1}{N}\sum_{j=1}^N \gamma_j f_z(\zeta_j)\,, \qquad S(z)= \E\left[f_z(\zeta_1)\gamma_1\right] =\mu_1\int_{\alpha}^\beta \frac{s\rho(s)}{s-z}\di s\,.
\end{equation}
We notice that since $z\in \sigma$ is separated from $(\alpha,\beta)$, there exist two constants $K,L>0$ depending only on $\sigma$ such that
\begin{equation}
\label{eq:upper_bound_prob}
    \sup_{x\in(\alpha,\beta),\,z\in\sigma} |f_z(x)|\leq K \,, \qquad \sup_{x\in(\alpha,\beta),\,z\in\sigma} |f_z'(x)| \leq L\,.
\end{equation}
Furthermore, defining the function
\begin{equation}
    D_N(x) = \frac{1}{N}\sum_{j=1}^N \left(\gamma_j \mathbf{1}_{\zeta_j<x} - \E\left[ \gamma_1 \mathbf{1}_{\zeta_1<x}\right] \right) = \frac{1}{N}\sum_{j=1}^N \gamma_j \mathbf{1}_{\zeta_j<x} - \mu_1 \int_\alpha^x \rho(s)\di s\,,
\end{equation}
and using integration by parts, we can rewrite 
\begin{equation}
    S_N(z) - S(z) = f_z(\beta)D_N(\beta) - \int_\alpha^\beta f_z'(x)D_N(x)\di x\,.
\end{equation}
Using \eqref{eq:upper_bound_prob}, we deduce that
\begin{equation}
    \sup_{z\in\sigma}\left\vert  S_N(z) - S(z)\right\vert \leq K|D_N(\beta)| + L\int_\alpha^\beta|D_N(x)|\di x\,.
\end{equation}
We notice that the right-hand side is independent of $z\in\sigma$. Therefore, we obtain
\begin{equation}
\begin{split}
    \mathbb{P}\left(  \sup_{z\in\sigma}\left\vert  S_N(z) - S(z)\right\vert > N^{-\varepsilon_\nu} \right)& \leq \mathbb{P}\left(K|D_N(\beta)| + L\int_\alpha^\beta|D_N(x)|\di x > N^{-\varepsilon_\nu}\right) \\ &\leq  \mathbb{P}\left(\left(K|D_N(\beta)| + L\int_\alpha^\beta|D_N(x)|\di x \right)^\nu> N^{-\nu\varepsilon_\nu}\right) \,.
\end{split}
\end{equation}
By Jensen's inequality, there exists a constant $C_\nu>0$ depending only on $\nu,\alpha,\beta$ such that
\begin{equation}
    \mathbb{P}\left(\left(K|D_N(\beta)| + L\int_\alpha^\beta|D_N(x)|\di x \right)^\nu> N^{-\nu\varepsilon_\nu}\right) \leq \mathbb{P}\left(C_\nu\left(K^\nu|D_N(\beta)|^\nu + L^\nu\int_\alpha^\beta|D_N(x)|^\nu\di x\right)> N^{-\nu\varepsilon_\nu}\right)\,.
\end{equation}
Applying Markov's inequality and Fubini's theorem, we find that there exists a constant $\wt C_\nu>0$, depending on $L,K$, and therefore on $\sigma$, such that 
\begin{equation}
\label{eq:almost_there}
    \mathbb{P}\left(\sup_{z\in\sigma}\left\vert  S_N(z) - S(z)\right\vert > N^{-\varepsilon_\nu} \right) \leq \wt C_\nu N^{\nu \varepsilon_\nu} \sup_{x\in[\alpha,\beta]}\mathbb{E} \left[|D_N(x)|^\nu\right]\,.
\end{equation}
To bound $\mathbb{E}[|D_N(x)|^\nu]$, we set $Y_j(x) = \gamma_j \mathbf{1}_{\zeta_j < x}$ and $X_j(x) = Y_j(x) - \mathbb{E}[Y_j(x)]$. Note that $\{X_j(x)\}_{j=1}^N$ are i.i.d.\ centered random variables satisfying, by Jensen's inequality,
\begin{equation}
\label{eq:Xj_moment_bound}
    \mathbb{E}\left[|X_j(x)|^\nu\right] \leq 2^{\nu-1}\left(\mathbb{E}[|Y_j(x)|^\nu] + |\mathbb{E}[Y_j(x)]|^\nu\right) \leq 2^\nu \mathbb{E}[\gamma_1^\nu] = 2^\nu \mu_\nu\,.
\end{equation}
We now distinguish two cases according to the value of $\nu > 1$:

\medskip
\noindent\textbf{Case 1: $1 < \nu \leq 2$.} 
Applying the von Bahr--Esseen inequality \cite[Theorem 2]{vonBahr1965} to the sum of independent centered random variables $\sum_{j=1}^N X_j(x)$ and using \eqref{eq:Xj_moment_bound}, we obtain
\begin{equation}
\label{eq:bound_DN_case1}
    \mathbb{E}[|D_N(x)|^\nu] = \frac{1}{N^\nu}\mathbb{E}\left[\left|\sum_{j=1}^N X_j(x)\right|^\nu\right] \leq \frac{2}{N^\nu} \sum_{j=1}^N\mathbb{E}\left[|X_j(x)|^\nu\right] \leq 2^{\nu+1}\mu_\nu N^{1-\nu}\,.
\end{equation}
\noindent\textbf{Case 2: $\nu > 2$.}
When $\nu > 2$, we apply the Dharmadhikari--Fabian--Jogdeo inequality \cite{DharmadhikariFabianJogdeo1968}, obtaining that there exists a constant $B_\nu > 0$ depending only on $\nu$ such that
\begin{equation}
\label{eq:bound_DN_case2}
    \mathbb{E}[|D_N(x)|^\nu] = \frac{1}{N^\nu}\mathbb{E}\left[\left|\sum_{j=1}^N X_j(x)\right|^\nu\right] \leq B_\nu N^{-\frac{\nu}{2}}\left(\frac{1}{N}\sum_{j=1}^N \mathbb{E}\left[|X_j(x)|^\nu\right]\right) \stackrel{ \eqref{eq:Xj_moment_bound}}{\leq} 2^\nu B_\nu \mu_\nu N^{-\frac{\nu}{2}}\,.
\end{equation}
Combining \eqref{eq:bound_DN_case1} and \eqref{eq:bound_DN_case2}, we conclude that for all $\nu > 1$,
\begin{equation}
    \sup_{x\in[\alpha,\beta]} \mathbb{E}\left[|D_N(x)|^\nu\right] \leq C_{\nu,\mu_\nu} N^{-\kappa_\nu}\,, \qquad \kappa_\nu = \min\left(\nu-1, \frac{\nu}{2}\right)\,.
\end{equation}
Substituting this estimate into \eqref{eq:almost_there} yields the claimed bound.
\end{proof}

\begin{remark}
    We note that our results can be interpreted as establishing the rate of convergence of the weighted empirical measure $\frac{1}{N}\sum_{j=1}^N \gamma_j \delta_{\zeta_j}$ to its expectation with respect to the Dudley (or bounded-Lipschitz) metric, see \cite[Chapter 11]{Dudley2002}.
\end{remark}

We can now complete the proofs of Theorem \ref{thm:poly_continuum_approx} and Theorem \ref{thm:main_prob}.
\begin{proof}[Proof of Theorem \ref{thm:poly_continuum_approx}]
We start by establishing the uniform bound on the good set $\fW_N$.
Fix $\delta \in (0, \varepsilon_\nu)$. Theorem \ref{thm:m_2_q_approx} guarantees that conditioned on the good set $\fW_N$, we have for $z$ in an open neighborhood of the origin,
\begin{equation}
	m^{(2)}_1(z) =   q_1(z)(1 + \mathcal{O}(N^{-\delta})).
    \end{equation}
For the Flaschka variable $a_n(t)$, using Lemma \ref{lem:recovery} and Lemma \ref{lem:recovery_q}, we have:
\begin{equation}\label{eq:an_aqn}
    a_n(t) = \frac{1}{2}\frac{m_1^{(2)}(0, n, t)}{m_1^{(2)}(0, n+1, t)} = \frac{1}{2}\frac{q_1(0, n, t)}{q_1(0, n+1, t)}\big(1 + \mathcal{O}(N^{-\delta})\big) = a_n^{(q)}(t)\big(1 + \mathcal{O}(N^{-\delta})\big)\,.
\end{equation}
For the velocity variable $b_n(t) = B(n,t) - B(n-1,t)$, the recovery formulas require evaluating the logarithmic derivative at the origin:
\begin{equation}
    -2B(n-1,t) = \left. \frac{\mathrm{d}}{\mathrm{d}z}\ln m_1^{(2)}(z; n, t) \right|_{z=0}, \qquad -2B^{(q)}(n-1,t) = \left. \frac{\mathrm{d}}{\mathrm{d}z}\ln q_1(z; n, t) \right|_{z=0}.
\end{equation} 
Due to holomorphicity of the error, using Cauchy's Theorem, we have $\frac{d}{dz}\ln(1+\mathcal{O}(N^{-\delta})) = \frac{d}{dz}\mathcal{O}(N^{-\delta}) = \mathcal{O}(N^{-\delta})$, giving us (again on the good set $\fW_N$)
\begin{align}\label{eq:bn_bqn}
    b_n(t) = b_n^{(q)}(t) + \mathcal{O}(N^{-\delta}).
\end{align}
As for both asymptotic formulas \eqref{eq:an_aqn}, \eqref{eq:bn_bqn} the convergence is uniform on the good set $\fW_N$, this establishes \eqref{eq:good_set_poly_bound}.

\medskip
\noindent\textbf{$L^1(\mathbb{P})$ Convergence.}
Splitting the expected value into integrals over $\fW_N$ and $\fW_N^c$, and applying Lemma \ref{lem:measure_bad_set} and Lemma \ref{lem:apriori}, we obtain that there exist two constants $C_0,C_a>0$:
\begin{align}
    \mathbb{E}\left[ |a_n(t) - a_n^{(q)}(t)| \right] &= \int_{\fW_N} |a_n(t) - a_n^{(q)}(t)| \di\mathbb{P} + \int_{\fW_N^c} |a_n(t) - a_n^{(q)}(t)| \di\mathbb{P} \nonumber\\
    &\leq C_0 N^{-\delta} + C_{a} \mathbb{P}(\fW_N^c) \nonumber\\
    &\leq C_0 N^{-\delta} + C_{a} c_{\textrm{bad}} N^{-\kappa_\nu + \nu\varepsilon_\nu} \leq C N^{-\delta_1}\,,
\end{align}
where $\delta_1 = \min(\delta, \kappa_\nu - \nu\varepsilon_\nu) > 0$.
Identically for $b_n(t)$:
\begin{equation}
    \mathbb{E}\left[ |b_n(t) - b_n^{(q)}(t)| \right] \leq C_0 N^{-\delta} + C_{a} c_{\textrm{bad}} N^{-\kappa_\nu + \nu\varepsilon_\nu} \leq C N^{-\delta_1}\,,
\end{equation}
which establishes the convergence in expectation.

\medskip
\noindent\textbf{Almost Sure Convergence.}
When $\nu > 2$, $\kappa_\nu = \frac{\nu}{2} > 1$. Choosing $\varepsilon_\nu \in \left(0, \frac{1}{2} - \frac{1}{\nu}\right)$ ensures that $-\frac{\nu}{2} + \nu\varepsilon_\nu < -1$. By Lemma \ref{lem:measure_bad_set}:
\begin{equation}
    \sum_{N=1}^\infty \mathbb{P}(\fW_N^c) \leq c_{\textrm{bad}} \sum_{N=1}^\infty N^{-\frac{\nu}{2} + \nu\varepsilon_\nu} < +\infty\,.
\end{equation}
The first Borel--Cantelli lemma implies that $\mathbb{P}(\limsup_{N\to\infty} \fW_N^c) = 0$. Hence, for $\mathbb{P}$-almost every realization $\omega$ of $\{(\zeta_j, \gamma_j)\}_{j=1}^\infty$, there exists a finite random index $N_0(\omega)$ such that $(\boldsymbol{\zeta}, \boldsymbol{\gamma}) \in \fW_N$ for all $N \geq N_0(\omega)$. Combining this with the uniform bound \eqref{eq:good_set_poly_bound} on $\fW_N$ establishes \eqref{eq:almost_sure_poly}.
\end{proof}

\begin{proof}[Proof of Theorem \ref{thm:main_prob}]
By the deterministic Riemann--Hilbert asymptotic analysis established across Sections \ref{sec:small_norm}--\ref{sec:one_cut_modulated_regions} (Corollary \ref{cor:small_norm_sol}, Theorem \ref{thm:complete_one_cut_asymptotics}, Theorem \ref{thm:modulated_two_cuts_asymptotics}, Theorem \ref{thm:left_one_cut_asymp}, and Theorem \ref{thm:right_one_cut_asymp}), for any $(\eta, \tau)$ in a compact subset of $\mathbb{R}^2 \setminus \Sigma_{\mathrm{phase}}$, there exists a deterministic constant $C_{\mathrm{RHP}}>0$ such that:
\begin{equation}
\label{eq:continuum_to_theta_error}
    |a_n^{(q)}(t) - \fa(\eta, \tau)| \leq \frac{C_{\mathrm{RHP}}}{\ln N}\,, \qquad |b_n^{(q)}(t) - \fb(\eta, \tau)| \leq \frac{C_{\mathrm{RHP}}}{\ln N}\,.
\end{equation}
Applying the triangle inequality:
\begin{equation}
    |a_n(t) - \fa(\eta,\tau)| \leq |a_n(t) - a_n^{(q)}(t)| + |a_n^{(q)}(t) - \fa(\eta,\tau)|\,,
\end{equation}
and similarly for $b_n(t)$.
Conditioned on $\fW_N$, using Theorem \ref{thm:poly_continuum_approx} and \eqref{eq:continuum_to_theta_error}:
\begin{equation}
    |a_n(t) - \fa(\eta,\tau)| \leq C_0 N^{-\delta} + \frac{C_{\mathrm{RHP}}}{\ln N} \leq \frac{C}{\ln N}\,, \qquad |b_n(t) - \fb(\eta,\tau)| \leq C_0 N^{-\delta} + \frac{C_{\mathrm{RHP}}}{\ln N} \leq \frac{C}{\ln N}\,.
\end{equation}
Integrating over $\fW_N$ and bounding the integral over $\fW_N^c$ via Lemma \ref{lem:measure_bad_set} and the a priori bounds yields
\begin{equation}
    \mathbb{E}\left[ |a_n(t) - \fa(\eta,\tau)| \right] \leq \frac{C}{\ln N} + 2C_{a} c_{\textrm{bad}} N^{-\kappa_\nu + \nu\varepsilon_\nu} \xrightarrow[N\to\infty]{} 0\,,
\end{equation}
and identically for $b_n(t)$.
Finally, for $\nu > 2$, since $(\boldsymbol{\zeta}, \boldsymbol{\gamma}) \in \fW_N$ almost surely for sufficiently large $N$, the almost sure convergence \eqref{eq:almost_sure_conv} with rate $\cO(1/\ln N)$ follows immediately.
\end{proof}

\section{Nonlinear dispersion relations and ensemble averages}
\label{sec:NDR}
In this final section, we explain how our asymptotic results connect to the broader physical theory of Soliton Gases and Generalized Hydrodynamics (GHD). In particular, we show that our Riemann--Hilbert $g$-function construction naturally yields the \textit{Nonlinear Dispersion Relations} (NDR) \cite{El2003,ElTovbis2020,Congy2023} for a Toda soliton condensate.
We start by showing that in each of the asymptotic regimes ($\cU_c, \cU_\ell, \cU_r, \cU_2$), the $g$-function constructed in Sections \ref{sec:complete_one_cut}--\ref{sec:one_cut_modulated_regions} admits the following representation 
\begin{equation}
\label{eq:alternative_g}
    g(z,\eta,\tau)=\int_\Sigma\ln\left(\frac{z-\lambda}{\lambda z- 1 }\right)\varphi(\lambda,\eta,\tau)\di \lambda\,,
\end{equation}
where the support $\Sigma = \Sigma(\eta,\tau) \subset (0,1)$ is given by $\Sigma=(\alpha,\beta)$ for $(\eta,\tau)\in \cU_c$, $\Sigma=(\alpha,b(\eta,\tau))$ for $(\eta,\tau)\in \cU_\ell$, $\Sigma=(a(\eta,\tau),\beta)$ for $(\eta,\tau)\in \cU_r$, and $\Sigma=(\alpha,b(\eta,\tau))\cup(a(\eta,\tau),\beta)$ for $(\eta,\tau)\in \cU_2$. 
In fact, the function $\partial_z g(z, \eta, \tau)$ is holomorphic in $\mathbb C \setminus (\Sigma \cup \Sigma^{-1})$ and satisfies $\lim_{z\to \infty} \partial_z g(z, \eta, \tau) = 0$. The Sokhotski--Plemelj formula tells us that $\partial_z g(z, \eta, \tau)$ must have the representation 
\begin{align}\label{eq:repGpartial}
    \partial_z g(z, \eta, \tau) = \int_{\Sigma \cup \Sigma^{-1}} \frac{(-\varphi(\lambda, \eta, \tau))}{\lambda-z} d\lambda,
\end{align}
where $\varphi(\lambda, \eta, \tau) = -\frac{1}{2\pi i}\left(\partial_\lambda g_+(\lambda, \eta, \tau) - \partial_\lambda g_-(\lambda, \eta, \tau)\right) = -\frac{1}{\pi i}\Omega_+(\lambda, \eta, \tau)$. Due the symmetry $g(z^{-1}) = - g(z)$ we have that $g'_\pm(z^{-1}) = z^2g'_\mp(z)$, and therefore $\varphi(\lambda^{-1}, \eta, \tau) = -\lambda^2 \varphi(\lambda, \eta, \tau)$. It follows that \eqref{eq:repGpartial} can be written as
\begin{align*}
  \partial_z g(z, \eta, \tau) &= \int_\Sigma \left(\frac{1}{z-\lambda} - \frac{1}{z-\lambda^{-1}}\right) \varphi(\lambda, \eta, \tau) d\lambda. 
\end{align*}
After integrating with respect to $z$ we obtain the integral representation \eqref{eq:alternative_g} up to an additive constant. That this constant is $0$ follows from the fact that the integral by itself already satisfy the required $z \mapsto z^{-1}$ symmetry, and the addition of a nonzero constant would break this symmetry.

The characterizing feature of the $g$-function across all regimes is the jump condition:
\begin{equation}
    g_+(z,\eta,\tau) + g_-(z,\eta,\tau)=\begin{cases}
        -1-\tau\Phi(z,\eta,\tau)\,,  & z\in\Sigma\,, \\
        +1-\tau\Phi(z,\eta,\tau)\,,  & z\in\Sigma^{-1}\,,
    \end{cases}
\end{equation}
where $\Sigma^{-1} = \{\lambda^{-1} : \lambda \in \Sigma\}$ denotes the reciprocal support (e.g., $\Sigma^{-1}=(\beta^{-1},\alpha^{-1})$ when $\Sigma=(\alpha,\beta)$), with both contours oriented from left to right. Using the representation \eqref{eq:alternative_g}, this jump relation can be written as the integral equation
\begin{equation}
\label{eq:g_jump_integral}
    2\int_{\Sigma}\ln\left(\left\vert\frac{z-\lambda}{1-\lambda z}\right\vert\right)\varphi(\lambda,\eta,\tau)\di\lambda=\begin{cases}
        -1-\tau\Phi(z,\eta,\tau)\,,  & z\in\Sigma\,, \\
        +1-\tau\Phi(z,\eta,\tau)\,,  & z\in\Sigma^{-1}\,.
    \end{cases}
\end{equation}
Recall that we have the exact representation 
\begin{align*}
    \varphi(\lambda, \eta, \tau) = -\frac{1}{\pi i} \Omega_+(\lambda, \eta, \tau),
\end{align*}
where $\Omega(\lambda)$ changes among the various regimes (e.g.~$\Omega_2(\lambda), \Omega_\ell(\lambda)$ etc.), and is a smooth function of the parameters $(\eta,\tau)$ for $z\in\Sigma$, except at the transition boundary $\Sigma_{\textrm{phase}}$. Due to Lemma \ref{lem:whole_regime}, \ref{lem:off_decay_two_cuts_lens}, \ref{lem:off_decay_left} and \ref{lem:off_decay_right} we have that $\varphi(\lambda, \eta, \tau) > 0$ on $\Sigma$, so it can be viewed as a positive density. Moreover, from the explicit representations of $\Omega$ it follows that
\begin{align*}
    \varphi(\lambda, \eta, \tau) = \begin{cases}
        \cO(|\lambda - \gamma|^{-1/2}), \qquad &\gamma \in \lbrace \alpha, \beta\rbrace \quad (\text{if any is an edge of } \, \Sigma),
        \\
        \cO(|\lambda - c|^{1/2}), \qquad &c \in \lbrace a(\eta, \tau), b(\eta, \tau)\rbrace \quad (\text{if any is an edge of } \, \Sigma),
    \end{cases}
\end{align*}
agreeing with the typical behavior of equilibrium measures near a hard and soft edge.
We now differentiate \eqref{eq:g_jump_integral} for $z \in \Sigma$ with respect to the variables $\eta$ and $\tau$. Observe that no boundary terms arise from the endpoints of $\Sigma(\eta,\tau)$: the hard edges $\alpha$ and $\beta$ are independent of $(\eta,\tau)$, while at the moving soft edges $a(\eta,\tau)$ and $b(\eta,\tau)$ the density vanishes, $\varphi(a(\eta,\tau),\eta,\tau) = \varphi(b(\eta,\tau),\eta,\tau) = 0$, by construction. Recalling that $\tau\Phi(z,\eta,\tau) = \tau(z-z^{-1}) + 2\eta\ln(z)$, we deduce that for $z \in \Sigma$,
\begin{equation}
\label{eq:NDR_toda}
    \int_{\Sigma}\ln\left(\left\vert\frac{z-\lambda}{1-\lambda z}\right\vert\right)(-\partial_\eta \varphi(\lambda,\eta,\tau))\di\lambda = \ln(z)\,,\qquad \int_{\Sigma}\ln\left(\left\vert\frac{z-\lambda}{1-\lambda z}\right\vert\right)\partial_\tau \varphi(\lambda,\eta,\tau)\di\lambda= \frac{z^{-1}-z}{2}\,.
\end{equation}
Equations \eqref{eq:NDR_toda} are the Toda lattice analogue of the NDR for the KdV soliton gas \cite{El2003,Congy2023}:
\begin{align}
    &\int_\Gamma\ln\left(\left\vert\frac{z+\lambda}{z-\lambda}\right\vert\right)u(\lambda,x,t)\di\lambda +\sigma(z)u(z,x,t) = z\,, \\
    &\int_{\Gamma}\ln\left(\left\vert\frac{z+\lambda}{z-\lambda}\right\vert\right)v(\lambda,x,t)\di\lambda +\sigma(z)v(z,x,t)= 4z^3\,,
\end{align}
in the condensate limit $\sigma(z)\equiv 0$. For this reason, we call \eqref{eq:NDR_toda} the \textbf{NDR for the Toda soliton condensate}. Introducing  the \textit{density of states} (DOS) $u(\lambda,\eta,\tau) = -\partial_\eta\varphi(\lambda,\eta,\tau)$ and the \textit{flux density} $v(\lambda,\eta,\tau) = \partial_\tau\varphi(\lambda,\eta,\tau)$ one can rewrite them as

\begin{equation}
        \int_{\Sigma}\ln\left(\left\vert\frac{z-\lambda}{1-\lambda z}\right\vert\right)u(\lambda,\eta,\tau)\di\lambda = \ln(z)\,,\qquad \int_{\Sigma}\ln\left(\left\vert\frac{z-\lambda}{1-\lambda z}\right\vert\right)v (\lambda,\eta,\tau)\di\lambda= \frac{z^{-1}-z}{2}\,.
\end{equation}
Note that analogous equations have already appeared in the literature for the Toda chain in thermal equilibrium \cite{Spohn2019,Spohn2020}.
 For a general Toda soliton gas, they naturally generalize to 
\begin{align}
    &\int_{\Sigma}\ln\left(\left\vert\frac{z-\lambda}{1-\lambda z}\right\vert\right)u(\lambda,\eta,\tau)\di\lambda + \sigma(z)u(z,\eta,\tau) = \ln(z)\,,\\
    &\int_{\Sigma}\ln\left(\left\vert\frac{z-\lambda}{1-\lambda z}\right\vert\right)v(\lambda,\eta,\tau)\di\lambda + \sigma(z)v(z,\eta,\tau) = \frac{z^{-1}-z}{2}\,.
\end{align}
Introducing the effective soliton velocities $v^{\textrm{eff}}(\lambda, \eta, \tau) = \frac{v(\lambda, \eta, \tau)}{u(\lambda, \eta, \tau)}$ we can now verify the soliton gas kinetic equations \cite[Eq.~(3),(4)]{ElKamchatnov2005}
\begin{align}\label{eq:KineticEq}
\begin{split}
    &\partial_\tau u + \partial_\eta (v^{\textrm{eff}}u) = 0 \\
    &v^{\textrm{eff}}(z, \eta, \tau) = \frac{z^{-1}-z}{2\ln(z)} + \frac{1}{\ln(z)}\int_\Sigma \ln\left(\left\vert\frac{z-\lambda}{1-\lambda z}\right\vert\right) u(\lambda, \eta, \tau)\left[v^{\textrm{eff}}(z, \eta, \tau)-v^{\textrm{eff}}(\lambda, \eta, \tau)\right] \di \lambda,
\end{split}
\end{align}
where $(z^{-1} - z)/(2\ln(z))$ is the velocity of a free soliton, see \eqref{eq:bare_velocity}.

A fundamental property of the NDR equations is that their solutions directly determine the macroscopic ensemble averages of the wave field. For instance, in the KdV soliton gas \cite{El2003,Congy2023}, the ensemble average of the wave field $\psi(x,t)$ is given by the spectral moment, see \cite[eq. (2.14)]{Congy2023}
\begin{equation}
\label{eq:ensemble_avg_kdv}
    \langle \psi(x,t)\rangle = 4\int_\Gamma \lambda u(\lambda,x,t)\di \lambda\,.
\end{equation}
In our setting, the macroscopic background fields of the Toda lattice across the various regimes are precisely $a_{\textrm{bg}}(\eta,\tau)$ and $b_{\textrm{bg}}(\eta,\tau)$. Using \eqref{eq:alternative_g}, we find that as $z \to \infty$ and $z \to 0$,
\begin{equation}
    g_\infty(\eta,\tau) = -\int_\Sigma \ln(\lambda)\,\varphi(\lambda,\eta,\tau)\di\lambda\,, \qquad g'(0,\eta,\tau) = \int_\Sigma (\lambda - \lambda^{-1})\varphi(\lambda,\eta,\tau)\di\lambda\,.
\end{equation}
Consequently, as a corollary of our asymptotic and probabilistic theorems (see e.g.~\eqref{eq:thm_abg_explicit}--\eqref{eq:thm_bbg_explicit} and \eqref{eq:thm_abg_explicit_two_cuts}--\eqref{eq:thm_bbg_explicit_two_cuts}), the backgrounds $a_{\textrm{bg}}(\eta,\tau),\,b_{\textrm{bg}}(\eta,\tau)$ can be expressed in terms of the DOS as
\begin{align}
    a_{\textrm{bg}}(\eta,\tau) &= \frac{1}{2}\exp\left(-\partial_\eta g_\infty(\eta,\tau)\right) = \frac{1}{2}\exp\left(\int_\Sigma \ln(\lambda)\,\partial_\eta \varphi(\lambda,\eta,\tau)\di \lambda\right) = \frac{1}{2}\exp\left(-\int_\Sigma \ln(\lambda)\,u(\lambda,\eta,\tau)\di\lambda\right)\,, \label{eq:a_avg}\\
    b_{\textrm{bg}}(\eta,\tau) &= \frac{1}{2}\partial_\eta g'(0,\eta,\tau) = \frac{1}{2}\int_\Sigma (\lambda - \lambda^{-1}) \partial_\eta \varphi(\lambda,\eta,\tau)\di \lambda = \int_\Sigma \frac{\lambda^{-1}-\lambda}{2}\,u(\lambda,\eta,\tau)\di\lambda\,. \label{eq:b_avg}
\end{align}
Notice that \eqref{eq:a_avg}--\eqref{eq:b_avg} are the exact counterparts of \eqref{eq:ensemble_avg_kdv}. 
We conclude by emphasizing two key features of our work in view of this hydrodynamic interpretation:
\begin{enumerate}[leftmargin=*]
    \item  Usually, the NDR and the ensemble averages \eqref{eq:ensemble_avg_kdv} are derived via heuristic phase-shift arguments or formal thermodynamic limits of finite-gap potentials; Theorem \ref{thm:main_prob} establishes that \eqref{eq:a_avg}--\eqref{eq:b_avg} emerge almost surely as the self-averaged background of a truly random $N$-soliton ensemble with independent eigenvalues and random norming constants.
    \item In standard soliton condensates \cite{bertola2026arbitrarygenusdarksolitongases,GGJM} (see also \cite{Egorova2023, EPT} for the case of shock resp.~steplike initial data), the jump condition for $g$ is homogeneous in $(x,t)$, leading to self-similar evolution along straight rays $x/t = \text{const}$.  By contrast, in our scaling regime $n = \eta\ln N$, $t = \tau\ln N$, the inhomogeneous $\pm 1$ term in \eqref{eq:g_jump_integral} breaks this scaling invariance. As a result, the ensemble averages $a_{\textrm{bg}}(\eta,\tau)$ and $b_{\textrm{bg}}(\eta,\tau)$ modulate  in both space and time across nonlinear phase boundaries, while our $g$-function provides the solution to the Toda NDR as the condensate continuously interpolates between genus $0$, $1$, and $2$.
\end{enumerate}
\appendix

\section{Small norm theory and a vanishing lemma}\label{Appendix}
\subsection{Small norm theory}
\label{App:SmallNorm}
We now outline how to obtain a small norm Riemann--Hilbert problem in case of the unusual normalization in the Toda case. Let us denote the vector solution with $m$ (not necessarily related to any $m$ appearing in other sections). We assume some jump condition $m_+ = m_- J$ on the jump contour and the normalization
\begin{align*}
    m(0) = (1, *), \quad m(\infty) = (*, 1).
\end{align*}
Following \cite{Krger2008} we now need the modified Cauchy operator 
\begin{align}\label{eq:CauchyOPerator}
    (Cf)(z) = \frac{1}{2\pi i} \int_\Sigma f(s) \begin{pmatrix}
        \frac{1}{s-z} - \frac{1}{s} & 0
        \\
        0 & \frac{1}{s-z}
    \end{pmatrix} \di s.
\end{align}
Denote by $(C_{\pm}f)(s) = \lim_{z \to s\pm} (C f)(z)$, which defines a bounded linear operator from $L^2$ to itself (see e.g.~\cite{BC84}). One can check that $C_+ - C_- = \text{id}$ and $(Cf)(0) = (0, *)$, $(Cf)(\infty) = (*, 0)$.

Define $w = J - I$. Then we have $m_+(s) - m_-(s) = m_-(s)w(s)$ from which 
\begin{align*}
    m(z) = (1, 1) + C(m_-w)(z)
\end{align*}
follows. In particular, if we let $\mu = m_-$, then we have $(1-C_w)\mu = (1,1)$, where $C_w(f) = C_-(fw)$. On the other hand, if we have a solution $\mu$ of this equation, then $m(z) = (1,1) + C(\mu w)$ solves the RHP:
\begin{align*}
    m_+ &= (1, 1) + C_+(\mu w)
    \\
    &= (1, 1) + C_-(\mu w) + \mu w 
    \\
    & = (1, 1) + C_-(\mu w) + [(1, 1) + C_-(\mu w)] w
    \\
    &= [(1, 1) + C_-(\mu w)] J
    \\
    &= m_- J.
\end{align*}
Now to obtain a small-norm problem we need an outer parametrix matrix solution $M$, such that the jumps of $r = m M^{-1}$ are small in the $L^\infty$-norm. Furthermore, to write down a Neumann series expansion, we need to express this solution in terms of an equivalent singular integral equation, similar to above. To this end we need the normalization at $z = 0, \infty$ to be compatible with the modified Cauchy operator $C$. Thus we require
\begin{align*}
    r(0) = m(0)M^{-1}(0) = (1, *), \qquad r(\infty) = m(\infty) M^{-1}(\infty) = (*, 1). 
\end{align*}
Given the normalization of $m$, this implies that 
\begin{align*}
    M(0) = \begin{pmatrix}
        1 & *
        \\
        0 & *
    \end{pmatrix}, \qquad M(\infty) = \begin{pmatrix}
        * & 0
        \\
        * & 1
    \end{pmatrix}.
\end{align*}
Moreover, one can check that provided $M$ satisfies the symmetry condition $M(z^{-1}) = \sigma_1 M(z) \sigma_1$, then we have the equivalence
\begin{align*}
    r(z^{-1}) = r(z) \sigma_1 \qquad \Longleftrightarrow \qquad m(z^{-1}) = m(z) \sigma_1.
\end{align*}
In our applications, $M$ will have the form (away from the local parametrices)
\begin{align}\label{eq:MAnsatz}
    M(z) = \frac{1}{2} \begin{pmatrix}
        k_s(z) + \ell_a(z)
        \\
        k_s(z) - \ell_a(z)
    \end{pmatrix},
\end{align}
where $k_s(z^{-1}) = k_s(z)\sigma_1$, $\ell_a(z^{-1}) = -\ell_a(z)\sigma_1$ and $k_s(0) = \ell_a(0) = (1, \, *)$. It is easy to check that these conditions will guarantee the required symmetry and normalization of $M$. Additionally, the anti-symmetric vector $\ell_a(z)$ might have a pole at $z=1$. However, $r(z)$ will nonetheless be holomorphic around $z = 1$. In fact, due to the symmetry we have that $m(1) = (c, \, c)$, while in view of \eqref{eq:MAnsatz} there exist two constants $d^{(M)}_\pm$ such that

\begin{equation}
\label{eq:Minv}
    M_\pm(z) = \frac{d^{(M)}_\pm}{z-1} \pmtwo{1}{1}{-1}{-1} + \cO(1) \,, 
\end{equation}
implying that $r(z) = m(z)M^{-1}(z)$ remains bounded (and hence holomorphic) near $z = 1$.

Let us now assume that $r$ satisfies the vector Riemann--Hilbert problem $r_+(z) = r_-(z)J_r(z)$ with the normalization $r(0) = (1, \, *)$. If $\Vert w_r \Vert_\infty = \Vert J_r - I \Vert_\infty = O(\varepsilon)$ is sufficiently small, then it follows that $\Vert C_{w_r} \Vert_{L^2 \to L^2} < 1$, implying that $I-C_{w_r}$ is invertible and $(I - C_{w_r})^{-1} = I + O(\varepsilon)$. In particular $r_- = (I - C_{w_r})^{-1}(1, \, 1) = (1, \, 1) + O_{L^2}(\varepsilon)$, where the error is $L^2$-valued. From the representation 
\begin{align*}
r(z) = (1, \, 1) + \frac{1}{2\pi i} \int_{\Sigma_r} r_-(s)w_r(s)  \begin{pmatrix}
        \frac{1}{s-z} - \frac{1}{s} & 0
        \\
        0 & \frac{1}{s-z}
    \end{pmatrix} \di s
    \end{align*}
    it now follows that $r(z) = (1, \, 1) + \cO(\varepsilon)$ pointwise and locally uniformly away from the jump contour of $r$. Assuming the form of the global parametrix $M$ in \eqref{eq:MAnsatz}, it follows that 
\begin{align*}
    m(z) = r(z)M(z) = k_s(z) + \cO(\varepsilon),
\end{align*}
again pointwise locally uniformly away from the jump contours.
\begin{remark}
    Notice that the outer parametrix $M(z)$ is not assumed to be unique, and indeed is not unique in our application due to the nonuniqueness of the \emph{meromorphic} antisymmetric vector solution $\ell_a(z)$. This is not an issue as the leading order asymptotics $(1, 1) M(z) = k_s(z)$ is independent of the choice of $\ell_a(z)$, while the \emph{holomorphic} symmetric solution $k_s(z)$ is unique.
\end{remark}

\begin{remark}
    An alternative to the small-norm theory approach can be found \cite{Piorkowski2023}, where uniform invertibility of $1- C_w$ restricted to an appropriate symmetric $L^2$-subspace is shown directly. 
\end{remark}

\subsection{Unique solvability of RHPs \ref{RHP:q} \& \ref{RHP:q_hat}}\label{App:Fred}
In this section we establish unique solvability of RHP \ref{RHP:q}, or equivalently RHP \ref{RHP:q_hat}; we  use both formulations at different stages of the argument. While most arguments are standard, there are a few differences related to the symmetry and normalization in case of the Toda lattice. We partially follow the arguments given in \cite[App.~B]{KRGER2009}. For the KdV case, which has a similar symmetry condition, compare with \cite[Thm.~2.1]{EPT}, \cite[Sect.~5]{Piorkowski2023}.
\begin{lemma}\label{lem:vanishingLem}
    Assume $\widehat q_0$ satisfies the conditions of RHP \ref{RHP:q_hat} (including the symmetry condition), but the normalization condition is replaced by $\widehat q_{0,1}(0)\cdot \widehat q_{0,2}(0) = 0$. Then $\widehat q_0(z) \equiv 0$. 
\end{lemma}
Using Lemma \ref{lem:vanishingLem} and the Fredholm alternative we obtain
\begin{proposition}\label{Prop:Fred}
    The RHP \ref{RHP:q} (equivalently RHP \ref{RHP:q_hat}) is uniquely solvable. 
\end{proposition}
\begin{proof}[Proof of Lemma \ref{lem:vanishingLem}]
Assume a solution $\widehat q_0(z)$ is given. We need to show that $\widehat q_0 \equiv 0$. The jump condition for $\widehat q_0(z)$ reads:
\begin{align*}
    \widehat q_{0,+}(z) &= \widehat q_{0,-}(z) \begin{pmatrix}
        1 & 0
        \\
        2\pi i h(z) & 1
    \end{pmatrix}, \quad z \in (\alpha, \beta), 
    \\
    \widehat q_{0, +}(z) &= \widehat q_{0, -}(z) \begin{pmatrix}
        1 & -2\pi i h(z^{-1})
        \\
       0 & 1
    \end{pmatrix}, \quad z \in (\beta^{-1}, \alpha^{-1})\, ,
\end{align*}
with $h(z) = N\mu_1 e^{t\Phi(z,\eta,\tau)}z\rho(z) > 0$ on $z \in (\alpha, \beta) \cup (\beta^{-1}, \alpha^{-1})$. Let us define the function 
\begin{align*}
    F_0(z) = \wh q_0(z) \wh q_0(\overline{z}^{-1})^\dag = \wh q_{0,1}(z)\overline{\wh q_{0,1}(\overline{z}^{-1}}) + \wh q_{0,2}(z)\overline{\wh q_{0,2}(\overline{z}^{-1}}).
\end{align*}
For $s \in S^1$ we have $\overline{s}^{-1} = s$ and therefore $F_0(s) = \Vert \wh q_0(s) \Vert^2_2 \geq 0$. Additionally, notice that due to $\wh q_0(z^{-1}) = \wh q_0(z) \sigma_1$ and $\wh q_{0,1}(0) \cdot \wh q_{0,2}(0) = 0$, we have that 
\begin{align*}
    F_0(0) = 2 \Re(\wh q_{0,1}(0) \overline{\wh q_{0,2}(0)}) = 0.
\end{align*}
It follows that $F_0(z)/z$ is holomorphic in $\mathbb{C} \setminus \lbrace [\alpha, \beta]\cup[\beta^{-1}, \alpha^{-1}]\rbrace$. In particular, by deforming contours and noticing that $\frac{ds}{is}$ is a positive measure on  $S^{1}$, we obtain
\begin{align}\nonumber
    0 \leq \int_{S^1} \Vert \wh q_0(s) \Vert_2^2 \frac{ds}{is} = \int_{S^1} F_0(s) \frac{ds}{is} &= \int_\alpha^\beta [F_{0,-}(z) - F_{0,+}(z)] \frac{dz}{iz}
    \\\label{eq:intqhat}
    &= \int_\alpha^\beta \wh q_{0,-}(z) \begin{pmatrix}
        0 & 0 
        \\
        0 & -4\pi i h(z)
    \end{pmatrix}\wh q_{0,-}(z)^\dag\frac{dz}{iz}
    \\\nonumber
    &= \int_\alpha^\beta |\wh q_{0,2-}(z)|^2 \left(-\frac{4\pi h(z)}{z}\right)dz\,  \leq 0.
\end{align}
It follows that $\wh q_0(z) \equiv 0$ for $z \in S^1$, and therefore $\wh q_0 \equiv 0$ everywhere. This finishes the proof.
\end{proof}
\begin{proof}[Proof of Proposition \ref{Prop:Fred}]
We now switch to the equivalent RHP \ref{RHP:q} for $q(z)$. Recall that $q(z)$ satisfies the jump condition 
\begin{align*}
    q_+(z) = q_-(z) \begin{cases}
        \begin{pmatrix}
  			1 & 0 \\
  			N\mu_1 e^{t\Phi(z,\eta,\tau)}\int_\alpha^\beta \frac{s\rho(s)}{z-s}\mathrm{d} s& 1
  		\end{pmatrix} \quad &z \in \sigma, \\\\[5pt]
  		\begin{pmatrix}
  			1 &-N\mu_1 ze^{-t\Phi(z,\eta,\tau)} \int_{\alpha}^{\beta} \frac{\rho(s)}{z-s^{-1}} \mathrm{d} s\\
  			0 & 1
  		\end{pmatrix} \quad &z^{-1}\in\sigma.
        \end{cases}
\end{align*}
As the jump matrices $J_q(z)$ are continuous and defined on closed contours, standard Fredholm theory of Cauchy operators due to Zhou applies, see \cite[Prop.~4.1]{Zhou1989}. While Zhou's results deal with the standard Cauchy kernel $(s-z)^{-1}$ corresponding to a RHP normalized at infinity, the Fredholm property can be shown to hold for the modified Cauchy operator \eqref{eq:CauchyOPerator} as shown in \cite[Lem.~B.4]{KRGER2009}. In particular, the Cauchy operator $I - C_w$ introduced in Appendix \ref{App:SmallNorm}  with $w(z) = J_q(z) - I$ is Fredholm.

Next observe that $L^2(\sigma \cup \sigma^{-1}) = L^2_s(\sigma \cup \sigma^{-1}) \oplus L^2_a(\sigma \cup \sigma^{-1})$ where
\begin{align*}
    L^2_s(\sigma \cup \sigma^{-1}) &= \lbrace \mu(z) \in L^2(\sigma \cup \sigma^{-1}) \mid \mu(z) = \mu(z^{-1})\sigma_1 \rbrace
    \\
    L^2_a(\sigma \cup \sigma^{-1}) &= \lbrace \mu(z) \in L^2(\sigma \cup \sigma^{-1}) \mid \mu(z) = -\mu(z^{-1})\sigma_1 \rbrace. 
\end{align*}
Due to the symmetries of the jump matrix, the operator $1-C_w$ descents to the symmetric and antisymmetric subspaces:
\begin{align}
    1- C_w|_{L^2_s(\sigma \cup \sigma^{-1})} &\colon L^2_s(\sigma \cup \sigma^{-1}) \to L^2_s(\sigma \cup \sigma^{-1}),
    \\
    1- C_w|_{L^2_a(\sigma \cup \sigma^{-1})} &\colon L^2_a(\sigma \cup \sigma^{-1}) \to L^2_a(\sigma \cup \sigma^{-1}).
\end{align}
In fact if $\mu(z^{-1}) = \pm \mu(z)\sigma_1$, then one can compute
\begin{align*}
    C_w(\mu)(z^{-1}) =& \lim_{y \to z^{-1}-} \frac{1}{2\pi i}\int_{\sigma\cup \sigma^{-1}} \mu(s)w(s)\begin{pmatrix}
        \frac{1}{s-y} - \frac{1}{s} & 0
        \\
        0 & \frac{1}{s-y}
    \end{pmatrix} \di s
    \\
    =& \lim_{x \to z-} \frac{1}{2\pi i}\int_{\sigma \cup \sigma^{-1}} \mu(\zeta^{-1}) w(\zeta^{-1})\begin{pmatrix}
        \frac{1}{\zeta^{-1}-x^{-1}} - \frac{1}{\zeta^{-1}} & 0
        \\
        0 & \frac{1}{\zeta^{-1}-x^{-1}}
    \end{pmatrix} \left(-\frac{\di \zeta}{\zeta^{2}}\right)
    \\
    =& \lim_{x \to z-} \frac{1}{2\pi i} \int_{\sigma \cup \sigma^{-1}} \pm\mu(\zeta)w(\zeta)\sigma_1\begin{pmatrix}
        \frac{1}{\zeta-x} & 0
        \\
        0 & \frac{1}{\zeta-x} - \frac{1}{\zeta}
    \end{pmatrix} \di \zeta
    \\
    =&\pm \lim_{x \to z-} \frac{1}{2\pi i}\int_{\sigma \cup \sigma^{-1}} \mu(\zeta)w(\zeta) \begin{pmatrix}
        \frac{1}{\zeta-x} - \frac{1}{\zeta} & 0
        \\
        0 & \frac{1}{\zeta-x}
    \end{pmatrix} \di \zeta \sigma_1 = \pm C_w(\mu)(z)\sigma_1\,.
\end{align*}
Here we used that the orientation of $\sigma\cup\sigma^{-1}$ is invariant under $z \mapsto z^{-1}$. Let us introduce $\displaystyle 1 - S_w = \left.1- C_w\right|_{L^2_s(\sigma \cup \sigma^{-1})}$. As $1 - C_w$ is a Fredholm operator, the same is true of $1- S_{w}$. Moreover, the operator $1 - S_{\lambda w}$ for any real scaling parameter $\lambda$ will be Fredholm by the same arguments, and choosing $\lambda = 0$, i.e. $1 - S_{0\cdot w} = \text{id}$, we see that the Fredholm index is $0$. Due to the homotopy invariance of the Fredholm index and the continuous dependence of $1- S_{\lambda w}$ on $\lambda$, the Fredholm index of $1-S_w$ is $0$.

One can show, using a similar calculation as found in Appendix \ref{App:SmallNorm}, that solutions of $(1-S_w)\mu = (0, 0)$ correspond bijectively to solutions of RHP \ref{RHP:q} with the normalization replaced by $q_0(z) = (0, \, *)$, $q_0(\infty) = (*, \, 0)$; the symmetry condition is enforced automatically by the domain restriction of $1-S_w$. The identification is $q_0(z) = C(\mu w)(z)$ and $\mu(z) = q_{0-}(z)$. Due to the previous Lemma \ref{lem:vanishingLem} which holds for symmetric vanishing solutions, and the equivalence of the (vanishing) RHPs for $q_0$ and $\wh q_0$, we deduce that no nontrivial symmetric vanishing solution $q_0$ exists. It follows that $1- S_w$ is injective, and as it's Fredholm index is $0$, it is in fact bijective.

We can now conclude that there is a unique solution $\mathfrak{q}(z)$ to RHP \ref{RHP:q}, i.e.~also satisfying the symmetry $\mathfrak{q}(z) = \mathfrak{q}(z^{-1})\sigma_1$, with the normalization
\begin{align*}
    \mathfrak{q}(0) = (1, \, *), \qquad  \mathfrak{q}(\infty) = (*, \, 1),
\end{align*}
given by $\mathfrak{q}(z) = (1, \, 1) + C(\mu w)(z)$, where $\mu(z)$ is the unique solution of $(1-S_w) \mu= (1, 1)$. It remains to show that $\mathfrak{q}_2(0) > 0$, so that we can renormalize and obtain a unique solution $q(z)$ with $q_1(0) \cdot q_2(0) = 1$ and $q_1(0) > 0$. To this end we consider the unique solution $\mu^{(\lambda)}$ of $(1-S_{\lambda w})\mu^{(\lambda)} = (1, \, 1)$ with $\lambda \geq 0$, giving rise to the symmetric Riemann--Hilbert solution $\mathfrak{q}^{(\lambda)}(z)$. This solution will be unique whenever $\lambda \geq 0$, as the proof of the vanishing lemma goes through unchanged, e.g.~absorb $\lambda > 0$ into $\rho(z)$. By continuity of $S_{\lambda w}$ in $\lambda$, it follows that $\mathfrak{q}_2^{(\lambda)}(0)$ is a continuous function of $\lambda$ with $\mathfrak{q}_2^{(0)}(0) = 1$. As $\mathfrak{q}_2^{(\lambda)}(0)$ is real by uniqueness and Schwarz symmetry of the jump matrix $J_q^{(\lambda)}(z)$, i.e.~$\overline{\mathfrak{q}^{(\lambda)}(\overline{z})} = \mathfrak{q}^{(\lambda)}(z)$ (assuming $\sigma \cup \sigma^{-1}$ is chosen Schwarz symmetric) and cannot vanish by the vanishing Lemma \ref{lem:vanishingLem}, it follows that $\mathfrak{q}_2^{(1)}(0) = \mathfrak{q}_2(0) > 0$. Hence we obtain the unique solution $q(z) = \mathfrak{q}(z)/\sqrt{\mathfrak{q}_2(0)}$, finishing the proof.
\end{proof}

\section*{Declaration on the Use of Artificial Intelligence}
The authors acknowledge the use of generative artificial intelligence tools for drafting code for illustrations, manuscript auditing, language refinement, and assisting with intermediate calculations. One key result, Lemma~\ref{lem:measure_bad_set}, is attributed to an AI: while the authors had previously established an independent proof of Lemma~\ref{lem:measure_bad_set} under stronger assumptions, the probabilistic argument presented here was suggested by GPT 6.1 Sol, which allowed us to obtain a more general result. All mathematical statements, proofs, and bibliographic references were independently scrutinized and verified by the authors, who assume full responsibility for the entirety of this work.

\section*{Acknowledgments}
Both authors acknowledge the support of the Royal Swedish Academy of
Sciences through the Mittag--Leffler Institute (grant no. 2021-06594) where this project was started
while the author attended the special program on ``Random Matrices and Scaling
Limits''  in the Fall of 2024. M.P. was supported by the starting grant from the Ragnar S\"oderbergs Foundation and the European Union’s Horizon Europe research and innovation programme under the Marie Skłodowska-Curie grant agreement No.~101272799 (Aztec RHP).
\par\medskip \noindent \begin{minipage}[t]{0.75\linewidth} \vspace{0pt} Funded by the European Union. Views and opinions expressed are however those of the author(s) only
and do not necessarily reflect those of the European Union or the European Research Executive Agency (REA). Neither the European Union nor the granting authority can be held responsible for them.\end{minipage}\hfill \begin{minipage}[t]{0.2\linewidth} \vspace{0pt} \raggedleft \includegraphics[width=\linewidth]{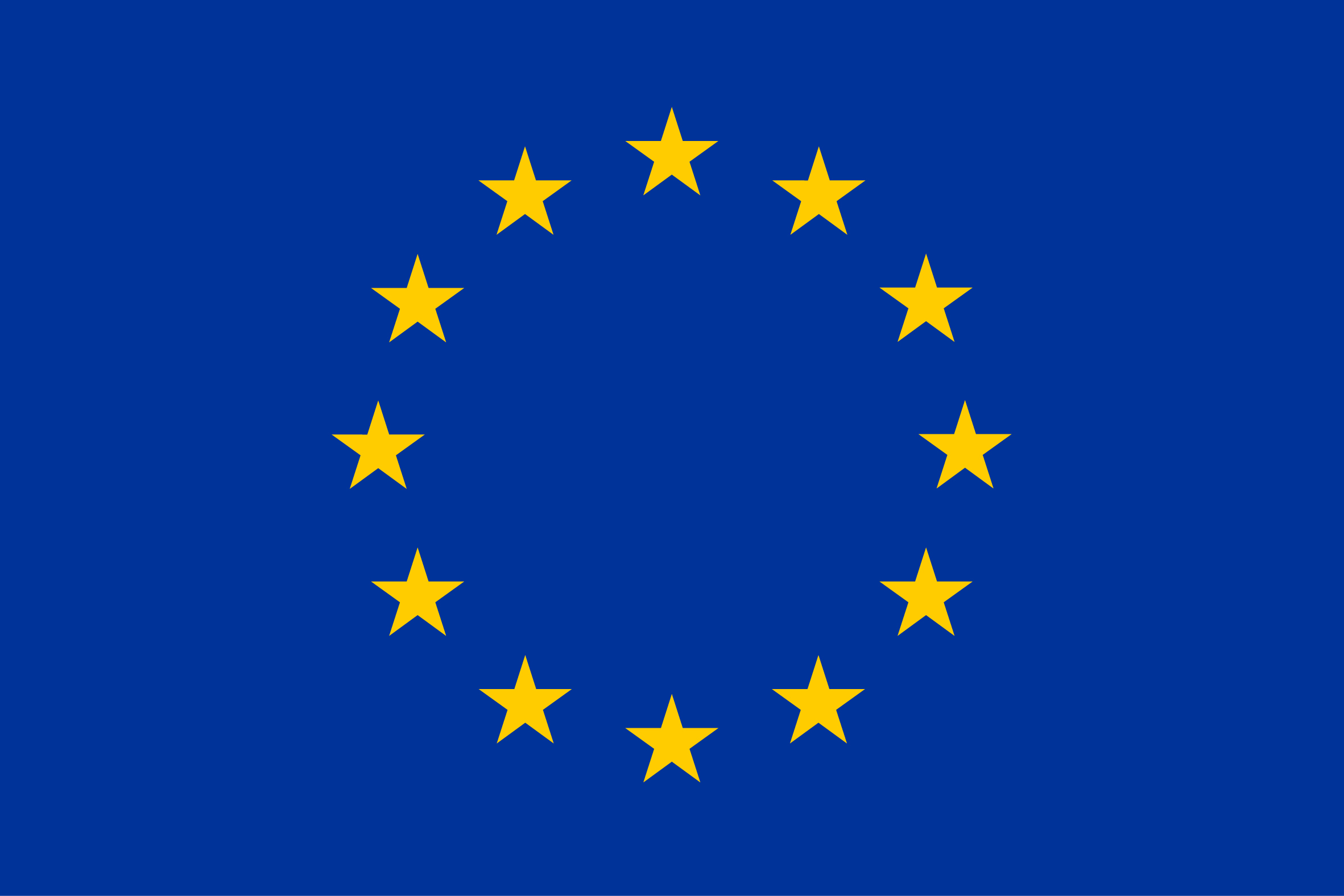} \end{minipage} \par\medskip
\bibliographystyle{siam}
\bibliography{mybib}

\end{document}